\documentclass[rivista]{cimento} 
\usepackage{graphicx}  % got figures? uncomment this
\usepackage[nosort]{cite}
\usepackage{amsmath}

\newcommand\kp{\boldmath{$k \cdot p$}\ }

\title{The $\!$\kp $\!$method $\!$and $\!$its $\!$application $\!$to $\!$graphene,
$\!$carbon $\!$nanotubes and graphene nanoribbons: the Dirac equation}
\shorttitle{The \protect\kp method and its application
to graphene-related materials}
\author{P.~Marconcini %\from{iet.unipi}
\thanks{E-mail: {\texttt p.marconcini@iet.unipi.it}}
        \atque
M.~Macucci %\from{iet.unipi}
\thanks{E-mail: {\texttt m.macucci@mercurio.iet.unipi.it}}}
\instlist{\inst %{iet.unipi} 
Dipartimento di Ingegneria dell'Informazione, Universit\`a di Pisa,\\
Via G. Caruso 16, I-56122 Pisa, Italy
}

\PACSes{
\PACSit{71.15.-m}{Methods of electronic structure calculations.}
\PACSit{73.22.-f}{Electronic structure of nanoscale materials:
clusters, nanoparticles, nanotubes, and nanocrystals.}
\PACSit{73.22.Pr}{Electronic structure of graphene.}}

\DeclareRobustCommand{\SkipTocEntry}[4]{}

\begin{document}

\setcounter{page}{489}

\issue{34}{8-9}{2011}
\checkin{il 9 Gennaio 2011}
%\onlinepub{YY YY 2011}

\DOI{10.1393/ncr/i2011-10068-1}

\maketitle

\def\vec#1{{\mathchoice
{\hbox{\boldmath{$\displaystyle{#1}$}}}
{\hbox{\boldmath{$\textstyle{#1}$}}}
{\hbox{\boldmath{$\scriptstyle{#1}$}}}
{\hbox{\boldmath{$\scriptscriptstyle{#1}$}}}}}

\begin{abstract}
The $\vec k \cdot \vec p$ method is a semi-empirical approach which allows
to extrapolate the band structure of materials from the knowledge of a
restricted set of parameters evaluated in correspondence of a single point of
the reciprocal space. In the first part of this review article we give a
general description of this method, both in the case of homogeneous
crystals (where we consider a formulation based on the standard
perturbation theory, and Kane's approach) and in the case of non-periodic
systems (where, following Luttinger and Kohn, we describe the single-band
and multi-band envelope function method and its application to
heterostructures). The following part of our review is completely
devoted to the application of the $\vec k \cdot \vec p$ method to graphene
and graphene-related materials. Following Ando's approach,
we show how the application of this method to graphene results in a
description of its properties in terms of the Dirac equation.
Then we find general expressions for the probability density and the
probability current density in graphene and we compare this formulation
with alternative existing representations. Finally, applying proper boundary
conditions, we extend the treatment to carbon nanotubes and graphene
nanoribbons, recovering their fundamental electronic properties.
\end{abstract}

\newpage

\tableofcontents

\section{Introduction}

To understand the physical properties of semiconductors it is necessary to
know their electronic band structure, {\em i.e.} the behavior of energy 
as a function of the wave vector $\vec k$ in the reciprocal lattice of 
the crystal. Several numerical methods can be successfully applied to find the 
band structure and the corresponding wave functions, such as the tight 
binding, the pseudopotential, the orthogonalized plane wave, the augmented
plane wave, the Green's function and the cellular methods
\cite{callaway1,grosso,martin}. These methodologies can yield the desired
results throughout the {$\vec k$}-space.

Many phenomena, for example in the study of electrical transport
(due to both electrons and holes) and of optical properties (such as
absorption or gain due to electronic transitions caused by an incident
optical wave), involve only the top of the valence band and the bottom
of the conduction band. Indeed, low-energy electrons and holes are situated
in these regions and also electronic transitions occur near the band edges
of direct band gap semiconductors. Therefore a technique to obtain the band
structure in such regions is of great interest.

The $\vec k \cdot \vec p$ method \cite{voon,wenckebach,chuang,tsidilkovski,
yu,li,kane1,kane2,kane3,pidgeon,bir,ivchenko,anselm,callaway2,singh,long,
kroemer,bassani,enderlein,haug,harrison,winkler,datta1,marconcini1}
allows to extrapolate the band structure of materials from the knowledge of
a restricted set of parameters (a limited number of energy gaps and of 
momentum matrix elements between band lattice functions), evaluated in
correspondence of a single point $\vec k_0$ of the reciprocal space,
which are generally treated as fitting parameters, that can be obtained
from experiments or {\em ab initio} calculations.
Even though, considering quite a large number of bands and thus of
parameters, the $\vec k \cdot \vec p$ method can be used to obtain the
band structure all over the Brillouin zone of the material 
\cite{cardona,cavasillas,radhia,richard,michelini},
its primary use is to explore with great detail the dispersion relations
around the considered point $\vec k_0$. In particular, it allows to
obtain the band structure of materials in the regions of the reciprocal
space near the band extrema, expanding the eigenvalues and eigenvectors
of the single-electron Hamiltonian as a function of $\vec k$ around the
wave vector $\vec k_0$ corresponding to the band maximum or minimum.
It has been shown to be very useful to study structures containing a large
number of atoms, for which atomistic approaches would be computationally
too expensive.

This method, first introduced by J.~Bardeen \cite{bardeen} and F.~Seitz
\cite{seitz}, was developed and adopted by W.~Shockley \cite{shockley}
and G.~Dresselhaus, A.~F.~Kip and C.~Kittel \cite{dresselhaus} in well-known
papers on the energy band structures of semiconductors. It received a general
formulation with E.~O.~Kane \cite{kane1,kane2,kane3,kane4,kane5,kane6}
and with J.~M.~Luttinger and W.~Kohn \cite{luttinger1,luttinger2}.
It was later applied to strained materials (by G.~E.~Pikus and G.~L.~Bir
\cite{bir}) and to heterostructures (for example by G.~Bastard 
\cite{bastard1,bastard2,bastard3}, M.~Altarelli 
\cite{altarelli1,altarelli2,altarelli3} and M.~G. Burt
\cite{burt1,burt2,burt3,burt4,burt5,foreman}), proving to be a very useful 
and straightforward way to study the local properties of materials. 

In the last few years, a significant theoretical and experimental effort
has been devoted to the study of graphene and graphene-related materials,
which appear as promising candidates for many technological applications
and are characterized by very unusual and interesting physical properties.
In particular, the application of the $\vec k \cdot \vec p$ method to the
study of the electronic properties of graphene, sistematically pursued
by T.~Ando \cite{ajiki,ando1,ando2} and other authors, results in a
description of the graphene properties in terms of the Dirac equation,
the same relation that describes the relativistic behavior of elementary
spin-(1/2) particles. This is at the basis of the experiments aiming to
observe in graphene, at non-relativistic speeds, the analogous of purely
relativistic quantum phenomena
\cite{katsnelson1,katsnelson2,katsnelson3,geim,beenakker}.
The application of proper boundary conditions to the relations found for
a sheet of graphene allows to obtain the electronic properties of
carbon nanotubes and graphene nanoribbons, materials which (contrary
to unconfined graphene)\break
can exibit (depending on their geometrical details) a non-zero energy gap.

The first part of this review is a short introduction to the
$\vec k \cdot \vec p$ method in some of its most common formulations.

In particular, sect.~{\bf 2} describes the application of the
$\vec k \cdot \vec p$ method to homogeneous crystals, where, due to
the periodicity of the material, the electron wave functions are Bloch
functions and thus the unperturbed Bloch lattice functions are adopted
as a basis for the method. We first describe (following W. T. Wenckebach
\cite{wenckebach}) how the $\vec k \cdot \vec p$ approach can be derived
by just applying the standard perturbation theory to the Schr\"odinger-Bloch
equation and how this formulation can be adopted to study the
dispersion relations of semiconductors with the diamond or zincblende
structure. Then we briefly summarize the alternative formulation by Kane,
consisting in the exact diagonalization of the Schr\"odinger-Bloch
Hamiltonian for a subset of bands, and in the inclusion of the effect
of the other energy bands with the L\"owdin perturbation theory.

In sect.~{\bf 3}, instead, we describe how the $\vec k \cdot \vec p$
method can be applied to the case of non-periodic systems, where the
phase factor (involving the wave vector measured from the considered
extremum point) of the Bloch lattice functions
has to be replaced by properly defined envelope functions. Following
J.~M.~Luttinger and W.~Kohn, we derive the single-band and multi-band
envelope function equations, and then we briefly outline the main
approaches followed in the application of the envelope function theory to
the study of heterostructures.

The second part of the review is devoted to the application of the 
$\vec k \cdot \vec p$ method, and in particular of the envelope function
approach, to graphene, carbon nanotubes and graphene nanoribbons.

In sect.~{\bf 4}, following T. Ando, we perform a first-order expansion
of a simple tight-binding description of graphene, obtaining the Dirac
equation for the envelope functions (corresponding to the
two degeneration points of graphene) in the presence of a generic external
potential, and we analytically solve this equation for the case of null
potential. Starting from this formulation, we also derive
general expressions for the probability density and for the probability
current density in graphene, and we compare them with those\break
used, adopting a valley-isotropic representation, by C.~W.~J.~Beenakker
{\sl et al.} \cite{akhmerov1,beenakker}.

In sect.~{\bf 5} we extend the previous treatment to the study
of carbon nanotubes, enforcing a periodic boundary condition along the chiral
vector, that univocally characterizes these tubules. In particular, we show
how this periodic condition on the overall wave function translates in terms
of the envelope functions, and we analytically solve the\break
\newpage
\noindent
Dirac problem in the
absence of an external potential, obtaining the conditions for which nanotubes
have a semiconducting or a metallic behavior.

Finally, in sect.~{\bf 6} we discuss the case of graphene nanoribbons.
Adapting the approach adopted by L.~Brey and H.~A.~Fertig \cite{brey1,brey2}
to the mathematical formulation of graphene proposed by T. Ando, we study
both zigzag and armchair nanoribbons, obtaining an analytical solution in
the absence of an external potential, and recovering the fundamental
properties of these structures.

\section{The $\vec k \cdot \vec p$ method in homogeneous crystals:
derivation based on the standard perturbation theory and Kane's model}

We begin our overview of the $\vec k \cdot \vec p$ method describing its
formulation in the case of homogeneous crystals.

In a pure crystal an electron is subject to a periodic potential energy
\begin{equation}
U_L(\vec r)=U_L(\vec r+\vec R),
\end{equation}
with $\vec R$ any linear combination of the lattice vectors, and thus also 
the Hamiltonian is invariant under translation by the lattice vectors.
Therefore, if $\psi^n_{\vec k} (\vec r)$ is the wave function of an electron 
moving in the crystal, also $\psi^n_{\vec k} (\vec r+\vec R)$ will be 
a solution of the Schr\"odinger equation and therefore will coincide with
$\psi^n_{\vec k} (\vec r)$, apart from a constant with unit modulus
(otherwise the wave function could grow to infinity, if we repeated the 
translation $\vec R$ indefinitely). Thus the
general form of the electron wave functions\break
will be 
\begin{equation}
\psi^n_{\vec k} (\vec r)=e^{i\,\vec k \cdot \vec r}u^n_{\vec k} (\vec r),
\end{equation}
where $\psi^n_{\vec k} (\vec r)$ is usually called ``Bloch function'',
while $u^n_{\vec k} (\vec r)$ is called ``Bloch lattice function'' and is
periodic with the lattice periodicity
\begin{equation}
u^n_{\vec k} (\vec r+\vec R)=u^n_{\vec k} (\vec r)
\end{equation}
(Bloch's theorem) \cite{bloch}.

Starting from the Schr\"odinger equation (in the absence of a magnetic field)
for $\psi^n_{\vec k} (\vec r)$
\begin{equation}
H^{(0)}\psi^n_{\vec k} (\vec r)=
E^n_{\vec k}\psi^n_{\vec k} (\vec r),
\end{equation}
with (in the absence of a magnetic field)
\begin{equation}
H^{(0)}=-\frac{\hbar^2}{2\,m_e}\nabla^2+U_L(\vec r)
\end{equation}
(where $m_e$ is the electron mass and $\hbar$ is the reduced Planck 
constant) and substituting $\psi^n_{\vec k} ( \vec r)$ with the generic 
expression of the Bloch function, we obtain
\begin{eqnarray}
&&\left(-\frac{\hbar^2}{2\,m_e}\nabla^2+U_L(\vec r)\right)
e^{i\,\vec k \cdot \vec r}u^n_{\vec k} (\vec r)=\\
&&\quad -\frac{\hbar^2}{2\,m_e}\vec\nabla\cdot
\left(e^{i\,\vec k \cdot \vec r}(\vec\nabla u^n_{\vec k} (\vec r))+
(\vec\nabla e^{i\,\vec k \cdot \vec r})u^n_{\vec k} (\vec r)\right)+
U_L(\vec r)e^{i\,\vec k \cdot \vec r}u^n_{\vec k} (\vec r)=\nonumber\\
&&\quad -\frac{\hbar^2}{2\,m_e}\vec\nabla\cdot\left(e^{i\,\vec k \cdot \vec r}
(\vec\nabla u^n_{\vec k} (\vec r)+i\,\vec k u^n_{\vec k} (\vec r))\right)+
U_L(\vec r)e^{i\,\vec k \cdot \vec r}u^n_{\vec k} (\vec r)=\nonumber\\
&&\quad-\frac{\hbar^2}{2\,m_e}\Big(e^{i\,\vec k \cdot \vec r}\vec\nabla\cdot
(\vec\nabla u^n_{\vec k} (\vec r)+i\,\vec k u^n_{\vec k} (\vec r))\nonumber\\
&&\qquad {}+(\vec\nabla e^{i\,\vec k \cdot \vec r})\cdot 
(\vec\nabla u^n_{\vec k} (\vec r)+i\,\vec k u^n_{\vec k} (\vec r))\Big)+
U_L(\vec r)e^{i\,\vec k \cdot \vec r}u^n_{\vec k} (\vec r)=\nonumber\\
&&\quad -\frac{\hbar^2}{2\,m_e}\Big(e^{i\,\vec k \cdot \vec r}
(\nabla^2 u^n_{\vec k} (\vec r)+i\vec k \cdot \vec\nabla u^n_{\vec k} 
(\vec r))\nonumber\\
&&\qquad {}+(i\,\vec k\, e^{i\,\vec k \cdot \vec r})\cdot 
(\vec\nabla u^n_{\vec k} (\vec r)+i\,\vec k u^n_{\vec k} (\vec r))\Big)+
U_L(\vec r)e^{i\,\vec k \cdot \vec r}u^n_{\vec k} (\vec r)=\nonumber\\
&&\quad -\frac{\hbar^2}{2\,m_e}e^{i\,\vec k \cdot \vec r}
\!\left(\nabla^2 u^n_{\vec k} 
(\vec r)+i\vec k \cdot \vec\nabla u^n_{\vec k} (\vec r) 
+i\vec k \cdot\vec\nabla u^n_{\vec k} (\vec r)-k^2u^n_{\vec k} (\vec r)\right)\nonumber\\
&&\qquad {}+U_L(\vec r)e^{i\,\vec k \cdot \vec r}u^n_{\vec k} (\vec r)\!=\nonumber\\
&&\quad e^{i\,\vec k \cdot \vec r}\left(\left(-\frac{\hbar^2}{2\,m_e}
\nabla^2 +U_L(\vec r)\right)
-\frac{i\,\hbar^2}{m_e}\vec k \cdot\vec\nabla+
\frac{\hbar^2 k^2}{2\,m_e}\right)u^n_{\vec k}(\vec r)=\nonumber\\
&&\quad e^{i\,\vec k \cdot \vec r}(H^{(0)}+H^{(1)}) u^n_{\vec k}(\vec r)=
e^{i\,\vec k \cdot \vec r}E^n_{\vec k}u^n_{\vec k}(\vec r)\nonumber
\end{eqnarray}
and thus
\begin{equation}
H u^n_{\vec k}(\vec r)=
(H^{(0)}+H^{(1)})u^n_{\vec k}(\vec r)=
E^n_{\vec k}u^n_{\vec k}(\vec r),
\end{equation}
with
\begin{equation}
H^{(1)}=-\frac{i\,\hbar^2}{m_e}\vec k \cdot\vec\nabla+
\frac{\hbar^2 k^2}{2\,m_e}
\end{equation}
(where $k=|\vec k|$). What we have just obtained is clearly an equation for
the Bloch lattice functions (the Schr\"odinger-Bloch equation), which needs
to be solved only for a single primitive cell with the boundary condition
that the function $u^n_{\vec k}(\vec r)$ must be periodic with the lattice
periodicity. For each value of $\vec k$ this equation has a periodic solution
only for selected values $E^n_{\vec k}$ of the energy $E$. Noting
that $H^{(1)}(\vec r)$ reduces to zero when $\vec k$ approaches $0$
and thus that this part of the Hamiltonian can be treated as a perturbation
around $\vec k=0$, we can locally solve this equation using 
the time-independent perturbation theory, assuming to know the eigenfunctions 
and eigenvalues of $H^{(0)}(\vec r)$, {\em i.e.} the Bloch lattice functions 
and the energy band values for $\vec k=0$.

For most of the semiconductors the maximum of the valence band is in the 
\hbox{$\Gamma$-point} (the center of the first Brillouin zone represented with
the Wigner-Seitz method) and therefore corresponds to $\vec k=0$; 
the minimum of the conduction band instead is for $\vec k=0$ only for
direct-gap semiconductors. When the extremum point of the energy band
(and thus the interesting region) is for a generic $\vec k_0$, we can easily
extend this argument observing that, if we define the value of $H$ in
$\vec k_0$ as
\begin{equation}
H_{\vec k_0}=H^{(0)}-\frac{i\,\hbar^2}{m_e}\vec k_0 
\cdot\vec\nabla+\frac{\hbar^2 {k_0}^2}{2\,m_e},
\end{equation}
we have that the value of $H$ in $\vec k$ is
\begin{eqnarray}
& H= & H^{(0)}+H^{(1)}=
H_{\vec k_0}+\left[-\frac{i\,\hbar^2}{m_e}(\vec k-\vec k_0) 
\cdot\vec\nabla+\frac{\hbar^2 (k^2-{k_0}^2)}{2\,m_e}\right]=\\
& H_{\vec k_0} & +\Big[-\frac{i\,\hbar^2}{m_e}(\vec k-\vec k_0) 
\!\cdot\!\vec\nabla+\!\frac{\hbar^2 (k^2-{k_0}^2)}{2\,m_e}\nonumber\\
&&+\frac{\hbar}{m_e}(\vec k-\vec k_0)\!\cdot\! \hbar \vec k_0-
\!\frac{\hbar}{m_e}(\vec k-\vec k_0)\!\cdot\! \hbar \vec k_0 \Big]\!=\nonumber\\
& H_{\vec k_0} & +\left[\frac{\hbar}{m_e}(\vec k-\vec k_0)
\cdot(\hbar \vec k_0-i\,\hbar\vec\nabla)+\frac{\hbar^2}{2\,m_e}
(k^2-{k_0}^2-2\vec k\cdot\vec k_0+2{k_0}^2)\right]=\nonumber\\
& H_{\vec k_0} & +\left[\frac{\hbar}{m_e}(\vec k-\vec k_0)
\cdot(\hbar \vec k_0-i\,\hbar\vec\nabla)+\frac{\hbar^2}{2\,m_e}
(k^2-2\vec k\cdot\vec k_0+{k_0}^2)\right]=\nonumber\\
& H_{\vec k_0} & +\left[\frac{\hbar}{m_e}(\vec k-\vec k_0)
\cdot(\hbar \vec k_0-i\,\hbar\vec\nabla)+\frac{\hbar^2}{2\,m_e}
(\vec k-\vec k_0)\cdot(\vec k-\vec k_0)\right]=\nonumber\\
& H_{\vec k_0} & +\left[\frac{\hbar}{m_e}(\vec k-\vec k_0)
\cdot(\hbar \vec k_0-i\,\hbar\vec\nabla)+\frac{\hbar^2}{2\,m_e}
|\vec k-\vec k_0|^2 \right]\nonumber
\end{eqnarray}
and for $\vec k$ near $\vec k_0$ the term between square brackets can be 
treated as a perturbation of $H_{\vec k_0}$ \cite{kane1}. 
For the sake of simplicity, in the following we will consider 
$\vec k_0=0$.

An important point to notice is that, for any selected $\vec k$, the 
functions $u^n_{\vec k}(\vec r)$ form an orthogonal and complete set (in 
the restricted sense that any function with the lattice periodicity can be 
expanded in terms of the Bloch lattice functions corresponding to the
selected $\vec k$).

To describe the main results of time-independent perturbation theory
\cite{perturbation,baym}, we have to distinguish the case in which the
unperturbed energy levels are non-degenerate from the case in which such
a degeneration exists (in the following we will use the notation of
W.~T.~Wenckebach \cite{perturbation}).
Let us begin from the first case. The problem we have to\break
solve is
\begin{equation}
[H^{(0)}+H^{(1)}]|n \rangle =E_n|n \rangle,
\end{equation}
where $H^{(0)}$ is the unperturbed Hamiltonian and $H^{(1)}$ the perturbation.
If we expand the eigenvalues $E_n$ and the eigenfunctions $|n \rangle $:
\begin{eqnarray}
E_n &=& E_n^{(0)}+E_n^{(1)}+E_n^{(2)}+\ldots,\\
|n \rangle &=& |n \rangle ^{(0)}+|n \rangle ^{(1)}+|n \rangle ^{(2)}+\ldots,\nonumber
\end{eqnarray}
we insert these expressions into the eigenvalue equation, and we enforce the
identity between terms of the same order, we find
\begin{eqnarray}
H^{(0)}|n \rangle ^{(0)} &=& E_n^{(0)}|n \rangle ^{(0)},\\
H^{(0)}|n \rangle ^{(1)}+H^{(1)}|n \rangle ^{(0)} &=&
E_n^{(0)}|n \rangle ^{(1)}+
E_n^{(1)}|n \rangle ^{(0)},\nonumber\\
H^{(0)}|n \rangle ^{(2)}+H^{(1)}|n \rangle ^{(1)} &=&
E_n^{(0)}|n \rangle ^{(2)}+
E_n^{(1)}|n \rangle ^{(1)}+E_n^{(2)}|n \rangle ^{(0)},\nonumber\\
&\ldots\ .\nonumber
\end{eqnarray}
The first equation corresponds to the unperturbed eigenvalue equation,
the solutions of which, $E_n^{(0)} \equiv E_0^n$ and
$|n \rangle ^{(0)} \equiv |n0 \rangle $, are assumed to be known.
From the other equations, instead, we can obtain the corrections to these 
values produced by the perturbation $H^{(1)}$. In particular, if we stop to 
the first-order corrections for the eigenfunctions and to the second-order
corrections for the eigenvalues we find
\begin{equation}
|n \rangle \simeq |n0 \rangle +|n \rangle ^{(1)}=|n0 \rangle +\sum_{m\ne n}
\left(|m0 \rangle \frac{\langle m0|H^{(1)}|n0 \rangle}{E_0^n-E_0^m}\right)
\end{equation}
(choosing $ \langle n0|n \rangle ^{(1)}=0$) and
\begin{eqnarray}
E_n\simeq E_0^n+E_n^{(1)}+E_n^{(2)} 
&=& E_0^n+ \langle n0|H^{(1)}|n0 \rangle\\
&&{}+\sum_{m\ne n}
\left(\frac{\langle n0|H^{(1)}|m0 \rangle  \langle m0|H^{(1)}|n0 \rangle}
{E_0^n-E_0^m} \right).\nonumber
\end{eqnarray}
When we examine degenerate unperturbed states, the expressions we have just 
found diverge and thus we have to modify our treatment. In particular, if the
degenerate energy level $E_0^n$ corresponds to a multiplet of degenerate
states $|na0 \rangle $ (with $a=1,2,\ldots,g_n$, where $g_n$ is the degeneracy) 
and we have to solve the perturbed problem 
\begin{equation}
H|\psi \rangle =[H^{(0)}+H^{(1)}]|\psi \rangle =E|\psi \rangle,
\end{equation}
we can express the new generic eigenfunction $|\psi \rangle $ as
\begin{equation}
|\psi \rangle =\sum_{a=1}^{g_n}|na \rangle  \langle na|\psi \rangle,
\end{equation}
where the $|na \rangle $'s are states which are related to the 
unperturbed eigenvectors $|na0 \rangle $'s by the perturbation matrix 
elements between different multiplets (as we will see in
eq.~(\ref{multiplet})).
If we define
\begin{equation}
H_{ab}^n= \langle na|H|nb \rangle = 
\langle na|[H^{(0)}+H^{(1)}]|nb \rangle,
\end{equation}
we can express our perturbed equation in the following way:
\begin{equation}
\sum_{b=1}^{g_n}H_{ab}^n \langle nb|\psi \rangle =
E \langle na|\psi \rangle.
\end{equation}
Noting that the definition of the $H_{ab}^n$'s can be equivalently expressed
in this way
\begin{equation}
[H^{(0)}+H^{(1)}]|nb \rangle =\sum_{a=1}^{g_n}|na \rangle H_{ab}^n,
\end{equation}
inserting into this equation the expansions
\begin{eqnarray}
H_{ab}^n &= (H_{ab}^n)^{(0)}+(H_{ab}^n)^{(1)}+(H_{ab}^n)^{(2)}+\ldots,\\
|na \rangle &= |na \rangle ^{(0)}+|na \rangle ^{(1)}+|na \rangle ^{(2)}+
\ldots,\nonumber
\end{eqnarray}
and enforcing the identity of the terms of the same order, we find
\begin{eqnarray}
H^{(0)}|nb \rangle ^{(0)} &=&
\sum_{a=1}^{g_n}|na \rangle ^{(0)}(H_{ab}^n)^{(0)},\\
H^{(0)}|nb \rangle ^{(1)}+H^{(1)}|nb \rangle ^{(0)} &=&
\sum_{a=1}^{g_n}|na \rangle ^{(1)}(H_{ab}^n)^{(0)}+
\sum_{a=1}^{g_n}|na \rangle ^{(0)}(H_{ab}^n)^{(1)},\nonumber\\
H^{(0)}|nb \rangle ^{(2)}+H^{(1)}|nb \rangle ^{(1)} &=&
\sum_{a=1}^{g_n}|na \rangle ^{(2)}(H_{ab}^n)^{(0)}\nonumber\\
&&+ \sum_{a=1}^{g_n}|na \rangle ^{(1)}(H_{ab}^n)^{(1)}+
\sum_{a=1}^{g_n}|na \rangle ^{(0)}(H_{ab}^n)^{(2)},\nonumber\\
\ldots\ .\nonumber
\end{eqnarray}
The first equation corresponds, noting that 
$(H_{ab}^n)^{(0)}=E_0^n \delta_{ab}$, to the unperturbed eigenvalue equation,
the solutions of which, $E_0^n$ and $|na \rangle ^{(0)}=|na0 \rangle $,
are assumed to be known.
From the other equations, instead, we can obtain the corrections to these 
values produced by the perturbation. In particular, if we 
stop to the first-order corrections for the eigenstates and to the second-order
corrections for the eigenvalues, we find
\begin{equation}
\label{multiplet}
|nb>\simeq |nb0 \rangle +|nb \rangle ^{(1)}=|nb0 \rangle +
\sum_{m\ne n}\sum_{c=1}^{g_m}
\left(|mc0 \rangle \frac{\langle mc0|H^{(1)}|nb0 \rangle}
{E_0^n-E_0^m}\right)
\end{equation}
(choosing $ \langle nc0|nb \rangle ^{(1)}=0$) and
\begin{eqnarray}
H_{cb}^n &\simeq& (H_{cb}^n)^{(0)}+(H_{cb}^n)^{(1)}+(H_{cb}^n)^{(2)}=
E_0^n\delta_{cb}+ \langle nc0|H^{(1)}|nb0 \rangle\\
&&{}+\sum_{m\ne n}\sum_{a=1}^{g_m}
\left(\frac{\langle nc0|H^{(1)}|ma0 \rangle  \langle ma0|H^{(1)}|nb0 \rangle}
{E_0^n-E_0^m} \right).\nonumber
\end{eqnarray}
Once the $H_{cb}^n$ have been found, we can obtain the energy levels $E$ 
solving the equation
\begin{equation}
\sum_{b=1}^{g_n}H_{ab}^n \langle nb|\psi \rangle =
E \langle na|\psi \rangle,
\end{equation}
or, equivalently, finding the eigenvalues of the matrix  
$H^n$ (matrix $g_n\times g_n$ with  elements $H_{ab}^n$)
by solving
\begin{equation}
\det\,(H^n-E I)=0
\end{equation}
(with  $I$ the $g_n\times g_n$ unit matrix).
We notice that, computing also the eigenvectors $ \langle na|\psi \rangle $ 
of such a matrix and combining such results with the $|nb \rangle $ that have  
been computed before up to the first order, it is also possible to know the 
eigenfunctions $|\psi \rangle $ of the perturbed problem.

In the case of the $\vec k \cdot \vec p$ Hamiltonian that we have found before 
\cite{wenckebach}, we can use the $u^n_0 (\vec r)$ ($u^n_{\vec k} (\vec r)$
for $\vec k=0$) as $|n0 \rangle $ and we have that
\begin{eqnarray}
\langle m0|H^{(1)}|n0 \rangle &=&
\langle m0|\left[-\frac{i\,\hbar^2}{m_e}(\vec k \cdot\vec\nabla)\right]
|n0 \rangle + \langle m0|\frac{\hbar^2 k^2}{2\,m_e}|n0 \rangle =\\
&&\frac{\hbar \vec k}{m_e} \cdot \langle m0|(-i\,\hbar \vec\nabla) |n0 \rangle
+ \langle m0|\frac{\hbar^2 k^2}{2\,m_e}|n0 \rangle .\nonumber
\end{eqnarray}
The second term clearly gives only diagonal matrix elements, because it is 
equal to $(\hbar^2 k^2/(2\,m_e)) \delta_{nm}$.
The first term, instead, gives only non-diagonal matrix elements 
because it is known \cite{velocity} that 
\begin{equation}
\langle n\vec k_0|(-i\hbar \vec\nabla)|n\vec k_0 \rangle +
\hbar \vec k_0=m_e \vec v_n=
\frac{m_e}{\hbar}\vec \nabla_{\vec k} E^n_{\vec k}
\end{equation}
(where $\vec v_n$ is the expectation value of the velocity of the Bloch 
waves, and in our considerations we are assuming $\vec k_0=0$) and
$\vec \nabla_{\vec k} E^n_{\vec k}=0$ in the band extrema.

Then, if the unperturbed energy bands are non-degenerate, we can write that
\begin{eqnarray}
E_{\vec k}^n &=& E_0^n+\frac{\hbar^2 k^2}{2\,m_e}+\frac{\hbar^2}{{m_e}^2}
\sum_{m\ne n}\frac{\langle n0|\vec k \cdot (-i\,\hbar\,\vec\nabla)|m0 \rangle 
 \langle m0|\vec k \cdot (-i\,\hbar\,\vec\nabla)|n0 \rangle}
{E_0^n-E_0^m}=\\
&& E_0^n+\frac{\hbar^2}{2}\sum_{\mu,\nu}\frac{k_{\mu}k_{\nu}}{m_{\mu\nu}^*},\nonumber
\end{eqnarray}
where $\mu,\nu=x,y,z$, while $m_{\mu\nu}^*$ is the effective-mass tensor
defined by
\begin{equation}
\frac{1}{m_{\mu\nu}^*}=\frac{1}{m_e}\delta_{\mu\nu}+
\frac{2}{{m_e}^2}\sum_{m\ne n}\frac{P_{\mu}^{nm}P_{\nu}^{mn}}
{E_0^n-E_0^m}
\end{equation}
and the momentum matrix elements at the band extremum are
\begin{equation}
P_{\mu}^{nm}= \langle n0|(-i\,\hbar\,\nabla_{\mu})|m0 \rangle.
\end{equation}

\begin{figure}               
\centering
\includegraphics[width=.45\textwidth,angle=0]{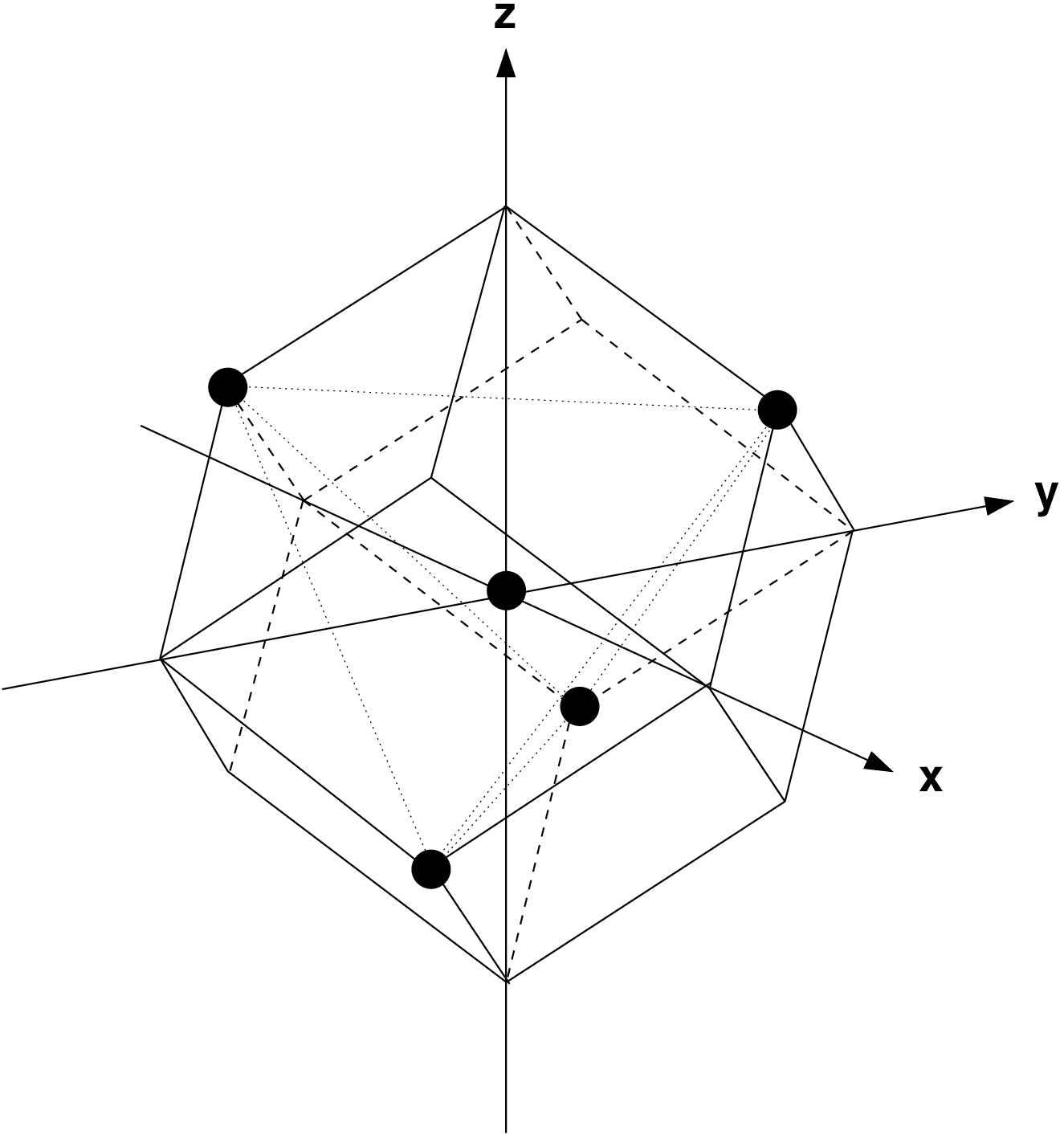}
\caption{Wigner-Seitz primitive cell for the diamond or zincblende 
structure (adapted from \cite{wenckebach}).}
\label{f1}
\end{figure}

\noindent
If the unperturbed energy bands are degenerate, instead, we have 
\begin{eqnarray}
(H_{\vec k}^n)_{cb} &=& E_0^n \delta_{cb}+\frac{\hbar^2 k^2}{2\,m_e}\delta_{cb}+
\frac{\hbar}{m_e} \langle nc0|\vec k \cdot (-i\,\hbar\,\vec\nabla)|nb0 \rangle 
\\
&&{}+\frac{\hbar^2}{{m_e}^2}\sum_{m\ne n}\sum_{a=1}^{g_m}
\frac{\langle nc0|\vec k \cdot (-i\,\hbar\,\vec\nabla)|ma0 \rangle 
\langle ma0|\vec k \cdot (-i\,\hbar\,\vec\nabla)|nb0 \rangle} 
{E_0^n-E_0^m}=\nonumber\\
&& E_0^n \delta_{cb}+\frac{\hbar}{m_e}\sum_{\mu}k_{\mu}(P_{\mu})_{cb}^{nn}+
\frac{\hbar^2}{2}\sum_{\mu,\nu}\frac{k_{\mu}k_{\nu}}{m_{\mu\nu}^{cb}},\nonumber
\end{eqnarray}
where $\mu,\nu=x,y,z$, while $m_{\mu\nu}^{cb}$ is the effective-mass tensor
defined by
\begin{equation}
\frac{1}{m_{\mu\nu}^{cb}}=\frac{1}{m_e}\delta_{cb}\delta_{\mu\nu}+
\frac{2}{{m_e}^2}\sum_{m\ne n}\sum_{a=1}^{g_m}
\frac{(P_{\mu})_{ca}^{nm}(P_{\nu})_{ab}^{mn}}
{E_0^n-E_0^m}
\end{equation}
and the momentum matrix elements at the band extremum are
\begin{equation}
(P_{\mu})_{cb}^{nm}= \langle nc0|(-i\,\hbar\,\nabla_{\mu})|mb0 \rangle.
\end{equation}
In most of the cases all the $(P_{\mu})_{cb}^{nn}=0$, and the linear term in
$k_{\mu}$ disappears.
The energy levels will be found solving 
\begin{equation}
\det\,(H_{\vec k}^n-E I)=0.
\end{equation}
Thus, in principle to perform a calculation of the energy bands we would have
to know the $| n0 \rangle$'s (the Bloch lattice functions at $\vec k=0$).
Since the Hamiltonian $H^{(0)}$ and its eigenfunctions $| n0 \rangle$ have
the periodicity of the lattice, the problem can be solved inside
a single primitive cell, enforcing periodic boundary conditions
at the surface of the cell. Most semiconductors of interest have the diamond
or zincblende crystal structure; for these materials we can choose as
lattice primitive cell a Wigner-Seitz cell centered around an atomic site
(the one with the strongest potential in the case of the zincblende
structure, characterized by atoms that are not all identical) and with, at
four vertices of the cell, four other atoms forming a tetrahedron with the
center coincident with the primitive cell center (fig.~\ref{f1}). 
We can use a central force model (the same results can be obtained
using group theory), considering the potential inside the primitive cell
as due only to the attraction of the nucleus of the central atom, shielded
by its electrons~\cite{wenckebach}. We find that the Bloch lattice
functions at $\vec k=0$ exhibit symmetry properties analogous to
those of atomic orbitals: we have completely symmetric $s$-type states
$\rho_{\nu s}(r)$, and $p$-type states antisymmetric with respect to a
coordinate and symmetric with respect to the others, {\em i.e.} of the form
$\rho_{\nu x}(r)x$, $\rho_{\nu y}(r)y$, and $\rho_{\nu z}(r)z$
(where $r=\sqrt{x^2+y^2+z^2}$).
Then, treating the electrostatic potential of the cores at the vertices
of the primitive cell as a perturbation, we see that, to first order, this
potential does not change the eigenfunctions but shifts the energy levels 
and in particular breaks the degeneracy between each $s$-type state and
the corresponding three $p$-type states (which remain mutually degenerate).
As a result, we find that at $\vec k=0$ the top of the valence band
can be described with three degenerate states: $|vx0 \rangle =\rho_v(r)x$,
$|vy0 \rangle =\rho_v(r)y$ and $|vz0 \rangle =\rho_v(r)z$,
while in most cases the bottom of the conductance band is described
by a non-degenerate symmetric state $|c0 \rangle =\rho_c(r)$ (with the
important exception of silicon, where at $\vec k=0$ also the bottom of
the conduction band is characterized by three states $|cx0 \rangle$, 
$|cy0 \rangle$ and $|cz0 \rangle$).
\noindent

Therefore, if we treat the conduction band as a non-degenerate band, we obtain
\begin{equation}
E_{\vec k}^c=
E_0^c+\frac{\hbar^2}{2}\sum_{\mu,\nu}\frac{k_{\mu}k_{\nu}}{m_{\mu\nu}^*},
\end{equation}
where $\mu,\nu=x,y,z$ and
\begin{equation}
\frac{1}{m_{\mu\nu}^*}=\frac{1}{m_e}\delta_{\mu\nu}+
\frac{2}{{m_e}^2}\sum_{m\ne n}
\frac{\langle c0|(-i\,\hbar\,\nabla_{\mu})|m0 \rangle 
 \langle m0|(-i\,\hbar\,\nabla_{\nu})|c0 \rangle}
{E_0^c-E_0^m}.
\end{equation}
The largest contribution to the sum comes from the bands $m$ for
which $| E_0^c-E_0^m |$ is smallest, {\em i.e.} from the three valence bands.
If we compute the momentum matrix elements between the valence bands and
the conduction band, we find that, due to the symmetry properties of the Bloch 
lattice functions, 
\begin{equation}
\langle v\mu 0|(-i\,\hbar\,\nabla_{\nu})|c0 \rangle =
- \langle c0|(-i\,\hbar\,\nabla_{\nu})|v\mu 0 \rangle =
-i\,\hbar\,P\,\delta_{\mu\nu}
\end{equation}
with $\mu,\nu=x,y,z$ and $P= \langle v\mu 0|\nabla_{\mu}|c0 \rangle $ a 
non-zero quantity, which multiplied by $\hbar$ has the dimensions of a
momentum.
Consequently, the effective mass in the conduction band that we find is
isotropic and equal to
\begin{equation}
\frac{1}{m_{\mu\nu}^*}=\frac{1}{m_c^*}\delta_{\mu\nu}=
\left(\frac{1}{m_e}+\frac{2\,\hbar^2 P^2}{m_e^2 E_g^0}\right)\delta_{\mu\nu},
\end{equation}
with $E_g^0=E_0^c-E_0^v$.

As to the valence band, we must use the degenerate perturbation theory
and, with a motivation analogous to that used in the study of the 
conduction band, we can consider only the interaction between the three
degenerate valence bands and the conduction band, which is the nearest 
energy band. Thus, using the previous results, we have that
\begin{equation}
(H_{\vec k}^v)_{\alpha\beta}=E_0^v\delta_{\alpha\beta}+
\frac{\hbar^2}{2}\sum_{\mu,\nu}
\frac{k_{\mu}k_{\nu}}{m_{\mu\nu}^{\alpha\beta}},
\end{equation}
with 
\begin{eqnarray}
\frac{1}{m_{\mu\nu}^{\alpha\beta}}\! &=& \!\frac{1}{m_e}\delta_{\alpha\beta}
\delta_{\mu\nu}+
\frac{2}{{m_e}^2}\!\sum_{m\ne v}\sum_{a=1}^{g_m}\!
\frac{\langle v \alpha 0|(-i\,\hbar\,\nabla_{\mu})|ma0 \rangle 
\langle ma0|(-i\,\hbar\,\nabla_{\nu})|v \beta 0 \rangle}
{E_0^v-E_0^m}\!=\\
&& \frac{1}{m_e}\delta_{\alpha\beta}
\delta_{\mu\nu}+\frac{2}{{m_e}^2}
\frac{\langle v \alpha 0|(-i\,\hbar\,\nabla_{\mu})|c0 \rangle 
 \langle c0|(-i\,\hbar\,\nabla_{\nu})|v \beta 0 \rangle}
{E_0^v-E_0^c}=\nonumber\\
&& \frac{1}{m_e}\delta_{\alpha\beta}\delta_{\mu\nu}
-\frac{2\,\hbar^2 P^2}{{m_e}^2 E_g^0}
\delta_{\alpha\mu}\delta_{\beta\nu}\nonumber
\end{eqnarray}
and thus the valence energy bands near the extremum can be obtained finding
the eigenvalues of the matrix 
\begin{equation}
H_{\vec k}^v=\left(E_0^v+\frac{\hbar^2 k^2}{2\,m_e}\right)I-
\frac{\hbar^4 P^2}{m_e^2 E_g^0}
\left[\begin{array}{ccc}
k_x^2 & k_x k_y & k_x k_z\\
k_y k_x & k_y^2 & k_y k_z\\
k_z k_x & k_z k_y & k_z^2
\end{array}\right].
\end{equation}

\begin{figure}
\centering
\includegraphics[width=0.7\textwidth,angle=0]{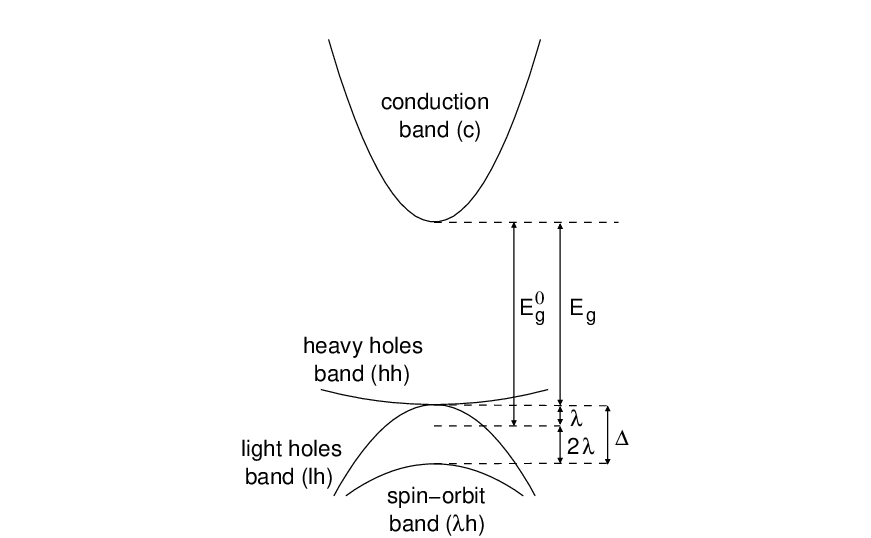}
\caption{Band structure near $\vec k=0$ obtained with the simplified 
model described in the text (the heavy-hole band has a wrong curvature)
(adapted from \cite{wenckebach}).} 
\label{f2}
\end{figure}

\noindent
Till now we have not considered the effect of the so-called spin-orbit 
interaction, which often has a non-negligible influence on the energy
bands. The physical phenomenon is the following~\cite{jackson,spinorbit}. 
An electron has an intrinsic magnetic moment 
\begin{equation}
\vec \mu=-\gamma_e \frac{\hbar}{2} \,\vec\sigma = 
-g_e \gamma_L \frac{\hbar}{2} \,\vec\sigma=
-g_e \frac{e}{2\,m_e} \frac{\hbar}{2} \,\vec\sigma=
-g_e \mu_B \,\frac{\vec \sigma}{2},
\end{equation}
where $e$ is the modulus of the electron charge, 
${\vec \sigma}$ is a vector with three components consisting of
the Pauli spin matrices:
\begin{equation}
\label{pauli}
\sigma_x=\left(\begin{array}{cc}
0 & 1\\
1 & 0
\end{array}\right),\quad 
\sigma_y=\left(\begin{array}{cc}
0 & -i\\
i & 0
\end{array}\right),\quad
\sigma_z=\left(\begin{array}{cc}
1 & 0\\
0 & -1
\end{array}\right),
\end{equation}
$\gamma_e$ is the intrinsic gyromagnetic ratio of the electron, 
$\gamma_L$ is its orbital gyromagnetic ratio, 
$g_e=\gamma_e / \gamma_L$ is its intrinsic g-factor 
and $\mu_B=e \hbar / (2\,m_e)$ is the Bohr magneton. When an electron moves in
a system (such as the atom) where the charge distribution (for example the 
nucleus charge) produces an electric field $\vec E$, for the theory of 
relativity this electric field will appear as a magnetic field in the frame
of reference of the electron. In particular if the motion of the electron
were uniform the equivalent magnetic field would be equal to 
$\vec B=-(\vec v \times \vec E)/c^2$. The fact that the electron (and its
frame of reference) is rotating halves such an equivalent magnetic
field~\cite{jackson,spinorbit}.
Thus the Hamiltonian of the electron will have an additional part
\begin{equation}
H_{SO}=\mu_B\,\vec\sigma \cdot 
\left(\frac{\vec E \times \vec v}{2\,c^2}\right)=
\frac{e\,\hbar}{4\,m_e c^2}\vec\sigma \cdot (\vec E \times \vec v)=
\frac{\hbar}{4\,m_e c^2}\vec\sigma \cdot ((\vec\nabla U_L) \times \vec v)
\end{equation}
(with $U_L$ the potential energy), 
which in the absence of an external magnetic field can be written also as
\begin{equation}
H_{SO}=\frac{\hbar}{4\,m_e^2 c^2}\vec\sigma \cdot 
((\vec\nabla U_L) \times \vec p).
\end{equation}
However, if we insert this additional term into the original Schr\"odinger 
equation for the wave function $\psi^n_{\vec k} (\vec r)=
e^{i\,\vec k \cdot \vec r}u^n_{\vec k} (\vec r)$, we obtain 
\begin{eqnarray}
&& H_{SO}\psi^n_{\vec k} (\vec r) =
\frac{\hbar}{4\,m_e^2 c^2}\vec\sigma \cdot 
\left((\vec\nabla U_L) \times (-i\,\hbar\,\vec\nabla)\right)
e^{i\,\vec k \cdot \vec r}u^n_{\vec k} (\vec r)=\\
&& \frac{\hbar}{4\,m_e^2 c^2}\vec\sigma \cdot 
\left((\vec\nabla U_L) \times \left((\hbar \vec k
e^{i\,\vec k \cdot \vec r})u^n_{\vec k} (\vec r)+
e^{i\,\vec k \cdot \vec r}
(-i\,\hbar\,\vec\nabla u^n_{\vec k} (\vec r))\right)\right)=\nonumber\\
&& e^{i\,\vec k \cdot \vec r}
\left(\frac{\hbar^2}{4\,m_e^2 c^2}\vec\sigma \cdot 
((\vec\nabla U_L) \times \vec k)+
\frac{\hbar}{4\,m_e^2 c^2}\vec\sigma \cdot 
((\vec\nabla U_L) \times (-i\,\hbar\,\vec\nabla))\right)
u^n_{\vec k} (\vec r).\nonumber
\end{eqnarray}
If we repeat the procedure used to move from the Schr\"odinger 
equation for the wave functions $\psi^n_{\vec k} (\vec r)$ to the
Schr\"odinger-Bloch equation for the Bloch lattice functions
$u^n_{\vec k} (\vec r)$, we obtain that in the Hamiltonian of this last
equation there will be two additional terms:
\begin{eqnarray}
&& \frac{\hbar^2}{4\,m_e^2 c^2}\vec\sigma \cdot 
((\vec\nabla U_L) \times \vec k)+
\frac{\hbar}{4\,m_e^2 c^2}\vec\sigma \cdot 
((\vec\nabla U_L) \times (-i\,\hbar\,\vec\nabla))=\\
&& \frac{\hbar^2}{4\,m_e^2 c^2}\vec\sigma \cdot 
((\vec\nabla U_L) \times \vec k)+H_{SO}.\nonumber
\end{eqnarray}
The first term near $\vec k=0$ is small compared with the other term; 
thus only the second term is usually considered.
The second term in the case of a potential energy with (locally) spherical
symmetry (and thus of a radial electric field) becomes
\begin{eqnarray}
H_{SO}&=&\frac{e\,\hbar}{4\,m_e^2 c^2}\vec\sigma \cdot (\vec E \times \vec p)=
\frac{e\,\hbar}{4\,m_e^2 c^2}\vec\sigma \cdot 
\frac{E_r}{r}(\vec r \times \vec p)=\\
&& -i\left(\frac{e\,\hbar^2 E_r}{4\,m_e^2 c^2 r}\right)\vec\sigma \cdot 
(\vec r \times \vec\nabla) \equiv
-i \frac{\Lambda}{2} \vec\sigma \cdot (\vec r \times \vec\nabla).\nonumber
\end{eqnarray}
\hfill\break
In order to calculate the influence that the spin-orbit term has on the
valence bands, we need to calculate the matrix elements on the basis states
$|vx0 \rangle $, $|vy0 \rangle $, $|vz0 \rangle $ and $|c0 \rangle $. 
Due to the symmetry proprieties of such states, we see that the only 
non-zero elements are the non-diagonal elements between valence band states
\begin{eqnarray}
\langle vy0|H_{SO}|vx0 \rangle &=& - \langle vx0|H_{SO}|vy0 \rangle =
i\,\lambda\sigma_z,\\
\langle vz0|H_{SO}|vy0 \rangle &=& - \langle vy0|H_{SO}|vz0 \rangle =
i\,\lambda\sigma_x,\nonumber\\
\langle vx0|H_{SO}|vz0 \rangle &=& - \langle vz0|H_{SO}|vx0 \rangle =
i\,\lambda\sigma_y,\nonumber
\end{eqnarray}
with $\lambda$ a non-zero quantity given by (if $V_c$ is the volume of the
lattice unit cell)
\begin{equation}
\lambda=\frac{\Lambda}{2}\frac{1}{V_c}
\int_{V_c} x^2 \rho_v^2 (r) d\,\vec r.
\end{equation}
Therefore, considering also the spin-orbit coupling, the matrix 
$H_{\vec k}^v$ becomes
\begin{eqnarray}
H_{\vec k}^v &=& \left(E_0^v+\frac{\hbar^2 k^2}{2\,m_e}\right)I-
\frac{\hbar^4 P^2}{m_e^2 E_g^0}
\left[\begin{array}{ccc}
k_x^2 & k_x k_y & k_x k_z\\
k_y k_x & k_y^2 & k_y k_z\\
k_z k_x & k_z k_y & k_z^2
\end{array}\right]\\
&&+i\,\lambda\left[\begin{array}{ccc}
0 & -\sigma_z & \sigma_y\\ 
\sigma_z & 0 & -\sigma_x\\
-\sigma_y & \sigma_x & 0
\end{array}\right],\nonumber
\end{eqnarray}
where $\sigma_x$, $\sigma_y$ and $\sigma_z$ are the Pauli spin matrices
(\ref{pauli}), which do not commute with one another.
If we consider the special case $\vec k \parallel \hbox{\boldmath{$\hat z$}}$
we can quite easily find the eigenvalues of this matrix, arriving at a
third-order equation in the energy, the solutions of which represent the
dispersion relations of the three valence 
bands, each one degenerate with respect to the spin. In particular, if
we make the approximation $(\hbar^4 P^2 k^2 /(m_e^2 E_g^0)) \ll \lambda$,
we find the solutions~\cite{wenckebach}
\begin{eqnarray}
E_{hh} &=& E_0^v+\lambda+\frac{\hbar^2}{2}\frac{1}{m_e}k^2,\\
E_{lh} &=& E_0^v+\lambda+\frac{\hbar^2}{2}\frac{1}{m_e}
\left(1-\frac{4\,\hbar^2\,P^2}{3\,m_e E_g^0}\right)k^2,\nonumber\\
E_{\lambda h} &=& E_0^v-2\,\lambda+\frac{\hbar^2}{2}\frac{1}{m_e}
\left(1-\frac{2\,\hbar^2\,P^2}{3\,m_eE_g^0}\right)k^2.\nonumber
\end{eqnarray}
Thus, considering the effect of the spin-orbit interaction, we have obtained
(fig.~\ref{f2}) two valence bands (the heavy-hole band and the light-hole 
band) degenerate at $\vec k=0$, where they have an energy 
$E_g=E_c^0-(E_v^0+\lambda )=E_g^0-\lambda$ 
lower than the conduction band, and one valence band (the spin-orbit band)
which for $\vec k=0$ has an energy $\Delta=3\,\lambda$ lower than 
the other two valence bands. We notice that,
while the light-hole band and the spin-orbit band have a negative effective
mass of the same order of magnitude as the effective mass of the electrons
in the conduction band, the heavy-hole band is characterized by a much larger 
effective mass (the fact that the obtained effective mass is positive instead 
disappears with a more refined treatment: obviously the effective mass of 
the electrons in the valence bands has to be negative, which corresponds to
a positive effective mass for the holes).
\noindent

This simplified model is amenable to several refinements.

As to the conduction band, we can include in the calculation the 
spin-orbit splitting of the valence band and the effect of the higher
conduction bands. In particular, with the first change we obtain
a better expression for the effective mass in the conduction band:
\begin{equation}
\frac{1}{m_c^*}=
\frac{1}{m_e}+\frac{2}{{m_e}^2}
\left[\frac{2\,\hbar^2P^2}{3\,E_g}+\frac{\hbar^2P^2}{3(E_g+\Delta)}\right],
\end{equation}
where $E_g=E_g^0-\lambda $.
\newpage

Also in the treatment of the valence bands we can consider the effect of the
higher conduction bands; one of the effects is that the resulting valence 
bands lose their isotropy and exhibit a complex orientation dependence in 
the reciprocal space (``band warping'').

It is important to notice that the expressions found for the band structure 
depend on a small number of parameters, for example $E_g$,
$\Delta$ and $m_c^*$ (from which we can calculate the parameter $P$ using 
the expression found for the effective mass of the conduction band). From
a practical point of view, these quantities are commonly obtained from
{\em a priori} band structure calculations or, better, experimentally:
in particular the bandgap values $E_g$ and $\Delta$ are accurately known
from optical experiments, while $m_c^*$ is known from cyclotron resonance
experiments.

The approach based on the ``traditional'' perturbation theory, that we
have reported in this first part following the description of
T.~Wenckebach \cite{wenckebach}, differs from the method proposed by
E.~O.~Kane \cite{kane1,kane2,kane3,kane4,kane5,kane6}.

Starting from the consideration that the Bloch lattice functions can
be expanded in terms of the complete, infinite set of the unperturbed Bloch
lattice functions, Kane computes this expansion in an approximate way,
considering only a finite set of bands. In particular he considers only
the three valence bands and the conduction band (not including the
effects of the other bands) and diagonalizes exactly the Hamiltonian
of the Schr\"odinger-Bloch equation in the presence of spin-orbit 
interaction \cite{kane5}, written taking as a basis the following set,
made up of a linear combination with constant coefficients of the
$u^n_0 (\vec r)$ considered in the absence of spin-orbit ({\em i.e.} of the
functions $|c0 \rangle $, $|vx0 \rangle $, $|vy0 \rangle $ and
$|vz0 \rangle $ taken with spin-up and spin-down):
\begin{eqnarray}
&& i |c0 \rangle |\!\downarrow \rangle,\quad
1/\sqrt{2}(|vx0 \rangle -i\,|vy0 \rangle )|\!\uparrow \rangle,\quad 
|vz0 \rangle |\!\downarrow \rangle,\quad 
-1/\sqrt{2}(|vx0 \rangle +i\,|vy0 \rangle )|\!\uparrow \rangle,\\
&& i |c0 \rangle |\!\uparrow \rangle,\quad
-1/\sqrt{2}(|vx0 \rangle +i\,|vy0 \rangle )|\!\downarrow \rangle,\quad
|vz0 \rangle |\!\uparrow \rangle,\quad
1/\sqrt{2}(|vx0 \rangle -i\,|vy0 \rangle )|\!\downarrow \rangle \nonumber
\end{eqnarray}
(where $|\!\uparrow \rangle$ and $|\!\downarrow \rangle$ are, respectively,
the spin-up and spin-down unit spinors).

From this diagonalization he finds, for small values of $k^2$, the
following expressions for the considered bands (choosing the zero
of energy at the top of the light-hole and heavy-hole bands and
defining the various quantities as before):
\begin{eqnarray}
E_c &=& E_g+\frac{\hbar^2}{2}\frac{1}{m_e}
\left(1+\frac{4\,\hbar^2\,P^2}{3\,m_e E_g}+
\frac{2\,\hbar^2\,P^2}{3\,m_e (E_g+\Delta)}\right)k^2,\\
E_{hh} &=& \frac{\hbar^2}{2}\frac{1}{m_e}k^2,\nonumber\\
E_{lh} &=& \frac{\hbar^2}{2}\frac{1}{m_e}
\left(1-\frac{4\,\hbar^2\,P^2}{3\,m_e E_g}\right)k^2,\nonumber\\
E_{\lambda h} &=& -\Delta+\frac{\hbar^2}{2}\frac{1}{m_e}
\left(1-\frac{2\,\hbar^2\,P^2}{3\,m_e (E_g+\Delta)}\right)k^2.\nonumber
\end{eqnarray}
These expressions are very similar to the expressions obtained with the 
previously described simplified model, but clearly show the dual effect 
that each reciprocal interaction has on the related couple of bands.
As before, these results give an incorrect effective mass for the
heavy-hole band.

From the diagonalization Kane also finds the Bloch lattice functions 
$u^n_{\vec k} (\vec r)$ that diagonalize the Hamiltonian of the
Schr\"odinger-Bloch equation in the presence of spin-orbit interaction
({\em i.e.} the eigenfunctions of this Hamiltonian) as linear combinations of the 
$u^n_0 (\vec r)$ considered in the absence of spin-orbit;
in particular for vanishing $k$ they are (in the simplest case in which
$\vec k \parallel \hbox{\boldmath{$\hat z$}}$):
\begin{eqnarray}
\label{pertu}
& i\,|c0 \rangle |\!\downarrow \rangle,\qquad
i\,|c0 \rangle |\!\uparrow \rangle, &\\
& -1/\sqrt{2}\,(|vx0 \rangle +i\,|vy0 \rangle )|\!\uparrow \rangle,\qquad
1/\sqrt{2}\,(|vx0 \rangle -i\,|vy0 \rangle )|\!\downarrow \rangle,& \nonumber\\
& 1/\sqrt{6}\,(|vx0 \rangle -i\,|vy0 \rangle )|\!\uparrow \rangle+
\sqrt{2/3}\,|vz0 \rangle |\!\downarrow \rangle,& \nonumber\\
& -1/\sqrt{6}\,(|vx0 \rangle +i\,|vy0 \rangle )|\!\downarrow \rangle+
\sqrt{2/3}\,|vz0 \rangle |\!\uparrow \rangle,& \nonumber\\
& 1/\sqrt{3}\,(|vx0 \rangle -i\,|vy0 \rangle )|\!\uparrow \rangle-
1/\sqrt{3}\,|vz0 \rangle |\!\downarrow \rangle,& \nonumber\\
& 1/\sqrt{3}\,(|vx0 \rangle +i\,|vy0 \rangle )|\!\downarrow \rangle+
1/\sqrt{3}\,|vz0 \rangle |\!\uparrow \rangle.& \nonumber
\end{eqnarray}
In order to take into account the effect of higher and lower bands on
the considered ones, Kane uses the L\"owdin perturbation theory
\cite{lowdin1,lowdin2}. Following this method, one can divide all the bands 
into two sets $A$ and $B$: $A$ is the set we want to treat exactly and $B$ 
contains all the other bands. At the lowest order of perturbation theory the
coupling between the set $A$ and the set $B$ can be removed introducing
the perturbed functions
\begin{equation}
u_i'=u_i+\sum_n^B \frac{H_{ni}u_{n}}{(H_{ii}-H_{nn})},
\end{equation}
where $i$ is in $A$ and $n$ is in $B$.
The renormalized interactions connecting $u_i'$ and $u_j'$ 
are given by
\begin{equation}
H_{ij}'=H_{ij}+\sum_n^B \frac{H_{in}H_{nj}}{\displaystyle
{\left(\frac{H_{ii}+H_{jj}}{2}-H_{nn}\right)}}
\end{equation}
(with $i$, $j$ in $A$).
In this way we can reduce the Hamiltonian matrix, which in principle 
connects all the possible bands, to a Hamiltonian matrix relating
only the bands of interest, but in which, however, the interactions with
the non-considered bands are included. The method is accurate as long
as $|H_{in}| \ll |H_{ii}-H_{nn}|$, with $i$ in $A$ and $n$ in $B$, and
thus the set $A$ has to be selected in order to satisfy this relation
(for example, also states degenerate with those in which we are
interested have to be considered inside the set $A$).
Note that the L\"owdin  perturbation theory reduces to the ordinary
perturbation theory when only a single band is considered in the set $A$.

Kane applies this perturbation method, starting from the Bloch lattice
functions $(\ref{pertu})$ of the set $A$ of considered conduction and valence
bands and from the unperturbed Bloch lattice functions of the set $B$
of the higher and lower bands, obtaining a better approximation of the
actual dispersion relations of the considered bands.

An exact diagonalization of the Hamiltonian has also been performed
(originally by M.~Cardona and F.~H.~Pollak \cite{cardona}, more
recently by other authors \cite{cavasillas,radhia,richard,michelini}),
extending the number of considered bands (and thus the number of
involved parameters) to reproduce the band structure all over the
Brillouin zone to a reasonable degree of accuracy (for example, in
their original paper M.~Cardona and F.~H.~Pollak consider 15 bands,
with 10 parameters, to reproduce the energy band structure of
germanium and silicon).

\section{The $\vec k \cdot \vec p$ method in non-periodic systems:
envelope function theory and application to heterostructures}

Till now, we have considered the applications of the $\vec k \cdot \vec p$
method to periodic, homogeneous crystals, using a description of the
electron wave function in terms of Bloch functions. However, in the
presence of a generic external potential the periodicity of the potential
inside the crystal breaks down and thus the electron wave functions are far
from periodic. Since the Bloch functions $|n\,\vec k \rangle =
e^{i\,\vec k \cdot \vec r} u^n_{\vec k} (\vec r)/\sqrt{(2\pi)^3}$,
considered as functions of $\vec r$ and $\vec k$, are a complete set of
orthonormal functions, also in this case the generic wave function could
be expanded on the basis of Bloch functions in this way
\begin{equation}
\psi (\vec r)=\sum_{n} \int d\vec k A_{n}(\vec k) |n\,\vec k \rangle,
\end{equation}
(where the sum over the number of bands together with the integral over
the Brillouin zone corresponds to an integral over all the reciprocal space).
However, in general a large number of Bloch functions, evaluated over a large
range of wave vectors, would be necessary in this expansion. Therefore in
this case it is convenient to replace the Bloch phase factor, involving the
wave vector measured from the reference extremum point, with an envelope
function, and thus to use a different formulation of the $\vec k \cdot \vec p$
method, based on the concept of envelope functions
\footnote{Notice that there is also an alternative approach to the 
envelope function theory using the definition of Wannier~\cite{wannier} 
and Slater~\cite{slater}, based on Wannier orbitals. See also
\cite{burt5,adams,young}.}.

In order to introduce this concept, we can make a very approximate calculation
\cite{envelope,mitin} in the hypothesis that the external potential energy
$U(\vec r)$ (``external'' here meaning ``not due to the periodic structure
of the lattice'') is slowly varying on an atomic scale and\break
the $n$-th energy
band that we are considering is non-degenerate (thus with unique independent
Bloch lattice function $u^n_{\vec k} (\vec r)$). In this case, the
Schr\"odinger equation (in the absence of a magnetic field) for the electron
wave function $\psi(\vec r)$
\begin{equation}
\label{es}
\left(-\frac{\hbar^2}{2\,m_e}\nabla^2+U_L (\vec r)\right)\psi (\vec r)+
U(\vec r)\psi (\vec r)=H^{(0)} \psi (\vec r)+
U(\vec r)\psi (\vec r)=E \psi (\vec r)
\end{equation}
(where $U_L (\vec r)$ is the periodic lattice potential energy and 
$H^{(0)}$ is the Hamiltonian in the absence of the external potential 
energy $U(\vec r)$) is equivalent to the equation
\begin{equation}
\label{efe}
E_n(-i\,\vec\nabla) F(\vec r)+
U(\vec r) F(\vec r)=E F(\vec r),
\end{equation}
where $E_n(-i\,\vec\nabla)$ represents the operator obtained replacing, 
in the dispersion relation $E_n(\vec k)$ describing the $n$-th
energy band in the absence of the external potential, each component of
$\vec k$ with the corresponding component of $-i\,\vec\nabla$,
and $F(\vec r)$ is the envelope function, a slowly varying 
function that, when we consider only the $n$-th band, multiplied by the 
fast varying Bloch lattice function $u_0^n(\vec r)$ 
(considered in $\vec k=0$) gives the electron wave function.

Indeed, if we expand $\psi(\vec r)$ in the orthogonal basis set 
$|\nu\vec k \rangle =
e^{i\,\vec k \cdot \vec r}u^{\nu}_{\vec k} (\vec r)/\sqrt{V}$
(with $V$ the crystal volume)
\begin{equation}
\psi(\vec r)=\sum_{\nu,\vec k}a_\nu (\vec k)|\nu\vec k \rangle,
\end{equation}
we can re-write the Schr\"odinger equation (\ref{es}) in matrix form using
the basis $|\nu\vec k \rangle $
\begin{eqnarray}
&& \sum_{\nu',\vec k'}\left( 
\langle \nu\vec k|H^{(0)}+U(\vec r)|\nu'\vec k' \rangle 
a_{\nu'}(\vec k')\right) = E a_{\nu} (\vec k) \Rightarrow\\
&& E_{\nu}(\vec k)a_{\nu} (\vec k)+
\sum_{\nu',\vec k'}
\left( \langle \nu\vec k|U(\vec r)|\nu'\vec k' \rangle 
a_{\nu'}(\vec k')\right) = E a_{\nu} (\vec k),\nonumber
\end{eqnarray}
where we have used the fact that (being $|\nu\vec k \rangle$ an
eigenfunction of $H^{(0)}$ with eigenvalue $E_{\nu}(\vec k)$)
\begin{equation}
\langle \nu\vec k|H^{(0)}|\nu'\vec k' \rangle =E_{\nu'}(\vec k') 
\langle \nu\vec k|\nu'\vec k' \rangle =
E_\nu(\vec k)\delta_{\nu\,\nu'}\delta_{\vec k\,\vec k'}.
\end{equation}
In particular, for $\nu=n$ we have that 
\begin{equation}
\label{mpschr}
E_n(\vec k)a_n (\vec k)+
\sum_{\nu',\vec k'}
\left( \langle n\vec k|U(\vec r)|\nu'\vec k' \rangle a_{\nu'}(\vec k')\right)=
E a_n (\vec k).
\end{equation}
If instead we expand the envelope function equation in the orthogonal 
set of plane waves $|\vec k \rangle =e^{i\,\vec k \cdot \vec r}/\sqrt{V}$
\begin{equation}
F(\vec r)=\sum_{\vec k}a(\vec k)|\vec k \rangle,
\end{equation}
we can re-write the envelope function equation (\ref{efe}) in matrix form using
the basis $|\vec k \rangle $
\begin{eqnarray}
\label{mpenv}
&& \sum_{\vec k'}\left( \langle \vec k|E_n(-i\,\vec\nabla)+
U(\vec r)|\vec k' \rangle 
a(\vec k')\right) = E a(\vec k)\Rightarrow\\
&& E_n(\vec k)a(\vec k)+
\sum_{\vec k'}\left( \langle \vec k|U(\vec r)|\vec k' \rangle a(\vec k')\right)
= E a(\vec k),\nonumber
\end{eqnarray}
using the fact that 
\begin{equation}
(-i\,\nabla_{\nu})^p |\vec k' \rangle =
(-i\,\nabla_{\nu})^p (e^{i\,\vec k' \cdot \vec r}/ \sqrt{V})=
(k'_{\nu})^p (e^{i\,\vec k' \cdot \vec r}/ \sqrt{V})=
(k'_{\nu})^p |\vec k' \rangle,
\end{equation}
with $\nu =x, y, z$ 
and thus
\begin{equation}
E_n(-i\,\vec\nabla)|\vec k' \rangle = E_n(\vec k')|\vec k' \rangle 
\end{equation}
(being $E_n(-i\,\vec\nabla)$ an operator
made up of operators of the type $(-i\,\nabla_{\nu})^p$)
and then exploiting the orthogonality relation 
$ \langle \vec k|\vec k' \rangle =\delta_{\vec k\,\vec k'}$.
The two equations (\ref{mpschr}) and (\ref{mpenv}), obtained from the
Schr\"odinger equation and from the envelope function equation are
exactly equal if 
\begin{equation}
\langle n\vec k|U(\vec r)|\nu'\vec k' \rangle =
\delta_{n\,\nu'} \langle \vec k|U(\vec r)|\vec k' \rangle,
\end{equation}
{\em i.e.} if the matrix elements of the external potential $U(\vec r)$
between states from different bands are negligible. This is what happens 
if $U$ is slowly varying on an atomic scale. Indeed, in this case we 
have that
\begin{eqnarray}
&& \langle n\vec k|U(\vec r)|\nu'\vec k' \rangle =
\frac{1}{V} \sum_{j=1}^N \int_{V_j}d \vec r\, {u_{\vec k}^n}^*(\vec r)
u_{\vec k'}^{\nu'}(\vec r)e^{i\,(\vec k'-\vec k) 
\cdot \vec r}U(\vec r)\simeq\\
&& \sum_{j=1}^N e^{i\,(\vec k'-\vec k) \cdot \vec r_j}U(\vec r_j)
\frac{1}{V} 
\int_{V_j} d \vec r \,{u_{\vec k}^n}^*(\vec r)u_{\vec k'}^{\nu'}(\vec r)
\simeq\nonumber\\
&& \sum_{j=1}^N e^{i\,(\vec k'-\vec k) \cdot \vec r_j}U(\vec r_j)
\delta_{n\,\nu'}\frac{1}{N}\simeq
\delta_{n\,\nu'}\int_V d \vec r \,
\frac{e^{i\,(\vec k'-\vec k)  \cdot \vec r}}{V} U(\vec r)=
\delta_{n\,\nu'} \langle \vec k|U(\vec r)|\vec k' \rangle,\nonumber
\end{eqnarray}
where $V$ the crystal volume, $V_j$ the volume of the $j$-th unit cell,
$\vec r_j$ the coordinate of its center and 
$N$ the number of unit cells. We have assumed that $U(\vec r)$
and $e^{i\,(\vec k'-\vec k) \cdot \vec r}$ are approximately constant over 
a unit cell and $u_{\vec k'}^{n}(\vec r)\simeq u_{\vec k}^{n}(\vec r)$ over
the range of values of $|\vec k'-\vec k|$ for which 
$ \langle \vec k|U(\vec r)|\vec k' \rangle $ is not negligible.

Note that usually for functions with the translation symmetry of the crystal
lattice the scalar product is defined as
\begin{equation}
\label{scalarproduct}
\langle \Psi_1 | \Psi_2 \rangle=
\frac{1}{V_c} \int_{V_c} d \vec r \,\Psi_1^* (\vec r) \Psi_2 (\vec r)
\end{equation}
(with $V_c$ the volume of the unit cell); in particular 
$u_{\vec k}^{\nu} (\vec r)$ and $e^{i\,\vec k \cdot \vec r}u^{\nu}_{\vec k}$ 
are normalized with respect to this scalar product.

If the two equations (\ref{mpschr}) and (\ref{mpenv}) are identical, they have 
the same solutions $a_{n}(\vec k)$ and $a(\vec k)$.
Thus (assuming that $a_\nu(\vec k)$ is non-zero only for the particular 
band $n$, coherently with our hypothesis that there is no mixing between 
the bands) we can write that 
\begin{eqnarray}
\psi(\vec r)&=&\sum_{\nu,\vec k}a_\nu (\vec k)|\nu\vec k \rangle =
\sum_{\vec k}a_n(\vec k)
\frac{e^{i\,\vec k \cdot \vec r}}{\sqrt{V}}
u^n_{\vec k} (\vec r)\simeq\\
&& u^n_0 (\vec r) \sum_{\vec k} a_n(\vec k)
\frac{e^{i\,\vec k \cdot \vec r}}{\sqrt{V}}=
u^n_0 (\vec r) \sum_{\vec k} a_n(\vec k) |\vec k \rangle =
u^n_0 (\vec r) F(\vec r),\nonumber
\end{eqnarray}
where we have assumed that $u^n_{\vec k}(\vec r)$ does not vary very much
with $\vec k$ (note that the main $\vec k$'s have to be quite close to
$\vec k_0=0$ for the previous derivation to be consistent).

We notice that if we express $E_n(\vec k)$ as 
\begin{equation}
E_n(\vec k)=E_0^n+\frac{\hbar^2}{2}\sum_{\mu ,\nu}
\frac{k_{\mu}k_{\nu}}{m_{\mu\,\nu}^*}
\end{equation}
(with $\mu,\nu=x,y,z$) the envelope function equation becomes
\begin{equation}
-\frac{\hbar^2}{2}\sum_{\mu ,\nu}
\frac{\nabla_{\mu}\nabla_{\nu}}{m_{\mu\,\nu}^*} 
F(\vec r)+(E_0^n+U(\vec r)) F(\vec r)=
E F(\vec r)
\end{equation}
and when the effective mass is isotropic 
($(1/m_{\mu\,\nu}^*)=(1/m^*)\delta_{\mu\,\nu}$) we have the well-known 
equation
\begin{equation}
-\frac{\hbar^2}{2\,m^*}\nabla^2 F(\vec r)
+(E_0^n+U(\vec r)) F(\vec r)=E F(\vec r).
\end{equation}
Luttinger and Kohn in a famous paper~\cite{luttinger1} have given an
alternative derivation of the single-band envelope function equation,
which has the advantage of being easily generalized to more complicated
cases. The starting equation is again the Schr\"odinger
equation $(H^{(0)}+U)\psi=E\psi$, with $H^{(0)}$ being the Hamiltonian of
the electron in the periodic lattice potential and $U$ an additional
potential which is assumed not to vary significantly over each unit cell.
They show that the fuctions $|n\,\vec k \rangle =\chi_{n \vec k}=
e^{i\,\vec k \cdot \vec r} u^n_0 (\vec r)/\sqrt{(2\pi)^3}$
(where $u^n_0 (\vec r)$ are the Bloch lattice functions in the absence of
the external potential, evaluated for $\vec k=0$)
are a complete orthonormal set, if considered as functions of
$\vec r$ and $\vec k$ (exactly as the functions 
$e^{i\,\vec k \cdot \vec r} u^n_{\vec k} (\vec r)/\sqrt{(2\pi)^3}$,
which, contrary to the $\chi_{n \vec k}$, are eigenfunctions of $H^{(0)}$).
This means that
\begin{equation}
\langle \chi_{n \vec k} | \chi_{n' \vec k'} \rangle=
\delta_{n n'}\delta(\vec k-\vec k').
\end{equation}
Therefore, they can expand the wave function $\psi$ over the complete 
orthonormal set of functions $|n\,\vec k \rangle$ in this way:
\begin{equation}
\label{eqp}
\psi=\sum_{n'} \int d\vec k' A_{n'}(\vec k')\chi_{n'\vec k'},
\end{equation}
and, considering this basis, they can rewrite the Schr\"odinger equation
in the following form:
\begin{equation}
\label{eqa}
\sum_{n'} \int d\vec k' \langle n \vec k | H^{(0)}+U | n' \vec k' \rangle
A_{n'}(\vec k')=E A_{n}(\vec k).
\end{equation}
After some calculations, they obtain that
\begin{eqnarray}
\langle n \vec k | H^{(0)} | n' \vec k' \rangle &=&
\left(E^n_0+\frac{\hbar^2k^2}{2\,m_e}\right)
\delta_{n n'}\delta(\vec k-\vec k')+
\sum_{\alpha=x,y,z}\frac{\hbar k_{\alpha} P_{\alpha}^{n n'}}{m_e}
\delta(\vec k-\vec k')\equiv\\
&&\langle n \vec k | H_a | n' \vec k' \rangle+
\langle n \vec k | H_b | n' \vec k' \rangle,\nonumber
\end{eqnarray}
where the momentum matrix elements at $\vec k=0$
\begin{equation}
P_{\alpha}^{n n'}=\frac{1}{V_c}\int_{V_c} {u^n_0}^* (-i\hbar \nabla_{\alpha})
u^{n'}_0 d\vec r
\end{equation}
are characterized by the following properties: $P_{\alpha}^{n n}=0$ if the
point $\vec k=0$ around which we are working is an extremum point of the
dispersion relations, and
$P_{\alpha}^{n n'}=P_{\alpha}^{n' n}=(P_{\alpha}^{n n'})^*$ if a center
of symmetry exists in the crystal.
Moreover, if $U$ is a ``gentle'' potential, with a very small variation over a
unit cell,
\begin{equation}
\langle n \vec k | U | n' \vec k' \rangle=
\mathcal{U}(\vec k-\vec k')\delta_{n n'},
\end{equation}
where $\mathcal{U}(\vec k)$ is the Fourier transform of $U$
\begin{equation}
\mathcal{U}(\vec k)=\frac{1}{(2\pi)^3}\int d\vec r 
e^{-i\vec k\cdot \vec r} U(\vec r).
\end{equation}
As a consequence, eq.~(\ref{eqa}) becomes
\begin{eqnarray}
\label{eqa1}
&& \left(E^n_0+\frac{\hbar^2 k^2}{2\,m_e}\right) A_{n}(\vec k)+\sum_{\alpha=x,y,z}
\sum_{\scriptstyle n' \atop \scriptstyle n'\ne n}
\frac{\hbar k_{\alpha} P_{\alpha}^{n n'}}{m_e} A_{n'}(\vec k)\\
&& {}+\int d\vec k' \mathcal{U}(\vec k-\vec k') A_{n}(\vec k')=E A_{n}(\vec k).\nonumber
\end{eqnarray}
In order to decouple the equation corresponding to the band $n$ from the
other bands, the terms involving $P_{\alpha}^{n n'}$, which couple the bands,
have to be removed to the first order. Luttinger and Kohn obtain this result
applying a proper canonical transformation $T$:
\begin{equation}
A_n(\vec k)=\sum_{n'}\int d \vec k' \langle n \vec k | T | n' \vec k' \rangle
B_{n'}(\vec k'),
\end{equation}
which corresponds, more abstractly, to $\vec A=T \vec B$.
Writing $T=e^S$ and applying this transformation to the equation~(\ref{eqa1}),
which can be rewritten as $H \vec A=E \vec A$, with
$H=H_a+H_b+U$, we obtain
$(e^{-S} H e^S) \vec B=E \vec B$. After some calculations, it can be proved
that, choosing $S$ in such a way that $H_b+[H_a,S]=0$ (the square brackets
denoting the commutator), {\em i.e.}
\begin{equation}
\langle n \vec k | S | n' \vec k' \rangle=
\left\{ \begin{array}{ll}
\displaystyle
-\frac{\hbar \vec k \cdot \vec P^{n n'} \delta(\vec k-\vec k')}
{m_e(E_0^n-E_0^{n'})},
&\textrm{if $n \ne n'$,}\\
\displaystyle
0, &\textrm{if $n=n'$,}
\end{array}\right.
\end{equation}
and neglecting the terms of order $k^3$ and higher and the terms which assume
very small values for a ``gentle'' potential $U$, this equation becomes
\begin{eqnarray}
&& \left(E_0^n+\frac{\hbar^2 k^2}{2\,m_e}+
\frac{\hbar^2}{{m_e}^2}\sum_{\alpha,\beta=x,y,z}
k_{\alpha}k_{\beta}\sum_{\scriptstyle n'' \atop \scriptstyle n''\ne n}
\frac{P_{\alpha}^{n n''} P_{\beta}^{n'' n}}{E_0^n-E_0^{n''}}\right)
B_n(\vec k)\\
&& {}+\int \mathcal{U}(\vec k-\vec k') B_n (\vec k') d\vec k'=E B_n (\vec k),\nonumber
\end{eqnarray}
which can be written more briefly in this form:
\begin{equation}
\label{eqb}
E_n (\vec k) B_n(\vec k)+\int \mathcal{U}(\vec k-\vec k')B_n (\vec k') 
d\vec k'=E B_n (\vec k),
\end{equation}
where $E_n (\vec k)$ is the dispersion relation in the absence of 
$U(\vec r)$ expanded to second order in $\vec k$.

Converting eq.~(\ref{eqb}) from the momentum space to the position space 
and defining the envelope function in this way
\begin{equation}
\label{eqf}
F_n(\vec r)=
\frac{1}{\sqrt{(2\pi)^3}}\int e^{i\vec k \cdot \vec r} B_n(\vec k) 
d\vec k,
\end{equation}
the single band envelope function equation is obtained 
\begin{equation}
\label{eqf1}
(E_n(-i\,\vec\nabla)+U(\vec r)) F_n(\vec r)=E F_n(\vec r),
\end{equation}
with $E_n(-i\,\vec\nabla)$ obtained expanding $E_n(\vec k)$ (the dispersion
relation in the absence of $U(\vec r)$) to second order in $\vec k$ around
$\vec k=0$ with non-degenerate perturbation theory and substituting
each component of $\vec k$ with the corresponding component of 
$-i\,\vec\nabla$.
Being $F_n(\vec r)$ a smooth function, it has significant Fourier components
only for small values of $\vec k$. Since for small values of $\vec k$ also
$S$ is small, for these components 
$A_n(\vec k)=e^S B_n(\vec k)\simeq B_n(\vec k)$
and thus, exploiting the eqs.~(\ref{eqp}) and (\ref{eqf}), we have
\begin{equation}
\psi\simeq \sum_n \int d\vec k B_n(\vec k)
e^{i\,\vec k \cdot \vec r} \frac{u^n_0 (\vec r)}{\sqrt{(2\pi)^3}}=
\sum_n F_n(\vec r) u^n_0 (\vec r)
\end{equation}
and, noting that eq.~(\ref{eqf1}) contains no interband coupling,
\begin{equation}
\psi=F_n(\vec r) u^n_0 (\vec r)
\end{equation}
(as already seen). If locally the external 
potential changes considerably within a cell, in such a region the equation
we have derived is no longer valid, but it continues to be valid in 
regions of space sufficiently distant from it.
\vskip2pt\noindent

Then Luttinger e Kohn adopt an analogous procedure starting from the
Schr\"odinger equation written in the presence of an external magnetic
field. In this way, they demonstrate that in such a case
the envelope function satisfies an equation similar to the
one in the absence of a magnetic field, the only difference being that
the new Hamiltonian is obtained replacing, in the expansion of
$E_n (\vec k)$ to quadratic terms, each $k_\alpha$ by the operator
$-i\nabla_{\alpha}+e A_{\alpha}/\hbar$ (using the MKS system of units)
with $A_{\alpha}$ the $\alpha$-th component of the vector potential. 
Moreover in the expansion of $E_n (\vec k)$ to the 
second order any arising product of non-commuting factors has to be 
interpreted as the symmetrized product. 

In the case in which the extremum is at $\vec k=\vec k_0 \ne 0$, the 
demonstrations (both with and without an external 
magnetic field) can be repeated by just replacing $u_0^n (\vec r)$ 
(the eigenfunctions of the Hamiltonian for $\vec k=0$ in the absence
of $U (\vec r)$ and of an external magnetic field) with
$\phi_{n \vec k_0}\equiv
e^{i\,\vec k_0 \cdot \vec r}u_{\vec k_0}^n (\vec r)$ 
(the eigenfunctions of the Hamiltonian for $\vec k=\vec k_0$ in the absence
of $U (\vec r)$ and of an external magnetic field). Indeed, it can be seen
that the functions $\varphi_{n \vec \kappa}\equiv 
e^{i\,\vec \kappa \cdot \vec r} (e^{i\,\vec k_0 \cdot \vec r}
u_{\vec k_0}^n (\vec r)/\sqrt{(2\pi)^3})$ have properties analogous to those
prevously seen for the $\chi_{n \vec k}=e^{i\,\vec k \cdot \vec r}
u^n_0 (\vec r)/\sqrt{(2\pi)^3}$: considered as functions of $\vec r$
and $\vec \kappa$, they are a complete orthonormal set of functions (such
that $\langle \varphi_{n \vec \kappa} | \varphi_{n' \vec \kappa'} \rangle=
\delta_{n n'}\delta(\vec \kappa-\vec \kappa')$)
and the momentum matrix elements computed in $\vec k_0$, defined as
\begin{equation}
P_{\alpha}^{n n'}=\frac{1}{V_c}\int_{V_c} {u^n_{\vec k_0}}^*
(\hbar k_{0_{\alpha}}-i\hbar \nabla_{\alpha})
u^{n'}_{\vec k_0} d\vec r,
\end{equation}
have properties analogous to those seen in the case in which 
$\vec k_0=0$. 
In this case the relation between the wave function and the
envelope function is
\begin{equation}
\label{k0}
\psi=F_n(\vec r)(e^{i\,\vec k_0 \cdot \vec r}u_{\vec k_0}^n (\vec r))
\end{equation}
and the envelope function equation is
\begin{equation}
\left[E_n(\vec k_0-i\vec\nabla)+U\right] F_n=E F_n,
\end{equation}
in the absence of magnetic field, and
\begin{equation}
\left[E_n \left(\vec k_0-i\vec\nabla+\frac{e \vec A}{\hbar}\right)+U\right]
F_n=E F_n,
\end{equation}
in the presence of magnetic field. As before, in these expressions
an expansion of $E_n$ around $\vec k_0$ to second-order terms in 
$-i\vec\nabla$ and in $-i\vec\nabla+e\vec A/\hbar$, respectively, is meant.

If there are extrema at several different values of $\vec k_0$ within the 
band, we obtain an envelope function equation for each of them; if the 
solutions corresponding to the different $\vec k_0$ values have different
energies, the corresponding wave functions represent independent solutions
of the Schr\"odinger equation; otherwise the correct wave function will be
a linear combination of those from the different extrema associated with
the same energy.

When the band of interest is degenerate, Luttinger and Kohn, using
a similar calculation, arrive at a set of coupled second-order equations
which correspond to the effective mass equation found in the case of 
non-degenerate bands. 
In particular (assuming for simplicity that the degeneracy occurs at
$\vec k=0$) they assume to have, at $\vec k=0$, $r$ unperturbed
degenerate Bloch lattice functions corresponding to the same unperturbed
energy $E_0^j$ (where ``unperturbed'' means for $\vec k=0$ and in
the absence of $U (\vec r)$ and of an external magnetic field) and they define
them as $\phi_j$ (with $j=1,\ldots , r$, where $r$ is the\break
degeneracy), {\em i.e.}
\begin{equation}
H^{(0)} \phi_j=E_0^j \phi_j
\end{equation}
(notice that the $\phi_j$'s, {\em i.e.} the $u^n_0$'s, can be seen as Bloch functions
$e^{i\,\vec k \cdot \vec r} u^n_{\vec k}$ for $\vec k=0$ and thus they
have to satisfy the Schr\"odinger equation for $\vec k=0$).
%(or equivalently they can be seen 
%as Bloch lattice functions $u^n_{\vec k}$ for $\vec k=0$ and so they 
%have to satisfy the Schr\"odinger-Bloch equation for $\vec k=0$). 
They instead indicate as $\phi_i$ (with $i\ne 1,\ldots , r$) the 
unperturbed Bloch lattice functions at $\vec k=0$ corresponding to
the other bands, that are not degenerate with the $\phi_j$'s.
If the crystal has a center of symmetry, it can be proved that the
momentum matrix elements between different $\phi_j$'s vanish, {\em i.e.}
$P_{\alpha}^{j j'}=0$.
Luttinger and Kohn introduce the complete set of functions 
$| n \vec k \rangle=\phi_{n \vec k}=e^{i\vec k \cdot \vec r}\phi_n
/\sqrt{(2\pi)^3}$
(where $\phi_n$ indicates both the $\phi_j$'s and the $\phi_i$'s).
Using this basis, they can expand the wave function in this way:
\begin{equation}
\psi=\sum_n \int d\vec k A_n(\vec k) \phi_{n \vec k}
\end{equation}
and rewrite the Schr\"odinger equation as
\begin{equation}
\sum_{n'} \int d\vec k' \langle n \vec k | H^{(0)}+U | n' \vec k' \rangle
A_{n'}(\vec k')=E A_{n}(\vec k),
\end{equation}
thus obtaining:
\vskip2pt\noindent
\begin{eqnarray}
&& \left(E_0^j+\frac{\hbar^2 k^2}{2\,m_e}\right) A_j(\vec k)+\sum_{\alpha=x,y,z}
\sum_i \frac{\hbar k_{\alpha} P_{\alpha}^{j i}}{m_e} A_i (\vec k)\\
&& {}+\int d\vec k' \mathcal{U}(\vec k-\vec k') A_j (\vec k')=E A_j (\vec k)\nonumber
\end{eqnarray}
\vskip2pt\noindent
(writing only the equations corresponding to the degenerate states $j$).
In order to decouple the equations corresponding to the states $j$ from those
of the states $i$, a proper canonical transformation $A=T B=e^S B$ is again
applied, with
\vskip2pt\noindent
\begin{equation}
\langle n \vec k | S | n' \vec k' \rangle=
\left\{ \begin{array}{ll}
\displaystyle
-\frac{\hbar \vec k \cdot \vec P^{n n'} \delta(\vec k-\vec k')}
{m_e (E_0^n-E_0^{n'})},
&\textrm{if $n$ or $n' \notin [1,r]$,}\\
\displaystyle
0, &\textrm{if $n$ and $n'\in [1,r]$.}
\end{array}\right.
\end{equation}
\vskip2pt\noindent
In this way Luttinger and Kohn obtain, to second-order terms in $k$, the
following set of equations for the $r$ degenerate states:
\begin{eqnarray}
&& \sum_{j'=1}^r \left(E_0^j \delta_{j j'}+
\sum_{\alpha,\beta=x,y,z} \left(D_{jj'}^{\alpha\beta}
k_{\alpha} k_{\beta}\right)\right) B_{j'} (\vec k)\\
&& {}+\int \mathcal{U}(\vec k-\vec k') B_j (\vec k') d\vec k'=E B_j (\vec k),\nonumber
\end{eqnarray}
with
\begin{equation}
D_{jj'}^{\alpha\beta}=\frac{\hbar^2}{2\,m_e}\delta_{j\,j'}
\delta_{\alpha\,\beta}+\frac{\hbar^2}{{m_e}^2}\sum_i
\frac{P_{\alpha}^{j\,i} P_{\beta}^{i\,j'}}{(E_0^j-E_0^i)}.
\end{equation}
Therefore, introducing again the envelope functions
\begin{equation}
F_j (\vec r)=
\frac{1}{\sqrt{(2\pi)^3}}\int e^{i\vec k \cdot \vec r} B_j (\vec k) 
d\vec k,
\end{equation}
Luttinger and Kohn arrive at the conclusion that the $r$ envelope
functions $F_j (\vec r)$ corresponding to the originally degenerate
energy bands satisfy the $r$ coupled differential equations
\begin{equation}
\label{coupled}
\sum_{j'=1}^r \left(E_0^j \delta_{j j'}+
\sum_{\alpha, \beta=x,y,z} \left(D_{jj'}^{\alpha\beta}
(-i\nabla_{\alpha})(-i\nabla_{\beta})\right)
+U (\vec r)\delta_{j\,j'}\right) F_{j'} (\vec r)=E F_j (\vec r)
\end{equation}
(if the energy zero is set at $E_0^j$ the term $E_0^j \delta_{j j'}$
disappears).

Analogously to what happens in the non-degenerate case, for small
values of $\vec k$, $A_n(\vec k)\simeq B_n(\vec k)$ and thus
\begin{equation}
\label{deg}
\psi\simeq \sum_n \int d\vec k B_n(\vec k)
e^{i\,\vec k \cdot \vec r} \frac{\phi_n (\vec r)}{\sqrt{(2\pi)^3}}=
\sum_n F_n(\vec r) \phi_n (\vec r)\simeq
\sum_{j=1}^r F_j (\vec r) \phi_j (\vec r),
\end{equation}
since in eq.~(\ref{coupled}) no coupling remains between
the states $j$ and the states $i$.
The numbers $D_{jj'}^{\alpha\beta}$ play the same role in the case
of degenerate bands as $\hbar^2 / ( 2\,m_{\alpha\beta}^*)$
for a non-degenerate band.

As before, in the presence of a magnetic field the components of
$-i\vec\nabla$ appearing in the envelope function equations will be
replaced with the corresponding components of $-i\vec\nabla+e\vec A/\hbar$.

In the presence of spin-orbit coupling, Luttinger and Kohn adopt the same
treatment, considering the spin-orbit contribution as
part of the unperturbed Hamiltonian (therefore the total unperturbed
Hamiltonian will be $H^{(0)}+H_{SO}$) and assuming the Bloch lattice functions 
and the corresponding energies for $\vec k=0$ of $H^{(0)}+H_{SO}$ as known
quantities. Thus the $u^n_0$ are replaced with the $\overline{u}^n_0$
(the spinorial Bloch lattice functions for $\vec k=0$ in\break
the presence of
spin-orbit interaction), $E_n (\vec k)$ by $\overline{E}_n (\vec k)$ (the
dispersion relations in the\break
\newpage
\noindent
presence of spin-orbit interaction) and the
$P_{\alpha}^{n\,n'}$ by 
\begin{equation}
(\pi_{\alpha})_{n\,n'}= \langle \overline{u}_0^n|
\left(-i\,\hbar\nabla_{\alpha}
+\frac{\hbar}{4\,m_ec^2}(\vec\sigma\times\vec\nabla V)_{\alpha}
\right)|\overline{u}_0^{n'} \rangle,
\end{equation}
where the extra term arises from the fact that the spin-orbit coupling 
contains the differential operator $\vec p$.
When we treat energy bands which are degenerate in the absence of spin-orbit
interaction, we have to remember that (as seen previously) the spin-orbit
coupling can lift, at least partially, the degeneracy. In such a case, we have
to consider that the validity of the adopted theory rests on
the assumption that the interband separations are large compared with the 
energies involved in the solution of the envelope function equation. Thus
we have to evaluate if the external potential $U$ or the magnetic field are
sufficiently small to produce no appreciable mixing of the bands, the 
degeneracy of which has been lifted by the spin-orbit coupling. If they
are sufficiently small, we can obtain a different set of coupled envelope 
function equations for each set of bands that have remained degenerate; 
otherwise we will have to deal with the full set of coupled equations for all
the bands that are degenerate in the absence of spin-orbit. 

We can introduce a matrix $D$, the elements of which are
\begin{equation}
D_{jj'}=\sum_{\alpha, \beta} D_{jj'}^{\alpha\beta}k_{\alpha}k_{\beta}.
\end{equation}
If in these matrix elements we replace each component of the vector
$\vec k$ with the corresponding component of the operator
$-i\,\vec\nabla+e \vec A/\hbar$, we obtain the terms which appear in the
envelope function coupled equations. 
In particular, the envelope function coupled equations written in
the absence of an external perturbation read (if we set the energy zero 
at $E_0^j$)
\begin{equation}
\sum_{j'=1}^r \sum_{\alpha, \beta}(D_{jj'}^{\alpha\beta}
(-i\nabla_{\alpha})(-i\nabla_{\beta})) 
F_{j'} (\vec r)=E F_j (\vec r).
\end{equation}
If we convert them from the position 
representation to the momentum representation, we obtain 
\begin{eqnarray}
\label{momentum}
&& \sum_{j'=1}^r \sum_{\alpha, \beta} (D_{jj'}^{\alpha\beta}
k_{\alpha}k_{\beta}) B_{j'} (\vec k) = E B_j (\vec k)\Rightarrow \\
&& \sum_{j'=1}^r D_{jj'}B_{j'} (\vec k) = E B_j (\vec k)\Rightarrow
D \vec B = E \vec B,\nonumber
\end{eqnarray}
from which it is evident that the dispersion relations $E(\vec k)$ near
the extremum can be obtained by finding the eigenvalues of the matrix $D$.
We notice that this clearly corresponds to what happens in the case 
of non-degeneracy, in which (as we have seen) the envelope function 
equation contains $E_n(-i\,\vec\nabla)$ (the dispersion
relation in the absence of external potential energy or magnetic field, 
where each component of $\vec k$ is replaced with the corresponding
component of $-i\,\vec\nabla$).

In order to determine the number of independent parameters which appear
in the matrix $D$, the symmetry properties of the considered lattices
are exploited.

In \cite{luttinger2} Luttinger proposes a different
way to obtain an explicit expression for $D$, based only on symmetry 
arguments. He writes this matrix for diamond-type semiconductors using
group theory, in particular considering that the Hamiltonian $D$ should
be invariant under the operations of the cubic group (so that the
Hamiltonian will give us results which transform correctly with respect
to the transformations of the cubic group, which is the symmetry group
of $\vec k$) and thus writing $D$ as a linear combination of the invariants
obtained combining angular momentum matrices and components of $\vec k$. 
The elements of such a matrix are polynomials in the components of
$\vec k$, at most of the second order, and involve
parameters characteristic of the materials, which have been experimentally 
found and are available for most common semiconductors \cite{lawaetz}.
For example, in the case of the $4\times 4$ matrix $D$ corresponding to the
light-hole and heavy-hole bands (the extra factor of $2$ coming from spin)
they are $\gamma_1$, $\gamma_2$, $\gamma_3$, $\kappa$ (which is useful 
in the presence of an external magnetic field) and $q$ (which approaches zero
as the spin-orbit coupling does).
%\hfill\break
%We notice that this method has been generalized, especially for 
%narrow band-gap semiconductors, to include also the conduction band in the 
%set of energy bands for which the matrix $D$ is computed.

Bir and Pikus \cite{bir} have shown that in uniformly strained semiconductors, 
such that the periodicity of the structure is preserved, the strain introduces
in the dispersion relation of non-degenerate bands an extra term of
the kind 
\begin{equation}
a_c(\epsilon_{xx}+\epsilon_{yy}+\epsilon_{zz})
\end{equation}
and in the Hamiltonian of degenerate bands additional terms of the form 
\begin{equation}
\sum_{\alpha ,\beta}\hat D_{j\,j'}^{\alpha\beta}
\epsilon_{\alpha\beta},
\end{equation}
where $\alpha ,\beta=x, y, z$ and $\epsilon_{\alpha\,\beta}$ is the generic
component of the strain matrix.
\begin{figure}
\centering
\includegraphics[width=.4\textwidth,angle=0]{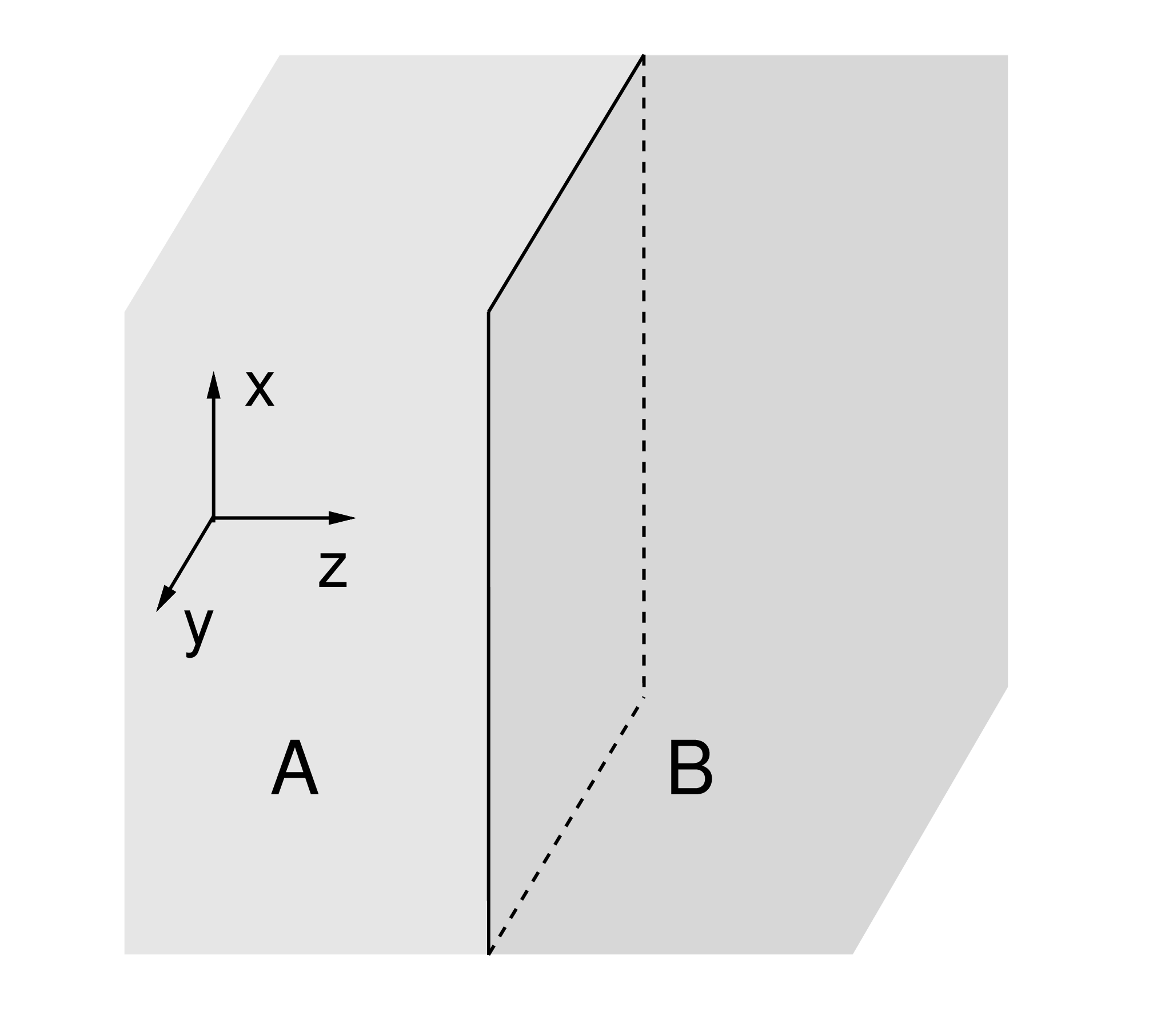}
\caption{Heterojunction between two semiconductors $A$ and $B$.} 
\label{f3}
\end{figure}\noindent

Bastard \cite{bastard1,bastard2,bastard3} uses the envelope function 
method to study heterostructures, for example made up of two materials $A$ 
and $B$ (fig.~\ref{f3}). 
In particular, he assumes that the two materials are perfectly lattice-matched 
and crystallize with the same crystallographic structure, so that the 
functions $u_0^n (\vec r)$ in the two materials can be considered identical. 
With this hypothesis, if in each material the wave functions are written as
\begin{equation}
\psi^{(A, B)}=\sum_n F_n^{(A, B)}(\vec r) u^n_0 (\vec r),
\end{equation}
it is evident that, since the $u^n_0$ are linearly independent and
the wave function has to be continuous at the interface, also the envelope
functions have to be continuous at the interface. For the derivative of
the envelope functions, Bastard finds, enforcing the continuity of the
probability current density at the interface, a general condition \cite{taylor},
which, in the simple case of two materials that are both characterized
by non-degenerate parabolic and isotropic bands but with different effective
masses $m^*_{(A)}$ and $m^*_{(B)}$, reduces to enforcing the continuity of
\begin{equation}
\frac{1}{m^*} \frac{\partial F_n}{\partial z}
\end{equation}
(where we have assumed the $\hbox{\boldmath{$\hat z$}}$ axis orthogonal to
the interface).
This can be easily obtained enforcing in this case the continuity of the
$z$ component of the probability current density, which is equal to
\begin{equation}
j_z=-\frac{i\,\hbar}{2\,m}\left(\psi^* \frac{\partial \psi}{\partial z}
- \psi \frac{\partial \psi^*}{\partial z}\right)
\end{equation}
and noting that the continuity of the envelope function has already been 
enforced. As to the asymptotic behavior of the envelope functions far
from the interface, it depends on the heterostructure under consideration.
For example, for superlattices the $z$-dependent part of the envelope 
function will be a Bloch wave, due to the periodicity of the structure in
that direction, while for the bound states of a quantum well it should
tend to zero for large $z$. Thus the envelope functions in the overall
structure can be found solving the envelope function equations in
the different materials, knowing the asymptotic behavior far from the
interface and enforcing the correct boundary conditions at the interface.
Bastard has also made an extensive analysis of the applications 
of this method \cite{bastard4}.

Also M. Altarelli has given important contributions to the 
development of the envelope function method \cite{altarelli4} and to
its applications to the study of heterostructures \cite{altarelli1,
altarelli2,altarelli3}.

M. G. Burt \cite{burt1,burt2,burt3,burt4,burt5} has pointed out the
errors deriving from the assumption, normally made in the application
of the envelope function method to heterostructures, that the
$u^n_0 (\vec r)$ in the two materials are the same and from the
boundary condition enforced on the derivative of the envelope function 
at the interface. In a series of interesting and detailed articles he
has developed an alternative envelope function theory
expanding the wave function in the overall structure on the same periodic
basis functions $U_n (\vec r)$ throughout, even though they are not 
necessarily eigenstates of the constituent crystals, without making 
any hypothesis about the real eigenstates $u^n_0 (\vec r)$
\begin{equation}
\psi (\vec r)=\sum_n F_n (\vec r) U_n (\vec r).
\end{equation}
The envelope functions $F_n (\vec r)$ univocally defined in this way
and all their derivatives are certainly continuous everywhere,
including at the interface. Using this approach, he has first derived exact 
envelope function equations, then, for local potentials and slowly varying
envelope functions (but without any assumption on the rate of variation 
of the composition), he has formulated approximate envelope function
equations, and finally, with the assumption of the dominance 
of one envelope function, he has arrived at an effective-mass equation that 
includes also the effect of the differences in the $u^n_0 (\vec r)$ between 
the two materials. At each step the associated approximations are accurately
described, so that it is possible to estimate the error.

A more detailed description of the applications of the $\vec k \cdot \vec p$
method to materials with a diamond, zincblende and wurtzite lattice, 
both in the periodic and in the non-periodic case, can be found (besides
in the other books and in the original publications reported in the list of
references of this review) in the recent book by L.~C.~Lew Yan Voon and
M.~Willatzen~\cite{voon}.

\section{Application of the $\vec k \cdot \vec p$ method to graphene}

In the last years the $\vec k \cdot \vec p$ method, and in particular the
formulation (described in the last section) based on the envelope functions,
has been successfully applied to the analysis of the electronic
properties of graphene and graphene-related stuctures, such as carbon
nanotubes and graphene nanoribbons.

In this section we will begin the description of this particular application 
deriving the $\vec k \cdot \vec p$ relations for a simple sheet of graphene.
\begin{figure}
\centering
\includegraphics[width=.65\textwidth,angle=0]{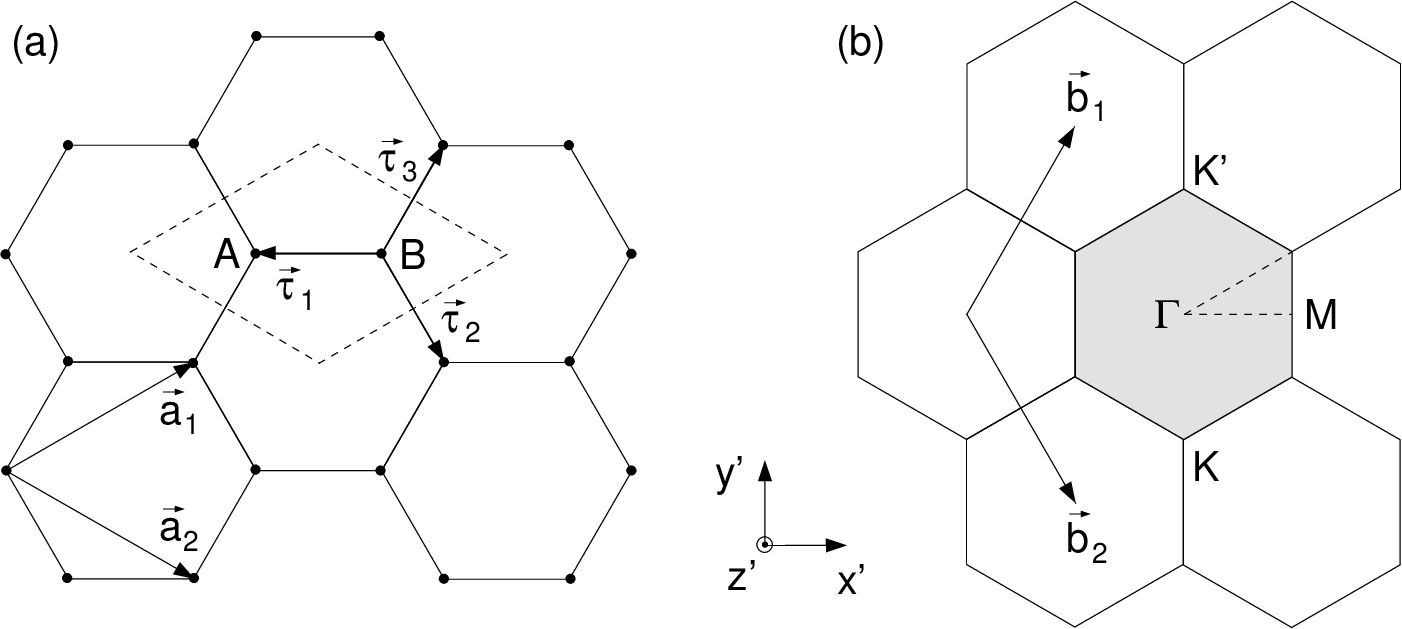}
\caption{The graphene lattice in the real space (a) and in the reciprocal 
space (b).}
\label{f4}
\end{figure}\noindent

\begin{figure}
\centering
\includegraphics[width=.55\textwidth,angle=0]{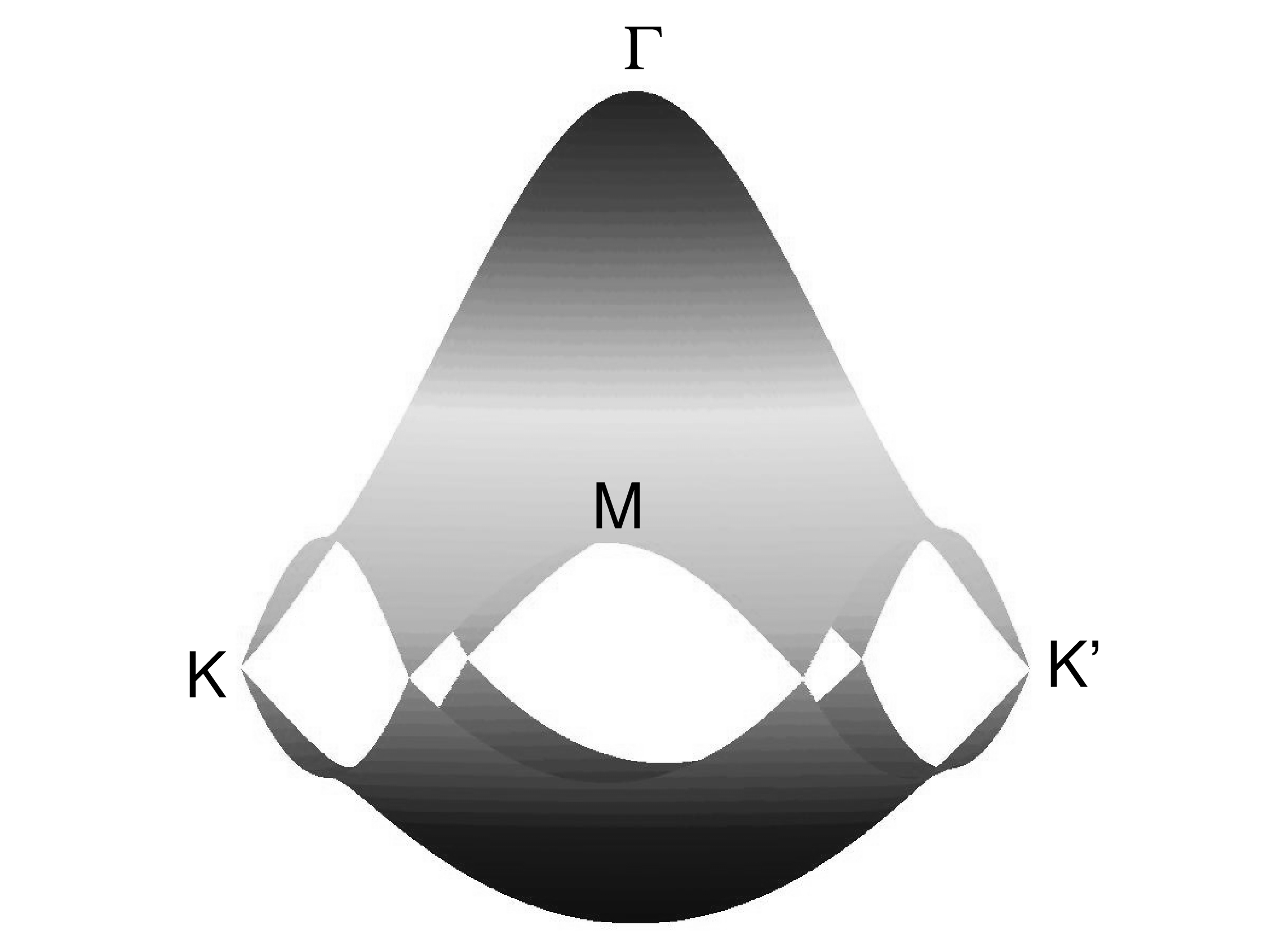}
\caption{The energy dispersion relations of graphene inside its 
hexagonal Brillouin zone.}
\label{f5}
\end{figure}

A graphene sheet is a hexagonal lattice of carbon atoms. In fig.~\ref{f4}(a) 
we show its structure in the real space and, in particular, its unit cell
as a dashed rhombus, containing two inequivalent carbon atoms $A$ and $B$,
while in fig.~\ref{f4}(b) we show the lattice in the reciprocal space with
the Brillouin zone as a shaded hexagon. The lattice unit vectors are 
$\vec a_1$ and $\vec a_2$ in the real space, and $\vec b_1$ and $\vec b_2$ 
in the reciprocal space. If we define 
$a=|\vec a_1|=|\vec a_2|=a_{C-C}\,\sqrt{3}$
(with $a_{C-C}$ the distance between nearest-neighbor carbon atoms),
the coordinates of these vectors in the right-hand reference frame
$\Sigma'=(\hbox{\boldmath{$\hat x$}}',\hbox{\boldmath{$\hat y$}}',
\hbox{\boldmath{$\hat z$}}')$ are (observe that we have taken
$\hbox{\boldmath{$\hat x$}}'$ along the vector $\vec a_1+\vec a_2$)
\begin{equation}
\vec a_1 \mathrel{\mathop\equiv_{\Sigma'}} \left[\begin{array}{c}
\displaystyle \frac{\sqrt{3}}{2}a\\
\noalign{\vskip3pt}
\displaystyle \frac{a}{2}\\
\noalign{\vskip3pt}
0
\end{array}\right]
,\quad
\vec a_2 \mathrel{\mathop\equiv_{\Sigma'}} \left[\begin{array}{c}
\displaystyle \frac{\sqrt{3}}{2}a\\
\noalign{\vskip3pt}
\displaystyle -\frac{a}{2}\\
\noalign{\vskip3pt}
0
\end{array}\right]
,\quad
\vec b_1 \mathrel{\mathop\equiv_{\Sigma'}} \left[\begin{array}{c}
\displaystyle \frac{2\pi}{\sqrt{3}a}\\ 
\noalign{\vskip3pt}
\displaystyle \frac{2\pi}{a}\\
\noalign{\vskip3pt}
0
\end{array}\right]
,\quad
\vec b_2 \mathrel{\mathop\equiv_{\Sigma'}} \left[\begin{array}{c}
\displaystyle \frac{2\pi}{\sqrt{3}a}\\
\noalign{\vskip3pt}
\displaystyle -\frac{2\pi}{a}\\
\noalign{\vskip3pt}
0
\end{array}\right]
\end{equation}
(following the conventions used by R.~Saito, G.~Dresselhaus and 
M.~S.~Dresselhaus \cite{saito}), which (being 
$\vec b_1=2\pi (\vec a_2 \times \hbox{\boldmath{$\hat z$}}')/
(\vec a_1 \cdot (\vec a_2 \times \hbox{\boldmath{$\hat z$}}'))$ and
$\vec b_2=2\pi (\hbox{\boldmath{$\hat z$}}' \times \vec a_1)/
(\vec a_1 \cdot (\vec a_2 \times \hbox{\boldmath{$\hat z$}}'))$)
fulfill the well-know relation 
$\vec a_i \cdot \vec b_j=2 \pi \delta_{ij}$
between lattice unit vectors in the real space and in the reciprocal space.
Note that the letter written under the symbol ``$\equiv$'' indicates the
adopted reference frame.
The most relevant graphene dispersion relations for transport and other
solid-state properties are the two $\pi$-bands (an upper anti-bonding band 
and a lower bonding band), which are degenerate at the points
(considering the point $\Gamma$ at the center of the hexagonal Brillouin zone 
of graphene as the origin of the reciprocal space)
\begin{equation}
\label{diracpoints}
\vec K=\frac{1}{3}(\vec b_2-\vec b_1) \mathrel{\mathop\equiv_{\Sigma'}}
\frac{4\pi}{3a}
\left[\begin{array}{c}
0\\
-1\\ 
0
\end{array}\right]
\quad\hbox{and}\quad
\vec K'=\frac{1}{3}(\vec b_1-\vec b_2) \mathrel{\mathop\equiv_{\Sigma'}}
\frac{4\pi}{3a}
\left[\begin{array}{c}
0\\
1\\
0
\end{array}\right]
\end{equation}
and obviously at their equivalents in the reciprocal space
(as we can see from fig.~\ref{f5}, which has been obtained with a 
nearest-neighbor tight-binding approach limited to the $2 p^z$ atomic 
orbitals, with nonzero nearest-neighbor overlap integral).

Thus we can use the $\vec k \cdot \vec p$ method to find the dispersion 
relations of graphene near these extrema points (called Dirac points),
following T.~Ando's approach \cite{ajiki,ando1,ando2}.
However, in our description we will continue to use the conventions of
ref.~\cite{saito} and we will consider the pair (\ref{diracpoints}) of
Dirac points (which will simplify the treatment of zigzag and armchair
graphene nanoribbons in the last section of this review).
Other articles where a $\vec k \cdot \vec p$ treatment of graphene is
introduced are refs.~\cite{wallace,mcclure,slonczewski,divincenzo,
semenoff,kanemele}.
\noindent

We start by using a simple tight-binding model, in which we use as basis
functions the $2 p^z$ orbitals of all the carbon atoms of the graphene
sheet, which are the orbitals leading to the $\pi$-bonds and
thus to the above-mentioned two $\pi$-bands.
The generic eigenfunction in the material can be expressed \cite{ando1,ando2}
as a linear combination (with coefficients $\psi_A (\vec R_A)$ and
$\psi_B (\vec R_B)$) of these atomic orbitals $\varphi(\vec r -\vec R_A)$ and
$\varphi(\vec r -\vec R_B)$ (centered on atoms of type $A$ and $B$,
respectively)
\begin{equation}
\label{wavefunction}
\psi (\vec r)=
\sum_{\vec R_A}\psi_A (\vec R_A)\varphi(\vec r -\vec R_A)+
\sum_{\vec R_B}\psi_B (\vec R_B)\varphi(\vec r -\vec R_B),
\end{equation}
where the first (second) sum spans over all the positions of the atoms of
type $A$ ($B$) in the lattice.

Using the definition of the Hamiltonian operator
\begin{equation}
\label{hamiltonian}
H |\psi \rangle =E |\psi \rangle,
\end{equation}
we have that \cite{saito}
\begin{equation}
\langle \psi| H |\psi \rangle =E  \langle \psi |\psi \rangle
\end{equation}
and thus (using $j$ and $j'$ to indicate the type of the atoms and $n$ and
$m$ to specify the particular atoms)
\begin{eqnarray}
\label{defe}
\qquad E &=& \frac{\langle \psi| H |\psi \rangle}
{\langle \psi |\psi \rangle }=\\
&& \frac{\displaystyle
\left<\sum_{j=A, B}\sum_{\vec R_{jn}}\psi_j (\vec R_{jn})
\varphi(\vec r -\vec R_{jn}) \Bigg| 
H \Bigg|\sum_{j'=A, B}\sum_{\vec R_{j'm}}
\psi_{j'} (\vec R_{j'm})\varphi(\vec r -\vec R_{j'm})\right>}
{\displaystyle
\left<\sum_{j=A, B}\sum_{\vec R_{jn}}\psi_j (\vec R_{jn})
\varphi(\vec r -\vec R_{jn}) \Bigg|
\sum_{j'=A, B}\sum_{\vec R_{j'm}}
\psi_{j'} (\vec R_{j'm})\varphi(\vec r -\vec R_{j'm})\right>}=\nonumber\\
&& \frac{\displaystyle
\sum_{j,j'=A, B}\sum_{\vec R_{jn}}\sum_{\vec R_{j'm}}
\psi_j^* (\vec R_{jn})\psi_{j'} (\vec R_{j'm})
\langle \varphi(\vec r -\vec R_{jn})| H |\varphi(\vec r -\vec R_{j'm}) \rangle}
{\displaystyle
\sum_{j,j'=A, B}\sum_{\vec R_{jn}}\sum_{\vec R_{j'm}}
\psi_j^* (\vec R_{jn})\psi_{j'} (\vec R_{j'm})
 \langle \varphi(\vec r -\vec R_{jn})|\varphi(\vec r -\vec R_{j'm}) \rangle }=\nonumber\\
&& \frac{\displaystyle
\sum_{j,j'=A, B}\sum_{\vec R_{jn}}\sum_{\vec R_{j'm}}
\psi_j^* (\vec R_{jn})\psi_{j'} (\vec R_{j'm}) h_{\vec R_{jn},\vec R_{j'm}}}
{\displaystyle
\sum_{j,j'=A, B}\sum_{\vec R_{jn}}\sum_{\vec R_{j'm}}
\psi_j^* (\vec R_{jn})\psi_{j'} (\vec R_{j'm})
s_{\vec R_{jn},\vec R_{j'm}}},\nonumber
\end{eqnarray}
where we have introduced the transfer integrals $h_{\vec R_{jn},\vec R_{j'm}}$
and the overlap integrals $s_{\vec R_{jn},\vec R_{j'm}}$ between atomic
orbitals. Now we can minimize $E$ (to obtain the actual physical state)
enforcing (for each coefficient, and thus for each atom)
\begin{eqnarray}
\frac{\partial E}{\partial \psi_j^* (\vec R_{jn})} &=&
\frac{\displaystyle
\sum_{j'=A, B}\sum_{\vec R_{j'm}} \psi_{j'} (\vec R_{j'm}) 
h_{\vec R_{jn},\vec R_{j'm}}}
{\displaystyle
\sum_{j,j'=A, B}\sum_{\vec R_{jn}}\sum_{\vec R_{j'm}}
\psi_j^* (\vec R_{jn})\psi_{j'} (\vec R_{j'm}) s_{\vec R_{jn},\vec R_{j'm}}}\\
&&{}-\frac{\displaystyle
\sum_{j,j'=A, B}\sum_{\vec R_{jn}}\sum_{\vec R_{j'm}}
\psi_j^* (\vec R_{jn})\psi_{j'} (\vec R_{j'm}) h_{\vec R_{jn},\vec R_{j'm}}}
{\displaystyle
\Big( \sum_{j,j'=A, B}\sum_{\vec R_{jn}}\sum_{\vec R_{j'm}}
\psi_j^* (\vec R_{jn})\psi_{j'} (\vec R_{j'm})
s_{\vec R_{jn},\vec R_{j'm}}\Big)^2}\nonumber\\
&&{}\cdot\sum_{j'=A, B}\sum_{\vec R_{j'm}}
\psi_{j'} (\vec R_{j'm}) s_{\vec R_{jn},\vec R_{j'm}}=0.\nonumber
\end{eqnarray}
Multiplying both members by the denominator of eq.~(\ref{defe}) and
rearranging, we find:
\begin{eqnarray}
&& \sum_{j'=A, B}\sum_{\vec R_{j'm}} \psi_{j'} (\vec R_{j'm}) 
h_{\vec R_{jn},\vec R_{j'm}}=\\
&& \frac{\displaystyle
\sum_{j,j'=A, B}\sum_{\vec R_{jn}}\sum_{\vec R_{j'm}}
\psi_j^* (\vec R_{jn})\psi_{j'} (\vec R_{j'm}) h_{\vec R_{jn},\vec R_{j'm}}}
{\displaystyle
\sum_{j,j'=A, B}\sum_{\vec R_{jn}}\sum_{\vec R_{j'm}}
\psi_j^* (\vec R_{jn})\psi_{j'} (\vec R_{j'm}) s_{\vec R_{jn},\vec R_{j'm}}}
\sum_{j'=A, B}\sum_{\vec R_{j'm}}
\psi_{j'} (\vec R_{j'm}) s_{\vec R_{jn},\vec R_{j'm}}\nonumber
\end{eqnarray}
and recognizing that the fraction in the right-hand side is the expression
of $E$, we have
\begin{equation}
\sum_{j'=A, B}\sum_{\vec R_{j'm}} \psi_{j'} (\vec R_{j'm}) 
h_{\vec R_{jn},\vec R_{j'm}}
=E \sum_{j'=A, B}\sum_{\vec R_{j'm}}
\psi_{j'} (\vec R_{j'm}) s_{\vec R_{jn},\vec R_{j'm}}.
\end{equation}
Let us expand this result for the coefficients (and thus for the atoms)
with $j=A$ and for those with $j=B$
\begin{equation}
\left\{ \begin{array}{l}
\displaystyle
\sum_{\vec R_{Am}} \psi_A (\vec R_{Am}) h_{\vec R_{An},\vec R_{Am}}+
\sum_{\vec R_{Bm}} \psi_B (\vec R_{Bm}) h_{\vec R_{An},\vec R_{Bm}}=\\
\displaystyle
\qquad
E \left(\sum_{\vec R_{Am}}\psi_A (\vec R_{Am}) s_{\vec R_{An},\vec R_{Am}}
\!+\!\sum_{\vec R_{Bm}}\psi_B (\vec R_{Bm})
s_{\vec R_{An},\vec R_{Bm}}\right);\\
\displaystyle
\sum_{\vec R_{Am}} \psi_A (\vec R_{Am}) h_{\vec R_{Bn},\vec R_{Am}}+
\sum_{\vec R_{Bm}} \psi_B (\vec R_{Bm}) h_{\vec R_{Bn},\vec R_{Bm}}=\\
\displaystyle
\qquad
E \left(\sum_{\vec R_{Am}}\psi_A (\vec R_{Am}) s_{\vec R_{Bn},\vec R_{Am}}
\!+\!\sum_{\vec R_{Bm}}\psi_B (\vec R_{Bm}) s_{\vec R_{Bn},\vec R_{Bm}}\right).
\end{array}\right.
\end{equation}
We consider non-negligible only the integrals between each atom and
itself and between each atom and its nearest neighbors (which are the
nearest three $B$ atoms for an $A$ atom, while they are the nearest
three $A$ atoms for a $B$ atom).
Therefore, if (in order to simplify the notation) we rename $\vec R_{An}$ as
$\vec R_A$ and $\vec R_{Bn}$ as $\vec R_B$ and we use the index $l$ to
indicate the nearest three atoms, we can rewrite these equations in the
following way:
\begin{equation}
\left\{ \begin{array}{l}
\displaystyle
\psi_A (\vec R_A) h_{\vec R_A,\vec R_A}+
\sum_{l=1}^3 \psi_B (\vec R_{B_l}) h_{\vec R_A,\vec R_{B_l}}=\\
\displaystyle
\qquad\qquad\qquad
E \left(\psi_A (\vec R_A) s_{\vec R_A,\vec R_A}
+\sum_{l=1}^3\psi_B (\vec R_{B_l}) s_{\vec R_A,\vec R_{B_l}}\right),\\
\displaystyle
\sum_{l=1}^3 \psi_A (\vec R_{A_l}) h_{\vec R_B,\vec R_{A_l}}+
\psi_B (\vec R_B) h_{\vec R_B,\vec R_B}=\\
\displaystyle
\qquad\qquad\qquad
E \left(\sum_{l=1}^3 \psi_A (\vec R_{A_l}) s_{\vec R_B,\vec R_{A_l}}
+\psi_B (\vec R_B) s_{\vec R_B,\vec R_B}\right).
\end{array}\right.
\end{equation}
In particular, we consider
\begin{eqnarray}
\label{integr}
\quad h_{\vec R_{jn},\vec R_{j'm}}&=&\left\{ \begin{array}{ll}
\displaystyle
\epsilon_{\vec R_{jn}}=u(\vec R_{jn}) &
\textrm{if $\vec R_{jn}=\vec R_{j'm}$,}\\
\displaystyle
-\gamma_0 &\textrm{if $\vec R_{jn}\ne \vec R_{j'm}$ and
$\vec R_{jn}$ and $\vec R_{j'm}$ are}\\
&\qquad\qquad\qquad\qquad\quad\textrm{nearest neighbors,}\\
\displaystyle
0 &\textrm{otherwise,}
\end{array}\right.\\
s_{\vec R_{jn},\vec R_{j'm}}&=&\left\{ \begin{array}{ll}
\displaystyle
1 &\textrm{if $\vec R_{jn}=\vec R_{j'm}$,}\\
\displaystyle
0 &\textrm{if $\vec R_{jn}\ne \vec R_{j'm}$.}
\end{array}\right.\nonumber
\end{eqnarray}
Here $\gamma_0$ is the modulus of the nearest-neighbor transfer integral.
Instead $\epsilon_{\vec R_{jn}}$ is the onsite energy, that we take as zero of
the energy in the absence of an external ({\em i.e.} not due to the periodic
structure of the lattice) potential energy; if the external
potential energy is not zero, we have to consider the term $u(\vec R_{jn})$,
which represents the value of this external potential energy in the position
$\vec R_{jn}$.

Note that the reason for the values of the overlap integrals reported in
eq.~(\ref{integr}) is that we consider atomic orbitals orthonormalized
using the L\"owdin procedure~\cite{lowdin3,slaterkoster,datta2}.

Thus the tight-binding relations become
\begin{equation}
\left\{ \begin{array}{l}
\displaystyle
-\gamma_0 \sum_{l=1}^3 \psi_B (\vec R_{B_l})=
(E-u(\vec R_A))\,\psi_A (\vec R_A),\\
\displaystyle
-\gamma_0 \sum_{l=1}^3 \psi_A (\vec R_{A_l})=
(E-u(\vec R_B))\,\psi_B (\vec R_B).
\end{array}\right.
\end{equation}
If we introduce the vectors (fig.~\ref{f4}(a))
\begin{equation}
\vec\tau_1 \mathrel{\mathop\equiv_{\Sigma'}}
\frac{a}{\sqrt{3}}\left[\begin{array}{c}
-1\\
\noalign{\vskip3pt}
0\\
\noalign{\vskip3pt}
0
\end{array}\right]
,\quad
\vec\tau_2 \mathrel{\mathop\equiv_{\Sigma'}}
\frac{a}{\sqrt{3}}\left[\begin{array}{c}
\displaystyle \frac{1}{2}\\
\noalign{\vskip3pt} 
\displaystyle -\frac{\sqrt{3}}{2}\\
\noalign{\vskip3pt} 
0
\end{array}\right]
,\quad
\vec\tau_3 \mathrel{\mathop\equiv_{\Sigma'}}
\frac{a}{\sqrt{3}}\left[\begin{array}{c}
\displaystyle \frac{1}{2}\\
\noalign{\vskip3pt}
\displaystyle \frac{\sqrt{3}}{2}\\
\noalign{\vskip3pt}
0
\end{array}\right]
\end{equation}
(with respect to the frame $\Sigma'=(\hbox{\boldmath{$\hat x$}}',
\hbox{\boldmath{$\hat y$}}',\hbox{\boldmath{$\hat z$}}')$),
we can write the positions of the nearest-neighbor atoms in this way:
\begin{eqnarray}
\vec R_{B_1} &= \vec R_A-\vec \tau_1,\\
\vec R_{B_2} &= \vec R_A-\vec \tau_2,\nonumber\\
\vec R_{B_3} &= \vec R_A-\vec \tau_3,\nonumber\\
\vec R_{A_1} &= \vec R_B+\vec \tau_1,\nonumber\\
\vec R_{A_2} &= \vec R_B+\vec \tau_2,\nonumber\\
\vec R_{A_3} &= \vec R_B+\vec \tau_3,\nonumber
\end{eqnarray}
and thus we can rewrite the tight-binding relations in the following form:
\begin{equation}
\label{tightbinding}
\left\{ \begin{array}{l}
\displaystyle
-\gamma_0 \sum_{l=1}^3 \psi_B (\vec R_A-\vec \tau_l)=
(E-u(\vec R_A))\,\psi_A (\vec R_A),\\
\displaystyle
-\gamma_0 \sum_{l=1}^3 \psi_A (\vec R_B+\vec \tau_l)=
(E-u(\vec R_B))\,\psi_B (\vec R_B).
\end{array} \right.
\end{equation}
%%%%%%%%%%%%%%%%%%%%%%%%%%%%%%%%%%%%%%%%%%%%%%%%%%%%%%%%%%%%%%%%%%%%%%%%%

Now let us consider what happens near the points $\vec K$ and $\vec K'$.

Let us assume that we can write
\begin{equation}
\label{assumptions}
\left\{ \begin{array}{l}
\displaystyle
\psi_A (\vec R_A)=e^{i \vec K\cdot \vec R_A} F_A^{\vec K}(\vec R_A)
-i \, e^{i \theta'} e^{i \vec K'\cdot \vec R_A} F_A^{\vec K'}(\vec R_A),\\
\\
\displaystyle
\psi_B (\vec R_B)=i \, e^{i \theta'} 
e^{i \vec K\cdot \vec R_B} F_B^{\vec K} (\vec R_B)+
e^{i \vec K'\cdot \vec R_B} F_B^{\vec K'} (\vec R_B)
\end{array} \right.
\end{equation}
(the angle $\theta'$ will be properly chosen later).
If $\vec k$ is the wave vector of $\psi_A$ and $\psi_B$, the functions
$F_A^{\vec K}$ and $F_B^{\vec K}$ have a wave vector $\vec\kappa=\vec k-\vec K$
and thus are slowly-varying functions (with small $\vec\kappa$) near the
point $\vec K$; analogously the functions $F_A^{\vec K'}$ and $F_B^{\vec K'}$
have a wave vector $\vec\kappa=\vec k-\vec K'$ and thus are slowly-varying
functions (with small $\vec\kappa$) near the point $\vec K'$ (note that
in the overall review we use $\vec k$ for the total wave vector and
$\vec\kappa$ for the wave vector measured from the reference extremum point).

Incidentally, with these assumptions, if we define $\alpha_A^{\vec K}=1$, 
$\alpha_A^{\vec K'}=-i \, e^{i \theta'}$,
$\alpha_B^{\vec K}=i \, e^{i \theta'}$,
and $\alpha_B^{\vec K'}=1$, we have that
\begin{eqnarray}
\psi (\vec r) &=&
\sum_{i=A,B}\sum_{\vec R_i} 
\psi_i (\vec R_i)\varphi(\vec r -\vec R_i)=\\
&& \sum_{i=A,B}\sum_{\vec R_i}\sum_{\vec K_j=\vec K,\vec K'}
\alpha_i^{\vec K_j}\,e^{i \vec K_j\cdot \vec R_i} 
F_i^{\vec K_j}(\vec R_i) \varphi(\vec r -\vec R_i)\simeq\nonumber\\
&& \sum_{i=A,B}\sum_{\vec R_i}\sum_{\vec K_j=\vec K,\vec K'}
\alpha_i^{\vec K_j}\,e^{i \vec K_j\cdot \vec R_i}
F_i^{\vec K_j}(\vec r) \varphi(\vec r -\vec R_i)=\nonumber\\
&& \sum_{i=A,B}\sum_{\vec K_j=\vec K,\vec K'}
F_i^{\vec K_j}(\vec r)\, e^{i \vec K_j\cdot \vec r}
\left[\alpha_i^{\vec K_j}\sum_{\vec R_i} \varphi(\vec r -\vec R_i)\,
e^{-i \vec K_j\cdot (\vec r-\vec R_i)} \right]=\nonumber\\
&& \sum_{i=A,B}\sum_{\vec K_j=\vec K,\vec K'}
F_i^{\vec K_j}(\vec r)\, e^{i \vec K_j\cdot \vec r}\,
\tilde u^i_{\vec K_j}(\vec r),\nonumber
\end{eqnarray}
where we have substituted $F_i^{\vec K_j}(\vec r)$ to
$F_i^{\vec K_j}(\vec R_i)$ using the fact that $F_i^{\vec K_j}$ is a
slowly varying function of $\vec r$ near $\vec K_j$, while the atomic
orbital $\varphi$ has significant values only near the corresponding atom.
The quantity between square brackets (that we have called here 
$\tilde u^i_{\vec K_j}$) is periodic with the periodicity of the lattice,
since, if $\vec a_\ell$ is a lattice unit vector (and thus, if $\vec R_i$ is
the position of a lattice point, also $\vec R_i^0=\vec R_i-\vec a_\ell$ is the
position of\break
a lattice point), then
\begin{eqnarray}
\tilde u^i_{\vec K_j} (\vec r+\vec a_\ell) &=&
\alpha_i^{\vec K_j}\sum_{\vec R_i} \varphi((\vec r+\vec a_\ell)-\vec R_i)\,
e^{-i \vec K_j\cdot ((\vec r+\vec a_\ell)-\vec R_i)}=\\
&& \alpha_i^{\vec K_j}\sum_{\vec R_i} \varphi(\vec r-(\vec R_i-\vec a_\ell))\,
e^{-i \vec K_j\cdot (\vec r-(\vec R_i-\vec a_\ell))}=\nonumber\\
&& \alpha_i^{\vec K_j}\sum_{\vec R_i^0} \varphi(\vec r-\vec R_i^0)\,
e^{-i \vec K_j\cdot (\vec r-\vec R_i^0)}=
\tilde u^i_{\vec K_j} (\vec r).\nonumber
\end{eqnarray}
Therefore, since $\tilde u^i_{\vec K_j}$ has the lattice periodicity
and $\vec K_j$ is an extremum point (different from $0$) of
the dispersion relations, from the relation between $\psi (\vec r)$, 
$\tilde u^i_{\vec K_j}(\vec r)$ and $F_i^{\vec K_j}(\vec r)$ we conclude 
that the 4 functions $F_i^{\vec K_j}$ can be seen as the electron
envelope functions corresponding to the 2 extremum points $\vec K_j$ where
the 2 considered bands of graphene are degenerate (see eq.~(\ref{k0}),
the related discussion, and eq.~(\ref{deg})).

Let us point out that this whole procedure does not need a particular choice
of scalar product and of normalization: these have just to be chosen coherently
with each other.
However, one could find desirable to normalize the periodic function
$\tilde u^i_{\vec K_j}$ according to the scalar product defined in 
(\ref{scalarproduct}), as is generally done in the envelope function theory.
Following this particular criterion, one should have (if $\Omega_0$ is the
area of a graphene unit cell, while $\Omega$ is the area of the overall
graphene sheet)
\begin{eqnarray}
\label{prenorm}
\quad 1 &=& \langle\tilde u^i_{\vec K_j}(\vec r)|\tilde u^i_{\vec K_j}(\vec r)\rangle=
\frac{1}{\Omega_0}
\int_{\Omega_0} |\tilde u^i_{\vec K_j}(\vec r)|^2 d \vec r=
\frac{1}{\Omega}
\int_{\Omega} |\tilde u^i_{\vec K_j}(\vec r)|^2 d \vec r=\\
&& \frac{1}{\Omega} \int_{\Omega}
\Big|\alpha_i^{\vec K_j}\sum_{\vec R_i} \varphi(\vec r -\vec R_i)\,
e^{-i \vec K_j\cdot (\vec r-\vec R_i)}\Big|^2 d \vec r=\nonumber\\
&& \frac{1}{\Omega}
\int_{\Omega}\Big(\sum_{\vec R_i} \varphi(\vec r -\vec R_i)\,
e^{-i \vec K_j\cdot (\vec r-\vec R_i)}\Big)^*
\Big(\sum_{\vec R_i'} \varphi(\vec r -\vec R_i')\,
e^{-i \vec K_j\cdot (\vec r-\vec R_i')}\Big) d \vec r=\nonumber\\
&& \frac{1}{\Omega}
\int_{\Omega} \sum_{\vec R_i} |\varphi(\vec r -\vec R_i)|^2 d \vec r\nonumber\\
&&\quad{}+\frac{1}{\Omega}\sum_{\scriptstyle \vec R_i,\vec R_i' \atop 
\scriptstyle \vec R_i \ne \vec R_i'}
\left[\int_{\Omega} \varphi^*(\vec r -\vec R_i)\varphi(\vec r -\vec R_i')
d \vec r \right] e^{i \vec K_j\cdot (\vec R_i'-\vec R_i)}=\nonumber\\
&& \frac{1}{\Omega}
\int_{\Omega} \sum_{\vec R_i} |\varphi(\vec r -\vec R_i)|^2 d \vec r=
\frac{1}{\Omega} \sum_{\vec R_i}
\int_{\Omega} |\varphi(\vec r -\vec R_i)|^2 d \vec r \simeq\nonumber\\
&& \frac{1}{\Omega} \frac{\Omega}{\Omega_0}
\int_{\Omega} |\varphi(\vec r -\vec R_i)|^2 d \vec r=
\frac{1}{\Omega_0}
\int_{\Omega} |\varphi(\vec r -\vec R_i)|^2 d \vec r.\nonumber
\end{eqnarray}
Here we have exploited the following properties of the involved
functions. First of all, integrating a function with the lattice
periodicity over the whole graphene sheet and dividing the result by its
area is equivalent to integrating it over
the lattice unit cell and dividing by the corresponding area.
Moreover, each atomic orbital $\varphi$ (orthonormalized using the L\"owdin
procedure) has a non-zero overlap only with itself. Finally, since each\break
\newpage
\noindent
atomic orbital has significative values only near the corresponding atom,
the integral of the square modulus over the whole graphene sheet is nearly
the same for all the considered atomic orbitals, and thus the sum
of all the integrals is approximately equal to a single integral 
multiplied by the number $\Omega/\Omega_0$ of orbitals.

Therefore, adopting this particular normalization for $\tilde u^i_{\vec K_j}$,
the atomic orbital $\varphi$ should be normalized in such a way that
\begin{equation}
\label{norm}
\frac{1}{\Omega_0} 
\int_{\Omega}|\varphi(\vec r -\vec R_i)|^2 d \vec r =1 \Rightarrow
\int_{\Omega}|\varphi(\vec r -\vec R_i)|^2 d \vec r = \Omega_0,
\end{equation}
and thus we should consider atomic orbitals $\sqrt{\Omega_0}$ times greater
than those deriving from the usual normalization over the whole graphene sheet.

The corresponding scalar product
\begin{equation}
\label{scalarproduct1}
\langle \varphi_1 |\varphi_2 \rangle=
\frac{1}{\Omega_0} \int_{\Omega} \varphi_1^* (\vec r) \varphi_2 (\vec r)\,
d \vec r
\end{equation}
should be used in all the calculations involving atomic orbitals.

\begin{figure}
\centering
\includegraphics[width=.8\textwidth,angle=0]{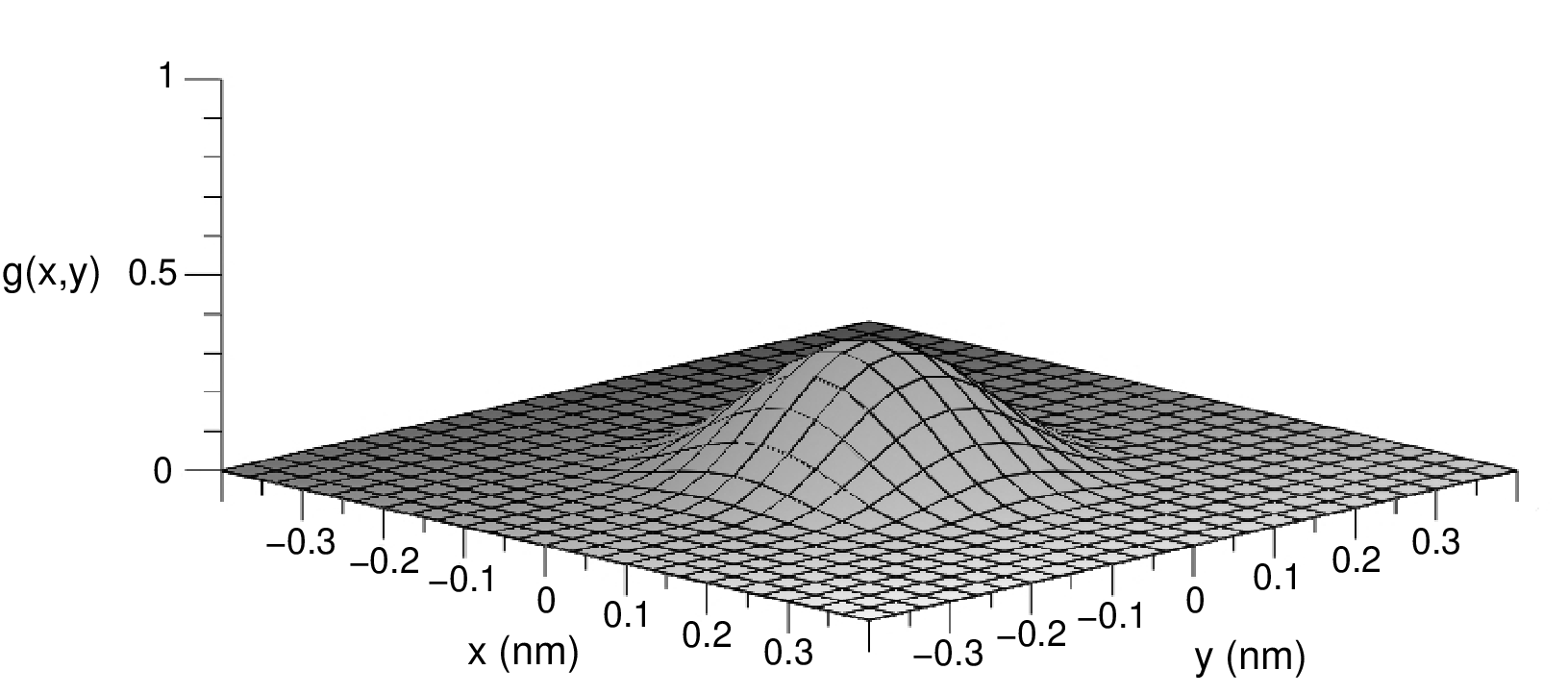}
\caption{A candidate smoothing function $g(\vec r)$.}
\label{f6}
\end{figure}

If we introduce the assumptions (\ref{assumptions}) into the tight-binding
equations (\ref{tightbinding}), we obtain
\begin{equation}
\label{tightbinding1}
\left\{ \begin{array}{l}
\displaystyle
(E-u(\vec R_A))\,\left[e^{i \vec K\cdot \vec R_A} F_A^{\vec K}(\vec R_A)
-i \, e^{i \theta'} e^{i \vec K'\cdot \vec R_A} F_A^{\vec K'}(\vec R_A)
\right]=\\
\displaystyle
-\gamma_0 \sum_{l=1}^3 \left[i \, e^{i \theta'} 
e^{i \vec K\cdot (\vec R_A-\vec \tau_l)} F_B^{\vec K} (\vec R_A-\vec \tau_l)+
e^{i \vec K'\cdot (\vec R_A-\vec \tau_l)} F_B^{\vec K'} (\vec R_A-\vec \tau_l)
\right];\\
\displaystyle
(E-u(\vec R_B))\,\left[i \, e^{i \theta'} 
e^{i \vec K\cdot \vec R_B} F_B^{\vec K} (\vec R_B)+
e^{i \vec K'\cdot \vec R_B} F_B^{\vec K'} (\vec R_B)\right]=\\
\displaystyle
-\gamma_0 \sum_{l=1}^3 \left[
e^{i \vec K\cdot (\vec R_B+\vec \tau_l)} F_A^{\vec K}(\vec R_B+\vec \tau_l)
-i \, e^{i \theta'} e^{i \vec K'\cdot (\vec R_B+\vec \tau_l)}
F_A^{\vec K'}(\vec R_B+\vec \tau_l)\right].
\end{array} \right.
\end{equation}
It is useful to introduce \cite{ando1,ando2} a smoothing function $g(\vec r)$,
{\em i.e.} a real function which varies smoothly around the point around which it is
centered, has non-negligible values only in a range of a few lattice
constants around the center, and then decays rapidly for larger distances.
This function (point-symmetric around its center) is chosen in such a way as to
satisfy the conditions
\begin{equation}
\label{sum}
\sum_{\vec R_A} g(\vec r-\vec R_A)=\sum_{\vec R_B} g(\vec r-\vec R_B)=1
\end{equation}
and
\begin{equation}
\int_{\Omega} d\vec r \, g(\vec r-\vec R_A)=
\int_{\Omega} d\vec r \, g(\vec r-\vec R_B)=\Omega_0
\end{equation}
(where $\Omega_0=\sqrt{3} a^2 /2$ is the area of a graphene unit cell,
while $\Omega$ is the area of the overall graphene sheet); moreover it has
to satisfy the relations
\begin{equation}
\label{phase}
\sum_{\vec R_A} g(\vec r-\vec R_A) e^{i (\vec K'-\vec K)\cdot \vec R_A}
=\sum_{\vec R_B} g(\vec r-\vec R_B) e^{i (\vec K'-\vec K)\cdot \vec R_B}
\simeq 0.
\end{equation}
Due to its locality, when this function is multiplied by a generic smooth
function $f(\vec r)$ (such as the envelope functions $F$ we have defined),
we clearly have that 
\begin{equation}
\label{smooth}
f(\vec r) g(\vec r-\vec R)\simeq f(\vec R) g(\vec r-\vec R)
\end{equation}
(for positions $\vec r$ for which $g(\vec r-\vec R)$ is not
negligible, the smooth function $f(\vec r)$ is approximately equal to
$f(\vec R)$, while for positions $\vec r$, further away from $\vec R$, for
which $f(\vec r)$ significantly differs from $f(\vec R)$, the function
$g(\vec r-\vec R)$ is null).
In fig.~\ref{f6} we show a possible smoothing function $g(\vec r)$,
which approximately satisfies all the previous relations \footnote
{In detail, we have represented the function defined as $106.5307 \, 
\exp(-\frac{5.7677}{1-(|\vec r|/(0.355~{\rm nm}))^2})$
for $|\vec r|<0.355~{\rm nm}$, and $0$ for $|\vec r| \ge 0.355~{\rm nm}$,
but better approximations for the smoothing function $g(\vec r)$ can be found.}.
\noindent

\begin{figure}
\centering
\includegraphics[width=.7\textwidth,angle=0]{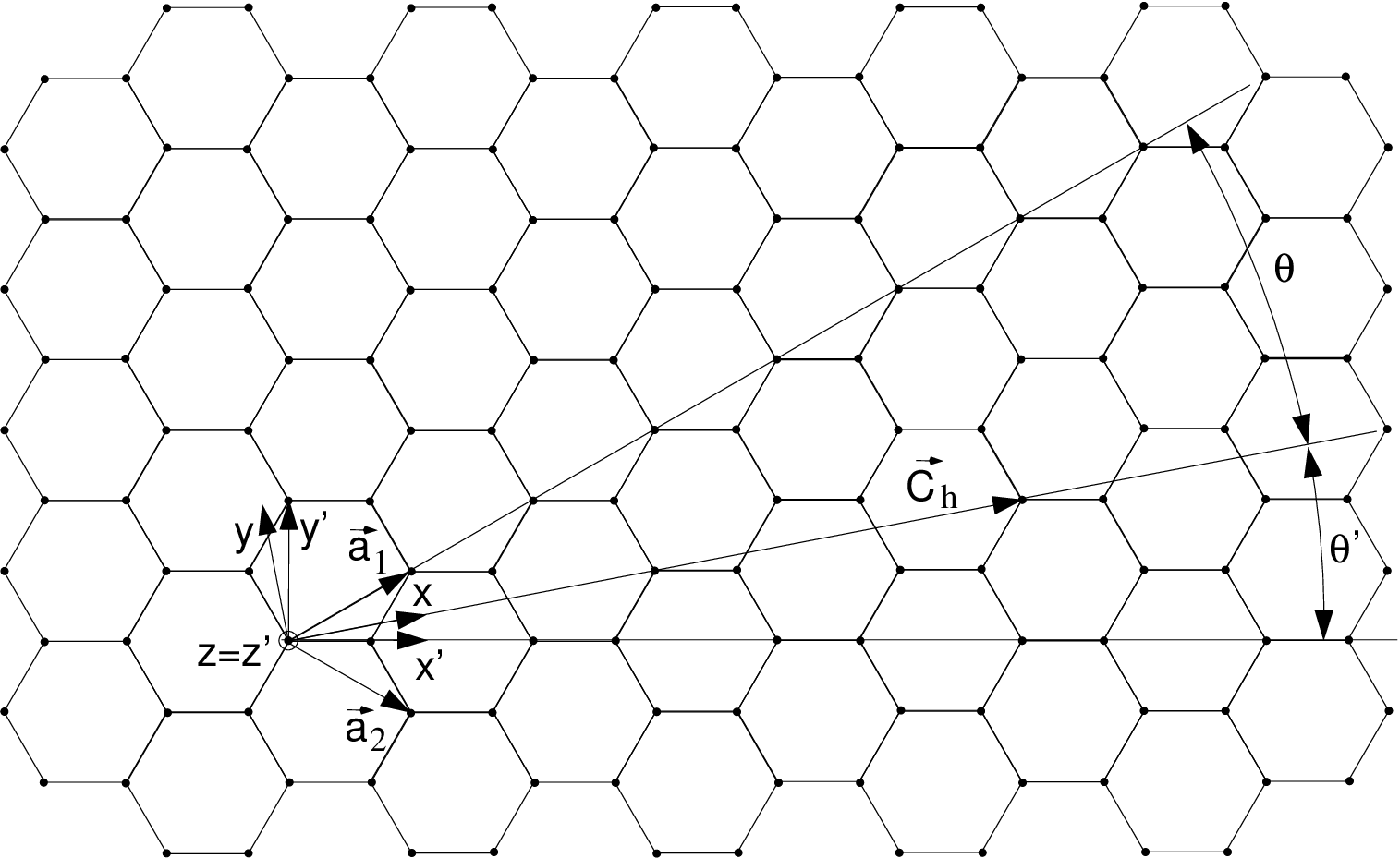}
\caption{The reference frames used in the calculations ($\vec C_h$ and
$\theta$ will be used for carbon nanotubes in the next section: this
figure corresponds to a $(4,2)$ nanotube).}
\label{f7}
\end{figure}

%%%%%% 1 %%%%%%
If we multiply the first of the tight-binding equations
(\ref{tightbinding1}) by
$g(\vec r-\vec R_A)e^{-i \vec K\cdot \vec R_A}$
and we sum it over $\vec R_A$, we find
\begin{eqnarray}
&& E\,\sum_{\vec R_A} g(\vec r-\vec R_A) F_A^{\vec K}(\vec R_A)\\
&&\qquad -E\,i \, e^{i \theta'}\,\sum_{\vec R_A} g(\vec r-\vec R_A)
e^{i (\vec K'-\vec K)\cdot \vec R_A} F_A^{\vec K'}(\vec R_A)\nonumber\\
&&\qquad -\sum_{\vec R_A} g(\vec r-\vec R_A) u(\vec R_A) F_A^{\vec K}(\vec R_A)\nonumber\\
&&\qquad +i \, e^{i \theta'}\,\sum_{\vec R_A} g(\vec r-\vec R_A)
e^{i (\vec K'-\vec K)\cdot \vec R_A}
u(\vec R_A) F_A^{\vec K'}(\vec R_A)=\nonumber\\
&&\qquad -\gamma_0 \,i \, e^{i \theta'}
\sum_{l=1}^3 e^{-i \vec K\cdot \vec\tau_l}
\sum_{\vec R_A} g(\vec r-\vec R_A) F_B^{\vec K}(\vec R_A-\vec\tau_l)\nonumber\\
&&\qquad -\gamma_0 \sum_{l=1}^3 e^{-i \vec K'\cdot \vec\tau_l}
\sum_{\vec R_A} g(\vec r-\vec R_A)
e^{i (\vec K'-\vec K)\cdot \vec R_A} F_B^{\vec K'}(\vec R_A-\vec\tau_l);\nonumber
\end{eqnarray}
\newpage
\noindent
exploiting the property (\ref{smooth}) it becomes
\vskip-5pt\noindent
\begin{eqnarray}
&& E\,\left[\sum_{\vec R_A} g(\vec r-\vec R_A)\right] F_A^{\vec K}(\vec r)-
E\,i \, e^{i \theta'}\,\left[\sum_{\vec R_A} g(\vec r-\vec R_A)
e^{i (\vec K'-\vec K)\cdot \vec R_A}\right] F_A^{\vec K'}(\vec r)\\
&& -\left[\sum_{\vec R_A} g(\vec r-\vec R_A) u(\vec R_A)\right]
F_A^{\vec K}(\vec r)\nonumber\\
&& +i \, e^{i \theta'}\,\left[\sum_{\vec R_A} g(\vec r-\vec R_A)
e^{i (\vec K'-\vec K)\cdot \vec R_A}
u(\vec R_A)\right] F_A^{\vec K'}(\vec r)=\nonumber\\
&& -\gamma_0 \,i \, e^{i \theta'}
\sum_{l=1}^3 e^{-i \vec K\cdot \vec\tau_l}
\left[\sum_{\vec R_A} g(\vec r-\vec R_A)\right]
F_B^{\vec K}(\vec r-\vec\tau_l)\nonumber\\
&& -\gamma_0 \sum_{l=1}^3 e^{-i \vec K'\cdot \vec\tau_l}
\left[\sum_{\vec R_A} g(\vec r-\vec R_A)
e^{i (\vec K'-\vec K)\cdot \vec R_A}\right] F_B^{\vec K'}(\vec r-\vec\tau_l).\nonumber
\end{eqnarray}
\vskip-5pt\noindent
For the quantities in the square brackets, we can use the properties
(\ref{sum}) and (\ref{phase}), together with the definitions
\vskip-5pt\noindent
\begin{equation}
\label{ua}
u_A(\vec r)=\sum_{\vec R_A} g(\vec r-\vec R_A) u(\vec R_A),\qquad
u'_A(\vec r)=\sum_{\vec R_A} g(\vec r-\vec R_A)
e^{i (\vec K'-\vec K)\cdot \vec R_A} u(\vec R_A),
\end{equation}
\vskip-5pt\noindent
obtaining
\vskip-5pt\noindent
\begin{eqnarray}
\label{first}
&& E\,F_A^{\vec K}(\vec r)-u_A(\vec r)\,F_A^{\vec K}(\vec r)+
i \, e^{i \theta'}\,u'_A(\vec r)\,F_A^{\vec K'}(\vec r)=\\
&& -\gamma_0 \,i \, e^{i \theta'}
\sum_{l=1}^3 e^{-i \vec K\cdot \vec\tau_l}
F_B^{\vec K}(\vec r-\vec\tau_l).\nonumber
\end{eqnarray}
\vskip-5pt\noindent
Expanding the smooth quantity $F_B^{\vec K} (\vec r-\vec \tau_l)$ to the
first order in $\vec \tau_l$, we have that
\vskip-5pt\noindent
\begin{eqnarray}
&& \sum_{l=1}^3 e^{-i \vec K\cdot \vec\tau_l}
F_B^{\vec K}(\vec r-\vec\tau_l)\simeq
\sum_{l=1}^3 e^{-i \vec K\cdot \vec \tau_l} 
\left[F_B^{\vec K} (\vec r)-
\left(\vec \tau_l\cdot\frac{\partial}{\partial\vec r}
\right) F_B^{\vec K} (\vec r)\right]=\\
&& \left\{\left(\sum_{l=1}^3 e^{-i \vec K\cdot \vec \tau_l}\right) 
F_B^{\vec K} (\vec r)-\left[\sum_{l=1}^3 e^{-i \vec K\cdot \vec \tau_l}
\left(\vec \tau_l\cdot\frac{\partial}{\partial\vec r}\right)\right]
F_B^{\vec K} (\vec r)\right\}.\nonumber
\end{eqnarray}
\vskip-5pt\noindent
Let us now calculate the value of the sums which appear in the previous 
expression
\vskip-5pt\noindent
\begin{eqnarray}
&& \sum_{l=1}^3 e^{-i \vec K\cdot \vec \tau_l}=
1+e^{-i\frac{2\pi}{3}}+e^{i\frac{2\pi}{3}}=0;\\
&& \sum_{l=1}^3 e^{-i \vec K\cdot \vec \tau_l}
\left(\vec \tau_l\cdot\frac{\partial}{\partial\vec r}\right)=
1 \frac{a}{\sqrt{3}} \left(-\frac{\partial}{\partial x'}\right)\nonumber\\
&& +e^{-i\frac{2\pi}{3}} \frac{a}{\sqrt{3}} 
\left(\frac{1}{2}\frac{\partial}{\partial x'}-\frac{\sqrt{3}}{2}
\frac{\partial}{\partial y'}\right)+
e^{i\frac{2\pi}{3}} \frac{a}{\sqrt{3}}
\left(\frac{1}{2}\frac{\partial}{\partial x'}+\frac{\sqrt{3}}{2}
\frac{\partial}{\partial y'}\right)=\nonumber\\
&& \frac{a}{\sqrt{3}}
\left(\left(-1+\frac{1}{2}e^{-i\frac{2\pi}{3}}+
\frac{1}{2}e^{i\frac{2\pi}{3}}\right)
\frac{\partial}{\partial x'}+\left(-\frac{\sqrt{3}}{2}e^{-i\frac{2\pi}{3}}+
\frac{\sqrt{3}}{2}e^{i\frac{2\pi}{3}} \right)
\frac{\partial}{\partial y'}\right).\nonumber
\end{eqnarray}
Since \ $\displaystyle 
-1+\frac{1}{2}e^{-i\frac{2\pi}{3}}+\frac{1}{2}e^{i\frac{2\pi}{3}}=
-\frac{3}{2}$ \ and \ 
$\displaystyle
-\frac{\sqrt{3}}{2}e^{-i\frac{2\pi}{3}}+\frac{\sqrt{3}}{2}e^{i\frac{2\pi}{3}}=
i\frac{3}{2}$, \ we have that
\begin{eqnarray}
&& \sum_{l=1}^3 e^{-i \vec K\cdot \vec \tau_l}
\left(\vec \tau_l\cdot\frac{\partial}{\partial\vec r}\right)=
-\frac{a}{\sqrt{3}}\frac{3}{2}
\left(\frac{\partial}{\partial x'}-i\frac{\partial}{\partial y'}\right)=\\
&& -\frac{\sqrt{3}}{2}a (i\hat\kappa_{x'}+\hat\kappa_{y'})=
-i\frac{\sqrt{3}}{2}a (\hat\kappa_{x'}-i\hat\kappa_{y'}),\nonumber
\end{eqnarray}
where we have defined
${\vec{\hat\kappa}}=-i\vec \nabla$ and thus
\begin{equation}
\quad
{\hat\kappa}_{x'}=-i\frac{\partial}{\partial x'}
\qquad \hbox{and} \qquad
{\hat\kappa}_{y'}=-i\frac{\partial}{\partial y'}.
\end{equation}
Substituting these results, eq.~(\ref{first}) becomes
\begin{eqnarray}
\label{first1}
&& E\,F_A^{\vec K}(\vec r)-u_A(\vec r)\,F_A^{\vec K}(\vec r)+
i \, e^{i \theta'}\,u'_A(\vec r)\,F_A^{\vec K'}(\vec r) \simeq\\
&& -\gamma_0 \,i \, e^{i \theta'}
\left(i\frac{\sqrt{3}}{2}a (\hat\kappa_{x'}-i\hat\kappa_{y'})
F_B^{\vec K}(\vec r)\right)=\nonumber\\
&& \frac{\sqrt{3}}{2}\gamma_0 a\, e^{i \theta'} 
(\hat\kappa_{x'}-i\hat\kappa_{y'}) F_B^{\vec K}(\vec r)=
\gamma ({\hat\kappa}_x-i{\hat\kappa}_y) F_B^{\vec K} (\vec r),\nonumber
\end{eqnarray}
where we have passed from the original reference frame
$\Sigma'=(\hbox{\boldmath{$\hat x$}}', \hbox{\boldmath{$\hat y$}}',
\hbox{\boldmath{$\hat z$}}')$ to a new frame
$\Sigma=(\hbox{\boldmath{$\hat x$}}, \hbox{\boldmath{$\hat y$}},
\hbox{\boldmath{$\hat z$}})$,
rotated, in the plane $(\hbox{\boldmath{$\hat x$}}', 
\hbox{\boldmath{$\hat y$}}')$, around the origin by an angle
$\theta'$ (positive in the counterclockwise direction) with respect to the
original one (fig.~\ref{f7}) and we have used the fact that
\begin{eqnarray}
\label{diffcoord}
&& e^{i\theta'}({\hat\kappa}_{x'}-i{\hat\kappa}_{y'})=
(\cos\theta'+i\sin\theta')({\hat\kappa}_{x'}-i{\hat\kappa}_{y'})=\\
&& (\cos\theta' {\hat\kappa}_{x'}+\sin\theta' {\hat\kappa}_{y'})-
i(\cos\theta' {\hat\kappa}_{y'}-\sin\theta' {\hat\kappa}_{x'})=
{\hat\kappa}_x-i{\hat\kappa}_y\nonumber
\end{eqnarray}
(due to the relations between old and new coordinates), with
\begin{equation}
\label{kappa}
\quad
{\hat\kappa}_{x}=-i\frac{\partial}{\partial x}
\qquad \hbox{and} \qquad
{\hat\kappa}_{y}=-i\frac{\partial}{\partial y}.
\end{equation}
Indeed, it is a well-known result that, for a rotation by $\theta'$ of
the reference frame, the relations between the new and the old coordinates
are $x=x'\cos\theta'+y'\sin\theta'$
and $y=y'\cos\theta'-x'\sin\theta'$. Therefore we have that
\begin{equation}
\frac{\partial F(x,y)}{\partial x'} =
\frac{\partial F(x,y)}{\partial x} \frac{\partial x}{\partial x'}+
\frac{\partial F(x,y)}{\partial y} \frac{\partial y}{\partial x'}=
\frac{\partial F(x,y)}{\partial x} \cos\theta'-
\frac{\partial F(x,y)}{\partial y} \sin\theta'
\end{equation}
and that
\begin{equation}
\frac{\partial F(x,y)}{\partial y'} =
\frac{\partial F(x,y)}{\partial x} \frac{\partial x}{\partial y'}+
\frac{\partial F(x,y)}{\partial y} \frac{\partial y}{\partial y'}=
\frac{\partial F(x,y)}{\partial x} \sin\theta'+
\frac{\partial F(x,y)}{\partial y} \cos\theta'.
\end{equation}
As a consequence, we have that
\begin{eqnarray}
\qquad (\cos\theta' {\hat\kappa}_{x'}&+&\sin\theta' {\hat\kappa}_{y'}) F(x,y)=
\cos\theta' \left(-i\frac{\partial F(x,y)}{\partial x'}\right)+
\sin\theta' \left(-i\frac{\partial F(x,y)}{\partial y'}\right)=\\
&-& i\bigg[\frac{\partial F(x,y)}{\partial x} \cos^2\theta'-
\frac{\partial F(x,y)}{\partial y} \cos\theta'\sin\theta'\nonumber\\
&&\quad +\frac{\partial F(x,y)}{\partial x} \sin^2\theta'+
\frac{\partial F(x,y)}{\partial y} \sin\theta'\cos\theta'
\bigg]=\nonumber\\
&-& i\frac{\partial F(x,y)}{\partial x}
(\cos^2\theta'+\sin^2\theta')=
-i\frac{\partial F(x,y)}{\partial x}=\hat\kappa_x F(x,y)\nonumber
\end{eqnarray}
and that
\begin{eqnarray}
\qquad (\cos\theta' {\hat\kappa}_{y'}&-&\sin\theta' {\hat\kappa}_{x'}) F(x,y)=
\cos\theta' \left(-i\frac{\partial F(x,y)}{\partial y'}\right)-
\sin\theta' \left(-i\frac{\partial F(x,y)}{\partial x'}\right)=\\
&-& i\bigg[
\frac{\partial F(x,y)}{\partial x} \sin\theta'\cos\theta'+
\frac{\partial F(x,y)}{\partial y} \cos^2\theta'-\nonumber\\
&&\quad -\frac{\partial F(x,y)}{\partial x} \cos\theta'\sin\theta'+
\frac{\partial F(x,y)}{\partial y} \sin^2\theta'
\bigg]=\nonumber\\
&-& i\frac{\partial F(x,y)}{\partial y}
(\cos^2\theta'+\sin^2\theta')=
-i\frac{\partial F(x,y)}{\partial y}=\hat\kappa_y F(x,y),\nonumber
\end{eqnarray}
from which we obtain eq.~(\ref{diffcoord}).
\noindent

$\theta'$ is the angle, taken counterclockwise, from the vector
$\vec a_1+\vec a_2$ to the axis $\hbox{\boldmath{$\hat x$}}$ of the new frame.
We have also defined the quantity $\gamma=(\sqrt{3}/2)\gamma_0 a$.

Note that in the new reference frame 
$\Sigma=(\hbox{\boldmath{$\hat x$}}, \hbox{\boldmath{$\hat y$}},
\hbox{\boldmath{$\hat z$}})$
\begin{eqnarray}
&\vec a_1 \mathrel{\mathop\equiv_{\Sigma}} \displaystyle \frac{a}{2}
\left[\begin{array}{c}
\displaystyle \sqrt{3}\cos\theta'+\sin\theta'\\[3pt]
\displaystyle \cos\theta'-\sqrt{3}\sin\theta'\\[3pt]
0
\end{array}\right],\quad&
\vec a_2 \mathrel{\mathop\equiv_{\Sigma}} \displaystyle \frac{a}{2}
\left[\begin{array}{c}
\displaystyle \sqrt{3}\cos\theta'-\sin\theta'\\[3pt]
\displaystyle -\cos\theta'-\sqrt{3}\sin\theta'\\[3pt]
0
\end{array}\right],\\
&\vec b_1 \mathrel{\mathop\equiv_{\Sigma}} \displaystyle \frac{2\pi}{\sqrt{3}a}
\left[\begin{array}{c}
\displaystyle \cos\theta'+\sqrt{3}\sin\theta'\\ [3pt]
\displaystyle \sqrt{3}\cos\theta'-\sin\theta'\\[3pt]
0
\end{array}\right],\quad&
\vec b_2 \mathrel{\mathop\equiv_{\Sigma}} \displaystyle \frac{2\pi}{\sqrt{3}a}
\left[\begin{array}{c}
\displaystyle \cos\theta'-\sqrt{3}\sin\theta'\\[3pt]
\displaystyle -\sqrt{3}\cos\theta'-\sin\theta'\\[3pt]
0
\end{array}\right],\nonumber\\
&\vec K \mathrel{\mathop\equiv_{\Sigma}} \displaystyle \frac{4\pi}{3a}
\left[\begin{array}{c}
-\sin\theta'\\
-\cos\theta'\\
0
\end{array}\right],\quad&
\vec K' \mathrel{\mathop\equiv_{\Sigma}} \displaystyle \frac{4\pi}{3a}
\left[\begin{array}{c}
\sin\theta'\\
\cos\theta'\\
0
\end{array}\right].\nonumber
\end{eqnarray}

%%%%%% 2 %%%%%%
Analogously, if we multiply the second of the tight-binding equations
(\ref{tightbinding1}) by
$g(\vec r-\vec R_B)(-i\, e^{-i\theta'} e^{-i \vec K\cdot \vec R_B})$
and we sum it over $\vec R_B$, using again the properties
(\ref{smooth}), (\ref{sum}) and (\ref{phase}), together with the definitions
\begin{equation}
\label{ub}
u_B(\vec r)=\sum_{\vec R_B} g(\vec r-\vec R_B) u(\vec R_B),\qquad 
u'_B(\vec r)=\sum_{\vec R_B} g(\vec r-\vec R_B)
e^{i (\vec K'-\vec K)\cdot \vec R_B} u(\vec R_B),
\end{equation}
we obtain~\cite{supplem}
\begin{eqnarray}
\label{second}
&& E\,F_B^{\vec K} (\vec r)-u_B(\vec r)\,F_B^{\vec K} (\vec r)+
i\,e^{-i\theta'}\, u'_B(\vec r)\,F_B^{\vec K'} (\vec r)=\\
&& \gamma_0 \,i\, e^{-i\theta'} \sum_{l=1}^3  
e^{i \vec K \cdot \vec \tau_l} F_A^{\vec K} (\vec r+\vec \tau_l).\nonumber
\end{eqnarray}
Expanding the smooth quantity $F_A^{\vec K} (\vec r+\vec \tau_l)$ to the
first order in $\vec \tau_l$, we have that
\begin{eqnarray}
&& \sum_{l=1}^3 e^{i \vec K \cdot \vec \tau_l}
F_A^{\vec K} (\vec r+\vec \tau_l)\simeq
\sum_{l=1}^3 e^{i \vec K \cdot \vec \tau_l}
\left[F_A^{\vec K} (\vec r)+
\left(\vec \tau_l\cdot\frac{\partial}{\partial\vec r}
\right) F_A^{\vec K} (\vec r)\right]=\\
&& \left(\sum_{l=1}^3 e^{i \vec K\cdot \vec \tau_l}\right) 
F_A^{\vec K} (\vec r)+\left[\sum_{l=1}^3 e^{i \vec K\cdot \vec \tau_l}
\left(\vec \tau_l\cdot\frac{\partial}{\partial\vec r}\right)\right]
F_A^{\vec K} (\vec r).\nonumber
\end{eqnarray}
Since 
\begin{equation}
\sum_{l=1}^3 e^{i \vec K\cdot \vec \tau_l}=0\quad\hbox{and}\quad
\sum_{l=1}^3 e^{i \vec K\cdot \vec \tau_l}
\left(\vec \tau_l\cdot\frac{\partial}{\partial\vec r}\right)=
-i\frac{\sqrt{3}}{2}a ({\hat\kappa}_{x'}+i{\hat\kappa}_{y'}),
\end{equation}
eq.~(\ref{second}) becomes
\begin{eqnarray}
\label{second1}
&& E\,F_B^{\vec K} (\vec r)-u_B(\vec r)\,F_B^{\vec K} (\vec r)+
i\,e^{-i\theta'}\, u'_B(\vec r)\,F_B^{\vec K'} (\vec r)\simeq\\
&& \gamma_0 \,i\, e^{-i\theta'} 
\left( -i\frac{\sqrt{3}}{2}a ({\hat\kappa}_{x'}+i{\hat\kappa}_{y'})
\right) F_A^{\vec K} (\vec r)=\nonumber\\
&& \frac{\sqrt{3}}{2}\gamma_0 a\, e^{-i \theta'} 
(\hat\kappa_{x'}+i\hat\kappa_{y'}) F_A^{\vec K}(\vec r)=
\gamma ({\hat\kappa}_x+i{\hat\kappa}_y) F_A^{\vec K} (\vec r),\nonumber
\end{eqnarray}
where we have made use of the relation
\begin{eqnarray}
\label{sumcoord}
&& e^{-i\theta'}({\hat\kappa}_{x'}+i{\hat\kappa}_{y'})=
(\cos\theta'-i\sin\theta')({\hat\kappa}_{x'}+i{\hat\kappa}_{y'})=\\
&& (\cos\theta' {\hat\kappa}_{x'}+\sin\theta' {\hat\kappa}_{y'})+
i(\cos\theta' {\hat\kappa}_{y'}-\sin\theta' {\hat\kappa}_{x'})=
{\hat\kappa}_x+i{\hat\kappa}_y.\nonumber
\end{eqnarray}

%%%%%% 3 %%%%%%
Instead, if we multiply the first of the tight-binding equations
(\ref{tightbinding1}) by
$g(\vec r-\vec R_A)\times (i\, e^{-i\theta'} e^{-i \vec K'\cdot \vec R_A})$
and we sum it over $\vec R_A$, we obtain (exploiting the properties
(\ref{smooth}), (\ref{sum})\break
and (\ref{phase}))~\cite{supplem}
\begin{eqnarray}
\label{third}
&& E F_A^{\vec K'} (\vec r)-i\, e^{-i\theta'} {u'_A}^*(\vec r)
F_A^{\vec K} (\vec r)-u_A(\vec r) F_A^{\vec K'} (\vec r)=\\
&& -\gamma_0\,i\, e^{-i\theta'} \sum_{l=1}^3 
e^{-i \vec K' \cdot \vec \tau_l} F_B^{\vec K'} (\vec r-\vec \tau_l).\nonumber
\end{eqnarray}
Expanding the smooth quantity $F_B^{\vec K'} (\vec r-\vec \tau_l)$ to the
first order in $\vec \tau_l$, we have that
\begin{eqnarray}
&& \sum_{l=1}^3 e^{-i \vec K'\cdot \vec \tau_l} 
F_B^{\vec K'} (\vec r-\vec \tau_l)\simeq
\sum_{l=1}^3 e^{-i \vec K'\cdot \vec \tau_l} 
\left[F_B^{\vec K'} (\vec r)-
\left(\vec \tau_l\cdot\frac{\partial}{\partial\vec r}
\right) F_B^{\vec K'} (\vec r)\right]=\\
&& \left(\sum_{l=1}^3 e^{-i \vec K'\cdot \vec \tau_l}\right) 
F_B^{\vec K'} (\vec r)-\left[\sum_{l=1}^3 e^{-i \vec K'\cdot \vec \tau_l}
\left(\vec \tau_l\cdot\frac{\partial}{\partial\vec r}\right)\right]
F_B^{\vec K'} (\vec r),\nonumber
\end{eqnarray}
with
\begin{equation}
\sum_{l=1}^3 e^{-i \vec K'\cdot \vec \tau_l}=0\quad\hbox{and}\quad
\sum_{l=1}^3 e^{-i \vec K'\cdot \vec \tau_l}
\left(\vec \tau_l\cdot\frac{\partial}{\partial\vec r}\right)=
-i\frac{\sqrt{3}}{2}a (\hat\kappa_{x'}+i\hat\kappa_{y'}).
\end{equation}
Therefore eq.~(\ref{third}) becomes
\begin{eqnarray}
\label{third1}
&& E F_A^{\vec K'} (\vec r)-i\, e^{-i\theta'} {u'_A}^*(\vec r)
F_A^{\vec K} (\vec r)-u_A(\vec r) F_A^{\vec K'} (\vec r)\simeq\\
&& -\gamma_0\,i\, e^{-i\theta'} 
\left(i\frac{\sqrt{3}}{2}a (\hat\kappa_{x'}+i\hat\kappa_{y'})
F_B^{\vec K'} (\vec r)\right)=\nonumber\\
&& \frac{\sqrt{3}}{2}\gamma_0 a\, e^{-i \theta'} 
(\hat\kappa_{x'}+i\hat\kappa_{y'}) F_B^{\vec K'}(\vec r)=
\gamma ({\hat\kappa}_x+i{\hat\kappa}_y) F_B^{\vec K'} (\vec r),\nonumber
\end{eqnarray}
where we have exploited the relation (\ref{sumcoord}).

%%%%%% 4 %%%%%%
Finally, if we multiply the second of the tight-binding equations
(\ref{tightbinding1}) by
$g(\vec r-\vec R_B) \times e^{-i \vec K'\cdot \vec R_B}$
and we sum it over $\vec R_B$, we obtain (using the properties
(\ref{smooth}), (\ref{sum})\break
and (\ref{phase}))~\cite{supplem}
\begin{eqnarray}
\label{fourth}
&& E\,F_B^{\vec K'} (\vec r)-i\,e^{i\theta'}\,
{u'_B}^*(\vec r)\,F_B^{\vec K} (\vec r)-
u_B(\vec r)\,F_B^{\vec K'} (\vec r)=\\
&& \gamma_0 \,i\, e^{i\theta'} \sum_{l=1}^3  
e^{i \vec K' \cdot \vec \tau_l} F_A^{\vec K'} (\vec r+\vec \tau_l).\nonumber
\end{eqnarray}
Expanding the smooth quantity $F_A^{\vec K'} (\vec r+\vec \tau_l)$ to the
first order in $\vec \tau_l$, we have that
\begin{eqnarray}
&& \sum_{l=1}^3 e^{i \vec K' \cdot \vec \tau_l}
F_A^{\vec K'} (\vec r+\vec \tau_l)\simeq
\sum_{l=1}^3 e^{i \vec K' \cdot \vec \tau_l}
\left[F_A^{\vec K'} (\vec r)+
\left(\vec \tau_l\cdot\frac{\partial}{\partial\vec r}
\right) F_A^{\vec K'} (\vec r)\right]=\\
&& \left(\sum_{l=1}^3 e^{i \vec K'\cdot \vec \tau_l}\right) 
F_A^{\vec K'} (\vec r)+\left[\sum_{l=1}^3 e^{i \vec K'\cdot \vec \tau_l}
\left(\vec \tau_l\cdot\frac{\partial}{\partial\vec r}\right)\right]
F_A^{\vec K'} (\vec r).\nonumber
\end{eqnarray}
Since
\begin{equation}
\sum_{l=1}^3 e^{i \vec K'\cdot \vec \tau_l}=0\quad\hbox{and}\quad
\sum_{l=1}^3 e^{i \vec K'\cdot \vec \tau_l}
\left(\vec \tau_l\cdot\frac{\partial}{\partial\vec r}\right)=
-i\frac{\sqrt{3}}{2}a (\hat\kappa_{x'}-i\hat\kappa_{y'}),
\end{equation}
eq.~(\ref{fourth}) becomes
\begin{eqnarray}
\label{fourth1}
&& E\,F_B^{\vec K'} (\vec r)-i\,e^{i\theta'}\,
{u'_B}^*(\vec r)\,F_B^{\vec K} (\vec r)-
u_B(\vec r)\,F_B^{\vec K'} (\vec r)=\\
&& \gamma_0 \,i\, e^{i\theta'} 
\left( -i\frac{\sqrt{3}}{2}a (\hat\kappa_{x'}-i\hat\kappa_{y'})
\right) F_A^{\vec K'} (\vec r)=\nonumber\\
&& \frac{\sqrt{3}}{2}\gamma_0 a\, e^{i \theta'}
(\hat\kappa_{x'}-i\hat\kappa_{y'}) F_A^{\vec K'}(\vec r)=
\gamma ({\hat\kappa}_x-i{\hat\kappa}_y) F_A^{\vec K'} (\vec r),\nonumber
\end{eqnarray}
where the relation (\ref{diffcoord}) has been used.

In this way, we have obtained the four equations (\ref{first1}),
(\ref{second1}), (\ref{third1}) and (\ref{fourth1}), that we can
summarize
\begin{equation}
\left\{ \begin{array}{l}
\displaystyle
u_A(\vec r)\,F_A^{\vec K}(\vec r)+
\gamma ({\hat\kappa}_x-i{\hat\kappa}_y) F_B^{\vec K} (\vec r)-
i\, e^{i \theta'}\,u'_A(\vec r)\,F_A^{\vec K'}(\vec r)=
E\,F_A^{\vec K}(\vec r),\\[5pt]
\displaystyle
\gamma ({\hat\kappa}_x+i{\hat\kappa}_y) F_A^{\vec K} (\vec r)+
u_B(\vec r)\,F_B^{\vec K} (\vec r)-
i\, e^{-i\theta'}\, u'_B(\vec r)\,F_B^{\vec K'} (\vec r)=
E\,F_B^{\vec K} (\vec r),\\[5pt]
\displaystyle
i\, e^{-i\theta'} {u'_A}^*(\vec r) F_A^{\vec K} (\vec r)+
u_A(\vec r) F_A^{\vec K'} (\vec r)+
\gamma ({\hat\kappa}_x+i{\hat\kappa}_y) F_B^{\vec K'} (\vec r)=
E F_A^{\vec K'} (\vec r),\\[5pt]
\displaystyle
i\, e^{i\theta'}\,{u'_B}^*(\vec r)\,F_B^{\vec K} (\vec r)+
\gamma ({\hat\kappa}_x-i{\hat\kappa}_y) F_A^{\vec K'} (\vec r)+
u_B(\vec r)\,F_B^{\vec K'} (\vec r)=
E\,F_B^{\vec K'} (\vec r),
\end{array} \right.
\end{equation}
and write in matrix form
\begin{eqnarray}
\label{diracpot}
\quad && \left[\begin{array}{cccc}
u_A(\vec r) &
\gamma ({\hat\kappa}_x-i{\hat\kappa}_y) &
-i \, e^{i \theta'}\,u'_A(\vec r) &
0 \\[3pt]
\gamma ({\hat\kappa}_x+i{\hat\kappa}_y) &
u_B(\vec r) &
0 &
-i\, e^{-i\theta'}\, u'_B(\vec r) \\[3pt]
i\, e^{-i\theta'}\,{u'_A}^*(\vec r) &
0 &
u_A(\vec r) &
\gamma ({\hat\kappa}_x+i{\hat\kappa}_y) \\[3pt]
0 &
i\, e^{i\theta'}\,{u'_B}^*(\vec r) &
\gamma ({\hat\kappa}_x-i{\hat\kappa}_y) &
u_B(\vec r)
\end{array}\right]
\left[\begin{array}{c}
F_A^{\vec K} (\vec r)\\[3pt]
F_B^{\vec K} (\vec r)\\[3pt]
F_A^{\vec K'} (\vec r)\\[3pt]
F_B^{\vec K'} (\vec r)
\end{array}\right]=\\
&& E \left[\begin{array}{c}
F_A^{\vec K} (\vec r)\\[3pt]
F_B^{\vec K} (\vec r)\\[3pt]
F_A^{\vec K'} (\vec r)\\[3pt]
F_B^{\vec K'} (\vec r)
\end{array}\right],\nonumber
\end{eqnarray}
which is the $\vec k \cdot \vec p$ equation of graphene.

Incidentally, if we repeat all the previous calculations considering
the following different pair of reference Dirac points: 
\begin{equation}
\vec K=\left[\begin{array}{c}
\displaystyle \frac{2\pi}{\sqrt{3}a}\\
\noalign{\vskip3pt}
\displaystyle \frac{2\pi}{3a}\\
\noalign{\vskip3pt} 
0
\end{array}
\right]
,\quad
\vec K'=\left[\begin{array}{c}
\displaystyle \frac{2\pi}{\sqrt{3}a}\\
\noalign{\vskip3pt}
\displaystyle -\frac{2\pi}{3a}\\
\noalign{\vskip3pt} 
0
\end{array}
\right]
\end{equation}
(equivalent, in the reciprocal space, to the pair (\ref{diracpoints})
of Dirac points), we have to replace (\ref{assumptions}) with
\begin{equation}
\left\{ \begin{array}{l}
\displaystyle
\psi_A (\vec R_A)=e^{i \vec K\cdot \vec R_A} F_A^{\vec K}(\vec R_A)
+e^{i\eta} e^{i \vec K'\cdot \vec R_A} F_A^{\vec K'}(\vec R_A),\\[5pt]
\displaystyle
\psi_B (\vec R_B)=-e^{i\frac{2\pi}{3}}\,e^{i\eta}
e^{i \vec K\cdot \vec R_B} F_B^{\vec K} (\vec R_B)+
e^{i \vec K'\cdot \vec R_B} F_B^{\vec K'} (\vec R_B),
\end{array} \right.
\end{equation}
where $\eta=\pi/6+\theta'$, and we obtain (instead of
eq.~(\ref{diracpot}))
\begin{eqnarray}
\qquad && \left[\begin{array}{@{\ }c@{\ }c@{\ \ \ }c@{\ }c@{\ }}
u_A(\vec r) &
\gamma ({\hat\kappa}_x-i{\hat\kappa}_y) &
e^{i\eta}\,u'_A(\vec r) &
0 \\[3pt]
\gamma ({\hat\kappa}_x+i{\hat\kappa}_y) &
u_B(\vec r) &
0 &
-e^{-i\frac{2\pi}{3}}\,e^{-i\eta}\, u'_B(\vec r) \\[3pt]
e^{-i\eta}\,{u'_A}^*(\vec r) &
0 &
u_A(\vec r) &
\gamma ({\hat\kappa}_x+i{\hat\kappa}_y) \\[3pt]
0 &
-e^{i\frac{2\pi}{3}}\,e^{i\eta}\,{u'_B}^*(\vec r) &
\gamma ({\hat\kappa}_x-i{\hat\kappa}_y) &
u_B(\vec r)
\end{array}\right]
\left[\begin{array}{c}
F_A^{\vec K} (\vec r)\\[3pt]
F_B^{\vec K} (\vec r)\\[3pt]
F_A^{\vec K'} (\vec r)\\[3pt]
F_B^{\vec K'} (\vec r)
\end{array}\right]=\\
&& E \left[\begin{array}{c}
F_A^{\vec K} (\vec r)\\[3pt]
F_B^{\vec K} (\vec r)\\[3pt]
F_A^{\vec K'} (\vec r)\\[3pt]
F_B^{\vec K'} (\vec r)
\end{array}\right],\nonumber
\end{eqnarray}
as found by Ando~\cite{ando1,ando2}.

Summarizing, we have that the overall wave function is given by
(see (\ref{wavefunction}))
\begin{equation}
\psi (\vec r)=
\sum_{\vec R_A}\psi_A (\vec R_A)\varphi(\vec r -\vec R_A)+
\sum_{\vec R_B}\psi_B (\vec R_B)\varphi(\vec r -\vec R_B),
\end{equation}
with (see (\ref{assumptions}))
\begin{equation}
\label{assumptions2}
\left\{ \begin{array}{l}
\displaystyle
\psi_A (\vec r)=e^{i \vec K\cdot \vec r} F_A^{\vec K}(\vec r)
-i \, e^{i \theta'} e^{i \vec K'\cdot \vec r} F_A^{\vec K'}(\vec r),\\
\displaystyle
\psi_B (\vec r)=i \, e^{i \theta'} 
e^{i \vec K\cdot \vec r} F_B^{\vec K} (\vec r)+
e^{i \vec K'\cdot \vec r} F_B^{\vec K'} (\vec r),
\end{array} \right.
\end{equation}
where the envelope functions $F$ satisfy eq.~(\ref{diracpot}).

We $\!$can $\!$treat $\!$two $\!$limiting $\!$cases $\!$for $\!$the $\!$external $\!$potential, $\!$depending $\!$on $\!$its$\!$ range~$\!\!$\cite{shon,ando3}.

If the potential range is much smaller than the lattice constant
(short-range case), we can consider the external potential as different from
zero only on one carbon atom.

If it is non-zero only on an atom of type A (in position $\vec R_{A_0}$), {\em i.e.}
$u (\vec R_{A_0}) \ne 0$, $u (\vec R_A)=0$ for $\vec R_A \ne \vec R_{A_0}$ and
$u (\vec R_B)=0$ for every $\vec R_B$, recalling eq.~(\ref{ua}) and
(\ref{ub}),\break
\newpage
\noindent
we have that
\vskip-3pt\noindent
\begin{eqnarray}
u_A(\vec r) &=& \sum_{\vec R_A} g(\vec r-\vec R_A) u(\vec R_A)=
g(\vec r-\vec R_{A_0}) u(\vec R_{A_0}),\\
u'_A(\vec r) &=& \sum_{\vec R_A} g(\vec r-\vec R_A)
e^{i (\vec K'-\vec K)\cdot \vec R_A} u(\vec R_A)=\nonumber\\
&& g(\vec r-\vec R_{A_0})
e^{i (\vec K'-\vec K)\cdot \vec R_{A_0}} u(\vec R_{A_0})=
u_A(\vec r) e^{i (\vec K'-\vec K)\cdot \vec R_{A_0}},\nonumber\\
u_B(\vec r) &=& \sum_{\vec R_B} g(\vec r-\vec R_B) u(\vec R_B)=0,\nonumber\\
u'_B(\vec r) &=& \sum_{\vec R_B} g(\vec r-\vec R_B)
e^{i (\vec K'-\vec K)\cdot \vec R_B} u(\vec R_B)=0.\nonumber
\end{eqnarray}
\vskip-3pt\noindent
Instead, if it is nonzero only on an atom of type B (in position
$\vec R_{B_0}$), i.e. 
$u (\vec R_{B_0}) \ne 0$, $u (\vec R_B)=0$ for $\vec R_B \ne \vec R_{B_0}$ and
$u (\vec R_A)=0$ for every $\vec R_A$, we have that
\vskip-3pt\noindent
\begin{eqnarray}
u_A(\vec r) &=& \sum_{\vec R_A} g(\vec r-\vec R_A) u(\vec R_A)=0,\\
u'_A(\vec r) &=& \sum_{\vec R_A} g(\vec r-\vec R_A)
e^{i (\vec K'-\vec K)\cdot \vec R_A} u(\vec R_A)=0,\nonumber\\
u_B(\vec r) &=& \sum_{\vec R_B} g(\vec r-\vec R_B) u(\vec R_B)=
g(\vec r-\vec R_{B_0}) u(\vec R_{B_0}),\nonumber\\
u'_B(\vec r) &=& \sum_{\vec R_B} g(\vec r-\vec R_B)
e^{i (\vec K'-\vec K)\cdot \vec R_B} u(\vec R_B)=\nonumber\\
&& g(\vec r-\vec R_{B_0})
e^{i (\vec K'-\vec K)\cdot \vec R_{B_0}} u(\vec R_{B_0})=
u_B(\vec r) e^{i (\vec K'-\vec K)\cdot \vec R_{B_0}}.\nonumber
\end{eqnarray}
\vskip-3pt\noindent
If instead the potential range is much larger than the lattice constant
(long-range case), using eq.~(\ref{sum}), (\ref{phase}) and (\ref{smooth}),
we have that
\vskip-3pt\noindent
\begin{eqnarray}
\label{longrange}
\qquad\quad u_A(\vec r)\! &=& \!\sum_{\vec R_A}\! g(\vec r-\vec R_A) u(\vec R_A)\!\simeq\!
\sum_{\vec R_A}\! g(\vec r-\vec R_A) u(\vec r)\!=\!
\left[\sum_{\vec R_A} g(\vec r-\vec R_A)\right]\! u(\vec r)\!=\!u(\vec r),\\
u'_A(\vec r) &=& \sum_{\vec R_A} g(\vec r-\vec R_A)
e^{i (\vec K'-\vec K)\cdot \vec R_A} u(\vec R_A)\simeq
\sum_{\vec R_A} g(\vec r-\vec R_A)
e^{i (\vec K'-\vec K)\cdot \vec R_A} u(\vec r)=\nonumber\\
&& \left[\sum_{\vec R_A} g(\vec r-\vec R_A)
e^{i (\vec K'-\vec K)\cdot \vec R_A}\right] u(\vec r)=0,\nonumber\\
u_B(\vec r) &=& \sum_{\vec R_B} g(\vec r-\vec R_B) u(\vec R_B)\simeq\nonumber\\
&& \sum_{\vec R_B} g(\vec r-\vec R_B) u(\vec r)=
\left[\sum_{\vec R_B} g(\vec r-\vec R_B)\right] u(\vec r)=u(\vec r)=
u_A(\vec r),\nonumber\\
u'_B(\vec r) &=& \sum_{\vec R_B} g(\vec r-\vec R_B)
e^{i (\vec K'-\vec K)\cdot \vec R_B} u(\vec R_B)\simeq
\sum_{\vec R_B} g(\vec r-\vec R_B)
e^{i (\vec K'-\vec K)\cdot \vec R_B} u(\vec r)=\nonumber\\
&& \left[\sum_{\vec R_B} g(\vec r-\vec R_B)
e^{i (\vec K'-\vec K)\cdot \vec R_B}\right] u(\vec r)=0.\nonumber
\end{eqnarray}
Here we have used first (exploiting the hypothesis that the external potential
is a very smooth function in comparison with $g(\vec r)$) the property
(\ref{smooth}) and then (for the quantities inside the square brackets)
the properties (\ref{sum}) and (\ref{phase}) of the function $g(\vec r)$.
In this last case the effect of the external potential on the Hamiltonian
matrix is only to sum the same quantity, $u(\vec r)$, to all the
diagonal elements of the matrix, as expected from the $\vec k \cdot \vec p$
theory (see eq.~(\ref{coupled}), where the external potential was
assumed slowly variable)
\vskip5pt\noindent
\begin{eqnarray}
\label{longrange2}
&& \left[\begin{array}{cccc}
u(\vec r) &
\gamma ({\hat\kappa}_x-i{\hat\kappa}_y) &
0 &
0 \\[6pt]
\gamma ({\hat\kappa}_x+i{\hat\kappa}_y) &
u(\vec r) &
0 &
0 \\[6pt]
0 &
0 &
u(\vec r) &
\gamma ({\hat\kappa}_x+i{\hat\kappa}_y) \\[6pt]
0 &
0 &
\gamma ({\hat\kappa}_x-i{\hat\kappa}_y) &
u(\vec r)
\end{array}\right]
\left[\begin{array}{c}
F_A^{\vec K} (\vec r)\\[6pt]
F_B^{\vec K} (\vec r)\\[6pt]
F_A^{\vec K'} (\vec r)\\[6pt]
F_B^{\vec K'} (\vec r)
\end{array}\right]=\\
&& E \left[\begin{array}{c}
F_A^{\vec K} (\vec r)\\[6pt]
F_B^{\vec K} (\vec r)\\[6pt]
F_A^{\vec K'} (\vec r)\\[6pt]
F_B^{\vec K'} (\vec r)
\end{array}\right].\nonumber
\end{eqnarray}
\vskip5pt\noindent
Let us note that by reordering the elements of the envelope function vector,
we can rewrite this equation in the form
\vskip5pt\noindent
\begin{eqnarray}
&& \left[\begin{array}{cccc}
u(\vec r) &
0 &
0 &
\gamma ({\hat\kappa}_x-i{\hat\kappa}_y) \\[6pt]
0 &
u(\vec r) &
\gamma ({\hat\kappa}_x+i{\hat\kappa}_y) &
0 \\[6pt]
0 &
\gamma ({\hat\kappa}_x-i{\hat\kappa}_y) &
u(\vec r) &
0 \\[6pt]
\gamma ({\hat\kappa}_x+i{\hat\kappa}_y) &
0 &
0 &
u(\vec r)
\end{array}\right]
\left[\begin{array}{c}
F_A^{\vec K} (\vec r)\\[6pt]
F_A^{\vec K'} (\vec r)\\[6pt]
F_B^{\vec K'} (\vec r)\\[6pt]
F_B^{\vec K} (\vec r)
\end{array}\right]=\\
&& E \left[\begin{array}{c}
F_A^{\vec K} (\vec r)\\[6pt]
F_A^{\vec K'} (\vec r)\\[6pt]
F_B^{\vec K'} (\vec r)\\[6pt]
F_B^{\vec K} (\vec r)
\end{array}\right],\nonumber
\end{eqnarray}
\vskip5pt\noindent
which can be more compactly written as
\vskip5pt\noindent
\begin{equation}
\left[\begin{array}{cccc}
u(\vec r)I &
\gamma \vec\sigma\cdot\vec{\hat\kappa}\\[6pt]
\gamma \vec\sigma\cdot\vec{\hat\kappa} &
u(\vec r)I
\end{array}\right]
\left[\begin{array}{c}
F_A^{\vec K} (\vec r)\\[6pt]
F_A^{\vec K'} (\vec r)\\[6pt]
F_B^{\vec K'} (\vec r)\\[6pt]
F_B^{\vec K} (\vec r)
\end{array}\right]=
E \left[\begin{array}{c}
F_A^{\vec K} (\vec r)\\[6pt]
F_A^{\vec K'} (\vec r)\\[6pt]
F_B^{\vec K'} (\vec r)\\[6pt]
F_B^{\vec K} (\vec r)
\end{array}\right]
\end{equation}
\vskip5pt\noindent
(where $I$ is the $2 \times 2$ identity matrix and $\vec\sigma$ is the
vector having as components the Pauli spin matrices $\sigma_x$ and $\sigma_y$
(\ref{pauli})).
This equation is analitically equivalent to the Dirac equation for
massless particles (Weyl's equation) of relativistic quantum
mechanics~\footnotemark{}; therefore eq.~(\ref{longrange2}) is commonly called
the Dirac equation for graphene. Since charge carriers in graphene obey a 
relation identical to that describing the relativistic behavior of elementary
massless spin-(1/2) particles, transport in graphene exhibits many phenomena,
such as Klein's tunneling, analogous to those predicted in relativistic
quantum mechanics~\cite{katsnelson1,katsnelson2,katsnelson3,geim,beenakker}.
\footnotetext{For example, compare this equation with eq.~(3.62) of
ref.~\cite{sakurai}, with $m=0$, $\vec A=0$,
$e A_0=u(\vec r)$, $c$ substituted by $v_F=\gamma/\hbar$,
$\vec \psi_A$ substituted by $[F_A^{\vec K}, \ F_A^{\vec K'}]^{\, T}$,
and $\vec \psi_B$ substituted by $[F_B^{\vec K'}, \ F_B^{\vec K}]^{\, T}$.}

Note that in the presence of a magnetic field the operator 
$\vec{\hat\kappa}=-i\vec\nabla$ which appears in the equation has to be
replaced by $-i\vec\nabla+e\vec A/\hbar$, as we have shown in the general
introduction on the $\vec k \cdot \vec p$ method.

In the absence of an external potential, the quantities $u_A$, $u'_A$,
$u_B$ and $u'_B$ are null and thus the matrix equation becomes
\begin{eqnarray}
\label{absence}
&& \left[\begin{array}{cccc}
0 &
\gamma ({\hat\kappa}_x-i{\hat\kappa}_y) &
0 &
0 \\[3pt]
\gamma ({\hat\kappa}_x+i{\hat\kappa}_y) &
0 &
0 &
0 \\[3pt]
0 &
0 &
0 &
\gamma ({\hat\kappa}_x+i{\hat\kappa}_y) \\[3pt]
0 &
0 &
\gamma ({\hat\kappa}_x-i{\hat\kappa}_y) &
0
\end{array}\right]
\left[\begin{array}{c}
F_A^{\vec K} (\vec r)\\[3pt]
F_B^{\vec K} (\vec r)\\[3pt]
F_A^{\vec K'} (\vec r)\\[3pt]
F_B^{\vec K'} (\vec r)
\end{array}\right]=\\
&& E \left[\begin{array}{c}
F_A^{\vec K} (\vec r)\\[3pt]
F_B^{\vec K} (\vec r)\\[3pt]
F_A^{\vec K'} (\vec r)\\[3pt]
F_B^{\vec K'} (\vec r)
\end{array}\right].\nonumber
\end{eqnarray}
Since in this case the part of equation corresponding to the point $\vec K$
is decoupled from that corresponding to the point $\vec K'$, we can consider
the two parts separately.
%%%%%%%%%%%%%%%%%%%%%%%%%%%%%%%%%%%%%%%%%%%%%%%%%%%%%%%%%%%%%%%%%%%%%%%%%%

In particular, the part of equation corresponding to the point $\vec K$ is
\begin{equation}
\left[\begin{array}{cc}
0 & \gamma ({\hat\kappa}_x-i{\hat\kappa}_y)\\[3pt]
\gamma ({\hat\kappa}_x+i{\hat\kappa}_y) & 0
\end{array}\right]
\left[\begin{array}{cc}
F_A^{\vec K} (\vec r)\\[3pt]
F_B^{\vec K} (\vec r)
\end{array}\right]=
E \left[\begin{array}{cc}
F_A^{\vec K} (\vec r)\\[3pt]
F_B^{\vec K} (\vec r)
\end{array}\right],
\end{equation}
or (using the Pauli spin matrices (\ref{pauli}))
\begin{equation}
\gamma({\hat\kappa}_x \sigma_x+
{\hat\kappa}_y \sigma_y)\vec F^{\vec K} (\vec r)=
\gamma({\vec{\hat\kappa}}\cdot\vec\sigma)
\vec F^{\vec K} (\vec r)=E\vec F^{\vec K} (\vec r).
\end{equation}
This $\vec k \cdot \vec p$ Hamiltonian matrix, converted into the momentum
representation (see eq.~(\ref{momentum})), has as eigenvalues the dispersion
relations of the two degenerate energy bands\break
$E_s^{\vec K}(\vec \kappa)$ and
as eigenvectors the corresponding electron envelope functions 
$F_{s \vec\kappa}^{\vec K} (\vec r)$.

In particular, if we set
\begin{equation}
\det\left\{\left[\begin{array}{cc}
0 & \gamma (\kappa_x-i\kappa_y)\\[3pt]
\gamma (\kappa_x+i\kappa_y) & 0
\end{array}\right]-E
\left[\begin{array}{cc}
1 & 0\\[3pt]
0 & 1
\end{array}\right]\right\}=0,
\end{equation}
we find the dispersion relations
\begin{equation}
E_s^{\vec K}(\vec\kappa)=s \gamma \sqrt{\kappa_x^2+\kappa_y^2}=
s \gamma |\vec \kappa|,
\end{equation}
where $s$ can assume the values $+1$ or $-1$.

If we define the angle $\alpha$ in such a way that
\begin{equation}
\label{alpha}
\kappa_x+i\kappa_y=|\vec\kappa|e^{i(\frac{\pi}{2}+\alpha)}=
i |\vec\kappa| e^{i \alpha}
\end{equation}
and thus
\begin{equation}
\kappa_x-i\kappa_y=(\kappa_x+i\kappa_y)^*=
|\vec\kappa|e^{i(-\frac{\pi}{2}-\alpha)}=
-i |\vec\kappa| e^{-i \alpha},
\end{equation}
we have that the corresponding envelope functions (properly normalized,
as we will see), are
\begin{equation}
\label{fk}
\vec F_{s \vec\kappa}^{\vec K}(\vec r)=
\frac{1}{\sqrt{2\Omega}}e^{i\vec\kappa\cdot\vec r}
e^{i\phi_s (\vec\kappa)} R(-\alpha (\vec\kappa)) |s \rangle,
\end{equation}
with
\begin{equation}
|s \rangle =\frac{1}{\sqrt{2}}\left[\begin{array}{c}
-is\\ 
1
\end{array}\right],
\end{equation}
where $\Omega$ is the considered surface area, $\phi_s (\vec\kappa)$ is an
arbitrary phase factor and $R(\alpha)$ is a spin-rotation operator, given by
\begin{equation}
R(\alpha)=\left[\begin{array}{cc}
e^{i\frac{\alpha}{2}} & 0\\
0 & e^{-i\frac{\alpha}{2}}
\end{array}\right].
\end{equation}
This can be easily verified noting that
\begin{eqnarray}
\label{ver1}
&& \gamma \left[\begin{array}{cc}
0 & \hat\kappa_x-i\hat\kappa_y\\
\hat\kappa_x+i\hat\kappa_y & 0
\end{array}\right] \vec F_{s \vec\kappa}^{\vec K}(\vec r)=\\
&& \gamma \left[\begin{array}{cc}
0 & \hat\kappa_x-i\hat\kappa_y\\
\hat\kappa_x+i\hat\kappa_y & 0
\end{array}\right] 
\frac{1}{\sqrt{2\Omega}}e^{i\vec\kappa\cdot\vec r}
e^{i\phi_s} R(-\alpha (\vec\kappa)) |s \rangle=\nonumber\\
&& \gamma \left[\begin{array}{cc}
0 & \kappa_x-i\kappa_y\\
\kappa_x+i\kappa_y & 0
\end{array}\right]
\frac{1}{\sqrt{2\Omega}}e^{i\vec\kappa\cdot\vec r}
e^{i\phi_s} R(-\alpha (\vec\kappa)) |s \rangle=\nonumber\\
&& \gamma \left[\begin{array}{cc}
0 & -i|\vec\kappa|e^{-i \alpha}\\
i|\vec\kappa|e^{i\alpha} & 0
\end{array}\right]
\left(\frac{1}{\sqrt{2\Omega}}e^{i\vec\kappa\cdot\vec r}
e^{i\phi_s} \left[\begin{array}{cc}
e^{-i\frac{\alpha}{2}} & 0\\
0 & e^{i\frac{\alpha}{2}}
\end{array}\right]
\frac{1}{\sqrt{2}}\left[\begin{array}{c}
-is\\
 1
\end{array}\right] \right)=\nonumber\\
&& \frac{1}{2\sqrt{\Omega}}\gamma e^{i\vec\kappa\cdot\vec r}
e^{i\phi_s} \!\left[\begin{array}{cc}
0 & -i|\vec\kappa|e^{-i \frac{\alpha}{2}}\\ 
i|\vec\kappa|e^{i\frac{\alpha}{2}} & 0
\end{array}\right]\!
\left[\begin{array}{c}
-is\\ 
1
\end{array}\right]\!=\nonumber\\
&& \frac{1}{2\sqrt{\Omega}}\gamma e^{i(\vec\kappa\cdot\vec r+\phi_s)}
\!\left[\begin{array}{c}
-i |\vec\kappa| e^{-i \frac{\alpha}{2}}\\
|\vec\kappa| s e^{i\frac{\alpha}{2}}
\end{array}\right]\nonumber
\end{eqnarray}
and also
\begin{eqnarray}
\label{ver2}
&& E_s^{\vec K} \vec F_{s \vec\kappa}^{\vec K}(\vec r)=
s \gamma |\vec \kappa| \left(\frac{1}{\sqrt{2\Omega}}e^{i\vec\kappa\cdot\vec r}
e^{i\phi_s} \left[\begin{array}{cc}
e^{-i\frac{\alpha}{2}} & 0\\
0 & e^{i\frac{\alpha}{2}}
\end{array}\right]
\frac{1}{\sqrt{2}}\left[\begin{array}{c}
-is\\ 
1
\end{array}\right] \right)=\\
&& s \gamma |\vec \kappa| \frac{1}{2\sqrt{\Omega}}
e^{i(\vec\kappa\cdot\vec r+\phi_s)}
\left[\begin{array}{c}
-i s e^{-i \frac{\alpha}{2}}\\
e^{i\frac{\alpha}{2}}
\end{array}\right]=
\frac{1}{2\sqrt{\Omega}} \gamma e^{i(\vec\kappa\cdot\vec r+\phi_s)}
\left[\begin{array}{c}
-i s^2 |\vec \kappa| e^{-i \frac{\alpha}{2}}\\
|\vec \kappa| s e^{i\frac{\alpha}{2}}
\end{array}\right]=\nonumber\\
&& \frac{1}{2\sqrt{\Omega}}\gamma e^{i(\vec\kappa\cdot\vec r+\phi_s)}
\left[\begin{array}{c}
-i |\vec\kappa| e^{-i \frac{\alpha}{2}}\\
|\vec\kappa| s e^{i\frac{\alpha}{2}}
\end{array}\right]\nonumber
\end{eqnarray}
(where we have used the fact that $s^2=(\pm 1)^2=1$ ). 
%%%%%%%%%%%%%%%%%%%%%%%%%%%%%%%%%%%%%%%%%%%%%%%%%%%%%%%%%%%%%%%%%%%%%%%%%%

Instead, the part of equation corresponding to the point $\vec K'$ is
\begin{equation}
\left[\begin{array}{cc}
0 & \gamma (\hat\kappa_x+i\hat\kappa_y)\\
\gamma (\hat\kappa_x-i\hat\kappa_y) & 0
\end{array}\right]
\left[\begin{array}{c}
F_A^{\vec K'} (\vec r)\\
F_B^{\vec K'} (\vec r)
\end{array}\right]=
E \left[\begin{array}{c}
F_A^{\vec K'} (\vec r)\\
F_B^{\vec K'} (\vec r)
\end{array}\right],
\end{equation}
or equivalently (using the Pauli spin matrices (\ref{pauli}))
\begin{equation}
\gamma(\hat\kappa_x \sigma_x-\hat\kappa_y \sigma_y)\vec F^{\vec K'} (\vec r)=
\gamma\left(\left[\begin{array}{c}
\hat\kappa_x \\
-\hat\kappa_y \\
0
\end{array}\right]
\cdot\vec\sigma\right) \vec F^{\vec K'} (\vec r)=E\vec F^{\vec K'} (\vec r).
\end{equation}
If we move to the momentum representation (see eq.~(\ref{momentum}))
and enforce 
\begin{equation}
\det\left\{\left[\begin{array}{cc}
0 & \gamma (\kappa_x+i\kappa_y)\\
\gamma (\kappa_x-i\kappa_y) & 0
\end{array}\right]-E
\left[\begin{array}{cc}
1 & 0\\ 
0 & 1
\end{array}\right]\right\}=0,
\end{equation}
we find the dispersion relations
\begin{equation}
E_s^{\vec K'}(\vec\kappa)=s \gamma \sqrt{\kappa_x^2+\kappa_y^2}=
s \gamma|\vec \kappa|,
\end{equation}
where $s$ can assume the values $+1$ or $-1$.

The corresponding envelope functions are
\begin{equation}
\label{fk1}
\vec F_{s \vec\kappa}^{\vec K'}(\vec r)=
\frac{1}{\sqrt{2\Omega}}e^{i\vec\kappa\cdot\vec r}
e^{i\tilde\phi_s (\vec\kappa)} R(\alpha (\vec\kappa)) \tilde{| s \rangle},
\end{equation}
with $\tilde\phi_s (\vec\kappa)$ an arbitrary phase factor and
\begin{equation}
\tilde{| s \rangle} =\frac{1}{\sqrt{2}}\left[\begin{array}{c}
is\\ 
1
\end{array}\right].
\end{equation}
\noindent
This result is easily verified in a way completely analogous to
eqs.~(\ref{ver1})-(\ref{ver2})~\cite{supplem}.

From these functions $F_A^{\vec K}$, $F_B^{\vec K}$, $F_A^{\vec K'}$
and $F_B^{\vec K'}$, we can find the functions $\psi_A$ and $\psi_B$
and thus the electron wave function $\psi$ in the absence of an external
potential, using the relations (\ref{assumptions}) and (\ref{wavefunction}).

We notice that the energy dispersion relations that we have found in
this way near $\vec K$ and $\vec K'$ are identical to those one can obtain
by first computing the dispersion relations in the absence of an external
potential by using the nearest-neighbor tight-binding technique, and then
expanding them near the extrema points.
%%%%%%%%%%%%%%%%%%%%%%%%%%%%%%%%%%%%%%%%%%%%%%%%%%%%%%%%%%

Let us now find an expression for the probability density and for the
probability current density in graphene.

The probability to find an electron in a region of area $S$ is equal to
\begin{eqnarray}
&& \int_{S} |\psi(\vec r)|^2 d \vec r=
\int_{S} \psi^*(\vec r) \psi(\vec r) d \vec r=\\
&& \int_{S} \Big[\sum_{\vec R_A}\psi_A^*(\vec R_A)\varphi^*(\vec r -\vec R_A)+
\sum_{\vec R_B}\psi_B^*(\vec R_B)\varphi^*(\vec r -\vec R_B)\Big]\nonumber\\
&& \cdot\Big[\sum_{\vec R_A}\psi_A (\vec R_A)\varphi(\vec r -\vec R_A)+
\sum_{\vec R_B}\psi_B (\vec R_B)\varphi(\vec r -\vec R_B)\Big] d \vec r=\nonumber\\
&& \sum_{\vec R_A \in S} |\psi_A(\vec R_A)|^2 
\int_{S} |\varphi(\vec r -\vec R_A)|^2 d \vec r+
\sum_{\vec R_B \in S} |\psi_B(\vec R_B)|^2 
\int_{S} |\varphi(\vec r -\vec R_B)|^2 d \vec r\simeq\nonumber\\
&& \sum_{\vec R_A \in S} |\psi_A(\vec R_A)|^2 
\int_{\Omega} |\varphi(\vec r -\vec R_A)|^2 d \vec r+
\sum_{\vec R_B \in S} |\psi_B(\vec R_B)|^2 
\int_{\Omega} |\varphi(\vec r -\vec R_B)|^2 d \vec r \nonumber
\end{eqnarray}
($\Omega$ is the area of the whole graphene sheet),
where we have exploited the fact that each atomic orbital $\varphi$ 
has a non-zero overlap only with itself (since we use L\"owdin orthonormalized
atomic orbitals) and has significant values only
near the atom on which it is centered. If the atomic orbital $\varphi$ is
normalized according to (\ref{norm}), the integral of its square modulus on
$\Omega$ is equal to the unit cell area $\Omega_0$ (otherwise, if the usual
normalization for $\varphi$ is adopted, this integral is equal to 1 and
the following results just have to be divided by the constant $\Omega_0$).
Therefore, in this case we have that
\begin{equation}
\int_{S} |\psi(\vec r)|^2 d \vec r\simeq
\Omega_0 \sum_{\vec R_A \in S} |\psi_A(\vec R_A)|^2+
\Omega_0 \sum_{\vec R_B \in S} |\psi_B(\vec R_B)|^2.
\end{equation}
Using the relations (\ref{assumptions}), we have that
\begin{eqnarray}
&& \sum_{\vec R_A} |\psi_A(\vec R_A)|^2=
\sum_{\vec R_A} \psi_A^*(\vec R_A) \psi_A(\vec R_A)=\\
&& \sum_{\vec R_A}
\Big\{\Big[e^{-i \vec K\cdot \vec R_A} {F_A^{\vec K}}^*(\vec R_A)
+i \, e^{-i \theta'} e^{-i \vec K'\cdot \vec R_A}
{F_A^{\vec K'}}^*(\vec R_A)\Big]\nonumber\\
&& \cdot\Big[e^{i \vec K\cdot \vec R_A} F_A^{\vec K}(\vec R_A)
-i \, e^{i \theta'} e^{i \vec K'\cdot \vec R_A}
F_A^{\vec K'}(\vec R_A)\Big]\Big\}=\nonumber\\
&& \sum_{\vec R_A} |F_A^{\vec K}(\vec R_A)|^2
+\sum_{\vec R_A} |F_A^{\vec K'}(\vec R_A)|^2\nonumber\\
&& -i \, e^{i \theta'} \sum_{\vec R_A}
\Big[e^{i (\vec K'-\vec K) \cdot \vec R_A}
{F_A^{\vec K}}^*(\vec R_A) F_A^{\vec K'}(\vec R_A)\Big]\nonumber\\
&& +i \, e^{-i \theta'} \sum_{\vec R_A}
\Big[e^{-i (\vec K'-\vec K) \cdot \vec R_A}
{F_A^{\vec K'}}^*(\vec R_A) F_A^{\vec K}(\vec R_A)\Big]\nonumber
\end{eqnarray}
and that
\begin{eqnarray}
&& \sum_{\vec R_B} |\psi_B(\vec R_B)|^2=
\sum_{\vec R_B} \psi_B^*(\vec R_B) \psi_B(\vec R_B)=\\
&& \sum_{\vec R_B}
\Big\{\Big[-i \, e^{-i \theta'} 
e^{-i \vec K\cdot \vec R_B} {F_B^{\vec K}}^*(\vec R_B)+
e^{-i \vec K'\cdot \vec R_B} {F_B^{\vec K'}}^*(\vec R_B)
\Big]\nonumber\\
&& \cdot\Big[i \, e^{i \theta'} 
e^{i \vec K\cdot \vec R_B} F_B^{\vec K} (\vec R_B)+
e^{i \vec K'\cdot \vec R_B} F_B^{\vec K'} (\vec R_B)
\Big]\Big\}=\nonumber\\
&& \sum_{\vec R_B} |F_B^{\vec K}(\vec R_B)|^2
+\sum_{\vec R_B} |F_B^{\vec K'}(\vec R_B)|^2\nonumber\\
&& -i \, e^{-i \theta'} \sum_{\vec R_B}
\Big[e^{i (\vec K'-\vec K) \cdot \vec R_B}
{F_B^{\vec K}}^*(\vec R_B) F_B^{\vec K'}(\vec R_B)\Big]\nonumber\\
&& +i \, e^{i \theta'} \sum_{\vec R_B}
\Big[e^{-i (\vec K'-\vec K) \cdot \vec R_B}
{F_B^{\vec K'}}^*(\vec R_B) F_B^{\vec K}(\vec R_B)\Big].\nonumber
\end{eqnarray}
However the terms containing the phase factors
$e^{i (\vec K'-\vec K) \cdot \vec R_A}$,
$e^{i (\vec K'-\vec K) \cdot \vec R_B}$, or their
complex conjugates are negligible with respect to the others.

Indeed, using the smoothing function $g(\vec r)$, we know from
the property (\ref{sum}) with $\vec r=\vec R_A$ that 
$\sum_{\vec R_A'} g(\vec R_A-\vec R_A')=1$. Therefore we
can insert this sum into the term
\begin{equation}
\sum_{\vec R_A} \Big[e^{i (\vec K'-\vec K) \cdot \vec R_A}
{F_A^{\vec K}}^*(\vec R_A) F_A^{\vec K'}(\vec R_A)\Big],
\end{equation}
obtaining
\begin{equation}
\sum_{\vec R_A} \left\{\left[\sum_{\vec R_A'} g(\vec R_A-\vec R_A')\right]
e^{i (\vec K'-\vec K) \cdot \vec R_A}
{F_A^{\vec K}}^*(\vec R_A) F_A^{\vec K'}(\vec R_A)\right\},
\end{equation}
that can be rewritten, as a result of the point-symmetry of the function $g$
with respect to its center and thus of the fact that $g(\vec R_A-\vec R_A')=
g(-(\vec R_A-\vec R_A'))$, in this way:
\begin{equation}
\sum_{\vec R_A} \sum_{\vec R_A'} g(\vec R_A'-\vec R_A)
e^{i (\vec K'-\vec K) \cdot \vec R_A}
{F_A^{\vec K}}^*(\vec R_A) F_A^{\vec K'}(\vec R_A).
\end{equation}
If then we use the property (\ref{smooth}) with $\vec r=\vec R_A'$
and in particular the fact that
\begin{equation}
g(\vec R_A'-\vec R_A) {F_A^{\vec K}}^*(\vec R_A) F_A^{\vec K'}(\vec R_A)
=g(\vec R_A'-\vec R_A) {F_A^{\vec K}}^*(\vec R_A') F_A^{\vec K'}(\vec R_A')
\end{equation}
(due to the smoothness of the envelope functions), the term becomes
\begin{equation}
\sum_{\vec R_A'} \Big[ \sum_{\vec R_A} g(\vec R_A'-\vec R_A)
e^{i (\vec K'-\vec K) \cdot \vec R_A} \Big]
{F_A^{\vec K}}^*(\vec R_A') F_A^{\vec K'}(\vec R_A')
\end{equation}
and, by way of the property (\ref{phase}) with $\vec r=\vec R_A'$, we conclude
that the quantities between square brackets, and thus the overall term,
are very small.

Analogously, we can see that the terms
\begin{eqnarray}
&& \sum_{\vec R_A} \Big[e^{-i (\vec K'-\vec K) \cdot \vec R_A}
{F_A^{\vec K'}}^*(\vec R_A) F_A^{\vec K}(\vec R_A)\Big],\nonumber\\
&& \sum_{\vec R_B} \Big[e^{i (\vec K'-\vec K) \cdot \vec R_B}
{F_B^{\vec K}}^*(\vec R_B) F_B^{\vec K'}(\vec R_B)\Big]\nonumber
\end{eqnarray}
and
\begin{equation}
\sum_{\vec R_B} \Big[e^{-i (\vec K'-\vec K) \cdot \vec R_B}
{F_B^{\vec K'}}^*(\vec R_B) F_B^{\vec K}(\vec R_B)\Big]\nonumber
\end{equation}
are negligible~\cite{supplem}.
Since $g(\vec r)$ has non-negligible values only within
a few lattice constants from its center, the previous considerations are
approximately valid also if we limit the sums to the atoms contained in 
the area $S$.

We conclude that
\begin{eqnarray}
&& \int_{S} |\psi(\vec r)|^2 d \vec r \simeq
\Omega_0 \sum_{\vec R_A \in S} |\psi_A(\vec R_A)|^2+
\Omega_0 \sum_{\vec R_B \in S} |\psi_B(\vec R_B)|^2\simeq\\
&& \Omega_0 \sum_{\vec R_A \in S} |F_A^{\vec K}(\vec R_A)|^2+
\Omega_0 \sum_{\vec R_A \in S} |F_A^{\vec K'}(\vec R_A)|^2\nonumber\\
&& +\Omega_0 \sum_{\vec R_B \in S} |F_B^{\vec K}(\vec R_B)|^2+
\Omega_0 \sum_{\vec R_B \in S} |F_B^{\vec K'}(\vec R_B)|^2 \simeq \nonumber\\
&& \int_{S} \Big[ |F_A^{\vec K}(\vec r)|^2+|F_A^{\vec K'}(\vec r)|^2+
|F_B^{\vec K}(\vec r)|^2+|F_B^{\vec K'}(\vec r)|^2 \Big] d \vec r,\nonumber
\end{eqnarray}
where we have exploited the fact that the envelope functions $F$ are
smooth functions, which are nearly constant over a unit cell. Therefore
we can consider
\begin{equation}
\label{p}
P=|F_A^{\vec K}(\vec r)|^2+|F_A^{\vec K'}(\vec r)|^2+
|F_B^{\vec K}(\vec r)|^2+|F_B^{\vec K'}(\vec r)|^2
\end{equation}
as a probability density, and the correct normalization condition is
\begin{equation}
\label{normalization}
\int_{\Omega} \left( |F_A^{\vec K}(\vec r)|^2+|F_A^{\vec K'}(\vec r)|^2+
|F_B^{\vec K}(\vec r)|^2+|F_B^{\vec K'}(\vec r)|^2 \right) d \vec r=1.
\end{equation}
We now follow a procedure similar to that used in relativistic quantum
mechanics \cite{sakurai} to find the expression of the probability current
density.
Let us consider the envelope function equation in the case of long-range
external potential (eq.~(\ref{longrange2})), writing explicitly the operators
${\hat\kappa}_x$ and ${\hat\kappa}_y$ (see eq.~(\ref{kappa})). Let us
consider the time-dependent wave function $\psi(\vec r,t)$ and thus the
time-dependent envelope functions $\vec F(\vec r,t)$
($\vec F$ will be the column vector
$[F_A^{\vec K}, \ 
F_B^{\vec K}, \ 
F_A^{\vec K'}, \
F_B^{\vec K'}]^{\, T}$).
We now convert the 
time-independent envelope function equation into a time-dependent envelope
function equation, substituting in the r.h.s. of eq.(\ref{longrange2})
the quantity $E \vec F(\vec r)$ with
$i\,\hbar\,(\partial \vec F(\vec r,t) / \partial t)$
(for stationary states $\psi(\vec r,t)=\psi(\vec r) e^{-iEt/\hbar}$,
$\vec F(\vec r,t)=\vec F(\vec r) e^{-iEt/\hbar}$, and thus the time-dependent
equation is clearly equivalent to the time-independent one).
Therefore we can write
\begin{eqnarray}
&& \gamma \left[\begin{array}{cccc}
0 &
-i\,\frac{\partial}{\partial x}-\frac{\partial}{\partial y} &
0 &
0 \\[3pt]
-i\,\frac{\partial}{\partial x}+\frac{\partial}{\partial y} &
0 &
0 &
0 \\[3pt]
0 &
0 &
0 &
-i\,\frac{\partial}{\partial x}+\frac{\partial}{\partial y} \\[3pt]
0 &
0 &
-i\,\frac{\partial}{\partial x}-\frac{\partial}{\partial y} &
0
\end{array}\right]
\left[\begin{array}{c}
F_A^{\vec K}\\[3pt]
F_B^{\vec K}\\[3pt]
F_A^{\vec K'}\\[3pt]
F_B^{\vec K'}
\end{array}\right]\\
&& +u(\vec r)
\left[\begin{array}{c}
F_A^{\vec K}\\[3pt]
F_B^{\vec K}\\[3pt]
F_A^{\vec K'}\\[3pt]
F_B^{\vec K'}
\end{array}\right]=
i\,\hbar\,\frac{\partial}{\partial t}\left[\begin{array}{c}
F_A^{\vec K}\\[3pt]
F_B^{\vec K}\\[3pt]
F_A^{\vec K'}\\[3pt]
F_B^{\vec K'}
\end{array}\right].\nonumber
\end{eqnarray}
Dividing by $\gamma$ and using the Pauli matrices (\ref{pauli}), we can
rewrite the equation in this form (in the following we will indicate
with $I$ the $2 \times 2$ identity matrix):
\begin{eqnarray}
\label{eqcu1}
&& \left[\begin{array}{cc}
-i\,\sigma_x & 0 \\[3pt]
0 & -i\,\sigma_x
\end{array}\right] \left(\frac{\partial}{\partial x} \vec F \right)+
\left[\begin{array}{cc}
-i\,\sigma_y & 0 \\[3pt]
0 & i\,\sigma_y
\end{array}\right] \left(\frac{\partial}{\partial y} \vec F \right)\\
&& -\frac{i\,\hbar}{\gamma}
\left[\begin{array}{cc}
I & 0 \\[3pt]
0 & I
\end{array}\right] \left(\frac{\partial}{\partial t} \vec F \right)+
\frac{u(\vec r)}{\gamma}
\left[\begin{array}{cc}
I & 0 \\[3pt]
0 & I
\end{array}\right] \vec F =0\nonumber
\end{eqnarray}
that, if we define
\begin{equation}
A=\left[\begin{array}{cc}
i\,\sigma_x & 0 \\[3pt]
0 & i\,\sigma_x
\end{array}\right]
,\quad
B=\left[\begin{array}{cc}
i\,\sigma_y & 0 \\[3pt]
0 & -i\,\sigma_y
\end{array}\right],
\end{equation}
we can rewrite in this compact way:
\begin{equation}
-A \left(\frac{\partial}{\partial x} \vec F \right)
-B \left(\frac{\partial}{\partial y} \vec F \right)
-\frac{i\,\hbar}{\gamma}
\left(\frac{\partial}{\partial t} \vec F \right)
+\frac{u(\vec r)}{\gamma} \vec F =0.
\end{equation}
If we left-multiply this equation by the row vector $\vec F^\dagger$
(the conjugate transpose of $\vec F$), we obtain:
\begin{equation}
\label{eqcu2}
-\vec F^\dagger A \left(\frac{\partial}{\partial x} \vec F \right)
-\vec F^\dagger B \left(\frac{\partial}{\partial y} \vec F \right)
-\frac{i\,\hbar}{\gamma}
\vec F^\dagger \left(\frac{\partial}{\partial t} \vec F \right)
+\frac{u(\vec r)}{\gamma} \vec F^\dagger \vec F =0.
\end{equation}
Instead, if we consider the conjugate transpose of eq.~(\ref{eqcu1}) we
obtain
\begin{eqnarray}
&& \left(\frac{\partial}{\partial x} \vec F^\dagger \right)
\left[\begin{array}{cc}
i\,\sigma_x^\dagger & 0 \\[3pt]
0 & i\,\sigma_x^\dagger
\end{array}\right]+
\left(\frac{\partial}{\partial y} \vec F^\dagger \right)
\left[\begin{array}{cc}
i\,\sigma_y^\dagger & 0 \\[3pt]
0 & -i\,\sigma_y^\dagger
\end{array}\right]\\
&& +\frac{i\,\hbar}{\gamma}
\left(\frac{\partial}{\partial t} \vec F^\dagger \right)
\left[\begin{array}{cc}
I & 0 \\[3pt]
0 & I
\end{array}\right]+
\frac{u(\vec r)}{\gamma}
\vec F^\dagger
\left[\begin{array}{cc}
I & 0 \\[3pt]
0 & I
\end{array}\right] =0,\nonumber
\end{eqnarray}
which, since $\sigma_x^\dagger=\sigma_x$ and $\sigma_y^\dagger=\sigma_y$,
is equal to
\begin{equation}
\left(\frac{\partial}{\partial x} \vec F^\dagger \right) A+
\left(\frac{\partial}{\partial y} \vec F^\dagger \right) B+
\frac{i\,\hbar}{\gamma}
\left(\frac{\partial}{\partial t} \vec F^\dagger \right)+
\frac{u(\vec r)}{\gamma} \vec F^\dagger =0.
\end{equation}
If we right-multiply this equation by the column vector $\vec F$, we obtain
\begin{equation}
\label{eqcu3}
\left(\frac{\partial}{\partial x} \vec F^\dagger \right) A \vec F+
\left(\frac{\partial}{\partial y} \vec F^\dagger \right) B \vec F+
\frac{i\,\hbar}{\gamma}
\left(\frac{\partial}{\partial t} \vec F^\dagger \right)\vec F+
\frac{u(\vec r)}{\gamma} \vec F^\dagger \vec F=0.
\end{equation}
Subtracting (\ref{eqcu2}) from (\ref{eqcu3}), we find
\begin{eqnarray}
&& \left[\left(\frac{\partial}{\partial x} \vec F^\dagger \right) A \vec F+
\vec F^\dagger A \left(\frac{\partial}{\partial x} \vec F \right)\right]+
\left[\left(\frac{\partial}{\partial y} \vec F^\dagger \right) B \vec F+
\vec F^\dagger B \left(\frac{\partial}{\partial y} \vec F \right)\right]\\
&& +\frac{i\,\hbar}{\gamma}
\left[\left(\frac{\partial}{\partial t} \vec F^\dagger \right)\vec F+
\vec F^\dagger \left(\frac{\partial}{\partial t} \vec F \right)\right]=0
\Rightarrow\nonumber\\
&& \frac{\partial}{\partial x}(\vec F^\dagger A \vec F)+
\frac{\partial}{\partial y}(\vec F^\dagger B \vec F)+
\frac{i\,\hbar}{\gamma}
\frac{\partial}{\partial t}(\vec F^\dagger \vec F)=0.\nonumber
\end{eqnarray}
Since $\vec F^\dagger \vec F=P$ (probability density), we have that
(defining $v_F=\gamma/\hbar$)
\begin{eqnarray}
&& -\frac{\partial}{\partial t} P=
-\frac{\partial}{\partial t}(\vec F^\dagger \vec F)=
-i\left(\frac{\gamma}{\hbar}\right) 
\vec\nabla \cdot \left[(\vec F^\dagger A \vec F) \hbox{\boldmath{$\hat x$}}+
(\vec F^\dagger B \vec F) \hbox{\boldmath{$\hat y$}} \right]= \\
&& \vec\nabla \cdot \left[(-i\,v_F\,\vec F^\dagger A \vec F)
\hbox{\boldmath{$\hat x$}}+
(-i\,v_F\,\vec F^\dagger B \vec F) \hbox{\boldmath{$\hat y$}} \right]=
\vec\nabla\cdot\vec J,\nonumber
\end{eqnarray}
which is the well-known continuity equation, if we define as probability
current density the vector
\begin{equation}
\vec J=\left[\begin{array}{c}
J_x \\
J_y
\end{array}\right]=
\left[\begin{array}{c}
-i\,v_F\,\vec F^\dagger A \vec F \\
-i\,v_F\,\vec F^\dagger B \vec F
\end{array}\right].
\end{equation}
In particular, we have that
\begin{eqnarray}
\label{jx}
J_x &=&-i\,v_F\,\vec F^\dagger A \vec F=\\
&& -i\,v_F\,
\left[\begin{array}{cccc}
{F_A^{\vec K}}^* &
{F_B^{\vec K}}^* &
{F_A^{\vec K'}}^* &
{F_B^{\vec K'}}^* 
\end{array}\right]
\left[\begin{array}{cccc}
0 & i & 0 & 0 \\[3pt]
i & 0 & 0 & 0 \\[3pt]
0 & 0 & 0 & i \\[3pt]
0 & 0 & i & 0
\end{array}\right]
\left[\begin{array}{c}
F_A^{\vec K} \\[3pt]
F_B^{\vec K} \\[3pt]
F_A^{\vec K'} \\[3pt]
F_B^{\vec K'} 
\end{array}\right]= \nonumber\\
&& -i\,v_F\,
\left[\begin{array}{cccc}
{F_A^{\vec K}}^* &
{F_B^{\vec K}}^* &
{F_A^{\vec K'}}^* &
{F_B^{\vec K'}}^* 
\end{array}\right]
\left[\begin{array}{c}
i\,F_B^{\vec K} \\[3pt]
i\,F_A^{\vec K} \\[3pt]
i\,F_B^{\vec K'} \\[3pt]
i\,F_A^{\vec K'} 
\end{array}\right]= \nonumber\\
&& v_F\,\left({F_A^{\vec K}}^* F_B^{\vec K}+
{F_B^{\vec K}}^* F_A^{\vec K}+
{F_A^{\vec K'}}^* F_B^{\vec K'}+
{F_B^{\vec K'}}^* F_A^{\vec K'}\right) \nonumber
\end{eqnarray}
and that
\begin{eqnarray}
\label{jy}
\quad J_y &=& -i\,v_F\,\vec F^\dagger B \vec F=\\
&& -i\,v_F\,
\left[\begin{array}{cccc}
{F_A^{\vec K}}^* &
{F_B^{\vec K}}^* &
{F_A^{\vec K'}}^* &
{F_B^{\vec K'}}^* 
\end{array}\right]
\left[\begin{array}{cccc}
0 & 1 & 0 & 0 \\[3pt]
-1 & 0 & 0 & 0 \\[3pt]
0 & 0 & 0 & -1 \\[3pt]
0 & 0 & 1 & 0
\end{array}\right]
\left[\begin{array}{c}
F_A^{\vec K} \\[3pt]
F_B^{\vec K} \\[3pt]
F_A^{\vec K'} \\[3pt]
F_B^{\vec K'} 
\end{array}\right]= \nonumber\\
&& -i\,v_F\,
\left[\begin{array}{cccc}
{F_A^{\vec K}}^* &
{F_B^{\vec K}}^* &
{F_A^{\vec K'}}^* &
{F_B^{\vec K'}}^* 
\end{array}\right]
\left[\begin{array}{c}
F_B^{\vec K} \\[3pt]
-F_A^{\vec K} \\[3pt]
-F_B^{\vec K'} \\[3pt]
F_A^{\vec K'} 
\end{array}\right]= \nonumber\\
&& -i\,v_F\,\left({F_A^{\vec K}}^* F_B^{\vec K}-
{F_B^{\vec K}}^* F_A^{\vec K}-
{F_A^{\vec K'}}^* F_B^{\vec K'}+
{F_B^{\vec K'}}^* F_A^{\vec K'}\right).\nonumber
\end{eqnarray}
We note that a different ordering of the elements inside the envelope function
vector is often used~\cite{akhmerov1,beenakker}, in which, instead of
$\vec F$, the vector 
$\vec{\tilde F}=
[F_A^{\vec K} (\vec r), \ 
F_B^{\vec K} (\vec r),$
$F_B^{\vec K'} (\vec r), \
F_A^{\vec K'} (\vec r)]^{\, T}$ is considered.
Consequently, the $\vec k \cdot \vec p$ equation in the case of long-range
external potential (\ref{longrange2}) can be rewritten in this way:
\begin{eqnarray}
&& \left[\begin{array}{cccc}
u(\vec r) &
\gamma ({\hat\kappa}_x-i{\hat\kappa}_y) &
0 &
0 \\[3pt]
\gamma ({\hat\kappa}_x+i{\hat\kappa}_y) &
u(\vec r) &
0 &
0 \\[3pt]
0 &
0 &
u(\vec r) &
\gamma ({\hat\kappa}_x-i{\hat\kappa}_y) \\[3pt]
0 &
0 &
\gamma ({\hat\kappa}_x+i{\hat\kappa}_y) &
u(\vec r)
\end{array}\right]
\left[\begin{array}{c}
F_A^{\vec K} (\vec r)\\[3pt]
F_B^{\vec K} (\vec r)\\[3pt]
F_B^{\vec K'} (\vec r)\\[3pt]
F_A^{\vec K'} (\vec r)
\end{array}\right]=\\
&& E \left[\begin{array}{c}
F_A^{\vec K} (\vec r)\\[3pt]
F_B^{\vec K} (\vec r)\\[3pt]
F_B^{\vec K'} (\vec r)\\[3pt]
F_A^{\vec K'} (\vec r)
\end{array}\right],\nonumber
\end{eqnarray}
which is the so-called valley-isotropic representation of the Dirac equation,
characterized by two identical $2 \times 2$ submatrices corresponding to the
two valleys $\vec K$ and $\vec K'$.

Following this representation, the previously obtained expressions for
the probability current density can be compactly restated in this form:
\begin{equation}
\vec J=v_F\,\vec{\tilde F}^\dagger (I\otimes\vec\sigma) \vec{\tilde F},\nonumber
\end{equation}
where $I\otimes\vec\sigma$ is the Kronecker product between the $2 \times 2$
identity matrix $I$ and the vector $\vec\sigma$ of Pauli matrices.
Indeed, the resulting $x$ and $y$ components of $\vec J$ are
\begin{eqnarray}
J_x &=& v_F\,\vec{\tilde F}^\dagger (I\otimes\vec\sigma_x) \vec{\tilde F}=\\
&& v_F \,\left[\begin{array}{cccc}
{F_A^{\vec K}}^* &
{F_B^{\vec K}}^* &
{F_B^{\vec K'}}^* &
{F_A^{\vec K'}}^*
\end{array}\right]
\left[\begin{array}{cccc}
0 & 1 & 0 & 0 \\[6pt]
1 & 0 & 0 & 0 \\[6pt]
0 & 0 & 0 & 1 \\[6pt]
0 & 0 & 1 & 0
\end{array}\right]
\left[\begin{array}{c}
F_A^{\vec K} \\[6pt]
F_B^{\vec K} \\[6pt]
F_B^{\vec K'} \\[6pt]
F_A^{\vec K'}
\end{array}\right]= \nonumber\\[6pt]
&& v_F\,\left({F_A^{\vec K}}^* F_B^{\vec K}+
{F_B^{\vec K}}^* F_A^{\vec K}+
{F_B^{\vec K'}}^* F_A^{\vec K'}+
{F_A^{\vec K'}}^* F_B^{\vec K'}\right);\nonumber\\[6pt]
J_y &=& v_F\,\vec{\tilde F}^\dagger (I\otimes\vec\sigma_y) \vec{\tilde F}=\nonumber\\[6pt]
&& v_F \,\left[\begin{array}{cccc}
{F_A^{\vec K}}^* &
{F_B^{\vec K}}^* &
{F_B^{\vec K'}}^* &
{F_A^{\vec K'}}^*
\end{array}\right]
\left[\begin{array}{cccc}
0 & -i & 0 & 0 \\[6pt]
i & 0 & 0 & 0 \\[6pt]
0 & 0 & 0 & -i \\[6pt]
0 & 0 & i & 0
\end{array}\right]
\left[\begin{array}{c}
F_A^{\vec K} \\[6pt]
F_B^{\vec K} \\[6pt]
F_B^{\vec K'} \\[6pt]
F_A^{\vec K'}
\end{array}\right]= \nonumber\\[6pt]
&& -i\,v_F\,\left({F_A^{\vec K}}^* F_B^{\vec K}-
{F_B^{\vec K}}^* F_A^{\vec K}+
{F_B^{\vec K'}}^* F_A^{\vec K'}-
{F_A^{\vec K'}}^* F_B^{\vec K'}\right),\nonumber
\end{eqnarray}
\vskip6pt\noindent
which coincide with eqs.~(\ref{jx})-(\ref{jy}).

%%%%%%%%%%%%%%%%%%%%%%%%%%%%%%%%%%%%%%%%%%%%%%%%%%%%%%%%%%
It is useful to notice that the Dirac equation in the absence of an
external potential is not satisfied only by the eigenvector
$\vec F (\vec r)=[F_A^{\vec K} (\vec r), \ 
F_B^{\vec K} (\vec r), \ 
F_A^{\vec K'} (\vec r), \
F_B^{\vec K'} (\vec r)]^{\, T}$
with eigenvalue $E$ (as we see in (\ref{absence})), but is
satisfied also by the eigenvector
$\vec F_1 (\vec r)=[F_A^{\vec K} (\vec r), \ 
-F_B^{\vec K} (\vec r), \ 
F_A^{\vec K'} (\vec r), \
-F_B^{\vec K'} (\vec r)]^{\, T}$
with eigenvalue $-E$, since (\ref{absence}) is equivalent to
\vskip6pt\noindent
\begin{eqnarray}
&& \left[\begin{array}{cccc}
0 &
\gamma ({\hat\kappa}_x-i{\hat\kappa}_y) &
0 &
0 \\[6pt]
\gamma ({\hat\kappa}_x+i{\hat\kappa}_y) &
0 &
0 &
0 \\[6pt]
0 &
0 &
0 &
\gamma ({\hat\kappa}_x+i{\hat\kappa}_y) \\[6pt]
0 &
0 &
\gamma ({\hat\kappa}_x-i{\hat\kappa}_y) &
0
\end{array}\right]
\left[\begin{array}{c}
F_A^{\vec K} (\vec r)\\[6pt]
-F_B^{\vec K} (\vec r)\\[6pt]
F_A^{\vec K'} (\vec r)\\[6pt]
-F_B^{\vec K'} (\vec r)
\end{array}\right]=\\[6pt]
&& -E \left[\begin{array}{c}
F_A^{\vec K} (\vec r)\\[6pt]
-F_B^{\vec K} (\vec r)\\[6pt]
F_A^{\vec K'} (\vec r)\\[6pt]
-F_B^{\vec K'} (\vec r)
\end{array}\right].\nonumber
\end{eqnarray}
\vskip6pt\noindent
The wave functions $\psi (\vec r)$ and $\psi_1 (\vec r)$ corresponding
to the envelope functions $\vec F (\vec r)$ and $\vec F_1 (\vec r)$
therefore have opposite energies and thus, being
(see eq.~(\ref{hamiltonian})) eigenfunctions of the Hermitian operator $H$
corresponding to different eigenvalues, are orthogonal. But,\break
\newpage
\noindent
due to the
form of $\vec F (\vec r)$ and $\vec F_1 (\vec r)$ and to
eq.~(\ref{assumptions2}), we see that $\psi (\vec r)$ and $\psi_1 (\vec r)$
have the same $\psi_A (\vec r)$ but opposite $\psi_B (\vec r)$.
Therefore, if we write the orthogonality relation between $\psi (\vec r)$ and
$\psi_1 (\vec r)$, we have that
\begin{eqnarray}
\quad 0 &=& \int_\Omega \psi (\vec r)^* \psi_1 (\vec r) d \vec r=\\
&& \int_\Omega \Big[\sum_{\vec R_A}\psi_A (\vec R_A)\varphi(\vec r -\vec R_A)+
\sum_{\vec R_B}\psi_B (\vec R_B)\varphi(\vec r -\vec R_B)\Big]^* \nonumber\\
&& \cdot\Big[\sum_{\vec R_A}\psi_A (\vec R_A)\varphi(\vec r -\vec R_A)-
\sum_{\vec R_B}\psi_B (\vec R_B)\varphi(\vec r -\vec R_B)\Big] d \vec r=\nonumber\\
&& \sum_{\vec R_A} |\psi_A (\vec R_A)|^2
\int_\Omega |\varphi(\vec r -\vec R_A)|^2 d \vec r-
\sum_{\vec R_B} |\psi_B (\vec R_B)|^2
\int_\Omega |\varphi(\vec r -\vec R_B)|^2 d \vec r=\nonumber\\
&& \sum_{\vec R_A} |\psi_A (\vec R_A)|^2 \Omega_0-
\sum_{\vec R_B} |\psi_B (\vec R_B)|^2 \Omega_0 \Rightarrow\nonumber\\
&& \Omega_0 \sum_{\vec R_A} |\psi_A (\vec R_A)|^2 =
\Omega_0 \sum_{\vec R_B} |\psi_B (\vec R_B)|^2,\nonumber
\end{eqnarray}
where we have exploited the fact that each atomic wave function $\varphi$
has a non-zero overlap only with itself and has been normalized according to
(\ref{norm}).
Since (as we have seen)
\begin{eqnarray}
&& \Omega_0 \sum_{\vec R_A} |\psi_A (\vec R_A)|^2\simeq
\Omega_0 \sum_{\vec R_A} |F_A^{\vec K} (\vec R_A)|^2+
\Omega_0 \sum_{\vec R_A} |F_A^{\vec K'} (\vec R_A)|^2\simeq\\
&& \int_{\Omega} \left(
|F_A^{\vec K} (\vec r)|^2+|F_A^{\vec K'} (\vec r)|^2
\right)d \vec r, \nonumber\\
&& \Omega_0 \sum_{\vec R_B} |\psi_B (\vec R_B)|^2\simeq
\Omega_0 \sum_{\vec R_B} |F_B^{\vec K} (\vec R_B)|^2+
\Omega_0 \sum_{\vec R_B} |F_B^{\vec K'} (\vec R_B)|^2\simeq\nonumber\\
&& \int_{\Omega} \left(
|F_B^{\vec K} (\vec r)|^2+|F_B^{\vec K'} (\vec r)|^2
\right)d \vec r,\nonumber
\end{eqnarray}
we conclude that
\begin{equation}
\int_{\Omega} \left(
|F_A^{\vec K} (\vec r)|^2+|F_A^{\vec K'} (\vec r)|^2
\right)d \vec r=
\int_{\Omega} \left(
|F_B^{\vec K} (\vec r)|^2+|F_B^{\vec K'} (\vec r)|^2
\right)d \vec r
\end{equation}
and this means that in the absence of an external potential the
normalization (\ref{normalization}) is equivalent to
\begin{equation}
\left\{ \begin{array}{l}
\displaystyle
\int_{\Omega} \left(
|F_A^{\vec K} (\vec r)|^2+|F_A^{\vec K'} (\vec r)|^2
\right)d \vec r=\frac{1}{2},\\
\\
\displaystyle
\int_{\Omega} \left(
|F_B^{\vec K} (\vec r)|^2+|F_B^{\vec K'} (\vec r)|^2
\right)d \vec r=\frac{1}{2}
\end{array}\right.
\end{equation}
(the expressions of the envelope functions previously written for graphene 
in the absence of an external potential satisfy this normalization
criterion).
%%%%%%%%%%%%%%%%%%%%%%%%%%%%%%%%%%%%%%%%%%%%%%%%%%%%%%%%%%

\section{Application of the $\vec k \cdot \vec p$ method to carbon
nanotubes}

A single-wall carbon nanotube can be described as a graphite sheet rolled, 
along one of its lattice translational vectors (the vector $\vec C_h$
shown in fig.~\ref{f7}),
into a cylindrical shape~\cite{saito}. In particular, it is completely
specified by the so-called chiral vector $\vec C_h$, which corresponds 
to a section of the nanotube perpendicular to the nanotube axis and thus 
has a length equal to the nanotube circumference and connects two points 
of the graphene sheet which coincide in the nanotube. This vector can
be expressed as a linear combination of the real space unit vectors of 
graphene with integer coefficients $n$ and $m$
\begin{equation}
\vec C_h=n\vec a_1 +m\vec a_2 \mathrel{\mathop\equiv_{\Sigma'}}
n a \left[\begin{array}{c}
\displaystyle \frac{\sqrt{3}}{2}\\[7pt]
\noalign{\vskip3pt}
\displaystyle \frac{1}{2}\\[7pt]
\noalign{\vskip3pt}
0
\end{array}\right]+
m a \left[\begin{array}{c}
\displaystyle \frac{\sqrt{3}}{2}\\[7pt]
\noalign{\vskip3pt}
\displaystyle -\frac{1}{2}\\[7pt]
\noalign{\vskip3pt}
0
\end{array}\right]=
a \left[\begin{array}{c}
\displaystyle \frac{\sqrt{3}}{2}(n+m)\\[7pt]
\noalign{\vskip3pt}
\displaystyle \frac{1}{2}(n-m)\\[7pt]
\noalign{\vskip3pt}
0
\end{array}\right].
\end{equation}
The corresponding carbon nanotube will be indicated as $(n, m)$.

If we define the chiral angle of the nanotube $\theta$ (with 
$-\pi /6 < \theta \leq \pi /6$, due to the hexagonal symmetry of graphene 
lattice) as the angle (positive in the clockwise direction) between
$\vec a_1$ and $\vec C_h$ (see fig.~\ref{f7}) or, equivalently, as the tilt 
angle of the edges of the hexagons constituting the graphene sheet with
respect to the direction of the nanotube axis, such an angle can be found
from the values of $n$ and $m$ noting that
\begin{equation}
\cos \theta=\frac{\vec C_h \cdot \vec a_1}{|\vec C_h||\vec a_1|}=
\frac{2n+m}{2\sqrt{n^2+m^2+nm}}
\end{equation}
and
\begin{equation}
\sin \theta=\frac{(\vec C_h \times \vec a_1)\cdot\hbox{\boldmath{$\hat z$}}'}
{|\vec C_h||\vec a_1|}=\frac{\sqrt{3} m}{2\sqrt{n^2+m^2+nm}},
\end{equation}
where the right-hand reference frame $\Sigma'=
(\hbox{\boldmath{$\hat x$}}',\hbox{\boldmath{$\hat y$}}',
\hbox{\boldmath{$\hat z$}}')$ is
that already used in the calculations on graphene.
In the successive expressions we will identify the previously introduced 
angle $\theta'$ with $\theta'=(\pi / 6)-\theta$ (the angle between 
$\vec C_h$ and the axis $\hbox{\boldmath{$\hat x$}}'$), as shown in
fig.~\ref{f7}, and thus we will take the axis $\hbox{\boldmath{$\hat x$}}$
along $\vec C_h$.

Following Ando's approach~\cite{ando1,ando2},
the dispersion relations and the electron wave functions of a carbon
nanotube can be obtained from those of graphene, enforcing
for the electron wave function the following periodic boundary condition in
the circumferential direction:
\begin{equation}
\psi (\vec r+\vec C_h)=\psi (\vec r)
\end{equation}
(in the calculations we will not consider the curvature effects
\footnotemark{}).
\footnotetext{For the effects of the finite curvature on the electronic
properties of carbon nanotubes see, for example, ref.~\cite{reich} and the
references therein.}
Remembering that using the tight-binding technique the electron wave function 
can be expressed as
\begin{equation}
\psi (\vec r)=\sum_{\vec R_A}\psi_A (\vec R_A)\varphi(\vec r -\vec R_A)+
\sum_{\vec R_B}\psi_B (\vec R_B)\varphi(\vec r -\vec R_B),
\end{equation}
the boundary condition can be written as
\begin{eqnarray}
&& \psi (\vec r+\vec C_h)=\\
&& \sum_{\vec R_A}\psi_A (\vec R_A)\varphi((\vec r+\vec C_h) -\vec R_A)+
\sum_{\vec R_B}\psi_B (\vec R_B)\varphi((\vec r+\vec C_h) -\vec R_B)=\nonumber\\
&& \sum_{\vec R_A}\psi_A (\vec R_A)\varphi(\vec r -(\vec R_A-\vec C_h))+
\sum_{\vec R_B}\psi_B (\vec R_B)\varphi(\vec r -(\vec R_B-\vec C_h))=\nonumber\\
&& \!\!\sum_{\vec R_A}\!\psi_A ((\vec R_A-\vec C_h)+\vec C_h)
\varphi(\vec r -(\vec R_A-\vec C_h))\nonumber\\
&& +\!\sum_{\vec R_B}\!\psi_B ((\vec R_B-\vec C_h)+\vec C_h)
\varphi(\vec r -(\vec R_B-\vec C_h))\!\!=\nonumber\\
&& \sum_{\vec R^*_A}\psi_A (\vec R^*_A+\vec C_h) \varphi(\vec r -\vec R^*_A)+
\sum_{\vec R^*_B}\psi_B (\vec R^*_B+\vec C_h) \varphi(\vec r -\vec R^*_B)=\nonumber\\
&& \psi (\vec r)=
\sum_{\vec R^*_A}\psi_A (\vec R^*_A) \varphi(\vec r -\vec R^*_A)+
\sum_{\vec R^*_B}\psi_B (\vec R^*_B) \varphi(\vec r -\vec R^*_B)\nonumber
\end{eqnarray}
(where we have used the fact that, being $\vec C_h$ a linear combination
with integer coefficients of the real space lattice unit vectors, also
$\vec R_A-\vec C_h$ and $\vec R_B-\vec C_h$ are atomic positions, defined
$\vec R^*_A$ and $\vec R^*_B$). Thus the boundary condition is equivalent
to the two conditions
\begin{equation}
\left\{ \begin{array}{l}
\displaystyle
\psi_A (\vec R^*_A+\vec C_h)=\psi_A (\vec R^*_A),\\[7pt]
\displaystyle
\psi_B (\vec R^*_B+\vec C_h)=\psi_B (\vec R^*_B).
\end{array} \right.
\end{equation}
\noindent
If we use the expressions (\ref{assumptions}) for $\psi_A (\vec r)$
and $\psi_B (\vec r)$ (and we define again the generic atomic position
$\vec R_A$ and $\vec R_B$, instead of $\vec R^*_A$ and $\vec R^*_B$), these
conditions can be rewritten in the following form:
\begin{equation}
\label{conditions}
\left\{ \begin{array}{l}
\displaystyle
e^{i \vec K\cdot (\vec R_A+\vec C_h)} F_A^{\vec K}(\vec R_A+\vec C_h)
-i \, e^{i \theta'} e^{i \vec K'\cdot (\vec R_A+\vec C_h)} 
F_A^{\vec K'}(\vec R_A+\vec C_h)=\\[7pt]
\qquad\qquad\qquad\qquad e^{i \vec K\cdot \vec R_A} F_A^{\vec K}(\vec R_A)
-i \, e^{i \theta'} e^{i \vec K'\cdot \vec R_A}
F_A^{\vec K'}(\vec R_A),\\[7pt]
\displaystyle
i \, e^{i \theta'} 
e^{i \vec K\cdot (\vec R_B+\vec C_h)} F_B^{\vec K} (\vec R_B+\vec C_h)+
e^{i \vec K'\cdot (\vec R_B+\vec C_h)} F_B^{\vec K'} (\vec R_B+\vec C_h)=\\[7pt]
\qquad\qquad\qquad\qquad i \, e^{i \theta'} 
e^{i \vec K\cdot \vec R_B} F_B^{\vec K} (\vec R_B)+
e^{i \vec K'\cdot \vec R_B} F_B^{\vec K'} (\vec R_B).
\end{array} \right.
\end{equation}
%%%%%%%%%%%%%%%%%%%%%%%%%%%%%%%%%%%%%%%%%%%%%%%%%%%%%%%%%%%%%%%
\noindent
Multiplying the first equation of (\ref{conditions}) by
$g(\vec r-\vec R_A) e^{-i \vec K\cdot \vec R_A}$,
summing it over $\vec R_A$ and\break
\newpage
\noindent
then using the properties of the function
$\!g\!$ (defined in eqs.~$\!$(\ref{sum}), $\!$(\ref{phase}) and $\!$(\ref{smooth})),$\!$ we find
\begin{eqnarray}
&& e^{i \vec K\cdot \vec C_h} \sum_{\vec R_A} g(\vec r-\vec R_A)
F_A^{\vec K}(\vec R_A+\vec C_h)\\
&& -i \, e^{i \theta'} e^{i \vec K'\cdot \vec C_h} \sum_{\vec R_A}
g(\vec r-\vec R_A) e^{i (\vec K'-\vec K) \cdot \vec R_A}
F_A^{\vec K'}(\vec R_A+\vec C_h)=\nonumber\\
&& \sum_{\vec R_A} g(\vec r-\vec R_A) F_A^{\vec K}(\vec R_A)
-i \, e^{i \theta'} \sum_{\vec R_A} g(\vec r-\vec R_A)
e^{i (\vec K'-\vec K) \cdot \vec R_A}
F_A^{\vec K'}(\vec R_A) \Rightarrow\nonumber\\
&& e^{i \vec K\cdot \vec C_h} \left[\sum_{\vec R_A} g(\vec r-\vec R_A)\right]
F_A^{\vec K}(\vec r+\vec C_h)\nonumber\\
&& -i \, e^{i \theta'} e^{i \vec K'\cdot \vec C_h} \left[\sum_{\vec R_A}
g(\vec r-\vec R_A) e^{i (\vec K'-\vec K) \cdot \vec R_A}\right]
F_A^{\vec K'}(\vec r+\vec C_h)=\nonumber\\
&& \left[\sum_{\vec R_A} g(\vec r-\vec R_A)\right] F_A^{\vec K}(\vec r)
-i \, e^{i \theta'} \left[\sum_{\vec R_A} g(\vec r-\vec R_A)
e^{i (\vec K'-\vec K) \cdot \vec R_A}\right]
F_A^{\vec K'}(\vec r) \Rightarrow\nonumber\\
&& e^{i \vec K\cdot \vec C_h} F_A^{\vec K}(\vec r+\vec C_h)=
F_A^{\vec K}(\vec r).\nonumber
\end{eqnarray}
If we calculate the scalar product between $\vec K$ and $\vec C_h$
we obtain
\begin{equation}
\label{kch}
\vec K\cdot \vec C_h =
\frac{2\pi}{3}(m-n)=2\pi\tilde N+\frac{2\pi\nu}{3},
\end{equation}
where $m-n=3 \tilde N +\nu$, with $\nu=0$ or $\pm 1$ and $\tilde N$ a proper
integer. Therefore we have that
\begin{equation}
e^{i \vec K\cdot \vec C_h}=e^{i 2\pi\tilde N} e^{i \frac{2\pi\nu}{3}}=
e^{i \frac{2\pi\nu}{3}}
\end{equation}
and thus the first boundary condition near $\vec K$ is
\begin{equation}
e^{i \frac{2\pi\nu}{3}}F_A^{\vec K}(\vec r+\vec C_h)=F_A^{\vec K}(\vec r),
\end{equation}
or equivalently
\begin{equation}
F_A^{\vec K}(\vec r+\vec C_h)=e^{-i \frac{2\pi\nu}{3}}F_A^{\vec K}(\vec r).
\end{equation}
Multiplying the second equation of (\ref{conditions}) by
$g(\vec r-\vec R_B) (-i e^{-i\theta'}e^{-i \vec K\cdot \vec R_B})$,
summing it over $\vec R_B$ and then using the properties of the function
$g$, we find analogously~\cite{supplem}
\begin{equation}
e^{i \vec K\cdot \vec C_h} F_B^{\vec K}(\vec r+\vec C_h)=
F_B^{\vec K}(\vec r).
\end{equation}
Substituting the value of $e^{i \vec K\cdot \vec C_h}$, we can rewrite 
this boundary condition in the form
\begin{equation}
e^{i \frac{2\pi\nu}{3}}F_B^{\vec K}(\vec r+\vec C_h)=F_B^{\vec K}(\vec r),
\end{equation}
or, equivalently
\begin{equation}
F_B^{\vec K}(\vec r+\vec C_h)=e^{-i \frac{2\pi\nu}{3}} F_B^{\vec K}(\vec r).
\end{equation}
Thus the periodic boundary condition near $\vec K$ is
\begin{equation}
\left[\begin{array}{c}
F_A^{\vec K}(\vec r+\vec C_h)\\[5pt]
F_B^{\vec K}(\vec r+\vec C_h)
\end{array}\right]=e^{-i \frac{2\pi\nu}{3}}
\left[\begin{array}{c}
F_A^{\vec K}(\vec r)\\[5pt]
F_B^{\vec K}(\vec r)
\end{array}\right],
\end{equation}
which can be written in this compact way:
\begin{equation}
\vec F^{\vec K}(\vec r+\vec C_h)=e^{-i \frac{2\pi\nu}{3}}
\vec F^{\vec K}(\vec r).
\end{equation}
However, as we have previously seen (eq.~(\ref{fk})), in the absence of an
external potential the envelope functions have the following form:
\begin{eqnarray}
\vec F_{s \vec\kappa}^{\vec K}(\vec r) &=&
\frac{1}{\sqrt{2 L \ell}}e^{i\vec\kappa\cdot\vec r} 
e^{i\phi_s (\vec\kappa)} R(-\alpha (\vec\kappa)) |s \rangle =\\
&& \frac{1}{\sqrt{2 L \ell}}e^{i (\kappa_x x +\kappa_y y )}
e^{i\phi_s (\vec\kappa)} R(-\alpha (\vec\kappa)) |s \rangle,\nonumber
\end{eqnarray}
with the surface area $\Omega=L \ell$, where $L=|\vec C_h|$ and $\ell$ is the
length of the nanotube. Thus the periodic boundary condition 
becomes
\begin{equation}
\frac{1}{\sqrt{2 L \ell}}
e^{i\vec\kappa\cdot(\vec r+\vec C_h)}
e^{i\phi_s (\vec\kappa)} R(-\alpha (\vec\kappa)) |s \rangle =
e^{-i \frac{2\pi\nu}{3}}
\frac{1}{\sqrt{2 L \ell}}
e^{i\vec\kappa\cdot\vec r}
e^{i\phi_s (\vec\kappa)} R(-\alpha (\vec\kappa)) |s \rangle,
\end{equation}
or equivalently
\begin{equation}
e^{i\vec\kappa\cdot\vec C_h}=e^{-i \frac{2\pi\nu}{3}}.
\end{equation}
This condition can be written also in the following way:
\begin{equation}
e^{i \kappa_x L}=e^{-i \frac{2\pi\nu}{3}}1=
e^{-i \frac{2\pi\nu}{3}}e^{i 2\pi \tilde n},
\end{equation}
or, equivalently
\begin{equation}
\kappa_x L=-\frac{2\pi\nu}{3}+2\pi \tilde n
\end{equation}
and thus 
\begin{equation}
\kappa_x =\frac{2\pi}{L}\left(\tilde n-\frac{\nu}{3}\right)=
\kappa_{\nu} (\tilde n),
\end{equation}
with $\tilde n$ integer.

This condition on $\kappa_x$ can be obtained also in a different way,
enforcing the boundary condition on the overall wave vector $\vec k$.
In order to do this, we have to observe that, considering only the periodic
lattice potential inside the graphene sheet, the wave function
$\psi (\vec r)$ has to be a Bloch function
$u(\vec k, \vec r) e^{i \vec k\cdot \vec r}$, where
$u(\vec k, \vec r)$ has the periodicity of the lattice.

Thus the boundary condition
\begin{equation}
\psi (\vec r+\vec C_h)=\psi (\vec r)
\end{equation}
is equivalent to
\begin{equation}
u(\vec k, \vec r+\vec C_h) e^{i \vec k\cdot (\vec r+\vec C_h)}=
u(\vec k, \vec r) e^{i \vec k\cdot \vec r}.
\end{equation}
Since we know that $u(\vec k, \vec r)$ has the lattice periodicity
and thus $u(\vec k, \vec r+\vec C_h)=u(\vec k, \vec r)$
($\vec C_h$ being a linear combination with integer coefficients of the 
lattice unit vectors) the boundary condition can also be written as
\begin{equation}
e^{i \vec k\cdot \vec C_h}=1,
\end{equation}
or, equivalently
\begin{equation}
\vec k\cdot \vec C_h=2\pi \tilde m.
\end{equation}
Thus the boundary condition is 
(being $\hbox{\boldmath{$\hat C$}}_h=\vec C_h /|\vec C_h|=\vec C_h /L$)
\begin{equation}
\vec k\cdot \hbox{\boldmath{$\hat C$}}_h=\vec k\cdot 
\hbox{\boldmath{$\hat x$}}=
k_x=(\vec K)_x+\kappa_x=\frac{2\pi}{L} \tilde m
\end{equation}
and (using eq.~(\ref{kch}))
\begin{eqnarray}
\kappa_x &=& \frac{2\pi}{L} \tilde m-(\vec K)_x=
\frac{2\pi}{L} \tilde m-\frac{\vec K\cdot \vec C_h}{L}=
\frac{2\pi}{L} \tilde m-\frac{2\pi}{L}\tilde N-\frac{2\pi}{3 L}\nu=\\
&& \frac{2\pi}{L} \left(\tilde m-\tilde N-\frac{\nu}{3}\right)=
\frac{2\pi}{L} \left(\tilde n-\frac{\nu}{3}\right)=
\kappa_{\nu} (\tilde n)\nonumber
\end{eqnarray}
(with $\tilde n \equiv \tilde m-\tilde N$), 
which is equal to the previously found expression.

If we substitute this condition on $\kappa_x$ in the dispersion 
relations of graphene, we find
\begin{equation}
E_{s,\tilde n}^{\vec K} (\kappa_y)=s\gamma|\vec\kappa|=
s\gamma\sqrt{\kappa_x^2+\kappa_y^2}=
s\gamma\sqrt{\kappa_{\nu} (\tilde n)^2+\kappa_y^2},
\end{equation}
where $s=+1$ and $s=-1$ indicate the conduction and valence bands,
respectively.

We notice that now $k_y$ is the wave vector $k$ of the nanotube,
which, being a substantially unidimensional material, 
has a one-dimensional Brillouin zone with width $2 \pi /T$ (where $T$ 
is the length of the unit cell of the nanotube, along its axis, which
can be easily found from the numbers $n$ and $m$ characterizing the 
nanotube \cite{saito}). Correspondingly, $\kappa_y$ is the difference between 
the wave vector $k$ of the nanotube and the component of $\vec K$ 
along $y$.

As to the envelope functions near $\vec K$, if, starting from eq.~(\ref{fk}), 
we choose as value of the arbitrary phase $\phi_s=-\alpha /2$ and then
we enforce the condition on $\kappa_x$, we can write
\begin{eqnarray}
&& \vec F_{s \vec\kappa}^{\vec K}(\vec r)=
\frac{1}{\sqrt{2 L \ell}}e^{i\vec\kappa\cdot\vec r}
e^{i\phi_s} \left[\begin{array}{cc}
e^{-i\frac{\alpha}{2}} & 0\\
0 & e^{i\frac{\alpha}{2}}
\end{array}\right]
\frac{1}{\sqrt{2}}\left[\begin{array}{c}
-is\\
1
\end{array}\right]=\\
&& \frac{1}{2\sqrt{L \ell}}
e^{i(\kappa_x x+\kappa_y y)}e^{i\phi_s}
\left[\begin{array}{c}
-i s e^{-i \frac{\alpha}{2}}\\
e^{i\frac{\alpha}{2}}
\end{array}\right]=
\frac{1}{2\sqrt{L \ell}} e^{i(\kappa_x x+\kappa_y y)}
\left[\begin{array}{c}
-i s e^{-i \alpha}\\ 
1
\end{array}\right]=\nonumber\\
&& \frac{1}{2\sqrt{L \ell}} 
\left[\begin{array}{c}
s e^{-i (\frac{\pi}{2}+\alpha)}\\
1
\end{array}\right]
e^{i\kappa_x x+i\kappa_y y}=\nonumber\\
&& \frac{1}{2\sqrt{L \ell}}
\left[\begin{array}{c}
s b_{\nu}(\tilde n,\kappa_y)\\
1
\end{array}\right]
e^{i \kappa_{\nu}(\tilde n) x+i\kappa_y y}=
\vec F_{s \tilde n \kappa_y}^{\vec K}(\vec r).\nonumber
\end{eqnarray}
The function $b_{\nu}(\tilde n,\kappa_y)=e^{-i (\frac{\pi}{2}+\alpha)}$ 
can be found noting that $\alpha$ has been defined (see eq.~(\ref{alpha}))
in such a way that
\begin{equation}
\kappa_x+i \kappa_y=|\vec\kappa|e^{i\left(\frac{\pi}{2}+\alpha\right)};
\end{equation}
this means that
\begin{equation}
e^{i\left(\frac{\pi}{2}+\alpha\right)}=
\frac{{\kappa_x+i \kappa_y}}{\sqrt{\kappa_x^2+\kappa_y^2}}
\end{equation}
and thus
\begin{eqnarray}
b_{\nu}(\tilde n,\kappa_y) &=& e^{-i\left(\frac{\pi}{2}+\alpha\right)}=
\left(e^{i\left(\frac{\pi}{2}+\alpha\right)}\right)^*=\\
&& \left(\frac{\kappa_x+i \kappa_y}{\sqrt{\kappa_x^2+\kappa_y^2}}\right)^*=
\frac{\kappa_x-i \kappa_y}{\sqrt{\kappa_x^2+\kappa_y^2}}=
\frac{\kappa_{\nu}(\tilde n)-i \kappa_y}
{\sqrt{\kappa_{\nu}(\tilde n)^2+\kappa_y^2}}.\nonumber
\end{eqnarray}
%%%%%%%%%%%%%%%%%%%%%%%%%%%%%%%%%%%%%%%%%%%%%%%%%%%%%%%%%%%%%%%

We can proceed analogously for the boundary conditions near $\vec K'$.

Indeed, multiplying the first equation of (\ref{conditions}) by
$g(\vec r-\vec R_A) 
(i e^{-i\theta'} e^{-i \vec K'\cdot \vec R_A})$,
summing it over $\vec R_A$ and then using the properties of the function
$g$, we find~\cite{supplem}
\begin{equation}
e^{i \vec K'\cdot \vec C_h} F_A^{\vec K'}(\vec r+\vec C_h)=
F_A^{\vec K'}(\vec r).
\end{equation}
The scalar product between $\vec K'$ and $\vec C_h$ is equal to
\begin{equation}
\label{k1ch}
\vec K'\cdot \vec C_h=-\frac{2\pi}{3}(m-n)=
-2\pi\tilde N-\frac{2\pi\nu}{3},
\end{equation}
where we have used the previously introduced relation
$m-n=3 \tilde N +\nu$ with $\nu=0$ or $\pm 1$ and $\tilde N$ a proper 
integer. Thus we have that
\begin{equation}
e^{i \vec K'\cdot \vec C_h}=e^{-i 2\pi \tilde N} 
e^{-i \frac{2\pi\nu}{3}}=e^{-i \frac{2\pi\nu}{3}}$$
and consequently the boundary condition near $K'$ is
$$e^{-i \frac{2\pi\nu}{3}}F_A^{\vec K'}(\vec r+\vec C_h)=
F_A^{\vec K'}(\vec r),
\end{equation}
or, equivalently
\begin{equation}
F_A^{\vec K'}(\vec r+\vec C_h)=
e^{i \frac{2\pi\nu}{3}}F_A^{\vec K'}(\vec r).
\end{equation}
On the other hand, multiplying the second equation of (\ref{conditions}) by 
$g(\vec r-\vec R_B) e^{-i \vec K'\cdot \vec R_B}$,
summing it over $\vec R_B$ and then using the properties of the function
$g$, we find~\cite{supplem}
\begin{equation}
e^{i \vec K'\cdot \vec C_h} F_B^{\vec K'}(\vec r+\vec C_h)=
F_B^{\vec K'}(\vec r).
\end{equation}
Substituting the value of $e^{i \vec K'\cdot \vec C_h}$, we can rewrite this 
second boundary condition near $\vec K'$ in the form 
\begin{equation}
e^{-i \frac{2\pi\nu}{3}}F_B^{\vec K'}(\vec r+\vec C_h)=
F_B^{\vec K'}(\vec r),
\end{equation}
or, equivalently
\begin{equation}
F_B^{\vec K'}(\vec r+\vec C_h)=
e^{i \frac{2\pi\nu}{3}}F_B^{\vec K'}(\vec r).
\end{equation}
Thus the overall periodic boundary condition near $\vec K'$ is
\begin{equation}
\left[\begin{array}{c}
F_A^{\vec K'}(\vec r+\vec C_h)\\[5pt]
F_B^{\vec K'}(\vec r+\vec C_h)
\end{array}\right]=e^{i \frac{2\pi\nu}{3}}
\left[\begin{array}{c}
F_A^{\vec K'}(\vec r)\\[5pt]
F_B^{\vec K'}(\vec r)
\end{array}\right],
\end{equation}
which can be written in a compact form
\begin{equation}
\vec F^{\vec K'}(\vec r+\vec C_h)=e^{i \frac{2\pi\nu}{3}}
\vec F^{\vec K'}(\vec r).
\end{equation}
Substituting the form that, in the absence of an external potential,
the envelope functions have near $\vec K'$ (eq.~(\ref{fk1}))
\begin{equation}
\vec F_{s \vec\kappa}^{\vec K'}(\vec r)=
\frac{1}{\sqrt{2 L \ell}}e^{i\vec\kappa\cdot\vec r} 
e^{i\tilde\phi_s (\vec\kappa)} R(\alpha (\vec\kappa)) \tilde{| s \rangle} =
\frac{1}{\sqrt{2 L \ell}}e^{i (\kappa_x x +\kappa_y y )}
e^{i\tilde\phi_s (\vec\kappa)} R(\alpha (\vec\kappa)) \tilde{| s \rangle},
\end{equation}
the periodic boundary condition becomes
\begin{equation}
\frac{1}{\sqrt{2 L \ell}}
e^{i\vec\kappa\cdot(\vec r+\vec C_h)}
e^{i\tilde\phi_s (\vec\kappa)} R(\alpha (\vec\kappa)) \tilde{| s \rangle} =
e^{i \frac{2\pi\nu}{3}}
\frac{1}{\sqrt{2 L \ell}}
e^{i\vec\kappa\cdot\vec r}
e^{i\tilde\phi_s (\vec\kappa)} R(\alpha (\vec\kappa)) \tilde{| s \rangle},
\end{equation}
or, equivalently
\begin{equation}
e^{i\vec\kappa\cdot\vec C_h}=e^{i \frac{2\pi\nu}{3}}.
\end{equation}
This can be rewritten in the form
\begin{equation}
e^{i \kappa_x L}=e^{i \frac{2\pi\nu}{3}}1
=e^{i \frac{2\pi\nu}{3}}e^{i 2\pi \overline{n}},
\end{equation}
or, equivalently
\begin{equation}
\kappa_x L=\frac{2\pi\nu}{3}+2\pi \overline{n}
\end{equation}
and thus
\begin{equation}
\kappa_x =\frac{2\pi}{L}\left(\overline{n}+\frac{\nu}{3}\right)=
\tilde\kappa_{\nu} (\overline{n}),
\end{equation}
with $\overline{n}$ integer.

Analogously to what we have done near $\vec K$, this condition on 
$\kappa_x$ can be found~\cite{supplem} also setting
$\displaystyle e^{i \vec k\cdot \vec C_h}=1$.

If we substitute this condition on $\kappa_x$ in the dispersion
relations of graphene, we find
\begin{equation}
E_{s,\overline{n}}^{\vec K'} (\kappa_y)=s\gamma|\vec\kappa|=
s\gamma\sqrt{\kappa_x^2+\kappa_y^2}=
s\gamma\sqrt{\tilde\kappa_{\nu} (\overline{n})^2+\kappa_y^2},
\end{equation}
where $k_y$ now is the wave vector $k$ of the nanotube and 
$\kappa_y$ is the difference between the wave vector $k$ of the nanotube
and the component of $\vec K'$ along $y$.

On the other hand, if, starting from eq.~(\ref{fk1}), we choose as arbitrary
phase $\tilde\phi_s=\alpha /2$ and then we enforce the condition on $\kappa_x$,
we find~\cite{supplem} as envelope functions in the carbon nanotube near
$\vec K'$
\begin{equation}
\vec F_{s \vec\kappa}^{\vec K'}(\vec r)=
\frac{1}{2 \sqrt{L \ell}}
\left[\begin{array}{c}
s \tilde b_{\nu}(\overline{n},\kappa_y)\\
1
\end{array}\right]
e^{i \tilde\kappa_{\nu}(\overline{n}) x+i\kappa_y y}=
\vec F_{s \overline{n} \kappa_y}^{\vec K'}(\vec r),
\end{equation}
where (using the definition of the angle $\alpha$: see eq.~(\ref{alpha}))
\begin{equation}
\tilde b_{\nu}(\overline{n},\kappa_y)=e^{i\left(\frac{\pi}{2}+\alpha\right)}=
\frac{\kappa_x+i \kappa_y}{\sqrt{\kappa_x^2+\kappa_y^2}}=
\frac{\tilde\kappa_{\nu}(\overline{n})+i \kappa_y}
{\sqrt{\tilde\kappa_{\nu}(\overline{n})^2+\kappa_y^2}}.
\end{equation}
%%%%%%%%%%%%%%%%%%%%%%%%%%%%%%%%%%%%%%%%%%%%%%%%%%%%%%
\noindent
If $m-n$ is a multiple of 3 and thus $\nu=0$, for $\tilde n=0$ and
$\overline{n}=0$ we have that
$\kappa_{\nu} (\tilde n)=0$ and $\tilde\kappa_{\nu} (\overline{n})=0$,
and consequently $E_{s}=s \gamma |\kappa_y|$ ,
which vanishes for $\kappa_y=0$, so that $E_{+}=E_{-}=0$.
This means that when $m-n$ is a multiple of 3 the points $\vec K$
and $\vec K'$, where the upper and lower bands of graphene are degenerate, 
are among the values of $\vec k$ allowed by the periodic boundary condition,
and thus the nanotube is metallic.

Instead, if $m-n$ is not a multiple of 3 and thus $\nu=\pm 1$,
the allowed $\vec k$'s nearest to $\vec K$ and $\vec K'$
correspond to $\tilde n=0$ and $\overline{n}=0$, for which 
$\kappa_{\nu} (\tilde n)=\mp 2\pi /(3L)$ and
$\tilde\kappa_{\nu} (\overline{n})=\pm 2\pi /(3L)$,
and consequently 
\begin{equation}
E_{s}=s\gamma\sqrt{\left(\frac{2\pi}{3 L}\right)^2+\kappa_y^2}.
\end{equation}
In particular, the minimum and maximum values of the nanotube bands are 
obtained with the further position $\kappa_y=0$ and therefore are equal to
\begin{equation}
E_{s}=s\gamma \frac{2\pi}{3 L};
\end{equation}
thus the bandgap of the nanotube is
\begin{equation}
E_g=E_{+}-E_{-}=2 \gamma \frac{2\pi}{3 L}=\frac{4\pi\gamma}{3 L}=
\frac{4\pi}{3 L}\frac{\sqrt{3}a\gamma_0}{2}=
2 \frac{\pi}{L}\frac{a}{\sqrt{3}}\gamma_0=\frac{2\gamma_0\, a_{C-C}}{d_t},
\end{equation}
where $d_t=L / \pi$ is the nanotube diameter. Therefore we have that the
bandgap of the nanotube depends on the reciprocal nanotube diameter.

\begin{figure}
\centering
\includegraphics[width=.75\textwidth,angle=0]{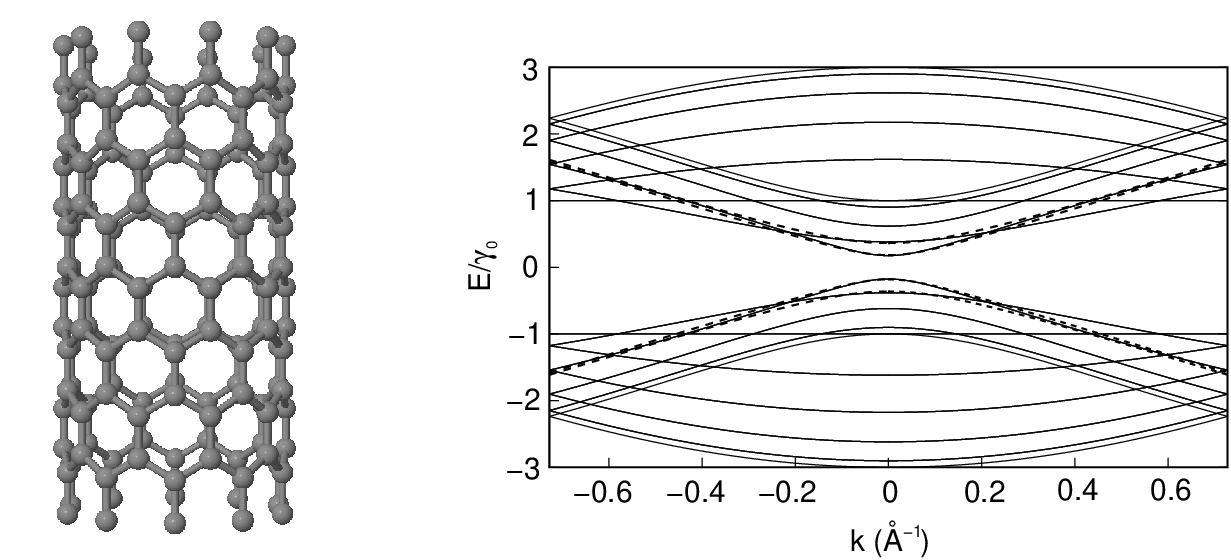}
\caption{The nanotube (10,0) and its dispersion relations, obtained both by 
means of the tight-binding method (solid lines) and (for the bands
corresponding to the smallest values of $|\kappa_{\nu} (\tilde n)|$ and
$|\tilde\kappa_{\nu} (\overline{n})|$) by means of the $\vec k\cdot\vec p$
method (dashed lines).}
\label{f8}
\vskip5pt
\centering
\includegraphics[width=.52\textwidth,angle=0]{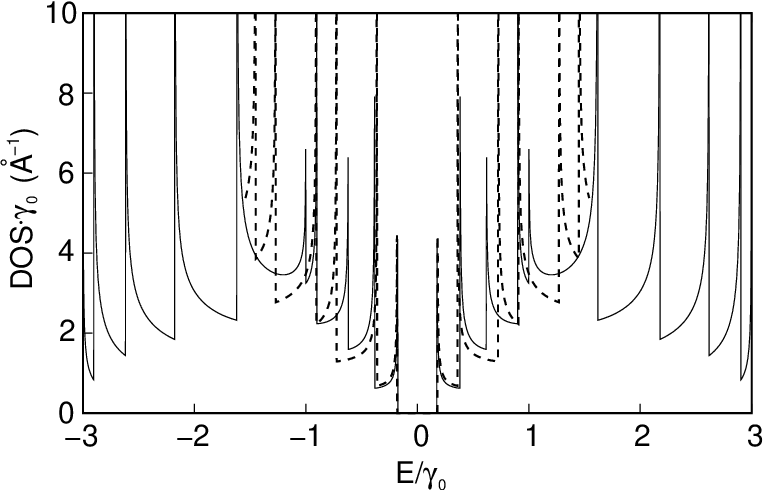}
\caption{The density of states per unit length of the nanotube (10,0), 
obtained both by means of the tight-binding method (solid lines) and
(in a smaller region around $E=0$) by means of the $\vec k\cdot\vec p$
method (dashed lines).}
\label{f9}
\end{figure}

We can observe that the approximate approach for the computation of the
density of states in carbon nanotubes proposed by J.~W.~Mintmire and
C.~T.~White~\cite{mintmire}, being based on a linear approximation of the 
dispersion relations of graphene near the extrema points, can be 
seen as a consequence of a $\vec k \cdot \vec p$ study of 
the nanotube energy bands.

In fig.~\ref{f8} we compare the dispersion relations that we have obtained 
for the same carbon nanotube using the nearest-neighbor tight-binding 
method and the $\vec k \cdot \vec p$ method (without considering curvature
effects)~\cite{marconcini1,marconcini2,marconcini3,marconcini4}. We see that
the $\vec k \cdot \vec p$ method gives a good approximation for the portions of
energy bands of the nanotube deriving from the graphene dispersion relations
around $\vec K$ and $\vec K'$.

In fig.~\ref{f9}, instead, for the same nanotube we show both the 
density of states that we have obtained by properly differentiating 
the tight-binding dispersion relations, and the
density of states deriving from the Mintmire-White approach
\cite{marconcini1,marconcini2}. We see that this last approximation gives
good results near $E=0$, thus in the region corresponding to the
graphene dispersion relations around $\vec K$ and $\vec K'$.

\section{Application of the $\vec k \cdot \vec p$ method to graphene
nanoribbons}

A graphene sheet can be laterally confined (along the $y$-direction) to
form a graphene nanoribbon (extending in the $x$-direction). The properties
of the nanorribbon strongly depend on the characteristics of the boundary.
Here we will consider nanoribbons with perfect zigzag and armchair edges, that
can be easily studied using the Dirac equation and enforcing the correct
boundary conditions~\cite{brey1,brey2,wakabayashi1,castroneto,wurm,tworzydlo,
wakabayashi2}.
An analysis of the boundary conditions that have to be enforced in
nanoribbons with more general terminations can be found in 
ref.~\cite{akhmerov2}.
In particular, we will perform the analytical calculations in the absence
of an external potential following Brey and Fertig's approach
\cite{brey1,brey2}, but using the representation adopted in the previous
sections. 
While inside the nanoribbon each atom has 3
nearest-neighbor atoms, for the atoms on the edges of the
ribbon some of the nearest-neighbor lattice sites are outside the
ribbon and thus are not occupied by a carbon atom. These lattice sites are
instead occupied by passivation atoms (such as hydrogen atoms), which saturate
the dangling bonds. The correct boundary condition to be enforced in
our calculations is the vanishing of the wave function in correspondence
of these lattice sites (let us call them ``boundary lattice sites'').

\begin{figure}[b]
\centering
\includegraphics[width=\textwidth,angle=0]{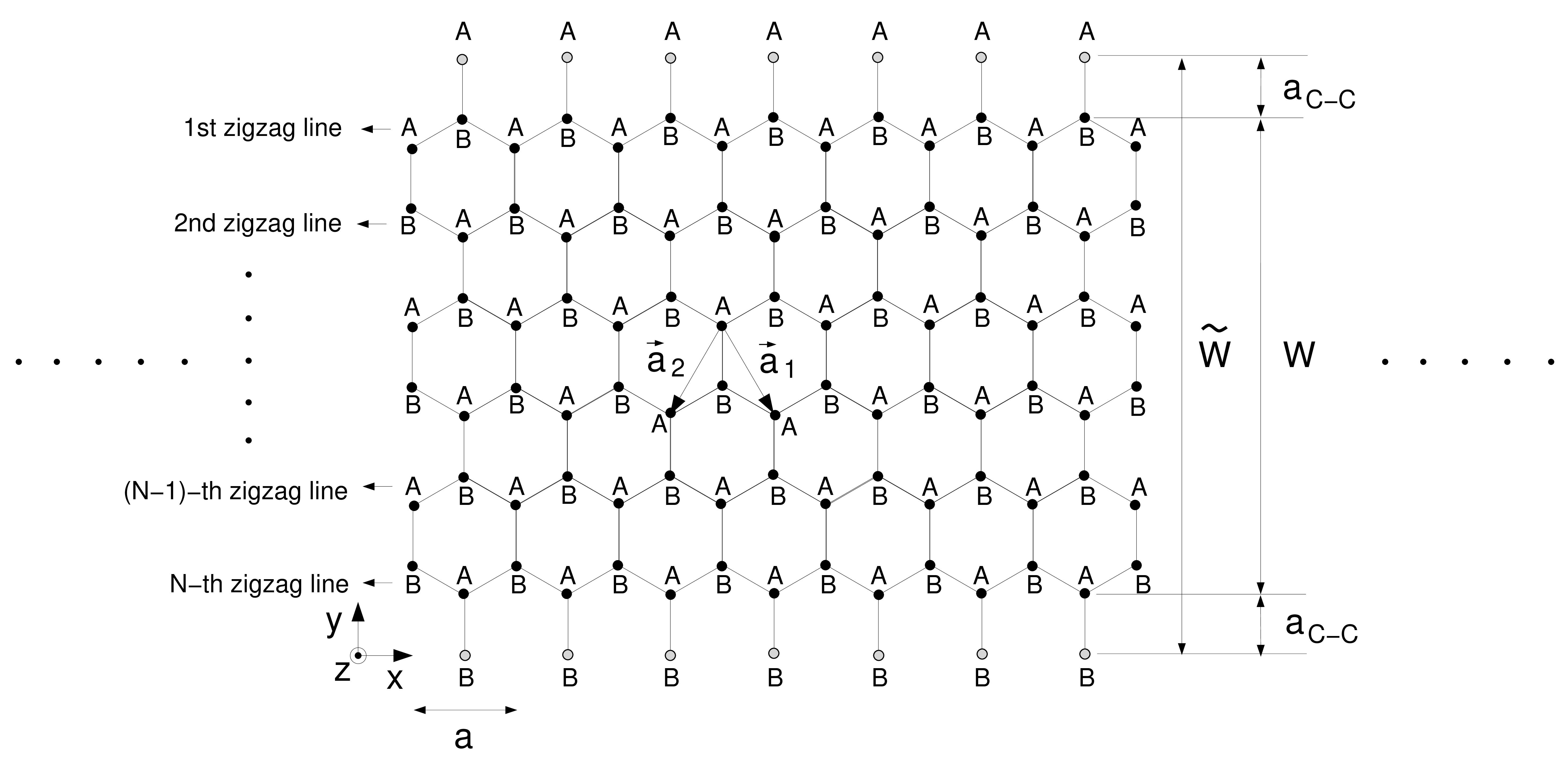}
\caption{Sketch of a zigzag nanoribbon with $N$ zigzag lines (the black
atoms are carbon atoms, while the grey atoms are passivation atoms).}
\label{f10}
\end{figure}

%%%%%%%%%% ZIGZAG %%%%%%%%%%
\subsection{Zigzag nanoribbons} 
In the case of zigzag nanoribbons (fig.~\ref{f10}), the graphene sheet has
been cut at an angle of $30^\circ$ with respect to the nearest-neighbor
carbon bonds, and therefore the edges have a zigzag shape. In order to
simplify the following calculations, we can choose (see fig.~\ref{f10})
the graphene lattice vectors in the real space $\vec a_1$ and $\vec a_2$
(and consequently those in the reciprocal space $\vec b_1$ and $\vec b_2$)
in this way (we express them in the reference frame
$\Sigma=(\hbox{\boldmath{$\hat x$}},\hbox{\boldmath{$\hat y$}},
\hbox{\boldmath{$\hat z$}})$):
\begin{eqnarray}
& \vec a_1 \mathrel{\mathop\equiv_{\Sigma}}
\left[\begin{array}{c}
\displaystyle \frac{a}{2}\\
\noalign{\vskip3pt}
\displaystyle -\frac{\sqrt{3}}{2}\,a\\
\noalign{\vskip3pt}
0
\end{array}\right]
,\quad
& \vec a_2 \mathrel{\mathop\equiv_{\Sigma}}
\left[\begin{array}{c}
\displaystyle -\frac{a}{2}\\
\noalign{\vskip3pt}
\displaystyle -\frac{\sqrt{3}}{2}\,a\\
\noalign{\vskip3pt}
0
\end{array}\right]
,\\
& \vec b_1 \mathrel{\mathop\equiv_{\Sigma}}
\left[\begin{array}{c}
\displaystyle \frac{2\pi}{a}\\ 
\noalign{\vskip3pt}
\displaystyle -\frac{2\pi}{\sqrt{3}a}\\
\noalign{\vskip3pt}
0
\end{array}\right]
,\quad
& \vec b_2 \mathrel{\mathop\equiv_{\Sigma}}
\left[\begin{array}{c}
\displaystyle -\frac{2\pi}{a}\\
\noalign{\vskip3pt}
\displaystyle -\frac{2\pi}{\sqrt{3}a}\\
\noalign{\vskip3pt}
0
\end{array}
\right]\nonumber
\end{eqnarray}
(which, being 
$\vec b_1=2\pi (\vec a_2 \times \hbox{\boldmath{$\hat z$}})/
(\vec a_1 \cdot (\vec a_2 \times \hbox{\boldmath{$\hat z$}}))$ and
$\vec b_2=2\pi (\hbox{\boldmath{$\hat z$}} \times \vec a_1)/
(\vec a_1 \cdot (\vec a_2 \times \hbox{\boldmath{$\hat z$}}))$,
fulfill the relation $\vec a_i \cdot \vec b_j=2 \pi \delta_{ij}$).
Consequently we have that
\begin{eqnarray}
\label{kz}
&& \vec K =\frac{1}{3}(\vec b_2-\vec b_1) \mathrel{\mathop\equiv_{\Sigma}}
\frac{4\pi}{3a}
\left[\begin{array}{c}
-1\\
\noalign{\vskip3pt}
0 \\
\noalign{\vskip3pt} 
0
\end{array}\right]=
\left[\begin{array}{c}
-K\\
\noalign{\vskip3pt}
0 \\
\noalign{\vskip3pt} 
0
\end{array}\right],\\
&& \vec K' =\frac{1}{3}(\vec b_1-\vec b_2) \mathrel{\mathop\equiv_{\Sigma}}
\frac{4\pi}{3a}
\left[\begin{array}{c}
1\\
\noalign{\vskip3pt}
0\\
\noalign{\vskip3pt} 
0
\end{array}\right]=
\left[\begin{array}{c}
K\\
\noalign{\vskip3pt}
0 \\
\noalign{\vskip3pt} 
0
\end{array}\right],\nonumber
\end{eqnarray}
where we have defined $K=4\pi/(3a)$. For our choice of $\vec a_1$
and $\vec a_2$, the angle $\theta'$ from the vector $\vec a_1+\vec a_2$ ({\em i.e.}
from the axis $\hbox{\boldmath{$\hat x$}}'$ used in previous calculations)
to the axis $\hbox{\boldmath{$\hat x$}}$ (taken in the longitudinal direction)
is equal to $\pi/2$.

Therefore the total wave function is given by (eq.~(\ref{wavefunction}))
\begin{equation}
\label{wavefunctionbisz}
\psi (\vec r)=
\sum_{\vec R_A}\psi_A (\vec R_A)\varphi(\vec r -\vec R_A)+
\sum_{\vec R_B}\psi_B (\vec R_B)\varphi(\vec r -\vec R_B).
\end{equation}
with (eq.~(\ref{assumptions2}) with $\theta'=\pi/2$)
\begin{equation}
\label{envelopez}
\left\{ \begin{array}{l}
\displaystyle
\psi_A (\vec r)=
e^{i \vec K\cdot \vec r} F_A^{\vec K}(\vec r)+
e^{i \vec K'\cdot \vec r} F_A^{\vec K'}(\vec r),\\[5pt]
\displaystyle
\psi_B (\vec r)=
-e^{i \vec K\cdot \vec r} F_B^{\vec K} (\vec r)+
e^{i \vec K'\cdot \vec r} F_B^{\vec K'} (\vec r),
\end{array} \right.
\end{equation}
where (using eq.~(\ref{kz})), if we write
$\displaystyle \vec r \mathrel{\mathop\equiv_{\Sigma}} [x,y,0]^T$ we
have that $\vec K\cdot \vec r=-Kx$ and that $\vec K'\cdot \vec r=Kx$.
In the absence of an external potential, the envelope functions
satisfy the usual Dirac equation (eq.~(\ref{absence}))
\begin{eqnarray}
&& \gamma
\left[\begin{array}{cccc}
0 &
-i\,\frac{\partial}{\partial x}-\frac{\partial}{\partial y} &
0 &
0 \\[3pt]
-i\,\frac{\partial}{\partial x}+\frac{\partial}{\partial y} &
0 &
0 &
0 \\[3pt]
0 &
0 &
0 &
-i\,\frac{\partial}{\partial x}+\frac{\partial}{\partial y} \\[3pt]
0 &
0 &
-i\,\frac{\partial}{\partial x}-\frac{\partial}{\partial y} &
0
\end{array}\right]
\left[\begin{array}{c}
F_A^{\vec K} (\vec r)\\[3pt]
F_B^{\vec K} (\vec r)\\[3pt]
F_A^{\vec K'} (\vec r)\\[3pt]
F_B^{\vec K'} (\vec r)
\end{array}\right]=\\[3pt]
&& E \left[\begin{array}{c}
F_A^{\vec K} (\vec r)\\[3pt]
F_B^{\vec K} (\vec r)\\[3pt]
F_A^{\vec K'} (\vec r)\\[3pt]
F_B^{\vec K'} (\vec r)
\end{array}\right].\nonumber
\end{eqnarray}
Due to the translational invariance along the $x$-direction, we can write
the envelope functions as the product of a propagating part along the
longitudinal direction $x$ and of a confined part along the transverse
direction $y$. Therefore we can assume that
\begin{equation}
\label{phiz}
\left[\begin{array}{c}
F_A^{\vec K} (\vec r)\\[5pt]
F_B^{\vec K} (\vec r)
\end{array}\right]=
e^{i \kappa_x x}
\left[\begin{array}{c}
\Phi_A^{\vec K} (y)\\[5pt]
\Phi_B^{\vec K} (y)
\end{array}\right],
\ \hbox{and that}\ 
\left[\begin{array}{c}
F_A^{\vec K'} (\vec r)\\[5pt]
F_B^{\vec K'} (\vec r)
\end{array}\right]=
e^{i \kappa'_x x}
\left[\begin{array}{c}
\Phi_A^{\vec K'} (y)\\[5pt]
\Phi_B^{\vec K'} (y)
\end{array}\right].
\end{equation}
We have to enforce that the overall wave function vanishes in correspondence
with the ``boundary lattice sites'' on the lower and upper edges of the ribbon.
Let us define as $W$ the real width of the nanoribbon, {\em i.e.} the distance
between the lowest row of carbon atoms (all of type $A$) and
the highest row of carbon atoms (all of type $B$); if the ribbon
has $N$ zigzag lines across its width, we have that $W=(3N-2)a_{C-C}/2$.
If we take $y=0$ in correspondence of the row of ``boundary lattice sites''
on the lower edge, the row of ``boundary lattice sites'' on the
upper edge will be for $y=\tilde W=W+2 a_{C-C}={(3N+2)a_{C-C}/2}$.
The proper boundary condition thus implies that, for every $x$,
$\psi (x,y=0)=\psi (x,y=\tilde W)=0$.
Since in the zigzag nanoribbon all the ``boundary lattice sites''
on the lower edge belong to the $B$ sublattice, while all those on the upper
edge belong to the $A$ sublattice, looking at
eq.~(\ref{wavefunctionbisz}) and observing that the atomic orbitals
$\varphi$ are strongly localized around the atom on which they are centered,
the boundary condition on the wave function is equivalent to setting,
for every $x$, $\psi_B (x,y=0)=\psi_A (x,y=\tilde W)=0$.
Using eq.~(\ref{envelopez}), we have that
\begin{eqnarray}
&& \psi_B (x,y=0)=0 \ \ \forall x \Rightarrow\ 
-e^{-iKx} F_B^{\vec K} (x,y=0)+e^{iKx} F_B^{\vec K'} (x,y=0)=\\[5pt]
&& -e^{-iKx} e^{i \kappa_x x} \Phi_B^{\vec K}(0)+
e^{iKx} e^{i \kappa'_x x} \Phi_B^{\vec K'}(0)=0\ \forall x 
\Rightarrow\nonumber\\[5pt]
&& \Phi_B^{\vec K}(0)=0,\quad\Phi_B^{\vec K'}(0)=0\nonumber
\end{eqnarray}
and that
\begin{eqnarray}
&& \psi_A (x,y=\tilde W)=0 \ \ \forall x \Rightarrow\ 
e^{-iKx} F_A^{\vec K} (x,y=\tilde W)+e^{iKx} F_A^{\vec K'} (x,y=\tilde W)=\\[5pt]
&& e^{-iKx} e^{i \kappa_x x} \Phi_A^{\vec K}(\tilde W)+
e^{iKx} e^{i \kappa'_x x} \Phi_A^{\vec K'}(\tilde W)=0\ \forall x 
\Rightarrow\nonumber\\[5pt]
&& \Phi_A^{\vec K}(\tilde W)=0,\quad
\Phi_A^{\vec K'}(\tilde W)=0.\nonumber
\end{eqnarray}
As we can see, in zigzag nanoribbons the boundary conditions do not couple
the envelope functions relative to the Dirac points $\vec K$ and $\vec K'$.

%%%%%%%%%% K %%%%%%%%%%
First let us make the calculation {\em around the point} {\boldmath $\vec K$}.
The corresponding part of the Dirac equation is:
\begin{eqnarray}
\label{systemza}
&& \gamma
\left[\begin{array}{cccc}
0 &
-i\,\frac{\partial}{\partial x}-\frac{\partial}{\partial y} \\[5pt]
-i\,\frac{\partial}{\partial x}+\frac{\partial}{\partial y} &
0
\end{array}\right]
\left[\begin{array}{c}
F_A^{\vec K} (\vec r)\\[5pt]
F_B^{\vec K} (\vec r)
\end{array}\right]=
E \left[\begin{array}{c}
F_A^{\vec K} (\vec r)\\[5pt]
F_B^{\vec K} (\vec r)
\end{array}\right]
\Rightarrow\\[5pt]
&& \gamma
\left[\begin{array}{cccc}
0 &
-i\,\frac{\partial}{\partial x}-\frac{\partial}{\partial y} \\[5pt]
-i\,\frac{\partial}{\partial x}+\frac{\partial}{\partial y} &
0
\end{array}\right]
\left[\begin{array}{c}
\Phi_A^{\vec K}(y)e^{i \kappa_x x}\\[5pt]
\Phi_B^{\vec K}(y)e^{i \kappa_x x}
\end{array}\right]=\nonumber\\[5pt]
&& \gamma 
\left[\begin{array}{c}
\kappa_x \Phi_B^{\vec K}(y) e^{i \kappa_x x}-
e^{i \kappa_x x}\frac{d}{d\,y}\Phi_B^{\vec K}(y)\\[5pt]
\kappa_x \Phi_A^{\vec K}(y) e^{i \kappa_x x}+
e^{i \kappa_x x}\frac{d}{d\,y} \Phi_A^{\vec K}(y)
\end{array}\right]=\nonumber\\[5pt]
&& \gamma
\left[\begin{array}{cccc}
0 &
\kappa_x-\frac{d}{d\,y} \\[5pt]
\kappa_x+\frac{d}{d\,y} &
0
\end{array}\right]
\left[\begin{array}{c}
\Phi_A^{\vec K}(y)\\[5pt]
\Phi_B^{\vec K}(y)
\end{array}\right] e^{i \kappa_x x}=\nonumber\\[5pt]
&& E \left[\begin{array}{c}
F_A^{\vec K} (\vec r)\\[5pt]
F_B^{\vec K} (\vec r)
\end{array}\right]=
E \left[\begin{array}{c}
\Phi_A^{\vec K}(y)\\[5pt]
\Phi_B^{\vec K}(y)
\end{array}\right] e^{i \kappa_x x}
\Rightarrow\nonumber\\[5pt]
&& \left[\begin{array}{cccc}
0 &
\kappa_x-\frac{d}{d\,y} \\[5pt]
\kappa_x+\frac{d}{d\,y} &
0
\end{array}\right]
\left[\begin{array}{c}
\Phi_A^{\vec K}(y)\\[5pt]
\Phi_B^{\vec K}(y)
\end{array}\right]=
\frac{E}{\gamma} \left[\begin{array}{c}
\Phi_A^{\vec K}(y)\\[5pt]
\Phi_B^{\vec K}(y)
\end{array}\right],\nonumber
\end{eqnarray}
which can be rewritten as
\begin{equation}
\label{systemzb}
\left\{ \begin{array}{l}
\displaystyle
\left(\kappa_x-\frac{d}{d\,y}\right)
\Phi_B^{\vec K}(y)=\frac{E}{\gamma}\Phi_A^{\vec K}(y),\\[10pt]
\displaystyle
\left(\kappa_x+\frac{d}{d\,y}\right)
\Phi_A^{\vec K}(y)=\frac{E}{\gamma}\Phi_B^{\vec K}(y).
\end{array} \right.
\end{equation}
Obtaining $\Phi_B^{\vec K}(y)$ from the second of (\ref{systemzb})
and then substituting $\Phi_A^{\vec K}(y)$ from the first of
(\ref{systemzb}), we find:
\begin{eqnarray}
\label{solaz}
&& \Phi_B^{\vec K}(y)=\frac{\gamma}{E}
\left(\kappa_x+\frac{d}{d\,y}\right)
\Phi_A^{\vec K}(y)=\left(\frac{\gamma}{E}\right)^2
\left(\kappa_x+\frac{d}{d\,y}\right)
\left(\kappa_x-\frac{d}{d\,y}\right)
\Phi_B^{\vec K}(y)=\\[5pt]
&& \left(\frac{\gamma}{E}\right)^2
\left(\kappa_x^2-\kappa_x\frac{d}{d\,y}
+\kappa_x\frac{d}{d\,y}-
\frac{d^2}{d\,y^2}\right)
\Phi_B^{\vec K}(y)=\nonumber\\[5pt]
&& \left(\frac{\gamma}{E}\right)^2
\left(\kappa_x^2-
\frac{d^2}{d\,y^2}\right)
\Phi_B^{\vec K}(y)\Rightarrow\nonumber\\[5pt]
&& \left(-\frac{d^2}{d\,y^2}+\kappa_x^2\right)
\Phi_B^{\vec K}(y)=\left(\frac{E}{\gamma}\right)^2
\Phi_B^{\vec K}(y),\nonumber
\end{eqnarray}
the solution of which is
\begin{eqnarray}
\label{solbz}
&& \Phi_B^{\vec K}(y)=A e^{zy}+B e^{-zy},\\
&& \hbox{with}\quad
z=\sqrt{\kappa_x^2-\left(\frac{E}{\gamma}\right)^2}
\quad\left(\;\hbox{and thus}\quad
E=\pm \gamma \sqrt{\kappa_x^2-z^2}\;\right).\nonumber
\end{eqnarray}
Substituting $\Phi_B^{\vec K}(y)$ back into the first of (\ref{systemzb}),
we obtain that
\begin{eqnarray}
\label{solcz}
\Phi_A^{\vec K}(y) &=& \frac{\gamma}{E}
\left(\kappa_x-\frac{d}{d\,y}\right)
\Phi_B^{\vec K}(y)=\\
&& \frac{\gamma}{E}\left(\kappa_x A e^{zy}+\kappa_x B e^{-zy}-
z A e^{zy}+z B e^{-zy}\right)=\nonumber\\
&& \frac{\gamma}{E}\left((\kappa_x-z) A e^{zy}+(\kappa_x+z) B e^{-zy}\right).\nonumber
\end{eqnarray}
Let us now enforce the boundary conditions on $\Phi_B^{\vec K}(y)$ and
$\Phi_A^{\vec K}(y)$
\begin{eqnarray}
\label{realk}
&& \Phi_B^{\vec K}(0)=0 \Rightarrow
A+B=0 \Rightarrow B=-A;\\
&& \Phi_A^{\vec K}(\tilde W)=0 \Rightarrow 
\frac{\gamma}{E}\left((\kappa_x-z) A e^{z \tilde W}+
(\kappa_x+z) B e^{-z \tilde W}\right)=0 \Rightarrow\nonumber\\
&& (\kappa_x-z) A e^{z \tilde W}-
(\kappa_x+z) A e^{-z \tilde W}=0 \Rightarrow\nonumber\\
&& (\kappa_x-z) A e^{z \tilde W}=
(\kappa_x+z) A e^{-z \tilde W} \Rightarrow\nonumber\\
&& e^{-2 z \tilde W}=\frac{\kappa_x-z}{\kappa_x+z}.\nonumber
\end{eqnarray}
As we can see, in zigzag nanoribbons the longitudinal and the transverse
wave vectors are coupled.

Incidentally, note that, instead of eq.~(\ref{realk}), an equivalent equation can be
used~\cite{wakabayashi2}; indeed, being $E=\pm \gamma \sqrt{\kappa_x^2-z^2}$
and thus $(E/\gamma)^2=\kappa_x^2-z^2$, we have that
\begin{eqnarray}
&& e^{-2 z \tilde W}=\frac{\kappa_x-z}{\kappa_x+z}=
\frac{(\kappa_x-z)(\kappa_x+z)}{(\kappa_x+z)^2}=
\frac{\kappa_x^2-z^2}{(\kappa_x+z)^2}=
\frac{(E/\gamma)^2}{(\kappa_x+z)^2}\Rightarrow\\
&& \frac{E}{\gamma}=\pm (\kappa_x+z) e^{-z \tilde W}.\nonumber
\end{eqnarray}
Here we consider real values of $\kappa_x$.

\begin{figure}
\centering
\includegraphics[width=.5\textwidth,angle=0]{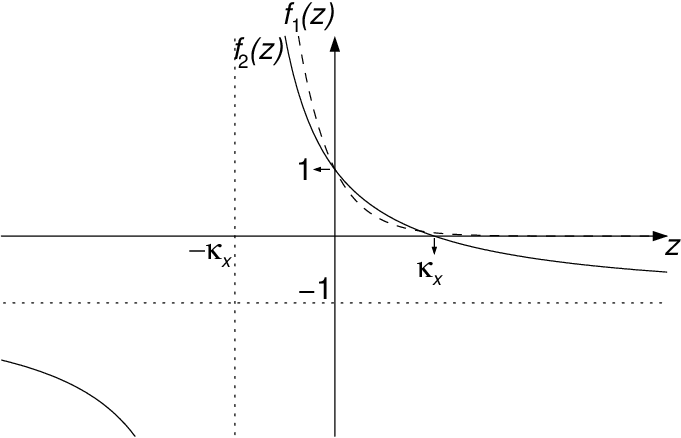}
\caption{Graphical solution (in the real domain) of eq.~(\ref{realk})
(the dotted lines are the asymptotes of $f_2 (z)$).}
\label{f11}
\end{figure}\noindent

If we graphically represent (fig.~\ref{f11}) the two functions
$f_1 (z)=e^{-2 z \tilde W}$ and
$f_2 (z)=(\kappa_x-z)/(\kappa_x+z)$, we see
that (apart from $z=0$, which corresponds to identically null $\Phi$'s)
there is an intersection between $f_1$ and $f_2$ for a
real value of $z$ (and thus eq.~(\ref{realk}) has a {\em real solution} $z$)
only if $\kappa_x>0$ and if $f_1 (z)$ is steeper than $f_2 (z)$ in $z=0$,
{\em i.e.} if
\begin{eqnarray}
&& \left|\left[\frac{d}{dz}f_1(z)\right]_{z=0}\right|>
\left|\left[\frac{d}{dz}f_2(z)\right]_{z=0}\right|\Rightarrow\\
&& \left|\left[-2 \tilde W e^{-2 z \tilde W}\right]_{z=0}\right|>
\left|\left[-\frac{1}{\kappa_x+z}-
\frac{\kappa_x-z}{(\kappa_x+z)^2}\right]_{z=0}\right|=\nonumber\\
&& \left|\left[-\frac{\kappa_x+z+\kappa_x-z}{(\kappa_x+z)^2}\right]_{z=0}\right|=
\left|\left[-\frac{2 \kappa_x}{(\kappa_x+z)^2}\right]_{z=0}\right|
\Rightarrow\nonumber\\
&& 2 \tilde W>\frac{2 \kappa_x}{\kappa_x^2}\Rightarrow
\tilde W>\frac{1}{\kappa_x}\Rightarrow
\kappa_x>\frac{1}{\tilde W}.\nonumber
\end{eqnarray}
\vskip-5pt\noindent
If instead $\kappa_x<1/\tilde W$, eq.~(\ref{realk}) does not have real
solutions $z$ (apart from $z=0$).

In the case of real $z$, from eq.~(\ref{realk}) we can find that
\vskip-5pt\noindent
\begin{eqnarray}
&& e^{-2 z \tilde W}=\frac{\kappa_x-z}{\kappa_x+z}\Rightarrow\\
&& \kappa_x e^{-2 z \tilde W}+z e^{-2 z \tilde W}=\kappa_x-z
\Rightarrow
\kappa_x (1-e^{-2 z \tilde W})=z (1+e^{-2 z \tilde W})
\Rightarrow\nonumber\\
&& \kappa_x =z\,\frac{1+e^{-2 z \tilde W}}{1-e^{-2 z \tilde W}}=
z\,\frac{e^{z \tilde W}+e^{-z \tilde W}}{e^{z \tilde W}-e^{-z \tilde W}}=
\frac{z}{\tanh(z \tilde W)}\nonumber
\end{eqnarray}
\vskip-5pt\noindent
($z=0$ does not have to be considered) and thus
\vskip-5pt\noindent
\begin{eqnarray}
\label{esinh}
&& \left(\frac{E}{\gamma}\right)^2=\kappa_x^2-z^2=
\frac{z^2}{\tanh^2(z \tilde W)}-z^2=
z^2 \left(\frac{\cosh^2(z \tilde W)}{\sinh^2(z \tilde W)}-1\right)=\\
&& z^2 \left(\frac{\cosh^2(z \tilde W)-\sinh^2(z \tilde W)}
{\sinh^2(z \tilde W)}\right)=
\frac{z^2}{\sinh^2(z \tilde W)} \Rightarrow
\left|\frac{E}{\gamma}\right|=\left|\frac{z}{\sinh(z \tilde W)}\right|.\nonumber
\end{eqnarray}
\vskip-5pt\noindent
Since (for the properties of the hyperbolic sine function)
$|\sinh(z \tilde W)|>|z \tilde W|=|z|\tilde W$, we see that in this case
\begin{equation}
\left|\frac{E}{\gamma}\right|<\frac{|z|}{|z| \tilde W}=
\frac{1}{\tilde W}.
\end{equation}
We can write (exploiting what we have found from the boundary
conditions) that
\begin{eqnarray}
\Phi_A^{\vec K}(y) &=&
\frac{\gamma}{E} \left((\kappa_x-z) A e^{zy}+(\kappa_x+z) B e^{-zy}\right)=\\
&& \frac{\gamma}{E} \left((\kappa_x-z) A e^{zy}-(\kappa_x+z) A e^{-zy}\right)=\nonumber\\
&& \frac{\gamma}{E} A \left(\kappa_x (e^{zy}-e^{-zy})-z (e^{zy}+e^{-zy})\right)=\nonumber\\
&& \frac{\gamma}{E} 2 A \left(\kappa_x \sinh(zy)-z \cosh(zy)\right)=\nonumber\\
&& 2 A \frac{\gamma}{E}
\left(\frac{z}{\tanh(z \tilde W)} \sinh(zy)-z \cosh(zy)\right)=\nonumber\\
&& 2 A \frac{\gamma}{E} z \,
\frac{\cosh(z \tilde W)\sinh(zy)-\sinh(z \tilde W)\cosh(zy)}
{\sinh(z \tilde W)}=\nonumber\\
&& -2 A \frac{\gamma}{E} z \,
\frac{\cosh(z \tilde W)\sinh(-zy)+\sinh(z \tilde W)\cosh(-zy)}
{\sinh(z \tilde W)}=\nonumber\\
&& -2 A \left(\frac{\gamma}{E} \frac{z}{\sinh(z \tilde W)}\right)
\sinh(z(\tilde W-y))=\nonumber\\
&& -2 A \,{\rm sign}\!\left(\frac{E}{\gamma} \frac{z}{\sinh(z \tilde W)}\right)
\sinh(z(\tilde W-y)),\nonumber
\end{eqnarray}
where in the last step we have taken advantage of the fact that, due to 
eq.~(\ref{esinh}), the product between $\gamma/E$ and $z/\sinh(z \tilde W)$
can only be equal to $+1$ (if the two quantities have the same sign) or
$-1$ (if they have opposite signs).

Moreover we have that
\begin{equation}
\Phi_B^{\vec K}(y)=A e^{zy}+B e^{-zy}=A e^{zy}-A e^{-zy}=A (e^{zy}-e^{-zy})=
2 A \sinh(zy).
\end{equation}
These are edge states, each one exponentially localized on one edge
of the ribbon.

These edge states correspond to bands flattened towards $E=0$, as we can
see both from the graphical solution of eq.~(\ref{realk}) (where we observe
that we have an intersection between $f_1$ and $f_2$ for a $z$ coordinate
very close to $\kappa_x$ and thus the energy
$E=\pm \gamma \sqrt{\kappa_x^2-z^2}$ has a very small value), and from our
previous analytical conclusion that $|E/\gamma|<1/\tilde W$ in this case. Since
the Dirac point $\vec K$, folded into the Brillouin zone $(-\pi/a,\pi/a)$
of the zigzag nanoribbon (the unit cell of which is of length $a$), corresponds
to $k_x=-4\pi/(3a)+2\pi/a=2\pi/(3a)$, the condition $\kappa_x>1/\tilde W$
(under which we have a real solution and thus the edge states) is equivalent to
$k_x=K_x+\kappa_x>2\pi/(3a)+1/\tilde W$ (note the difference between the total
wave vectors $k$ and the wave vectors $\kappa$ measured from the Dirac points).
Therefore in the region $2\pi/(3a)+1/\tilde W<k_x<\pi/a$ we have two bands
flattened towards $E=0$; this means that the zigzag nanoribbons are always
metallic~\cite{nakada}.
However, further studies~\cite{son,fujita1,fujita2} have shown that actual
zigzag nanoribbons have a non-zero gap deriving from a staggered sublattice
potential due to edge magnetization.

Let us now instead consider the {\em imaginary solutions} $z=i \kappa_n$ (with
$\kappa_n$ real) of\break
eq.~(\ref{realk}). In this case the dispersion
relation $E=\pm \gamma \sqrt{\kappa_x^2-z^2}$ becomes $E=$\break
$\pm \gamma \sqrt{\kappa_x^2+\kappa_n^2}$, from which we see more clearly
that $\kappa_x$ and $\kappa_n=-i z$ have the meaning of longitudinal and
transverse components of the wave vector, measured from the Dirac point. 
The solutions are given by
\begin{eqnarray}
&& e^{-2 z \tilde W}=\frac{\kappa_x-z}{\kappa_x+z}
\Rightarrow\\
&& e^{-i 2 \kappa_n \tilde W}=
\frac{\kappa_x-i \kappa_n}{\kappa_x+i \kappa_n}=
\frac{\sqrt{\kappa_x^2+\kappa_n^2}\,e^{-i \angle (\kappa_x+i \kappa_n)}}
{\sqrt{\kappa_x^2+\kappa_n^2}\,e^{i \angle (\kappa_x+i \kappa_n)}}=\nonumber\\
&& e^{-i 2 \angle (\kappa_x+i \kappa_n)}=
e^{-i 2 \angle (\kappa_x+i \kappa_n)} e^{i 2 \pi m}
\Rightarrow\nonumber\\
&& \kappa_n \tilde W=\angle (\kappa_x+i \kappa_n)-\pi m \Rightarrow
\tan (\kappa_n \tilde W)=\frac{\kappa_n}{\kappa_x}\Rightarrow
\kappa_x=\frac{\kappa_n}{\tan (\kappa_n \tilde W)}\nonumber
\end{eqnarray}
(with $m$ integer); $\kappa_n=0$ corresponds to identically null $\Phi$'s
and thus does not have to be considered. We have that
\begin{eqnarray}
\label{esin}
\left(\frac{E}{\gamma}\right)^2 &=& \kappa_x^2+\kappa_n^2=
\left(\frac{\kappa_n}{\tan (\kappa_n \tilde W)}\right)^2+\kappa_n^2=
\left(\frac{\cos^2 (\kappa_n \tilde W)}{\sin^2 (\kappa_n \tilde W)}+1\right)
\kappa_n^2=\\
&& \frac{\cos^2 (\kappa_n \tilde W)+\sin^2 (\kappa_n \tilde W)}
{\sin^2 (\kappa_n \tilde W)} \kappa_n^2=
\frac{\kappa_n^2}{\sin^2 (\kappa_n \tilde W)}\Rightarrow
\left|\frac{E}{\gamma}\right|=
\left|\frac{\kappa_n}{\sin(\kappa_n \tilde W)}\right|;\nonumber
\end{eqnarray}
since (for the properties of the sin function)
$|\sin (\kappa_n \tilde W)|<|\kappa_n \tilde W|=|\kappa_n|\tilde W$, 
we see that in this case
\begin{equation}
\left|\frac{E}{\gamma}\right|>\frac{|\kappa_n|}{|\kappa_n| \tilde W}=
\frac{1}{\tilde W}.
\end{equation}
We can write (exploiting what we have found from the boundary
conditions) that
\begin{eqnarray}
\Phi_A^{\vec K}(y) &=&
\frac{\gamma}{E} \left((\kappa_x-i\kappa_n) A e^{i\kappa_n y}+
(\kappa_x+i\kappa_n) B e^{-i\kappa_n y}\right)=\\
&& \frac{\gamma}{E} \left((\kappa_x-i\kappa_n) A e^{i\kappa_n y}-
(\kappa_x+i\kappa_n) A e^{-i\kappa_n y}\right)=\nonumber\\
&& \frac{\gamma}{E} A \left(\kappa_x (e^{i\kappa_n y}-e^{-i\kappa_n y})-
i\kappa_n (e^{i\kappa_n y}+e^{-i\kappa_n y})\right)=\nonumber\\
&& \frac{\gamma}{E} 2 i A 
\left(\kappa_x \sin(\kappa_n y)-\kappa_n \cos(\kappa_n y)\right)=\nonumber\\
&& 2 i A \frac{\gamma}{E}
\left(\frac{\kappa_n}{\tan(\kappa_n \tilde W)} \sin(\kappa_n y)-
\kappa_n \cos(\kappa_n y)\right)=\nonumber\\
&& 2 i A \frac{\gamma}{E} \kappa_n \,
\frac{\cos(\kappa_n \tilde W)\sin(\kappa_n y)-
\sin(\kappa_n \tilde W)\cos(\kappa_n y)}
{\sin(\kappa_n \tilde W)}=\nonumber\\
&& -2 i A \left(\frac{\gamma}{E} \frac{\kappa_n}{\sin(\kappa_n \tilde W)}\right)
\sin(\kappa_n (\tilde W-y))=\nonumber\\
&& -2 i A \,{\rm sign}\!\left(\frac{E}{\gamma} 
\frac{\kappa_n}{\sin(\kappa_n \tilde W)}\right)
\sin(\kappa_n (\tilde W-y)),\nonumber
\end{eqnarray}
where in the last step we have taken advantage of the fact that, due to 
eq.~(\ref{esin}), the product between $\gamma/E$ and 
$\kappa_n/\sin(\kappa_n \tilde W)$
can only be equal to $+1$ (if the two quantities have the same sign) or
$-1$ (if they have opposite signs).

Moreover we have that
\begin{eqnarray}
\Phi_B^{\vec K}(y) &=& A e^{i\kappa_n y}+B e^{-i\kappa_n y}=
A e^{i\kappa_n y}-A e^{-i\kappa_n y}=\\
&& A (e^{i\kappa_n y}-e^{-i\kappa_n y})=
A 2 i \sin(\kappa_n y).\nonumber
\end{eqnarray}
These are clearly confined states extending all over the ribbon.

%%%%%%%%%% K' %%%%%%%%%%
The calculations {\em around the point} {\boldmath $\vec K'$} are completely
analogous. 
The corresponding part of the Dirac equation is
\begin{eqnarray}
\label{systemz1a}
&& \gamma
\left[\begin{array}{cccc}
0 &
-i\,\frac{\partial}{\partial x}+\frac{\partial}{\partial y} \\
-i\,\frac{\partial}{\partial x}-\frac{\partial}{\partial y} &
0
\end{array}\right]
\left[\begin{array}{c}
F_A^{\vec K'} (\vec r)\\
F_B^{\vec K'} (\vec r)
\end{array}\right]=
E \left[\begin{array}{c}
F_A^{\vec K'} (\vec r)\\
F_B^{\vec K'} (\vec r)
\end{array}\right]
\Rightarrow\\
&& \gamma
\left[\begin{array}{cccc}
0 &
-i\,\frac{\partial}{\partial x}+\frac{\partial}{\partial y} \\
-i\,\frac{\partial}{\partial x}-\frac{\partial}{\partial y} &
0
\end{array}\right]
\left[\begin{array}{c}
\Phi_A^{\vec K'}(y)e^{i \kappa'_x x}\\
\Phi_B^{\vec K'}(y)e^{i \kappa'_x x}
\end{array}\right]=\nonumber\\
&& \gamma
\left[\begin{array}{c}
\kappa'_x \Phi_B^{\vec K'}(y) e^{i \kappa'_x x}+
e^{i \kappa'_x x}\frac{d}{d\,y}\Phi_B^{\vec K'}(y)\\
\kappa'_x \Phi_A^{\vec K'}(y) e^{i \kappa'_x x}-
e^{i \kappa'_x x}\frac{d}{d\,y} \Phi_A^{\vec K'}(y)
\end{array}\right]=\nonumber\\
&& \gamma
\left[\begin{array}{cccc}
0 &
\kappa'_x+\frac{d}{d\,y} \\
\kappa'_x-\frac{d}{d\,y} &
0
\end{array}\right]
\left[\begin{array}{c}
\Phi_A^{\vec K'}(y)\\
\Phi_B^{\vec K'}(y)
\end{array}\right] e^{i \kappa'_x x}=
E \left[\begin{array}{c}
F_A^{\vec K'} (\vec r)\\
F_B^{\vec K'} (\vec r)
\end{array}\right]=\nonumber\\
&& E \left[\begin{array}{c}
\Phi_A^{\vec K'}(y)\\
\Phi_B^{\vec K'}(y)
\end{array}\right] e^{i \kappa'_x x}
\Rightarrow\nonumber\\
&& \left[\begin{array}{cccc}
0 &
\kappa'_x+\frac{d}{d\,y} \\
\kappa'_x-\frac{d}{d\,y} &
0
\end{array}\right]
\left[\begin{array}{c}
\Phi_A^{\vec K'}(y)\\
\Phi_B^{\vec K'}(y)
\end{array}\right]=
\frac{E}{\gamma} \left[\begin{array}{c}
\Phi_A^{\vec K'}(y)\\
\Phi_B^{\vec K'}(y)
\end{array}\right],\nonumber
\end{eqnarray}
which can be rewritten as
\begin{equation}
\label{systemz1b}
\left\{ \begin{array}{l}
\displaystyle
\left(\kappa'_x+\frac{d}{d\,y}\right)
\Phi_B^{\vec K'}(y)=\frac{E}{\gamma}\Phi_A^{\vec K'}(y),\\
\displaystyle
\left(\kappa'_x-\frac{d}{d\,y}\right)
\Phi_A^{\vec K'}(y)=\frac{E}{\gamma}\Phi_B^{\vec K'}(y).
\end{array} \right.
\end{equation}
Obtaining $\Phi_B^{\vec K'}(y)$ from the second of (\ref{systemz1b})
and then substituting $\Phi_A^{\vec K'}(y)$ from the first of
(\ref{systemz1b}), we find
\begin{eqnarray}
\label{solaz1}
\quad \Phi_B^{\vec K'}(y) &=& \frac{\gamma}{E}
\left(\kappa'_x-\frac{d}{d\,y}\right)
\Phi_A^{\vec K'}(y)=\left(\frac{\gamma}{E}\right)^2
\left(\kappa'_x-\frac{d}{d\,y}\right)
\left(\kappa'_x+\frac{d}{d\,y}\right)
\Phi_B^{\vec K'}(y)=\\
&& \left(\frac{\gamma}{E}\right)^2
\left({\kappa'_x}^2+\kappa'_x\frac{d}{d\,y}
-\kappa'_x\frac{d}{d\,y}-
\frac{d^2}{d\,y^2}\right)
\Phi_B^{\vec K'}(y)=\nonumber\\
&& \left(\frac{\gamma}{E}\right)^2
\left({\kappa'_x}^2-
\frac{d^2}{d\,y^2}\right)
\Phi_B^{\vec K'}(y)\Rightarrow\nonumber\\
&& \left(-\frac{d^2}{d\,y^2}+{\kappa'_x}^2\right)
\Phi_B^{\vec K'}(y)=\left(\frac{E}{\gamma}\right)^2
\Phi_B^{\vec K'}(y),\nonumber
\end{eqnarray}
the solution of which is
\begin{eqnarray}
\label{solbz1}
&& \Phi_B^{\vec K'}(y)=C e^{z'y}+D e^{-z'y},\\
&& \hbox{with}\quad
z'=\sqrt{{\kappa'_x}^2-\left(\frac{E}{\gamma}\right)^2}
\quad\left(\;\hbox{and thus}\quad
E=\pm \gamma \sqrt{{\kappa'_x}^2-{z'}^2}\;\right).\nonumber
\end{eqnarray}
Substituting $\Phi_B^{\vec K'}(y)$ back into the first of 
(\ref{systemz1b}), we obtain that
\begin{eqnarray}
\label{solcz1}
\Phi_A^{\vec K'}(y) &=& \frac{\gamma}{E}
\left(\kappa'_x+\frac{d}{d\,y}\right)
\Phi_B^{\vec K'}(y)=\\
&& \frac{\gamma}{E}\left(\kappa'_x C e^{z'y}+\kappa'_x D e^{-z'y}+
z' C e^{z'y}-z' D e^{-z'y}\right)=\nonumber\\
&& \frac{\gamma}{E}
\left((\kappa'_x+z') C e^{z'y}+(\kappa'_x-z') D e^{-z'y}\right).\nonumber
\end{eqnarray}
Let us now enforce the boundary conditions on $\Phi_B^{\vec K'}(y)$ and
$\Phi_A^{\vec K'}(y)$:
\begin{eqnarray}
\label{realk1}
\Phi_B^{\vec K'}(0) &=& 0 \Rightarrow
C+D=0\quad\Rightarrow\quad D=-C;\\
\Phi_A^{\vec K'}(\tilde W) &=& 0 \Rightarrow
\frac{\gamma}{E}\left((\kappa'_x+z') C e^{z' \tilde W}+
(\kappa'_x-z') D e^{-z' \tilde W}\right)=0 \Rightarrow\nonumber\\
&& (\kappa'_x+z') C e^{z' \tilde W}-
(\kappa'_x-z') C e^{-z' \tilde W}=0 \Rightarrow\nonumber\\
&& (\kappa'_x+z') C e^{z' \tilde W}=
(\kappa'_x-z') C e^{-z' \tilde W} \Rightarrow\nonumber\\
e^{-2 z' \tilde W} &=& \frac{\kappa'_x+z'}{\kappa'_x-z'}=
\frac{(-\kappa'_x)-z'}{(-\kappa'_x)+z'},\nonumber
\end{eqnarray}
which is equal to eq.~(\ref{realk}) if we substitute $\kappa_x$ with
$-\kappa'_x$. Therefore the calculations are completely
analogous to those seen around the point $\vec K$.

We consider again real values of $\kappa'_x$.

We conclude~\cite{supplem} that (apart from $z'=0$,
which corresponds to identically null $\Phi$'s) eq.~(\ref{realk1}) has a
{\em real solution} $z'$ only if $\displaystyle -\kappa'_x>1/\tilde W$,
{\em i.e.} if $\displaystyle \kappa'_x<-1/\tilde W$.

If instead $\kappa'_x>-1/\tilde W$, eq.~(\ref{realk1}) does not have real
solutions $z'$ (apart from $z'=0$).

In the case of real $z'$, from eq.~(\ref{realk1}) we can find that
\cite{supplem}
\begin{equation}
\kappa'_x =-\frac{z'}{\tanh(z' \tilde W)}
\end{equation}
($z'=0$ does not have to be considered) and thus~\cite{supplem}
\begin{equation}
\label{esinh1}
\left(\frac{E}{\gamma}\right)^2={\kappa'_x}^2-{z'}^2=
\frac{{z'}^2}{\sinh^2(z' \tilde W)} \Rightarrow
\left|\frac{E}{\gamma}\right|=\left|\frac{z'}{\sinh(z' \tilde W)}\right|
<\frac{|z'|}{|z'| \tilde W}=\frac{1}{\tilde W}.
\end{equation}
The corresponding $\Phi$ functions are
\cite{supplem}
\begin{eqnarray}
\Phi_A^{\vec K'}(y) &=&
\frac{\gamma}{E} 
\left((\kappa'_x+z') C e^{z'y}+(\kappa'_x-z') D e^{-z'y}\right)=\\
&& 2 C \,{\rm sign}\!\left(\frac{E}{\gamma} \frac{z'}{\sinh(z' \tilde W)}\right)
\sinh(z'(\tilde W-y));\nonumber\\
\Phi_B^{\vec K'}(y) &=& C e^{z'y}+D e^{-z'y}=2 C \sinh(z'y).\nonumber
\end{eqnarray}
These are edge states, each one exponentially localized on one edge
of the ribbon.

Also in this case, these edge states correspond to bands flattened towards
$E=0$. Since the Dirac point $\vec K'$, folded into the Brillouin zone
$(-\pi/a,\pi/a)$ of the zigzag nanoribbon, corresponds to
$k_x=4\pi/(3a)-2\pi/a=-2\pi/(3a)$,
the condition $\kappa'_x<-1/\tilde W$ is equivalent to
$k'_x=K'_x+\kappa'_x<-2\pi/(3a)-1/\tilde W$. Therefore also in the region
$-\pi/a<k_x<-2\pi/(3a)-1/\tilde W$ we have two bands flattened towards $E=0$,
which confirms the metallic nature of zigzag nanoribbons.

Let us now instead consider the {\em imaginary solutions} $z'=i \kappa'_n$
(with $\kappa'_n$ real) of eq.~(\ref{realk1}). The dispersion
relation $E=\pm \gamma \sqrt{{\kappa'_x}^2-{z'}^2}$ becomes
$E=\pm \gamma \sqrt{{\kappa'_x}^2+{\kappa'_n}^2}$. 
The solutions are given by~\cite{supplem}
\begin{equation}
\kappa'_x=-\frac{\kappa'_n}{\tan (\kappa'_n \tilde W)}
\end{equation}
($\kappa'_n=0$ corresponds to identically null $\Phi$'s
and thus does not have to be considered) and thus~\cite{supplem}
\begin{equation}
\label{esin1}
\left(\frac{E}{\gamma}\right)^2={\kappa'_x}^2+{\kappa'_n}^2=
\frac{{\kappa'_n}^2}{\sin^2 (\kappa'_n \tilde W)}\Rightarrow
\left|\frac{E}{\gamma}\right|=
\left|\frac{\kappa'_n}{\sin(\kappa'_n \tilde W)}\right|
>\frac{|\kappa'_n|}{|\kappa'_n| \tilde W}=\frac{1}{\tilde W}.
\end{equation}
The corresponding $\Phi$ functions are~\cite{supplem}
\begin{eqnarray}
\Phi_A^{\vec K'}(y) &=&
\frac{\gamma}{E} \left((\kappa'_x+i\kappa'_n) C e^{i\kappa'_n y}+
(\kappa'_x-i\kappa'_n) D e^{-i\kappa'_n y}\right)=\\
&& 2 i C \,{\rm sign}\!\left(\frac{E}{\gamma} 
\frac{\kappa'_n}{\sin(\kappa'_n \tilde W)}\right)
\sin(\kappa'_n (\tilde W-y));\nonumber\\
\Phi_B^{\vec K'}(y) &=& C e^{i\kappa'_n y}+D e^{-i\kappa'_n y}=
C 2 i \sin(\kappa'_n y).\nonumber
\end{eqnarray}
These are confined states extending all over the ribbon.

Obviously, once the expressions of the functions $\Phi$ have been obtained,
the overall wave function is given by the equations (\ref{wavefunctionbisz}),
(\ref{envelopez}) and (\ref{phiz}).

\begin{figure}
\centering
\includegraphics[width=0.8\textwidth,angle=0]{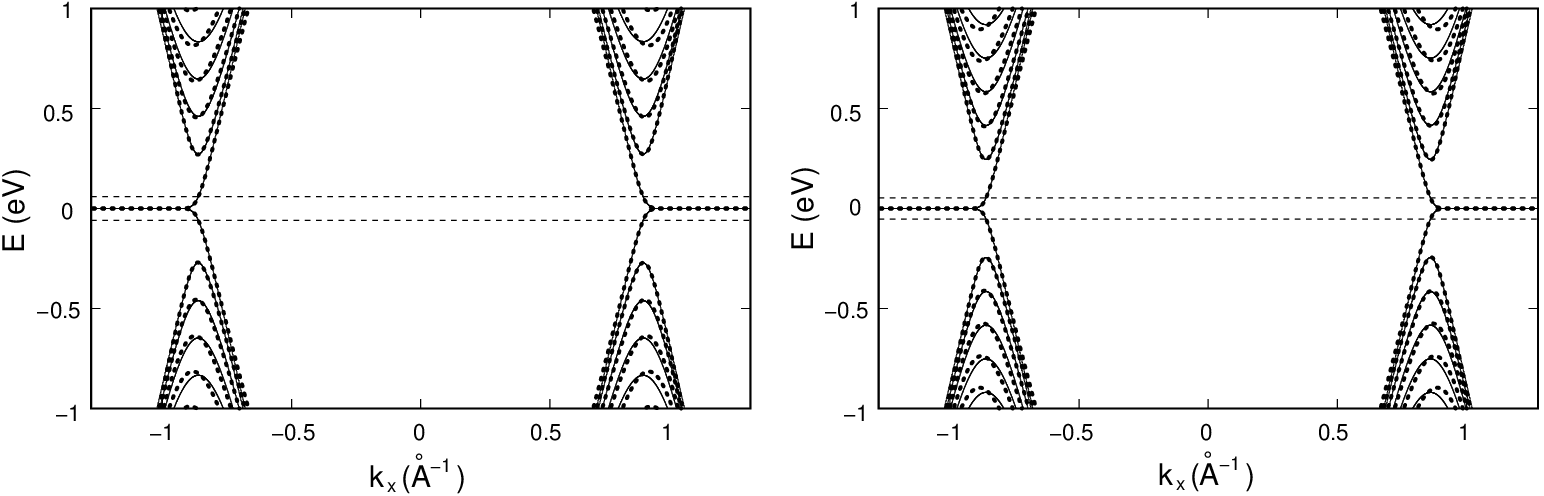}
\caption{Bands of a zigzag nanoribbon with $N=45$ zizag lines (a)
and with $N=50$ zigzag lines (b), computed both with a 
simple tight-binding model not including edge magnetization effects
(thick dotted lines) and with the $\vec k \cdot \vec p$ method
(thin solid lines). The two dashed lines correspond to the energy values
$\pm\gamma/\tilde W$; the dispersion relations in the region between the
two dashed lines are obtained for real values of $z$, while those outside
this region correspond to purely imaginary values of $z$.}
\label{f12}
\end{figure}\noindent

In fig.~\ref{f12} we show the bands of a zigzag nanoribbon with $N=45$ zigzag
lines and of a zigzag nanoribbon with $N=50$ zigzag lines, that we have 
computed both with a simple tight-binding model not including edge
magnetization effects (thick dotted lines) and with the $\vec k \cdot \vec p$
(Dirac equation) method (thin solid lines). For low energy values and for not
too narrow ribbons the results obtained with the two techniques are very
similar.
In both cases, the presence of the two bands flattened towards zero and
corresponding to the edge states can be clearly seen.

\begin{figure}[b]
\centering
\includegraphics[width=\textwidth,angle=0]{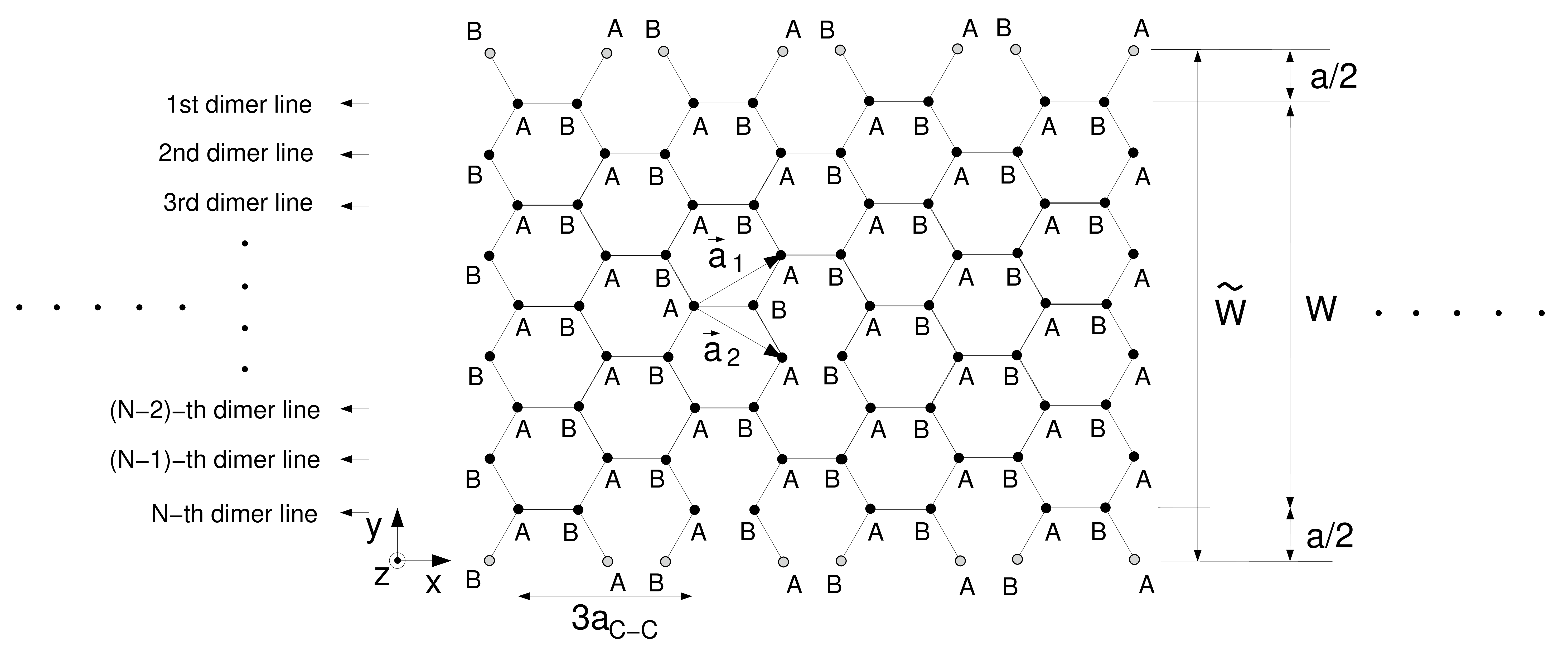}
\caption{Sketch of an armchair nanoribbon with $N$ dimer lines (the black
atoms are carbon atoms, while the grey atoms are passivation atoms).}
\label{f13}
\end{figure}\noindent

%%%%%%%%%% ARMCHAIR %%%%%%%%%%
\subsection{Armchair nanoribbons}
Instead, in the case of armchair nanoribbons (fig.~\ref{f13}), the graphene
sheet has been cut along the direction of the nearest-neighbor carbon bonds,
and therefore the edges have an armchair shape. In order to
simplify the following calculations, we can choose (see fig.~\ref{f13})
the graphene lattice vectors in the real space $\vec a_1$ and $\vec a_2$
(and consequently those in the reciprocal space $\vec b_1$ and $\vec b_2$)
in this way (we express them in the reference frame
$\Sigma=(\hbox{\boldmath{$\hat x$}},\hbox{\boldmath{$\hat y$}},
\hbox{\boldmath{$\hat z$}})$):
\begin{equation}
\vec a_1 \mathrel{\mathop\equiv_{\Sigma}}
\left[\begin{array}{c}
\displaystyle \frac{\sqrt{3}}{2}\,a \\[5pt]
\noalign{\vskip3pt}
\displaystyle \frac{a}{2} \\[5pt]
\noalign{\vskip3pt}
0
\end{array}\right]
,\quad
\vec a_2 \mathrel{\mathop\equiv_{\Sigma}}
\left[\begin{array}{c}
\displaystyle \frac{\sqrt{3}}{2}\,a \\[5pt]
\noalign{\vskip3pt}
\displaystyle -\frac{a}{2} \\[5pt]
\noalign{\vskip3pt}
0
\end{array}\right]
,\quad
\vec b_1 \mathrel{\mathop\equiv_{\Sigma}}
\left[\begin{array}{c}
\displaystyle \frac{2\pi}{\sqrt{3}a}\\ [5pt]
\noalign{\vskip3pt}
\displaystyle \frac{2\pi}{a}\\[5pt]
\noalign{\vskip3pt}
0
\end{array}\right]
,\quad
\vec b_2 \mathrel{\mathop\equiv_{\Sigma}}
\left[\begin{array}{c}
\displaystyle \frac{2\pi}{\sqrt{3}a}\\[5pt]
\noalign{\vskip3pt}
\displaystyle -\frac{2\pi}{a}\\[5pt]
\noalign{\vskip3pt}
0
\end{array}\right]
\end{equation}
(which, being 
$\vec b_1=2\pi (\vec a_2 \times \hbox{\boldmath{$\hat z$}})/
(\vec a_1 \cdot (\vec a_2 \times \hbox{\boldmath{$\hat z$}}))$ and
$\vec b_2=2\pi (\hbox{\boldmath{$\hat z$}} \times \vec a_1)/
(\vec a_1 \cdot (\vec a_2 \times \hbox{\boldmath{$\hat z$}}))$,
fulfill the relation $\vec a_i \cdot \vec b_j=2 \pi \delta_{ij}$).
Consequently we have that
\begin{eqnarray}
\label{ka}
&& \vec K =\frac{1}{3}(\vec b_2-\vec b_1) \mathrel{\mathop\equiv_{\Sigma}}
\frac{4\pi}{3a}
\left[\begin{array}{c}
0\\[6pt]
\noalign{\vskip3pt}
-1\\[6pt]
\noalign{\vskip3pt} 
0
\end{array}\right]=
\left[\begin{array}{c}
0\\[6pt]
\noalign{\vskip3pt}
-K\\[6pt]
\noalign{\vskip3pt} 
0
\end{array}\right],\\[6pt]
&& \vec K' =\frac{1}{3}(\vec b_1-\vec b_2) \mathrel{\mathop\equiv_{\Sigma}}
\frac{4\pi}{3a}
\left[\begin{array}{c}
0\\[6pt]
\noalign{\vskip3pt}
1\\[6pt]
\noalign{\vskip3pt} 
0
\end{array}\right]=
\left[\begin{array}{c}
0\\[6pt]
\noalign{\vskip3pt}
K\\[6pt]
\noalign{\vskip3pt} 
0
\end{array}\right],\nonumber
\end{eqnarray}
with $K=4\pi/(3a)$. For our choice of $\vec a_1$
and $\vec a_2$, the angle $\theta'$ from the vector $\vec a_1+\vec a_2$ (i.e.
from the axis $\hbox{\boldmath{$\hat x$}}'$ used in previous calculations)
to the axis $\hbox{\boldmath{$\hat x$}}$ (taken in the longitudinal direction)
is equal to $0$.

Therefore the total wave function is given by (eq.~\ref{wavefunction})
\begin{equation}
\label{wavefunctionbisa}
\psi (\vec r)=
\sum_{\vec R_A}\psi_A (\vec R_A)\varphi(\vec r -\vec R_A)+
\sum_{\vec R_B}\psi_B (\vec R_B)\varphi(\vec r -\vec R_B),
\end{equation}
with (eq.~(\ref{assumptions2}) with $\theta'=0$)
\begin{equation}
\label{envelopea}
\left\{ \begin{array}{l}
\displaystyle
\psi_A (\vec r)=
e^{i \vec K\cdot \vec r} F_A^{\vec K}(\vec r)-
i\,e^{i \vec K'\cdot \vec r} F_A^{\vec K'}(\vec r),\\[6pt]
\displaystyle
\psi_B (\vec r)=
i\,e^{i \vec K\cdot \vec r} F_B^{\vec K} (\vec r)+
e^{i \vec K'\cdot \vec r} F_B^{\vec K'} (\vec r),
\end{array} \right.
\end{equation}
where (using eq.~(\ref{ka})), if we write
$\displaystyle \vec r \mathrel{\mathop\equiv_{\Sigma}} [x,y,0]^T$ we
have that $\vec K\cdot \vec r=-Ky$ and that $\vec K'\cdot \vec r=Ky$.
In the absence of an external potential the envelope functions
satisfy the usual Dirac equation (eq.~(\ref{absence}))
\begin{eqnarray}
&& \gamma
\left[\begin{array}{cccc}
0 &
-i\,\frac{\partial}{\partial x}-\frac{\partial}{\partial y} &
0 &
0 \\[6pt]
-i\,\frac{\partial}{\partial x}+\frac{\partial}{\partial y} &
0 &
0 &
0 \\[6pt]
0 &
0 &
0 &
-i\,\frac{\partial}{\partial x}+\frac{\partial}{\partial y} \\[6pt]
0 &
0 &
-i\,\frac{\partial}{\partial x}-\frac{\partial}{\partial y} &
0
\end{array}\right]
\left[\begin{array}{c}
F_A^{\vec K} (\vec r)\\[6pt]
F_B^{\vec K} (\vec r)\\[6pt]
F_A^{\vec K'} (\vec r)\\[6pt]
F_B^{\vec K'} (\vec r)
\end{array}\right]=\\[6pt]
&& E \left[\begin{array}{c}
F_A^{\vec K} (\vec r)\\[6pt]
F_B^{\vec K} (\vec r)\\[6pt]
F_A^{\vec K'} (\vec r)\\[6pt]
F_B^{\vec K'} (\vec r)
\end{array}\right].\nonumber
\end{eqnarray}
Due to the translational invariance along the $x$-direction, we can write
the envelope functions as the product of a propagating part along the
longitudinal direction $x$ and of a confined part along the transverse
direction $y$. Here we have to consider the same longitudinal component
$\kappa_x$ for the wave vector measured from $\vec K$ and $\vec K'$
because in this case if we consider $\kappa'_x \ne \kappa_x$ the
boundary conditions are satisfied for every $x$ only by the identically
null wave function. Therefore we can assume that
\begin{equation}
\label{phia}
\left[\begin{array}{c}
F_A^{\vec K} (\vec r)\\
F_B^{\vec K} (\vec r)
\end{array}\right]=
e^{i \kappa_x x}
\left[\begin{array}{c}
\Phi_A^{\vec K} (y)\\
\Phi_B^{\vec K} (y)
\end{array}\right]
\ \hbox{and that}\ 
\left[\begin{array}{c}
F_A^{\vec K'} (\vec r)\\
F_B^{\vec K'} (\vec r)
\end{array}\right]=
e^{i \kappa_x x}
\left[\begin{array}{c}
\Phi_A^{\vec K'} (y)\\
\Phi_B^{\vec K'} (y)
\end{array}\right].
\end{equation}
We have to enforce that the overall wave function vanishes in correspondence
with the ``boundary lattice sites'' on the lower and upper edges of the ribbon.
Let us define as $W$ the real width of the nanoribbon, {\em i.e.} the distance between
the bottom row and the top row of carbon atoms of the ribbon; if the ribbon
has $N$ dimer lines across the ribbon width, we have that $W=(N-1)a/2$.
If we take $y=0$ in correspondence of the row of ``boundary lattice sites''
on the lower edge of the ribbon, the row of ``boundary lattice sites'' on the
upper edge of the ribbon will be at $y=\tilde W=W+2 \, a/2=W+a=(N+1)a/2$.
Therefore, for every $x$, we must have $\psi (x,y=0)=\psi (x,y=\tilde W)=0$.
We notice that in an armchair nanoribbon the ``boundary lattice sites''
on the lower and upper edges belong to both the $A$ and the $B$ sublattices.
Therefore, looking at eq.~(\ref{wavefunctionbisa}) and observing that
the atomic orbitals $\varphi$ are strongly localized around the atom on which
they are centered, the boundary condition on the wave function is equivalent
to setting, for every $x$, 
$\psi_A (x,y=0)=\psi_B (x,y=0)=\psi_A (x,y=\tilde W)=\psi_B (x,y=\tilde W)=0$. 
Using eq.~(\ref{envelopea}) we obtain the following 4 boundary conditions:
\begin{eqnarray}
\label{b1}
&& \psi_A (x,y=0)=0 \quad\forall x \Rightarrow
e^{-iK0} F_A^{\vec K} (x,y=0)-i e^{iK0} F_A^{\vec K'} (x,y=0)=\\
&& F_A^{\vec K} (x,y=0)-i F_A^{\vec K'} (x,y=0)=
e^{i \kappa_x x} \Phi_A^{\vec K}(0)-i e^{i \kappa_x x} \Phi_A^{\vec K'}(0)=
0\quad\forall x \Rightarrow\nonumber\\ 
&& \Phi_A^{\vec K}(0)-i \Phi_A^{\vec K'}(0)=0;\nonumber
\end{eqnarray}
\begin{eqnarray}
\label{b2}
&& \psi_B (x,y=0)=0 \quad\forall x \Rightarrow
i e^{-iK0} F_B^{\vec K} (x,y=0)+e^{iK0} F_B^{\vec K'} (x,y=0)=\\
&& i F_B^{\vec K} (x,y=0)+F_B^{\vec K'} (x,y=0)=
i e^{i \kappa_x x} \Phi_B^{\vec K}(0)+
e^{i \kappa_x x} \Phi_B^{\vec K'}(0)=0\quad\forall x 
\Rightarrow\nonumber\\
&& i \Phi_B^{\vec K}(0)+\Phi_B^{\vec K'}(0)=0;\nonumber
\end{eqnarray}
\begin{eqnarray}
\label{b3}
&& \psi_A (x,y=\tilde W)=0 \quad\forall x \Rightarrow
e^{-iK \tilde W} F_A^{\vec K} (x,y=\tilde W)-
i e^{iK \tilde W} F_A^{\vec K'} (x,y=\tilde W)=\\
&& e^{-iK \tilde W} e^{i \kappa_x x} \Phi_A^{\vec K}(\tilde W)-
i e^{iK \tilde W} e^{i \kappa_x x} \Phi_A^{\vec K'}(\tilde W)=
0\quad\forall x \Rightarrow\nonumber\\
&& e^{-iK \tilde W} \Phi_A^{\vec K}(\tilde W)-
i e^{iK \tilde W} \Phi_A^{\vec K'}(\tilde W)=0;\nonumber
\end{eqnarray}
\begin{eqnarray}
\label{b4}
&& \psi_B (x,y=\tilde W)=0 \quad\forall x \Rightarrow
i e^{-iK \tilde W} F_B^{\vec K} (x,y=\tilde W)+
e^{iK \tilde W} F_B^{\vec K'} (x,y=\tilde W)=\\
&& i e^{-iK \tilde W} e^{i \kappa_x x} \Phi_B^{\vec K}(\tilde W)+
e^{iK \tilde W} e^{i \kappa_x x} \Phi_B^{\vec K'}(\tilde W)=
0\quad\forall x \Rightarrow\nonumber\\
&& i e^{-iK \tilde W} \Phi_B^{\vec K}(\tilde W)+
e^{iK \tilde W} \Phi_B^{\vec K'}(\tilde W)=0.\nonumber
\end{eqnarray}
As we can see, in armchair nanoribbons the boundary conditions couple
the envelope functions relative to the Dirac points $\vec K$ and $\vec K'$.

We can solve the part of the Dirac equation around the point $\vec K$, that is
\begin{equation}
\label{systema}
\gamma
\left[\begin{array}{cccc}
0 &
-i\,\frac{\partial}{\partial x}-\frac{\partial}{\partial y} \\[3pt]
-i\,\frac{\partial}{\partial x}+\frac{\partial}{\partial y} &
0
\end{array}\right]
\left[\begin{array}{c}
F_A^{\vec K} (\vec r)\\[3pt]
F_B^{\vec K} (\vec r)
\end{array}\right]=
E \left[\begin{array}{c}
F_A^{\vec K} (\vec r)\\[3pt]
F_B^{\vec K} (\vec r)
\end{array}\right],
\end{equation}
repeating the calculations made for zigzag nanoribbons
(eqs.~(\ref{systemza})-(\ref{solcz})) and obtaining that
\begin{equation}
\left\{ \begin{array}{l}
\displaystyle
\Phi_A^{\vec K}(y)=
\frac{\gamma}{E}\left((\kappa_x-z) A e^{zy}+(\kappa_x+z) B e^{-zy}\right),\\[5pt]
\displaystyle
\Phi_B^{\vec K}(y)=A e^{zy}+B e^{-zy},
\end{array} \right.
\end{equation}
with $z=\sqrt{\kappa_x^2-\left(\frac{E}{\gamma}\right)^2}$
and thus $E=\pm \gamma \sqrt{\kappa_x^2-z^2}$.

Analogously, we can solve the part of the Dirac equation around the point
$\vec K'$, that is
\begin{equation}
\gamma
\left[\begin{array}{cccc}
0 &
-i\,\frac{\partial}{\partial x}+\frac{\partial}{\partial y} \\[3pt]
-i\,\frac{\partial}{\partial x}-\frac{\partial}{\partial y} &
0
\end{array}\right]
\left[\begin{array}{c}
F_A^{\vec K'} (\vec r)\\[3pt]
F_B^{\vec K'} (\vec r)
\end{array}\right]=
E \left[\begin{array}{c}
F_A^{\vec K'} (\vec r)\\[3pt]
F_B^{\vec K'} (\vec r)
\end{array}\right],
\end{equation}
repeating the calculations made for zigzag nanoribbons
(eqs.~(\ref{systemz1a})-(\ref{solcz1}), with the difference that $\kappa'_x$
and $z'$ here have to be replaced by $\kappa_x$ and $z$) and obtaining
that
\begin{equation}
\left\{ \begin{array}{l}
\displaystyle
\Phi_A^{\vec K'}(y)=
\frac{\gamma}{E}\left((\kappa_x+z) C e^{zy}+(\kappa_x-z) D e^{-zy}\right),\\[5pt]
\displaystyle
\Phi_B^{\vec K'}(y)=C e^{zy}+D e^{-zy},
\end{array} \right.
\end{equation}
with (as written before) 
$z=\sqrt{{\kappa_x}^2-\left(\frac{E}{\gamma}\right)^2}$
and thus $E=\pm \gamma \sqrt{{\kappa_x}^2-{z}^2}$.

Let us define $z=i \kappa_n$. In this case the dispersion relation becomes
$E=$\break
$\pm \gamma \sqrt{{\kappa_x}^2+{\kappa_n}^2}$; therefore $\kappa_x$ and
$\kappa_n=-i z$ are the longitudinal and transverse components of the
wave vector, measured from the Dirac point.

The functions $\Phi$ become
\begin{equation}
\label{phiaa}
\left\{ \begin{array}{l}
\displaystyle
\Phi_A^{\vec K}(y)=
\frac{\gamma}{E}\left((\kappa_x-i\kappa_n) A e^{i\kappa_n y}+
(\kappa_x+i\kappa_n) B e^{-i\kappa_n y}\right),\\[5pt]
\displaystyle
\Phi_B^{\vec K}(y)=A e^{i\kappa_n y}+B e^{-i\kappa_n y},\\[5pt]
\displaystyle
\Phi_A^{\vec K'}(y)=
\frac{\gamma}{E}\left((\kappa_x+i\kappa_n) C e^{i\kappa_n y}+
(\kappa_x-i\kappa_n) D e^{-i\kappa_n y}\right),\\[5pt]
\displaystyle
\Phi_B^{\vec K'}(y)=C e^{i\kappa_n y}+D e^{-i\kappa_n y}.
\end{array} \right.
\end{equation}
Now we can enforce the 4 boundary conditions (\ref{b1})-(\ref{b4}),
obtaining:
\begin{eqnarray}
\label{bc1}
\qquad && \Phi_A^{\vec K}(0)-i \Phi_A^{\vec K'}(0)=0 \Rightarrow\\[5pt]
&& (\kappa_x-i\kappa_n)A+(\kappa_x+i\kappa_n)B
-i(\kappa_x+i\kappa_n)C-i(\kappa_x-i\kappa_n)D=0 \Rightarrow\nonumber\\[5pt]
&& \kappa_x\,(A+B-iC-iD)+i\kappa_n\,(-A+B-iC+iD)=0;\nonumber\\[5pt]
\label{bc2}
&& i \Phi_B^{\vec K}(0)+\Phi_B^{\vec K'}(0)=0 \Rightarrow
i(A+B)+(C+D)=0 \Rightarrow\\[5pt]
&& A+B-iC-iD=0;\nonumber\\[5pt]
\label{bc3}
&& e^{-iK \tilde W} \Phi_A^{\vec K}(\tilde W)
-i e^{iK \tilde W} \Phi_A^{\vec K'}(\tilde W)=0 \Rightarrow\\[5pt]
&& e^{-iK \tilde W} (\kappa_x-i\kappa_n) A e^{i\kappa_n \tilde W}
+e^{-iK \tilde W} (\kappa_x+i\kappa_n) B e^{-i\kappa_n \tilde W}\nonumber\\[5pt]
&& -i e^{iK \tilde W} (\kappa_x+i\kappa_n) C e^{i\kappa_n \tilde W}
-i e^{iK \tilde W} (\kappa_x-i\kappa_n) D e^{-i\kappa_n \tilde W}=\nonumber\\[5pt]
&& \kappa_x 
\left(A e^{i(\kappa_n-K)\tilde W}+B e^{-i(\kappa_n+K)\tilde W}
-iC e^{i(\kappa_n+K)\tilde W}-iD e^{-i(\kappa_n-K)\tilde W}\right)\nonumber\\[5pt]
&& +i\kappa_n
\left(-A e^{i(\kappa_n-K)\tilde W}+B e^{-i(\kappa_n+K)\tilde W}
-iC e^{i(\kappa_n+K)\tilde W}+iD e^{-i(\kappa_n-K)\tilde W}\right)=0;\nonumber\\[5pt]
\label{bc4}
&& i e^{-iK \tilde W} \Phi_B^{\vec K}(\tilde W)
+e^{iK \tilde W} \Phi_B^{\vec K'}(\tilde W)=0 \Rightarrow\\[5pt]
&& i e^{-iK \tilde W} (A e^{i\kappa_n \tilde W}+B e^{-i\kappa_n \tilde W})
+e^{iK \tilde W} (C e^{i\kappa_n \tilde W}+D e^{-i\kappa_n \tilde W})=0 
\Rightarrow\nonumber\\[5pt]
&& iA e^{i(\kappa_n-K)\tilde W}+iB e^{-i(\kappa_n+K)\tilde W}
+C e^{i(\kappa_n+K)\tilde W}+D e^{-i(\kappa_n-K)\tilde W}=0 
\Rightarrow\nonumber\\[5pt]
&& A e^{i(\kappa_n-K)\tilde W}+B e^{-i(\kappa_n+K)\tilde W}
-iC e^{i(\kappa_n+K)\tilde W}-iD e^{-i(\kappa_n-K)\tilde W}=0.\nonumber
\end{eqnarray}
In the following we examine the different cases in which all of these 4
boundary conditions are satisfied.

\noindent
{\bf Case I}

\noindent
If $\kappa_n=0$ the condition (\ref{bc1}) is equivalent to the condition
(\ref{bc2}), and the condition (\ref{bc3}) is equivalent to the condition
(\ref{bc4}).

But the condition (\ref{bc2}) is satisfied if
\begin{equation}
A+B-iC-iD=0 \Rightarrow
A+B=iC+iD \Rightarrow
\left\{ \begin{array}{l}
A+B=G\\[5pt]
C+D=-iG
\end{array} \right.
\end{equation}
(where we have defined $A+B\equiv G$).

The condition (\ref{bc4}) instead is satisfied if (exploiting the fact that
$\kappa_n=0$)
\begin{eqnarray}
&& A e^{i(\kappa_n-K)\tilde W}+B e^{-i(\kappa_n+K)\tilde W}
-iC e^{i(\kappa_n+K)\tilde W}-iD e^{-i(\kappa_n-K)\tilde W}=0 
\Rightarrow\\[5pt]
&& A e^{-iK\tilde W}+B e^{-iK\tilde W}
-iC e^{iK\tilde W}-iD e^{iK\tilde W}=0
\Rightarrow\nonumber\\[5pt]
&& (A+B) e^{-iK\tilde W}-i(C+D) e^{iK\tilde W}=0
\Rightarrow
G e^{-iK\tilde W}-G e^{iK\tilde W}=0
\Rightarrow\nonumber\\[5pt]
&& -G (e^{iK\tilde W}-e^{-iK\tilde W})=0
\Rightarrow
-G 2 i \sin(K\tilde W)=0.\nonumber
\end{eqnarray}
Since in this case ($\kappa_n=0$) for $G=0$ all the $\Phi$ functions
(\ref{phiaa}) would become identically null and thus we have to consider
$G \ne 0$, this equation can be satisfied only if $\sin(K\tilde W)=0$.
But, since $K=4\pi/(3a)$ and $\tilde W=(N+1)a/2$, we have that
\begin{equation}
\sin(K\tilde W)=0 \Rightarrow
\sin\left(\frac{4\pi}{3a}(N+1)\frac{a}{2}\right)=0 \Rightarrow
\sin\left(\frac{2\pi}{3}(N+1)\right)=0
\end{equation}
and this is true only if $N+1$ is a multiple of 3, {\em i.e.} if $N+1=3M$ (with
$M$ integer) and thus $N=3M-1$. In this case we have that
\begin{eqnarray}
K\tilde W &=& \frac{2\pi}{3}(N+1)=\frac{2\pi}{3}(3M)=2\pi M \Rightarrow\\[5pt]
K &=& 2M\frac{\pi}{\tilde W}\Rightarrow
2M\frac{\pi}{\tilde W}-K=0(=\kappa_n)\nonumber
\end{eqnarray}
and the nanoribbon is metallic (as we will see). Being $\kappa_n=0$, the
$\Phi$ functions (\ref{phiaa}) are equal to
\begin{equation}
\left\{ \begin{array}{l}
\displaystyle
\Phi_A^{\vec K}(y)=
\frac{\gamma}{E}(\kappa_x A+\kappa_x B)=\frac{\gamma}{E}\kappa_x (A+B)=
\frac{\gamma}{E}\kappa_x G,\\[8pt]
\displaystyle
\Phi_B^{\vec K}(y)=A+B=G,\\[8pt]
\displaystyle
\Phi_A^{\vec K'}(y)=
\frac{\gamma}{E}(\kappa_x C+\kappa_x D)=
\frac{\gamma}{E} \kappa_x (C+D)=
-\frac{\gamma}{E} \kappa_x i G,\\[8pt]
\displaystyle
\Phi_B^{\vec K'}(y)=C+D=-iG.
\end{array} \right.
\end{equation}

\vskip10pt\noindent
\noindent
{\bf Case II}
\vskip5pt\noindent

\noindent
The other possibility is to satisfy the conditions (\ref{bc1})-(\ref{bc2})
in this way:
\begin{equation}
\label{firsttwo}
\left\{ \begin{array}{l}
A+B-iC-iD=0\\[5pt]
-A+B-iC+iD=0
\end{array} \right.
\Rightarrow
\left\{ \begin{array}{l}
2B-2iC=0\\[5pt]
2A-2iD=0
\end{array} \right.
\Rightarrow
\left\{ \begin{array}{l}
C=-iB\\[5pt]
D=-iA
\end{array} \right.
\end{equation}
(where in the first step we have summed and subtracted the two equations of
the system), and to satisfy the conditions (\ref{bc3})-(\ref{bc4}) enforcing
\begin{equation}
\left\{ \begin{array}{l} 
A e^{i(\kappa_n-K)\tilde W}+B e^{-i(\kappa_n+K)\tilde W}
-iC e^{i(\kappa_n+K)\tilde W}-iD e^{-i(\kappa_n-K)\tilde W}=0,\\[5pt]
-A e^{i(\kappa_n-K)\tilde W}+B e^{-i(\kappa_n+K)\tilde W}
-iC e^{i(\kappa_n+K)\tilde W}+iD e^{-i(\kappa_n-K)\tilde W}=0.
\end{array} \right.
\end{equation}
Using (\ref{firsttwo}), we can write these equations in the following form:
\begin{equation}
\left\{ \begin{array}{l} 
A e^{i(\kappa_n-K)\tilde W}+B e^{-i(\kappa_n+K)\tilde W}
-B e^{i(\kappa_n+K)\tilde W}-A e^{-i(\kappa_n-K)\tilde W}=0,\\[5pt]
-A e^{i(\kappa_n-K)\tilde W}+B e^{-i(\kappa_n+K)\tilde W}
-B e^{i(\kappa_n+K)\tilde W}+A e^{-i(\kappa_n-K)\tilde W}=0.
\end{array} \right.
\end{equation}
If now we separate the real and imaginary part of $\kappa_n$ 
({\em i.e.} we write $\kappa_n$ as $\kappa_n=\kappa_{nr}+i\kappa_{ni}$) we have
that
\begin{eqnarray}
&& \left\{ \begin{array}{l}
A e^{-\kappa_{ni} \tilde W} e^{i(\kappa_{nr}-K)\tilde W}
+B e^{\kappa_{ni} \tilde W} e^{-i(\kappa_{nr}+K)\tilde W}\\[4pt]
-B e^{-\kappa_{ni} \tilde W} e^{i(\kappa_{nr}+K)\tilde W}
-A e^{\kappa_{ni} \tilde W} e^{-i(\kappa_{nr}-K)\tilde W}=0,\\[4pt]
-A e^{-\kappa_{ni} \tilde W} e^{i(\kappa_{nr}-K)\tilde W}
+B e^{\kappa_{ni} \tilde W} e^{-i(\kappa_{nr}+K)\tilde W}\\[4pt]
-B e^{-\kappa_{ni} \tilde W} e^{i(\kappa_{nr}+K)\tilde W}
+A e^{\kappa_{ni} \tilde W} e^{-i(\kappa_{nr}-K)\tilde W}=0,
\end{array} \right.\Rightarrow\\[4pt]
&& \left\{ \begin{array}{l}
\big[A e^{-\kappa_{ni} \tilde W} \cos((\kappa_{nr}-K)\tilde W)
+B e^{\kappa_{ni} \tilde W} \cos((\kappa_{nr}+K)\tilde W)\\[4pt]
-B e^{-\kappa_{ni} \tilde W} \cos((\kappa_{nr}+K)\tilde W)
-A e^{\kappa_{ni} \tilde W} \cos((\kappa_{nr}-K)\tilde W)\big]\\[4pt]
+i\big[A e^{-\kappa_{ni} \tilde W} \sin((\kappa_{nr}-K)\tilde W)
-B e^{\kappa_{ni} \tilde W} \sin((\kappa_{nr}+K)\tilde W)\\[4pt]
-B e^{-\kappa_{ni} \tilde W} \sin((\kappa_{nr}+K)\tilde W)
+A e^{\kappa_{ni} \tilde W} \sin((\kappa_{nr}-K)\tilde W)\big]=0,\\[4pt]
\big[-A e^{-\kappa_{ni} \tilde W} \cos((\kappa_{nr}-K)\tilde W)
+B e^{\kappa_{ni} \tilde W} \cos((\kappa_{nr}+K)\tilde W)\\[4pt]
-B e^{-\kappa_{ni} \tilde W} \cos((\kappa_{nr}+K)\tilde W)
+A e^{\kappa_{ni} \tilde W} \cos((\kappa_{nr}-K)\tilde W)\big]\\[4pt]
+i\big[-A e^{-\kappa_{ni} \tilde W} \sin((\kappa_{nr}-K)\tilde W)
-B e^{\kappa_{ni} \tilde W} \sin((\kappa_{nr}+K)\tilde W)\\[4pt]
-B e^{-\kappa_{ni} \tilde W} \sin((\kappa_{nr}+K)\tilde W)
-A e^{\kappa_{ni} \tilde W} \sin((\kappa_{nr}-K)\tilde W)\big]=0,
\end{array} \right.\Rightarrow\nonumber\\[4pt]
&& \left\{ \begin{array}{l}
-(e^{\kappa_{ni} \tilde W}-e^{-\kappa_{ni} \tilde W})
\big[A \cos((\kappa_{nr}-K)\tilde W)-B \cos((\kappa_{nr}+K)\tilde W)\big]\\[4pt]
+i(e^{\kappa_{ni} \tilde W}+e^{-\kappa_{ni} \tilde W})
\big[A \sin((\kappa_{nr}-K)\tilde W)-B \sin((\kappa_{nr}+K)\tilde W)\big]=0,\\[4pt]
(e^{\kappa_{ni} \tilde W}-e^{-\kappa_{ni} \tilde W})
\big[A \cos((\kappa_{nr}-K)\tilde W)+B \cos((\kappa_{nr}+K)\tilde W)\big]\\[4pt]
-i(e^{\kappa_{ni} \tilde W}+e^{-\kappa_{ni} \tilde W})
\big[A \sin((\kappa_{nr}-K)\tilde W)+B \sin((\kappa_{nr}+K)\tilde W)\big]=0,
\end{array} \right.\Rightarrow\nonumber\\[4pt]
&& \left\{ \begin{array}{l}
-2\sinh(\kappa_{ni} \tilde W)
\big[A \cos((\kappa_{nr}-K)\tilde W)-B \cos((\kappa_{nr}+K)\tilde W)\big]\\[4pt]
+i2\cosh(\kappa_{ni} \tilde W)
\big[A \sin((\kappa_{nr}-K)\tilde W)-B \sin((\kappa_{nr}+K)\tilde W)\big]=0,\\[4pt]
2\sinh(\kappa_{ni} \tilde W)
\big[A \cos((\kappa_{nr}-K)\tilde W)+B \cos((\kappa_{nr}+K)\tilde W)\big]\\[4pt]
-i2\cosh(\kappa_{ni} \tilde W)
\big[A \sin((\kappa_{nr}-K)\tilde W)+B \sin((\kappa_{nr}+K)\tilde W)\big]=0.
\end{array} \right.\nonumber
\end{eqnarray}
If we sum and subtract the two equations, we obtain
\begin{eqnarray}
\label{secondtwo}
&& \left\{ \begin{array}{l}
4\sinh(\kappa_{ni} \tilde W) B \cos((\kappa_{nr}+K)\tilde W)\\[4pt]
-i4\cosh(\kappa_{ni} \tilde W) B \sin((\kappa_{nr}+K)\tilde W)=0,\\[4pt]
-4\sinh(\kappa_{ni} \tilde W) A \cos((\kappa_{nr}-K)\tilde W)\\[4pt]
+i4\cosh(\kappa_{ni} \tilde W) A \sin((\kappa_{nr}-K)\tilde W)=0,
\end{array} \right.\Rightarrow\\[4pt]
&& \left\{ \begin{array}{l}
B \big[\sinh(\kappa_{ni} \tilde W) \cos((\kappa_{nr}+K)\tilde W)
-i \cosh(\kappa_{ni} \tilde W) \sin((\kappa_{nr}+K)\tilde W)\big]=0,\\[4pt]
A \big[\sinh(\kappa_{ni} \tilde W) \cos((\kappa_{nr}-K)\tilde W)
-i \cosh(\kappa_{ni} \tilde W) \sin((\kappa_{nr}-K)\tilde W)\big]=0.
\end{array} \right.\nonumber
\end{eqnarray}
Apart from the case $A=B=0$, which (being also $C=-iB$ and $D=-iA$) gives
identically null functions $\Phi$, both of these two equations are satisfied
in 3 cases, that we will indicate with II-A, II-B and II-C.

\noindent
{\bf Case II-A}
\vskip5pt

\noindent
The eqs.~(\ref{secondtwo}) are satisfied if
\begin{equation}
\left\{ \begin{array}{l}
\sinh(\kappa_{ni} \tilde W) \cos((\kappa_{nr}+K)\tilde W)
-i \cosh(\kappa_{ni} \tilde W) \sin((\kappa_{nr}+K)\tilde W)=0,\\[6pt]
\sinh(\kappa_{ni} \tilde W) \cos((\kappa_{nr}-K)\tilde W)
-i \cosh(\kappa_{ni} \tilde W) \sin((\kappa_{nr}-K)\tilde W)=0.
\end{array} \right.
\end{equation}
If we separately equate to zero the real and imaginary parts, we find
\begin{equation}
\left\{ \begin{array}{l}
\sinh(\kappa_{ni} \tilde W) \cos((\kappa_{nr}+K)\tilde W)=0,\\[6pt]
\cosh(\kappa_{ni} \tilde W) \sin((\kappa_{nr}+K)\tilde W)=0,\\[6pt]
\sinh(\kappa_{ni} \tilde W) \cos((\kappa_{nr}-K)\tilde W)=0,\\[6pt]
\cosh(\kappa_{ni} \tilde W) \sin((\kappa_{nr}-K)\tilde W)=0.
\end{array} \right.
\end{equation}
Since the hyperbolic cosine is never equal to zero, these become
\begin{equation}
\left\{ \begin{array}{l}
\sinh(\kappa_{ni} \tilde W) \cos((\kappa_{nr}+K)\tilde W)=0,\\[6pt]
\sin((\kappa_{nr}+K)\tilde W)=0,\\[6pt]
\sinh(\kappa_{ni} \tilde W) \cos((\kappa_{nr}-K)\tilde W)=0,\\[6pt]
\sin((\kappa_{nr}-K)\tilde W)=0.
\end{array} \right.
\end{equation}
However, when the sine of an angle is equal to zero, the cosine of
that angle is certainly different from zero; therefore the previous
equations become
\begin{equation}
\left\{ \begin{array}{l}
\sinh(\kappa_{ni} \tilde W)=0,\\[6pt]
\sin((\kappa_{nr}+K)\tilde W)=0,\\[6pt]
\sin((\kappa_{nr}-K)\tilde W)=0.
\end{array} \right.
\end{equation}
Since the hyperbolic sine is null only when its argument is null, we
conclude that in this case:
\begin{equation}
\left\{ \begin{array}{l}
\kappa_{ni}=0,\\[6pt]
\sin((\kappa_{nr}+K)\tilde W)=0,\\[6pt]
\sin((\kappa_{nr}-K)\tilde W)=0,
\end{array} \right.
\Rightarrow
\left\{ \begin{array}{l}
\kappa_{n}\hbox{ real},\\[6pt]
\sin((\kappa_n+K)\tilde W)=0,\\[6pt]
\sin((\kappa_n-K)\tilde W)=0.
\end{array} \right.
\end{equation}
From the condition on $\sin((\kappa_n+K)\tilde W)$ it follows that
\begin{eqnarray}
\sin((\kappa_n+K)\tilde W)=0 &\Rightarrow&
(\kappa_n+K)\tilde W=n\pi \Rightarrow\\[6pt]
\kappa_n+K=n\frac{\pi}{\tilde W} &\Rightarrow&
\kappa_n=n\frac{\pi}{\tilde W}-K \nonumber
\end{eqnarray}
(with $n$ integer).
Then from the condition on $\sin((\kappa_n-K)\tilde W)$, substituting what
we have just found and then remembering that $K=4\pi/(3a)$
and that $\tilde W=(N+1)a/2$, we obtain that
\begin{eqnarray}
&& \sin((\kappa_n-K)\tilde W)=0 \Rightarrow 
\sin\left(\left(n\frac{\pi}{\tilde W}-K-K\right)\tilde W\right)=
\sin\left(n\pi-2K \tilde W\right)=\\
&& \sin\left(n\pi-2\frac{4\pi}{3a}(N+1)\frac{a}{2}\right)=
\sin\left(\pi\left(n-4\,\frac{N+1}{3}\right)\right)=0 \Rightarrow\nonumber\\
&& n-4\,\frac{N+1}{3}\ \hbox{is an integer.}\nonumber
\end{eqnarray}
This is true only if $N+1$ is a multiple of 3, {\em i.e.} if $N+1=3M$ (with $M$
integer), {\em i.e.} if $N=3M-1$; this means that the nanoribbon is metallic
(as we will see).
In this case the $\Phi$ functions (\ref{phiaa}) are equal to
\begin{equation}
\left\{ \begin{array}{r@{\ }l}
\displaystyle
\Phi_A^{\vec K}(y)=
& \displaystyle
\frac{\gamma}{E}\left((\kappa_x-i\kappa_n) A e^{i\kappa_n y}+
(\kappa_x+i\kappa_n) B e^{-i\kappa_n y}\right),\\[6pt]
\displaystyle
\Phi_B^{\vec K}(y)=
& \displaystyle
A e^{i\kappa_n y}+B e^{-i\kappa_n y},\\[6pt]
\displaystyle
\Phi_A^{\vec K'}(y)=
& \displaystyle
\frac{\gamma}{E}\left((\kappa_x+i\kappa_n) C e^{i\kappa_n y}+
(\kappa_x-i\kappa_n) D e^{-i\kappa_n y}\right)=\\[6pt]
& \displaystyle
-i\frac{\gamma}{E}\left((\kappa_x+i\kappa_n) B e^{i\kappa_n y}+
(\kappa_x-i\kappa_n) A e^{-i\kappa_n y}\right),\\[6pt]
\displaystyle
\Phi_B^{\vec K'}(y)=
& \displaystyle
C e^{i\kappa_n y}+D e^{-i\kappa_n y}=
-i\left(B e^{i\kappa_n y}+A e^{-i\kappa_n y}\right),
\end{array} \right.
\end{equation}
that can be written as a superposition of the modes
\begin{equation}
\left\{ \begin{array}{l}
\displaystyle
\!\!\Phi_A^{\vec K}(y)=
\frac{\gamma}{E}(\kappa_x+i\tilde\kappa_n) A e^{-i\tilde\kappa_n y},\\[6pt]
\displaystyle
\!\!\Phi_B^{\vec K}(y)=A e^{-i\tilde\kappa_n y},\\[6pt]
\displaystyle
\!\!\!\Phi_A^{\vec K'}(y)=
-\frac{\gamma}{E}(\kappa_x+i\tilde\kappa_n) iA e^{i\tilde\kappa_n y},\\[6pt]
\displaystyle
\!\!\Phi_B^{\vec K'}(y)=-iA e^{i\tilde\kappa_n y},
\end{array} \right.
\ \hbox{and}\ 
\left\{ \begin{array}{l}
\displaystyle
\!\!\Phi_A^{\vec K}(y)=
\frac{\gamma}{E}(\kappa_x+i\kappa_n) B e^{-i\kappa_n y},\\[6pt]
\displaystyle
\!\!\Phi_B^{\vec K}(y)=B e^{-i\kappa_n y},\\[6pt]
\displaystyle
\!\!\Phi_A^{\vec K'}(y)=
-\frac{\gamma}{E}(\kappa_x+i\kappa_n) iB e^{i\kappa_n y},\\[6pt]
\displaystyle
\!\!\Phi_B^{\vec K'}(y)=-iB e^{i\kappa_n y},
\end{array} \right.
\end{equation}
with $\kappa_n=(n\pi/\tilde W)-K$ and $\tilde\kappa_n=-\kappa_n$.
We notice that, since in this case $N+1=3M$, we have that
\begin{equation}
K\tilde W=\frac{4\pi}{3a}(N+1)\frac{a}{2}=\frac{2\pi}{3}(N+1)=
\frac{2\pi}{3}(3M)=2\pi M \Rightarrow
K=2M\frac{\pi}{\tilde W}
\end{equation}
and therefore $\tilde\kappa_n$ can be written as
\begin{eqnarray}
\tilde\kappa_n &=& -\kappa_n=-\left(n\frac{\pi}{\tilde W}-K\right)=
\left(-n\frac{\pi}{\tilde W}+2K\right)-K=\\
&& \left(-n\frac{\pi}{\tilde W}+4M\frac{\pi}{\tilde W}\right)-K=
(4M-n)\frac{\pi}{\tilde W}-K=\tilde n\frac{\pi}{\tilde W}-K,\nonumber
\end{eqnarray}
with $\tilde n=4M-n$ integer.
Clearly, if $\kappa_n$ satisfies 
$E=\pm \gamma \sqrt{{\kappa_x}^2+{\kappa_n}^2}$, also
$\tilde\kappa_n=-\kappa_n$ satisfies
$E=\pm \gamma \sqrt{{\kappa_x}^2+{\tilde\kappa_n}^2}$.

It can be observed that in the particular case in which $\kappa_n=0$
we find again Case I.

\vskip5pt\noindent
\noindent
{\bf Case II-B}

\noindent
Equations~(\ref{secondtwo}) are satisfied also if
\begin{equation}
\left\{ \begin{array}{l}
\sinh(\kappa_{ni} \tilde W) \cos((\kappa_{nr}+K)\tilde W)
-i \cosh(\kappa_{ni} \tilde W) \sin((\kappa_{nr}+K)\tilde W)=0,\\[5pt]
A=0.
\end{array} \right.
\end{equation}
If we separately equate to zero the real and imaginary parts of the first
equation, we find
\begin{equation}
\left\{ \begin{array}{l}
\sinh(\kappa_{ni} \tilde W) \cos((\kappa_{nr}+K)\tilde W)=0,\\[5pt]
\cosh(\kappa_{ni} \tilde W) \sin((\kappa_{nr}+K)\tilde W)=0,\\[5pt]
A=0.
\end{array} \right.
\end{equation}
Since the hyperbolic cosine is never equal to zero, these become
\begin{equation}
\left\{ \begin{array}{l}
\sinh(\kappa_{ni} \tilde W) \cos((\kappa_{nr}+K)\tilde W)=0,\\[5pt]
\sin((\kappa_{nr}+K)\tilde W)=0,\\[5pt]
A=0.
\end{array} \right.
\end{equation}
But when the sine of an angle is equal to zero, surely the cosine of
that angle is different from zero; therefore the previous equations become
\begin{equation}
\left\{ \begin{array}{l}
\sinh(\kappa_{ni} \tilde W)=0,\\[5pt]
\sin((\kappa_{nr}+K)\tilde W)=0,\\[5pt]
A=0,
\end{array} \right.
\end{equation}
Since the hyperbolic sine is null only when its argument is null, we
conclude that in this case:
\begin{equation}
\left\{ \begin{array}{l}
\kappa_{ni}=0,\\[5pt]
\sin((\kappa_{nr}+K)\tilde W)=0,\\[5pt]
A=0,
\end{array} \right.
\Rightarrow
\left\{ \begin{array}{l}
\kappa_{n}\hbox{ real},\\[5pt]
\sin((\kappa_n+K)\tilde W)=0,\\[5pt]
A=0.
\end{array} \right.
\end{equation}
Due to the fact that $A=0$, also $D=-iA=0$ (while $C=-iB$).

Instead the consequence of the condition on $\sin((\kappa_n+K)\tilde W)$ is
\begin{eqnarray}
\sin((\kappa_n+K)\tilde W)=0 &\Rightarrow&
(\kappa_n+K)\tilde W=n\pi \Rightarrow\\
\kappa_n+K=n\frac{\pi}{\tilde W} &\Rightarrow&
\kappa_n=n\frac{\pi}{\tilde W}-K.\nonumber
\end{eqnarray}
In this case the $\Phi$ functions (\ref{phiaa}) are equal to
\begin{equation}
\left\{ \begin{array}{r@{\ }l}
\displaystyle
\Phi_A^{\vec K}(y)=
& \displaystyle
\frac{\gamma}{E}\left((\kappa_x-i\kappa_n) A e^{i\kappa_n y}+
(\kappa_x+i\kappa_n) B e^{-i\kappa_n y}\right)=\\[8pt]
& \displaystyle
\frac{\gamma}{E}(\kappa_x+i\kappa_n) B e^{-i\kappa_n y},\\[8pt]
\displaystyle
\Phi_B^{\vec K}(y)=
& \displaystyle
A e^{i\kappa_n y}+B e^{-i\kappa_n y}=
B e^{-i\kappa_n y},\\[8pt]
\displaystyle
\Phi_A^{\vec K'}(y)=
& \displaystyle
\frac{\gamma}{E}\left((\kappa_x+i\kappa_n) C e^{i\kappa_n y}+
(\kappa_x-i\kappa_n) D e^{-i\kappa_n y}\right)=\\[8pt]
& \displaystyle
-\frac{\gamma}{E}(\kappa_x+i\kappa_n) iB e^{i\kappa_n y},\\[8pt]
\displaystyle
\Phi_B^{\vec K'}(y)=
& \displaystyle
C e^{i\kappa_n y}+D e^{-i\kappa_n y}=
-iB e^{i\kappa_n y}.
\end{array} \right.
\end{equation}

\noindent
{\bf Case II-C}

\noindent
Finally, eqs.~(\ref{secondtwo}) are satisfied also if
\begin{equation}
\left\{ \begin{array}{l}
B=0,\\[5pt]
\sinh(\kappa_{ni} \tilde W) \cos((\kappa_{nr}-K)\tilde W)
-i \cosh(\kappa_{ni} \tilde W) \sin((\kappa_{nr}-K)\tilde W)=0.
\end{array} \right.
\end{equation}
With calculations analogous to Case II-B, we conclude~\cite{supplem}
that in this case:
\begin{equation}
\left\{ \begin{array}{l}
B=0,\\[5pt]
\kappa_{n}\hbox{ real},\\[5pt]
\sin((\kappa_n-K)\tilde W)=0.
\end{array} \right.
\end{equation}
Due to the fact that $B=0$, also $C=-iB=0$ (while $D=-iA$).

Instead the consequence of the condition on $\sin((\kappa_n-K)\tilde W)$ is
\begin{eqnarray}
\sin((\kappa_n-K)\tilde W)=0 &\Rightarrow&
(\kappa_n-K)\tilde W=n\pi \Rightarrow\\
\kappa_n-K=n\frac{\pi}{\tilde W} &\Rightarrow&
\kappa_n=n\frac{\pi}{\tilde W}+K.\nonumber
\end{eqnarray}
In this case the $\Phi$ functions (\ref{phiaa}) are equal to
\begin{equation}
\left\{ \begin{array}{r@{\ }l}
\displaystyle
\Phi_A^{\vec K}(y)=
& \displaystyle
\frac{\gamma}{E}\left((\kappa_x-i\kappa_n) A e^{i\kappa_n y}+
(\kappa_x+i\kappa_n) B e^{-i\kappa_n y}\right)=\\[8pt]
& \displaystyle
\frac{\gamma}{E}(\kappa_x-i\kappa_n) A e^{i\kappa_n y}=
\frac{\gamma}{E}(\kappa_x+i\tilde\kappa_n) A e^{-i\tilde\kappa_n y},\\[8pt]
\displaystyle
\Phi_B^{\vec K}(y)=
& \displaystyle
A e^{i\kappa_n y}+B e^{-i\kappa_n y}=
A e^{i\kappa_n y}=A e^{-i\tilde\kappa_n y},\\[8pt]
\displaystyle
\Phi_A^{\vec K'}(y)=
& \displaystyle
\frac{\gamma}{E}\left((\kappa_x+i\kappa_n) C e^{i\kappa_n y}+
(\kappa_x-i\kappa_n) D e^{-i\kappa_n y}\right)=\\[8pt]
& \displaystyle
-\frac{\gamma}{E}(\kappa_x-i\kappa_n) iA e^{-i\kappa_n y}=
-\frac{\gamma}{E}(\kappa_x+i\tilde\kappa_n) iA e^{i\tilde\kappa_n y},\\[8pt]
\displaystyle
\Phi_B^{\vec K'}(y)=
& \displaystyle
C e^{i\kappa_n y}+D e^{-i\kappa_n y}=
-iA e^{-i\kappa_n y}=-iA e^{i\tilde\kappa_n y},
\end{array} \right.
\end{equation}
with
\begin{equation}
\tilde\kappa_n=-\kappa_n=-\left(n\frac{\pi}{\tilde W}+K\right)=
-n\frac{\pi}{\tilde W}-K=\tilde n\frac{\pi}{\tilde W}-K
\end{equation}
(where $\tilde n=-n$ is an integer).
Clearly, if $\kappa_n$ satisfies 
$E=\pm \gamma \sqrt{{\kappa_x}^2+{\kappa_n}^2}$, also 
$\tilde\kappa_n=-\kappa_n$ satisfies
$E=\pm \gamma \sqrt{{\kappa_x}^2+{\tilde\kappa_n}^2}$.

%%%%%%%%%% conclusion %%%%%%%%%%
In conclusion, in all the cases we have that
\begin{equation}
\label{concluphia}
\left\{ \begin{array}{l}
\displaystyle
\Phi_A^{\vec K}(y)=
\frac{\gamma}{E}(\kappa_x+i\kappa_n) A e^{-i\kappa_n y},\\[8pt]
\displaystyle
\Phi_B^{\vec K}(y)=A e^{-i\kappa_n y},\\[8pt]
\displaystyle
\Phi_A^{\vec K'}(y)=
-\frac{\gamma}{E}(\kappa_x+i\kappa_n) iA e^{i\kappa_n y},\\[8pt]
\displaystyle
\Phi_B^{\vec K'}(y)=-iA e^{i\kappa_n y},
\end{array} \right.
\end{equation}
with $A$ being a proper normalization constant, 
$\kappa_n=(n\pi/\tilde W)-K$ and 
$E=\pm \gamma \sqrt{{\kappa_x}^2+{\kappa_n}^2}$.
Consequently, for eq.~(\ref{envelopea}) we have that
\begin{eqnarray}
&& \psi_A (\vec r)=
e^{i \vec K\cdot \vec r} F_A^{\vec K}(\vec r)-
i\,e^{i \vec K'\cdot \vec r} F_A^{\vec K'}(\vec r)=\\[3pt]
&& e^{-iKy} \Phi_A^{\vec K}(y) e^{i\kappa_x x}-
i\,e^{iKy} \Phi_A^{\vec K'}(y) e^{i\kappa_x x}=\nonumber\\[3pt]
&& \left(e^{-iKy} \Phi_A^{\vec K}(y)-i\,e^{iKy} \Phi_A^{\vec K'}(y)\right)
e^{i\kappa_x x}=\nonumber\\[3pt]
&& \frac{\gamma}{E}
\left(e^{-iKy} (\kappa_x+i\kappa_n) A e^{-i\kappa_n y}+
i e^{iKy} (\kappa_x+i\kappa_n) iA e^{i\kappa_n y}\right)
e^{i\kappa_x x}=\nonumber\\[3pt]
&& -\frac{\gamma}{E}
(\kappa_x+i\kappa_n) A \left(e^{i(\kappa_n+K) y}-e^{-i(\kappa_n+K) y}\right)
e^{i\kappa_x x}=\nonumber\\[3pt]
&& -\frac{\gamma}{E}
(\kappa_x+i\kappa_n) A 2 i \sin\left((\kappa_n+K) y\right)e^{i\kappa_x x}\nonumber
\end{eqnarray}
and that
\begin{eqnarray}
&& \psi_B (\vec r)=
i\,e^{i \vec K\cdot \vec r} F_B^{\vec K}(\vec r)+
e^{i \vec K'\cdot \vec r} F_B^{\vec K'} (\vec r)=\\[3pt]
&& i\,e^{-iKy} \Phi_B^{\vec K}(y) e^{i\kappa_x x}+
e^{iKy} \Phi_B^{\vec K'}(y) e^{i\kappa_x x}=\nonumber\\[3pt]
&& \left(i\,e^{-iKy} \Phi_B^{\vec K}(y)+
e^{iKy} \Phi_B^{\vec K'}(y)\right) e^{i\kappa_x x}=\nonumber\\[3pt]
&& \left(i\,e^{-iKy} A e^{-i\kappa_n y}-
e^{iKy} iA e^{i\kappa_n y}\right) e^{i\kappa_x x}=\nonumber\\[3pt]
&& -iA \left(e^{i(\kappa_n+K) y}-e^{-i(\kappa_n+K) y}\right) e^{i\kappa_x x}=\nonumber\\[3pt]
&& -iA 2i \sin\left((\kappa_n+K) y\right) e^{i\kappa_x x}=\nonumber\\[3pt]
&& 2A \sin\left((\kappa_n+K) y\right) e^{i\kappa_x x}.\nonumber
\end{eqnarray}
We observe that in large ribbons the lowest-energy modes will have
values of $\kappa_n$ much smaller than $K$ and thus their wave functions
will be characterized by a transverse wave vector approximately equal to
$K$ and by a transverse wave length about equal to
$2\pi/K=2\pi\,(3a/(4\pi))=3a/2$, {\em i.e.} of the order of the lattice constant.

No edge state exists in armchair nanoribbons.

Using the relations $K=4\pi/(3a)$ and $\tilde W=(N+1)a/2$, we have that
\begin{equation}
\label{kn}
\kappa_n=n\frac{\pi}{\tilde W}-K=\frac{n 2 \pi}{(N+1)a}-\frac{4\pi}{3a}=
\frac{2\pi(3n-2(N+1))}{3(N+1)a}.
\end{equation}
Since $E_n=\pm \gamma \sqrt{{\kappa_x}^2+{\kappa_n}^2}$, we have a double
band degeneracy if, for any integer $n$, another integer $n'$ exists such
that $\kappa_{n'}=-\kappa_n$ and thus $E_{n'}=E_n$. This happens if
\begin{eqnarray}
3n'-2(N+1) &=& -(3n-2(N+1))\Rightarrow
3n'=-3n+4(N+1)\Rightarrow\\
n' &=& -n+\frac{4(N+1)}{3},\nonumber
\end{eqnarray}
with $n$ and $n'$ integer, which means that $N+1$ has to be a multiple
of 3, {\em i.e.} $N+1=3M$ with $M$ integer, or equivalently $N=3M-1$
(so that $n'=-n+4M$).

We also observe that among the allowed $\kappa_n$'s (given by eq.~(\ref{kn}))
we have $\kappa_n=0$ if an integer $n$ exists, such that
\begin{equation}
3n-2(N+1)=0\Rightarrow
n=\frac{2(N+1)}{3},
\end{equation}
which again means that $N+1$ has to be a multiple of 3, {\em i.e.} $N+1=3M$
with $M$ integer, or equivalently $N=3M-1$ (so that $n=2M$).

Therefore an armchair nanoribbon has a double band degeneracy and has
$\kappa_n=0$ among the allowed values of $\kappa_n$ only if it has a
number of dimer lines $N=3M-1$ (with $M$ an integer). In this case
for $\kappa_n=0$ we have $E=\pm \gamma |\kappa_x|$ which vanishes
for $\kappa_x\to0$ and thus the nanoribbon is metallic. Instead for
$N \ne 3M-1$ the armchair nanoribbon is not metallic and has non-degenerate
bands.

This conclusion is coherent with the fact that the dispersion relations of
an armchair nanoribbon can be obtained from those of graphene enforcing the
Dirichlet boundary conditions at $y=0$ and $y=\tilde W$; this means that there
has to be an integer number of transverse half-wavelengths $\lambda_y/2$
inside $\tilde W$; thus it must happen that
\begin{equation}
\tilde W=n \frac{\lambda_y}{2} \Rightarrow
k_y=\frac{2\pi}{\lambda_y}=n\frac{\pi}{\tilde W}
\end{equation}
(where $k_y$ is the transverse component of the  total wave vector, measured
from the origin of the reciprocal space). Therefore the bands of the ribbon
can be obtained by cross-sectioning those of graphene along the parallel lines
$k_y=n\pi/\tilde W$, and then folding them into the Brillouin zone 
$(-\pi/(\sqrt{3}a),\pi/(\sqrt{3}a))$ of the armchair nanoribbon (the unit cell
of which has a length $3a_{C-C}=\sqrt{3}a$). There are bands of the nanoribbon with
a zero gap, and thus the nanoribbon is metallic, only if some of the lines with
$k_y=n\pi/\tilde W$ pass through a Dirac point of graphene (where the
graphene dispersion relations have a zero gap). But, since
\begin{equation}
\tilde W=(N+1)\frac{a}{2} \Rightarrow
a=\frac{2 \tilde W}{N+1} \Rightarrow
K=\frac{4\pi}{3a}=\frac{4\pi}{3}\frac{N+1}{2 \tilde W}=
2\,\frac{N+1}{3}\frac{\pi}{\tilde W},
\end{equation}
this is possible only if $N+1$ is a multiple of 3, {\em i.e.} $N+1=3M$ with
$M$ integer, or equivalently $N=3M-1$.

A more exact tight-binding analysis (taking into consideration the reduction
of the carbon-carbon bond lengths parallel to dimer lines at the edges
with respect to the bond lengths in the core of the ribbon) leads to the
appearance of a small gap also in this subset of armchair nanoribbons,
that have to be more correctly considered as quasi-metallic
ones~\cite{son,fujita1,fujita2}.

\begin{figure}
\centering
\includegraphics[width=\textwidth,angle=0]{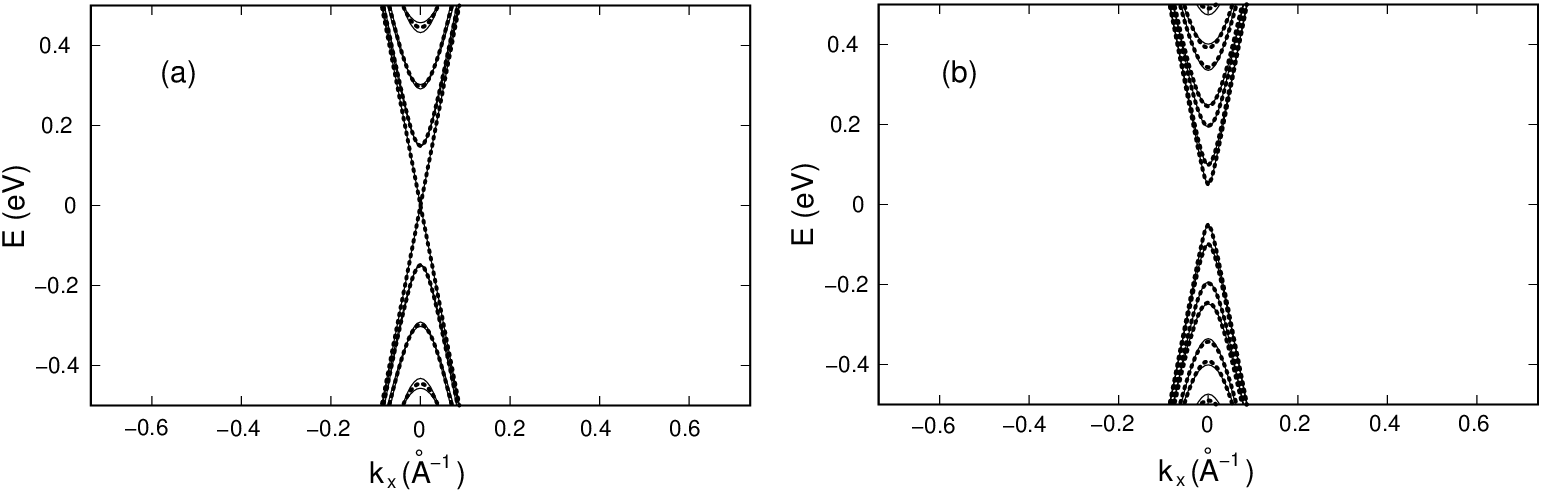}
\caption{Bands of an armchair nanoribbon with $N=98$ dimer lines (a)
and with $N=99$ dimer lines (b), computed both with a tight-binding
method not including the reduction of the bond lengths at the
edges (thick dotted lines) and with the $\vec k \cdot \vec p$ method
(thin solid lines).}
\label{f14}
\end{figure}

In fig.~\ref{f14} we show the bands of an armchair nanoribbon with
$N=98$ dimer lines (metallic) and of an armchair nanoribbon with
$N=99$ dimer lines (semiconductor), that we have computed both with a
tight-binding method not including the reduction of the bond lengths at the
edges (thick dotted lines) and with the $\vec k \cdot \vec p$ (Dirac equation)
method (thin solid lines). As we see, for low energy values and for not
too narrow ribbons they are nearly coincident.

\noindent

%%%%%%%%%% current %%%%%%%%%%
All previous considerations are valid both for real values of
$\kappa_x$ (propagating modes), and for purely imaginary values of
$\kappa_x$ (evanescent modes).

As an application of the relations (\ref{jx}) and (\ref{jy}) to
the case of an armchair nanoribbon in the absence of an external potential,
we can observe, using the (\ref{concluphia}) (with $\kappa_n$ real and
$\kappa_x$ real or purely imaginary), that
\begin{eqnarray}
J_x &=& v_F\,\left({F_A^{\vec K}}^* F_B^{\vec K}+
{F_B^{\vec K}}^* F_A^{\vec K}+
{F_A^{\vec K'}}^* F_B^{\vec K'}+
{F_B^{\vec K'}}^* F_A^{\vec K'}\right)=\\[5pt]
&& v_F\,\frac{\gamma}{E}\,\Big(
(\kappa_x^*-i\kappa_n) A^* e^{i\kappa_n y} e^{-i\kappa_x^* x}
A e^{-i\kappa_n y} e^{i\kappa_x x}\nonumber\\[5pt]
&& +A^* e^{i\kappa_n y} e^{-i\kappa_x^* x}
(\kappa_x+i\kappa_n) A e^{-i\kappa_n y} e^{i\kappa_x x}\nonumber\\[5pt]
&& +(\kappa_x^*-i\kappa_n) i A^* e^{-i\kappa_n y} e^{-i\kappa_x^* x}
(-i) A e^{i\kappa_n y} e^{i\kappa_x x}\nonumber\\[5pt]
&& +iA^* e^{-i\kappa_n y} e^{-i\kappa_x^* x}
(\kappa_x+i\kappa_n) (-i)A e^{i\kappa_n y} e^{i\kappa_x x}\Big)=\nonumber\\[5pt]
&& v_F\,\frac{\gamma}{E} |A|^2 e^{i(\kappa_x-\kappa_x^*) x}
(\kappa_x^*-i\kappa_n+\kappa_x+i\kappa_n+
\kappa_x^*-i\kappa_n+\kappa_x+i\kappa_n)=\nonumber\\[5pt]
&& 2 v_F\,\frac{\gamma}{E} |A|^2 e^{i(\kappa_x-\kappa_x^*) x}
(\kappa_x+\kappa_x^*),\nonumber
\end{eqnarray}
which if $\kappa_x$ is real (and thus $\kappa_x^*=\kappa_x$) is equal to
(remembering that $v_F=\gamma/\hbar$)
\begin{equation}
\label{jxa}
J_x=4\,v_F\,\frac{\gamma}{E}\,|A|^2 \kappa_x=
4|A|^2\,\frac{\gamma^2}{\hbar E}\,\kappa_x,
\end{equation}
while if $\kappa_x$ is purely imaginary (and thus $\kappa_x^*=-\kappa_x$)
is null.

Note that (using (\ref{p}) and (\ref{concluphia})) if $\kappa_x$ is real
the probability density is equal to
\begin{eqnarray}
P &=& |F_A^{\vec K}(\vec r)|^2+|F_A^{\vec K'}(\vec r)|^2+
|F_B^{\vec K}(\vec r)|^2+|F_B^{\vec K'}(\vec r)|^2=\\
&& \left(\frac{\gamma}{E}\right)^2 |\kappa_x+i \kappa_n|^2 |A|^2+|A|^2+
\left(\frac{\gamma}{E}\right)^2 |\kappa_x+i \kappa_n|^2 |A|^2+|A|^2=\nonumber\\
&& 2 |A|^2 
\left(1+\left(\frac{\gamma}{E}\right)^2 |\kappa_x+i \kappa_n|^2\right)=
2 |A|^2 \left(1+\frac{\gamma^2 (\kappa_x^2+\kappa_n^2)}{E^2}\right)=\nonumber\\
&& 2 |A|^2 \left(1+\frac{E^2}{E^2}\right)=2 |A|^2 2=4 |A|^2.\nonumber
\end{eqnarray}
Moreover, since in this case the energy dispersion relations are
$E=\pm \gamma \sqrt{{\kappa_x}^2+{\kappa_n}^2}$,
the mean velocity of the electrons is
\begin{eqnarray}
v_x &=& \frac{1}{\hbar}\frac{\partial E}{\partial k_x}=
\frac{1}{\hbar} \left(\pm \gamma \frac{1}{2}
\frac{1}{\sqrt{{\kappa_x}^2+{\kappa_n}^2}} 2 \kappa_x \right)=\\
&& \pm \frac{\gamma}{\hbar} \frac{\kappa_x}{\sqrt{{\kappa_x}^2+{\kappa_n}^2}}=
\frac{\gamma^2}{\hbar}
\frac{\kappa_x}{\pm \gamma \sqrt{{\kappa_x}^2+{\kappa_n}^2}}=
\frac{\gamma^2}{\hbar E} \kappa_x.\nonumber
\end{eqnarray}
Therefore if $\kappa_x$ is real we have that $J_x=P v_x$, as expected.

As to the transversal part of the probability current density, we have that
\begin{eqnarray}
\qquad J_y &=& -i\,v_F\,\left({F_A^{\vec K}}^* F_B^{\vec K}-
{F_B^{\vec K}}^* F_A^{\vec K}-
{F_A^{\vec K'}}^* F_B^{\vec K'}+
{F_B^{\vec K'}}^* F_A^{\vec K'}\right)=\\
&& -i\,v_F\,\frac{\gamma}{E}\,\Big(
(\kappa_x^*-i\kappa_n) A^* e^{i\kappa_n y} e^{-i\kappa_x^* x}
A e^{-i\kappa_n y} e^{i\kappa_x x}\nonumber\\
&& -A^* e^{i\kappa_n y} e^{-i\kappa_x^* x}
(\kappa_x+i\kappa_n) A e^{-i\kappa_n y} e^{i\kappa_x x}\nonumber\\
&& -(\kappa_x^*-i\kappa_n) iA^* e^{-i\kappa_n y} e^{-i\kappa_x^* x}
(-i) A e^{i\kappa_n y} e^{i\kappa_x x}\nonumber\\
&& +iA^* e^{-i\kappa_n y} e^{-i\kappa_x^* x}
(\kappa_x+i\kappa_n) (-i)A e^{i\kappa_n y} e^{i\kappa_x x}\Big)=\nonumber\\
&& -i\,v_F\,\frac{\gamma}{E}\,|A|^2\,e^{i(\kappa_x-\kappa_x^*) x}
(\kappa_x^*-i\kappa_n-\kappa_x-i\kappa_n
-\kappa_x^*+i\kappa_n+\kappa_x+i\kappa_n)=0,\nonumber
\end{eqnarray}
as expected (at least at the edges) in a transversally confined structure.

\section{Conclusion}

The $\vec k \cdot \vec p$ method and the related envelope function method are 
widely used to study the physical properties of materials within a continuum
approach, without having to resort to an atomistic analysis, which requires
(in the case of large structures) a prohibitive computational effort. These
methods have been developed in many and sometimes quite different ways by
several authors and have been successfully applied to a multitude of different 
problems. This explains the great variety and inhomogeneity of the related 
literature. In this review, we have briefly described the basics of these
methodologies, dwelling upon the treatments that we have considered more 
useful for an easy comprehension. For a detailed explanation of the different 
approaches, the interested reader can resort to the many papers and 
books on the topic, some of which are listed in the references.
In particular, we have focused on the application of the $\vec k \cdot \vec p$
method to graphene and graphene-related materials, where it results in a
description of the electronic properties in terms of the Dirac equation.
We have compared the different formulations adopted in the literature and
we have shown how this continuum approach allows to quickly obtain the
most relevant electrical properties of graphene, carbon nanotubes and
graphene nanoribbons.

\acknowledgments

We would like to thank Prof.~\textsc{T.~Ando}, Prof.~\textsc{P.~Lugli} and
Dr.~\textsc{G.~Scarpa} for useful discussions and suggestions.
We also gratefully acknowledge support from the EU\break
FP7 IST Project GRAND
(contract number 215752) via the IUNET consortium.

\addtocontents{toc}{\SkipTocEntry}


\begin{thebibliography}{0}

\setlength{\itemindent}{-1pt}
% numbers with 1 digit

\bibitem{callaway1} \BY{Callaway~J.}
\TITLE{Energy band theory} (Academic Press, New York) 1964.

\bibitem{grosso} \BY{Grosso~G. \atque Pastori Parravicini~G.}
\TITLE{Solid State Physics} (Academic Press, San\break
Diego) 2000.

\bibitem{martin} \BY{Martin~R.~M.}
\TITLE{Electronic Structure: Basic Theory and Practical Methods}
(Cambridge University Press, Cambridge) 2004.

\bibitem{voon}  \BY{Lew Yan Voon~L.~C. \atque Willatzen~M.}
\TITLE{The $\vec k \cdot \vec p$ Method: Electronic Properties of
Semiconductors} (Springer, Berlin) 2009.

\bibitem{wenckebach} \BY{Wenckebach~W.~T.}
\TITLE{Essentials of Semiconductor Physics} (J.~Wiley \& Sons, 
Chichester) 1999, with the addenda downloadable from
\texttt{http://tomwenckebach.com/}.

\bibitem{chuang} \BY{Chuang~S.~L.}
\TITLE{Physics of Optoelectronic Devices} (J.~Wiley \& Sons,
New York) 1995.

\bibitem{tsidilkovski} \BY{Tsidilkovski~I.~M.} 
\TITLE{Band Structure of Semiconductors, International 
Series on the Science of Solid State}, Vol.~{\bf 19} (Pergamon Press,
Oxford) 1982 (translation of 
\TITLE{Zonnaia\break
struktura poluprovodnikov}).

\bibitem{yu} \BY{Yu~P.~Y. \atque Cardona~M.}
\TITLE{Fundamentals of Semiconductors} (Springer-Verlag,
Berlin) 1995.

\bibitem{li} \BY{Li~Ming-Fu}
\TITLE{Modern Semiconductor Quantum Physics} (World Scientific,
Singapore)\break
1994.

\setlength{\itemindent}{-5pt}
% numbers with 2 digits

\bibitem{kane1} \BY{Kane~E.~O.}
\TITLE{Band structure of narrow gap semiconductors}, in
\TITLE{Narrow Gap Semicon-\break
ductors: Physics and Applications,
Springer Lect. Notes Phys.}, Vol.~{\bf 133}, edited by
\NAME{Zawadzki~W.} 
(Springer-Verlag, Berlin) 1980, p.~13.

\bibitem{kane2} \BY{Kane~E.~O.}
\TITLE{The $\vec k \cdot \vec p$ Method}, in
\TITLE{Semiconductors and Semimetals}, Vol.~{\bf 1}: \TITLE{Physics of
III-V Compounds}, edited by \NAME{Willardson~R.~K. \atque Beer~A.~C.}
(Academic Press, New\break
York) 1966, p.~75.

\bibitem{kane3} \BY{Kane~E.~O.}
\TITLE{Energy Band Theory}, in 
\TITLE{Handbook on Semiconductors}, Vol~{\bf 1}: \TITLE{Band Theory and
Transport Properties}, edited by \NAME{Moss~T.~S. \atque Paul~W.},
(North-Holland Publishing Company, Amsterdam) 1982, p.~193.

\bibitem{pidgeon} \BY{Pidgeon~C.~R.}
\TITLE{Free Carrier Optical Properties of Semiconductors}, in
\TITLE{Handbook on Semiconductors}, Vol.~{\bf 2}: \TITLE{Optical Properties of Solids},
edited by \NAME{Moss~T.~S. \atque Balkanski~M.},
(North-Holland Publishing Company, Amsterdam) 1980, p.~223.

\bibitem{bir} \BY{Bir~G.~L. \atque Pikus~G.~E.}
\TITLE{Symmetry and Strain-Induced Effects in Semiconductors}
(J.~Wiley \& Sons, New York / Keter Publishing House,
Jerusalem) 1974 (translation of 
\TITLE{Simmetriya i deformatsionnye effekty
v poluprovodnikakh}, Moskow, Izdatel'stvo ``Nauka'', 1972).

\bibitem{ivchenko} \BY{Ivchenko~E.~L. \atque Pikus~G.~E.}
\TITLE{Superlattices and Other
Heterostructures: Symmetry\break
and Optical Phenomena,
Springer Ser. Solid-State Sci.}, Vol.~{\bf 110}
(Springer-Verlag, Berlin) 1995.

\bibitem{anselm} \BY{Anselm~A.~I.}
\TITLE{Introduction to Semiconductor Theory} 
(MIR Publishers-Moscow, Moscow) 1978 
(English translation:Prentice Hall, London, 1981).

\bibitem{callaway2} \BY{Callaway~J.}
\TITLE{Quantum Theory of the Solid State} (Academic Press,
San Diego) 1974.

\bibitem{singh} \BY{Singh~J.}
\TITLE{Electronic and Optoelectronic Properties of Semiconductor
Structures} (Cam-\break
bridge University Press, Cambridge) 2003.

\bibitem{long} \BY{Long~D.}
\TITLE{Energy Bands in Semiconductors}
(J.~Wiley \& Sons, New York) 1968.

\bibitem{kroemer} \BY{Kroemer~H.}
\TITLE{Quantum Mechanics} (Prentice Hall, Englewood Cliffs) 1994.

\bibitem{bassani} \BY{Bassani~F. \atque Pastori~Parravicini~G.}
\TITLE{Electronic States and Optical Transitions in Solids}
(Pergamon Press, Oxford) 1975.

\bibitem{enderlein} \BY{Enderlein~R. \atque Horing~N.~J.~M.}
\TITLE{Fundamentals of Semiconductor Physics and\break
Devices} (World Scientific, Singapore) 1997.

\bibitem{haug} \BY{Haug~H. \atque Koch~S.~W.}
\TITLE{Quantum Theory of the Optical and Electronic 
Properties of Semiconductors}
(World Scientific, Singapore) 1990.

\bibitem{harrison} \BY{Harrison~W.~A.}
\TITLE{Electronic Structure and the Properties of Solids}
(Dover Publications, New York), 1980.

\bibitem{winkler} \BY{Winkler~R.}
\TITLE{Spin-orbit Coupling Effects in Two-Dimensional Electron
and Hole Systems, Springer Tracts Mod. Phys.}, Vol.~{\bf 191)}
(Springer-Verlag, Berlin) 2003.

\bibitem{datta1} \BY{Datta~S.} \TITLE{Quantum Phenomena} 
(Addison-Wesley, Reading), 1989.

\bibitem{marconcini1} \BY{Marconcini~P.}
\TITLE{Numerical Simulation of Transport and Noise in Nanoelectronic Devices},
Ph.D. thesis, University of Pisa (2006),
in particular Appendix and Par.~4.3.

\bibitem{cardona} \BY{Cardona~M. \atque Pollak~F.~H.}
\IN{Phys.~Rev.}{142}{1966}{530}.

\bibitem{cavasillas} \BY{Cavassilas~N., Aniel~F., Boujdaria~K. \atque
Fishman~G.} 
\IN{Phys.~Rev.~B}{64}{2001}{115207}.

\bibitem{radhia} \BY{Radhia~S.~Ben, Ridene~S., Boujdaria~K.,
Bouchriha~H. \atque Fishman~G.}
\IN{J. Appl. Phys.}{92}{2002}{4422}.

\bibitem{richard} \BY{Richard~S., Aniel~F. \atque Fishman~G.}
\IN{Phys.~Rev.~B}{70}{2004}{235204}.

\bibitem{michelini} \BY{Michelini~F., Cavassilas~N., Hayn~R. 
\atque Szczap~M.} 
\IN{Phys.~Rev.~B}{80}{2009}{245210}.

\bibitem{bardeen} \BY{Bardeen~J.}
\IN{J.~Chem.~Phys.}{6}{1938}{367}. 

\bibitem{seitz} \BY{Seitz~F.}
\TITLE{The Modern Theory of Solids} (McGraw~Hill, New York) 1940, p.~352.

\bibitem{shockley} \BY{Shockley~W.}
\IN{Phys.~Rev.}{78}{1950}{173}.

\bibitem{dresselhaus} \BY{Dresselhaus~G., Kip~A.~F. \atque Kittel~C.}
\IN{Phys.~Rev.}{98}{1955}{368}.

\bibitem{kane4} \BY{Kane~E.~O.}
\IN{J.~Phys.~Chem.~Solids}{1}{1956}{82}.

\bibitem{kane5} \BY{Kane~E.~O.}
\IN{J.~Phys.~Chem.~Solids}{1}{1957}{249}.

\bibitem{kane6} \BY{Kane~E.~O.}
\IN{J.~Phys.~Chem.~Solids}{8}{1959}{38}.

\bibitem{luttinger1} \BY{Luttinger~J.~M. \atque Kohn~W.}
\IN{Phys.~Rev.}{97}{1955}{869}.

\bibitem{luttinger2} \BY{Luttinger~J.~M.}
\IN{Phys.~Rev.}{102}{1956}{1030}.

\bibitem{bastard1} \BY{Bastard~G.}
\TITLE{Wave Mechanics Applied to Semiconductor 
Heterostructures} (Les editions de physique, Les Ulis Cedex), 1992.

\bibitem{bastard2} \BY{Bastard~G.}
\IN{Phys.~Rev.~B}{24}{1981}{5693}.

\bibitem{bastard3} \BY{Bastard~G.}
\IN{Phys.~Rev.~B}{25}{1982}{7584}.

\bibitem{altarelli1} \BY{Altarelli~M.}
\TITLE{Electronic Structure of Semiconductor Superlattices}, in
\TITLE{Application of High Magnetic Fields in Semiconductor Physics,
Springer Lect. Notes Phys.}, Vol.~{\bf 177}, edited by
\NAME{Araki~H., Ehlers~J., Hepp~K., Kippenhahn~R., Weidenm\"uller~H.~A.,\break
Zittartz~J.} (Springer-Verlag, Berlin) 1983, p.~174.

\bibitem{altarelli2} \BY{Altarelli~M.}
\IN{Phys.~Rev.~B}{28}{1983}{842}.

\bibitem{altarelli3} \BY{Altarelli~M.}
\IN{Physica}{117B \& 118B}{1983}{747}.

\bibitem{burt1} \BY{Burt~M.~G.}
\IN{Semicond.~Sci.~Technol.}{2}{1987}{460}.

\bibitem{burt2} \BY{Burt~M.~G.}
\IN{Semicond.~Sci.~Technol.}{3}{1988}{1224}.

\bibitem{burt3} \BY{Burt~M.~G.}
\IN{Appl.~Phys.~Lett.}{65}{1994}{717}.

\bibitem{burt4} \BY{Burt~M.~G.}
\IN{Semicond.~Sci.~Technol.}{3}{1988}{739}.

\bibitem{burt5} \BY{Burt~M.~G.}
\IN{J.~Phys.:~Condens.~Matter}{4}{1992}{6651}.

\newpage

\bibitem{foreman}
An application of Burt's theory for valence-band problems can be found in:
\BY{Foreman\break
B.~A.}
\IN{Phys.~Rev.~B}{48}{1993}{4964}.

\bibitem{ajiki} \BY{Ajiki~M. \atque Ando~T.}
\IN{J.~Phys.~Soc.~Jpn.}{62}{1993}{1255}.

\bibitem{ando1} \BY{Ando~T.}
\IN{J.~Phys.~Soc.~Jpn.}{74}{2005}{777}.

\bibitem{ando2} \BY{Ando~T.},
\TITLE{Electronic States and Transport in Carbon Nanotubes}, notes
of the lectures\break
held at the Kyushu University on September 28-30, 2004.

\bibitem{katsnelson1} \BY{Katsnelson~M.~I., Novoselov~K.~S. \atque Geim~A.~K.}
\IN{Nature Phys.}{2}{2006}{620}.

\bibitem{katsnelson2} \BY{Katsnelson~M.~I. \atque Novoselov~K.~S.}
\IN{Solid~State~Commun.}{143}{2007}{3}.

\bibitem{katsnelson3} \BY{Katsnelson~M.~I.}
\IN{Eur.~Phys.~J.~B}{51}{2006}{157}.

\bibitem{geim} \BY{Geim~A.~K.}
\IN{Science}{324}{2009}{1530}.

\bibitem{beenakker} \BY{Beenakker~C.~W.~J.}
\IN{Rev. Mod. Phys.}{80}{2008}{1337}.

\bibitem{akhmerov1} \BY{Akhmerov~A.~R. \atque Beenakker~C.~W.~J.}
\IN{Phys.~Rev.~Lett.}{98}{2007}{157003}.

\bibitem{brey1} \BY{Brey~L. \atque Fertig~H.~A.}
\IN{Phys.~Rev.~B}{73}{2006}{235411}.

\bibitem{brey2} \BY{Brey~L. \atque Fertig~H.~A.}
\IN{Phys.~Rev.~B}{73}{2006}{195408}.

\bibitem{bloch} \BY{Bloch~F.}
\IN{Z.~Phys.}{52}{1928}{555}.

\bibitem{perturbation} For a brief but very clear explanation of the
perturbation theory, also in the degenerate case, see the Appendix C of
ref.~\cite{wenckebach}.

\bibitem{baym} \BY{Baym~G.} 
\TITLE{Lectures On Quantum Mechanics
(Lecture Notes and Supplements in Physics)}
(Addison-Wesley, Redwood City) 1969.

\bibitem{velocity} See for example p.~434 of ref.~\cite{kroemer}.

\bibitem{jackson} \BY{Jackson~J.~D.}
\TITLE{Classical Electrodynamics} (Wiley, New York) 1998, section 11.8.

\bibitem{spinorbit}
For a simplified explanation see also the section 21.4 of ref.~\cite{kroemer}.

\bibitem{lowdin1} \BY{L\"owdin~P.}
\IN{J.~Chem.~Phys.}{19}{1951}{1396}.

\bibitem{lowdin2} See also pp.~68-70 of ref.~\cite{datta1} and section 3.6
of ref.~\cite{chuang}.

\bibitem{wannier} \BY{Wannier~G.~H.} 
\IN{Phys.~Rev.}{52}{1937}{191}.

\bibitem{slater} \BY{Slater~J.~C.}
\IN{Phys.~Rev.}{76}{1949}{1592}.

\bibitem{adams} \BY{Adams~E.~N.}
\IN{J.~Chem.~Phys.}{21}{1953}{2013}.

\bibitem{young} \BY{Young~K.}
\IN{Phys.~Rev.~B}{39}{1989}{13434}.

\bibitem{envelope} See Chapter 6 of ref.~\cite{datta1} and Appendix B of
ref.~\cite{chuang}.

\bibitem{mitin} \BY{Mitin~V.~V., Kochelap~V.~A. \atque Stroscio~M.~A.} 
\TITLE{Quantum Heterostructures} (Cambridge University Press, Cambridge) 1999.

\bibitem{lawaetz} \BY{Lawaetz~P.} 
\IN{Phys.~Rev.~B}{4}{1971}{3460}.

\bibitem{taylor} See also \BY{Taylor~R.~I. \atque Burt~M.~G.}
\IN{Semicond.~Sci.~Technol.}{2}{1987}{485}.

\bibitem{bastard4} \BY{Bastard~G., Brum~J.~A. \atque Ferreira~R.}
\TITLE{Electronic States in Semiconductor Heterostructures},
in \TITLE{Solid State Physics}, Vol.~{\bf 44}, edited by
\NAME{Ehrenreich~H. \atque Turnbull~D.}
(Academic Press, New York) 1991, p.~229.

\bibitem{altarelli4} \BY{Altarelli~M. \atque Bassani~F.}
\TITLE{Impurity States: Theoretical},
in \TITLE{Handbook on Semiconductors}, Vol~{\bf 1}:
\TITLE{Band Theory and Transport Properties}, edited by
\NAME{Moss~T.~S. \atque Paul~W.}
(North-Holland Publishing Company, Amsterdam) 1982, p.~269.

\bibitem{saito} \BY{Saito~R., Dresselhaus~G. \atque Dresselhaus~M.~S.}
\TITLE{Physical Properties of Carbon Nanotubes} 
(Imperial College Press, London) 1998.

\bibitem{wallace} \BY{Wallace~P.~R.}
\IN{Phys.~Rev.}{71}{1947}{622}.

\bibitem{mcclure} \BY{McClure~J.~W.}
\IN{Phys.~Rev.}{104}{1956}{666}.

\bibitem{slonczewski} \BY{Slonczewski~J.~C.\atque Weiss~P.~R.}
\IN{Phys.~Rev.}{109}{1958}{272}.

\bibitem{divincenzo} \BY{DiVincenzo~D.~P \atque Mele~E.~J.}
\IN{Phys.~Rev.~B}{29}{1984}{1685}.

\bibitem{semenoff} \BY{Semenoff~G.~W.}
\IN{Phys.~Rev.~Lett.}{53}{1984}{2449}.

\bibitem{kanemele} \BY{Kane~C.~L. \atque Mele~E.~J.}
\IN{Phys.~Rev.~Lett.}{78}{1997}{1932}.

\bibitem{lowdin3} \BY{L\"owdin~P.}
\IN{J.~Chem.~Phys.}{18}{1950}{365}.

\bibitem{slaterkoster} \BY{Slater~J.~C. \atque Koster~G.~F.} 
\IN{Phys.~Rev.}{94}{1954}{1498}.

\bibitem{datta2} \BY{Datta~S.}
\TITLE{Quantum Transport: Atom to Transistor}
(Cambridge University Press, Cambridge) 2005, p.~89.

\bibitem{supplem} Supplementary material, including an extended version
of this calculation, is available at
\texttt{http://brahms.iet.unipi.it/supplem/reviewgraph.html} .

\bibitem{shon} \BY{Shon~N.~H. \atque Ando~T.}
\IN{J.~Phys.~Soc.~Jpn.}{67}{1998}{2421}.

\bibitem{ando3} \BY{Ando~T. \atque Nakanishi~T.}
\IN{J.~Phys.~Soc.~Jpn.}{67}{1998}{1704}.

\bibitem{sakurai} \BY{Sakurai~J.~J.} \TITLE{Advanced Quantum Mechanics}
(Addison-Wesley, Reading) 1967.

\bibitem{reich} \BY{Reich~S., Thomsen~C. \atque Maultzsch~J.}
\TITLE{Carbon Nanotubes. Basic Concepts and\break
Physical Properties}
(Wiley-VCH, Berlin) 2004.

\bibitem{mintmire} \BY{Mintmire~J.~W. \atque White~C.~T.}
\IN{Phys.~Rev.~Lett.}{81}{1998}{2506}.

\bibitem{marconcini2} \BY{Marconcini~P. \atque Macucci~M.}
\IN{J.~Comput.~Electron.}{6}{2007}{211}.

\setlength{\itemindent}{-9pt}
% numbers with 3 digits

\bibitem{marconcini3} \BY{Marconcini~P. \atque Macucci~M.}
\IN{Carbon}{45}{2007}{1018}.

\bibitem{marconcini4} \BY{Marconcini~P. \atque Macucci~M.}
\IN{Phys.~Status~Solidi~A}{204}{2007}{1898}.

\bibitem{wakabayashi1} 
\BY{Wakabayashi~K., Takane~Y., Yamamoto~M. \atque Sigrist~M.}
\IN{New~J.~Phys.}{11}{2009}{095016}.

\bibitem{castroneto}
\BY{Castro~Neto~A.~H., Guinea~F., Peres~N.~M.~R., Novoselov~K.~S. \atque 
Geim~A.~K.}  
\IN{Rev.~Mod.~Phys.}{81}{2009}{109}.

\bibitem{wurm}
\BY{Wurm~J., Wimmer~M., Adagideli~{\.I}., Richter~K. \atque Baranger~H.~U.}
\IN{New~J.\break
Phys.}{11}{2009}{095022}.

\bibitem{tworzydlo}
\BY{Tworzyd{\l}o~J., Trauzettel~B., Titov~M., Rycerz~A. \atque 
Beenakker~C.~W.~J.}
\IN{Phys.~Rev.~Lett.}{96}{2006}{246802}.

\bibitem{wakabayashi2} \BY{Wakabayashi~K.},
\TITLE{Low-Energy Physical Properties of Edge States in Nano-Graphites},\break
PhD thesis, University of Tsukuba (2000).

\bibitem{akhmerov2} \BY{Akhmerov~A.~R. \atque Beenakker~C.~W.~J.}
\IN{Phys.~Rev.~B}{77}{2008}{085423}.

\bibitem{nakada}  
\BY{Nakada~K., Fujita~M., Dresselhaus~G. \atque Dresselhaus~M.~S.}
\IN{Phys.~Rev.~B}{54}{1996}{17954}.

\bibitem{son} \BY{Son~Y.-W., Cohen~M.~L. \atque Louie~S.~G.}
\IN{Phys.~Rev.~Lett.}{97}{2006}{216803}.

\bibitem{fujita1}
\BY{Fujita~M., Igami~M. \atque Nakada~K.}
\IN{J.~Phys.~Soc.~Jpn.}{66}{1997}{1864}.

\bibitem{fujita2}
\BY{Fujita~M., Wakabayashi~K., Nakada~K. \atque Kusakabe~K.}
\IN{J.~Phys.~Soc.~Jpn.}{65}{1996}{1920}.

\end{thebibliography}
\end{document}